\documentclass[a4paper,epsf]{cernrep}
\usepackage[spanish,english]{babel}
\usepackage{graphicx}
\usepackage{epsfig,rotating,feynmp,amssymb,calc}
\usepackage{amsmath}

\input axodraw.sty
\makeatletter
\def\citer{\@ifnextchar [{\@tempswatrue\@citexr}{\@tempswafalse\@citexr[]}}

\def\@citexr[#1]#2{\if@filesw\immediate\write\@auxout{\string\citation{#2}}\fi
  \def\@citea{}\@cite{\@for\@citeb:=#2\do
    {\@citea\def\@citea{--\penalty\@m}\@ifundefined
       {b@\@citeb}{{\bf ?}\@warning
       {Citation `\@citeb' on page \thepage \space undefined}}%
\hbox{\csname b@\@citeb\endcsname}}}{#1}}
\makeatother

\newcommand{\beq}{\begin{eqnarray}}
\newcommand{\eeq}{\end{eqnarray}}
\newcommand{\lessim}{\raisebox{-0.13cm}{~\shortstack{$<$ \\[-0.07cm] $\sim$}}~}
\newcommand{\gsim}{\raisebox{-0.13cm}{~\shortstack{$>$ \\[-0.07cm] $\sim$}}~}
\newcommand{\tgb}{\mbox{tg}\beta}
\newcommand{\ctgb}{\mbox{ctg}\beta}
\newcommand{\tg}{\mbox{tg}}
\newcommand{\non}{\nonumber}

\newcommand{\STS}{\vspace{1mm}}

\newcommand{\ee}{$e^+e^-$\ }

\newcommand{\demi}{1\! /\! 2}

\newcommand{\W}{{\it W} }

\newcommand{\SM}{Standard Model }

\newcommand{\ra}{\to }
\newcommand{\epem}{e^+e^- }
\newcommand{\cs}{cross section }

\newcommand{\Hs}{Higgs-strahlung }
\newcommand{\p}{particle }
\newcommand{\ps}{particles }
\newcommand{\cp}{coupling }
\newcommand{\cps}{couplings }

\newcommand{\ssy}{supersymmetric }

\begin{document}

\numberwithin{equation}{section}
\numberwithin{table}{section}
\numberwithin{figure}{section}
\selectlanguage{english}
\thispagestyle{empty}
\begin{flushright} 
{\footnotesize{
  CERN-PH-TH/2007-262 \\
  DESY 07-214      \\
  LAPTH-CONF-1223/2007    \\
  LPT-ORSAY 07-128 \\
  PITHA 07/20      \\
  PSI-PR-07-11     \\}}
\end{flushright} 
\vspace*{5mm}
{\Large\raggedright\noindent \bf\par
 Concepts of Electroweak Symmetry Breaking and Higgs Physics}\\[2ex]
\begin{enumerate}
      \item[]\normalsize\raggedright
{ \bf M. Gomez-Bock$^1$, M. Mondrag\'on$^2$, M. M\"uhlleitner$^{3,4}$, 
M. Spira$^5$, P.M. Zerwas$^{6,7,8}$}
\end{enumerate}
\begin{enumerate}
\item[]\rm
{ $^1$ Inst. de F\'{\i}sica ``LRT'', Benemerita Univ. Auton. de Puebla, 
72570 Puebla, Pue, Mexico \\
$^2$ Inst. de F\'{\i}sica, Universidad Nacional Autononoma de Mexico, 01000 Mexico D.F., 
Mexico \\
$^3$ Laboratoire d'Annecy-Le-Vieux de Physique Th\'eorique, LAPTH,
Annecy-Le-Vieux, France \\ 
$^4$ CERN TH Division, CERN, Geneva, Switzerland \\
$^5$ Paul Scherrer Institut, CH-5232 Villigen PSI, Switzerland \\
$^6$ Deutsches Elektronen-Synchrotron DESY, D-22603 Hamburg, Germany \\
$^7$ Inst. Theor. Physik E, RWTH Aachen, D-52074 Aachen, Germany}    \\
$^8$ Laboratoire de Physique Th\'eorique, U. Paris-Sud, F-91405 Orsay, France

\end{enumerate}

\begin{abstract}
We present an introduction to the basic concepts of electroweak symmetry 
breaking and Higgs physics within the Standard Model and its supersymmetric 
extensions. A brief overview will also be given on alternative mechanisms of 
electroweak symmetry breaking. In addition to the theoretical basis, the present 
experimental status of Higgs physics and prospects at the Tevatron,
the LHC and $\epem$ linear colliders are discussed.
\end{abstract}


\vspace*{4mm}
\section{Introduction}   

\vspace*{2mm}
{\bf 1.$\,$} Revealing  the physical mechanism 
which breaks the 
electroweak symmetries, is one of the key
problems in particle physics. If the fundamental
particles of the Standard Model -- leptons, quarks 
and gauge bosons -- remain weakly interacting up to
very high energies, potentially close to the 
Planck scale, the sector in which
the electroweak symmetry is broken
must contain one or more fundamental
scalar Higgs bosons with light masses
of the order of the symmetry-breaking
scale $v\simeq 
246$ GeV. The masses of the fundamental
particles are generated by the
interaction with the scalar 
Higgs field, which is  non-zero in the ground
state \cite{1}. Alternatively, the symmetry
breaking could be generated dynamically
by new strong forces characterized by
an interaction scale $\Lambda \sim
1$ TeV and beyond \cite{2}. If global symmetries 
of the strong interactions are broken
spontaneously, the associated Goldstone
bosons can be absorbed by the gauge
fields, generating the masses of the
gauge particles. The masses of leptons
and quarks can be generated by
interactions with the fermion condensate
of the new strong interaction theory.
In other strong-interaction scenarios,
$\Lambda > \mathcal{O}$(10 TeV),
the low-energy spectrum includes scalar Higgs fields
\cite{2A} which acquire light masses as 
pseudo-Goldstone bosons only by collective
symmetry breaking.
Other breaking mechanisms of the 
electroweak symmetries are associated
with the dynamics in extra space dimensions 
at low energies \cite{RI2}. The Higgs field 
may be identified with the fifth component 
of a vector field in $D=5$ dimensions, 
or no light Higgs field is realized 
in four dimensions. 

\vspace*{2mm}
\noindent
{\bf 2.$\,$} A simple mechanism for the breaking of
the electroweak symmetry is incorporated in 
the Standard Model (SM) \cite{3}. A complex 
isodoublet scalar field is introduced 
which acquires a non-vanishing vacuum expectation
value by self-interactions, 
breaking spontaneously the electroweak
symmetry SU(2)$_I\times$ U(1)$_Y$
down to the electromagnetic U(1)$_{EM}$
symmetry. The interactions of the gauge
bosons and fermions with the background
field generate the masses of these
particles. One scalar field component
is not absorbed in this process,
manifesting itself as the physical
Higgs particle $H$.

The mass of the Higgs boson is the
only unknown parameter in the symmetry-breaking 
sector of the Standard
Model, while all couplings are fixed
by the masses of the particles, a
consequence of the Higgs mechanism
{\it sui generis}. However, the mass of the Higgs
boson is constrained in two ways. Since
the quartic self-coupling of the Higgs
field grows indefinitely with rising
energy, an upper limit on the Higgs
mass can be derived from demanding
the SM particles to remain weakly
interacting up to a scale $\Lambda$
\cite{4}. On the other hand, stringent
lower bounds on the Higgs mass follow
from requiring the electroweak vacuum
to be stable \cite{5}. If the Standard
Model is valid up to scales near the 
Planck scale, the SM Higgs mass is
restricted to a narrow window between
130 and 190 GeV. For Higgs masses
either above or below this window, new
physical phenomena are expected to
occur at a scale $\Lambda$ between
$\sim 1$~TeV and the Planck scale. For Higgs
masses of order 1 TeV, the scale of
new strong interactions would be as
low as $\sim 1$ TeV \cite{4,6}. 

\begin{figure}[hbt]
\begin{center}
\hspace*{-1.1cm}
\epsfig{figure=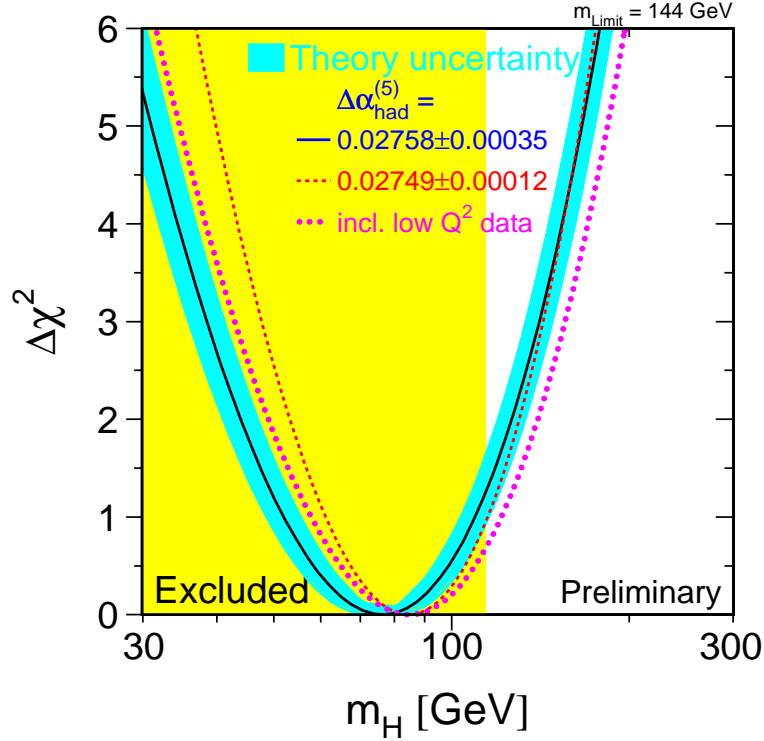,bbllx=17,bblly=36,bburx=564,bbury=564,width=10cm,clip=}
\end{center}
\vspace*{-0.4cm}

\caption[]{\label{fg:SMHiggs}\it The $\Delta\chi^2$ curve 
derived from high-$Q^2$ precision 
electroweak measurements, performed at LEP, SLC and Tevatron, as a 
function of the Higgs boson mass in the Standard Model.}
\end{figure}

The electroweak observables are affected
by the Higgs mass through radiative
corrections \cite{7}. Despite  the weak 
logarithmic dependence to leading order, 
the high-precision
electroweak data, cf. Fig.~\ref{fg:SMHiggs}, 
indicate a preference
for light Higgs masses close to
$\sim 100$ GeV \cite{8}. At the 95\% CL, they
require a value of the Higgs mass 
less than $\sim 144$~GeV.
By searching directly for the SM Higgs particle,
the LEP experiments have set a lower
limit of $M_H\gsim 114$ GeV
on the Higgs mass \cite{9}. Since the
Higgs boson has not been found at LEP2, 
the search will continue at the
Tevatron, which may reach masses up to $\sim 140$ GeV
\cite{11}. The proton collider LHC can sweep
the entire canonical Higgs mass range
of the Standard Model \cite{12,12A}. 
While first steps in analyzing the properties
of the Higgs particle can be taken at the LHC,
a comprehensive and high-resolution picture
of the Higgs mechanism can be established 
experimentally by performing very accurate analyses at
$e^+e^-$ linear colliders \cite{13}.  

\vspace*{2mm}
\noindent
{\bf 3.$\,$} If the Standard Model is embedded in a 
Grand Unified Theory (GUT) at high energies,
the scale of electroweak symmetry
breaking would naively be expected close to the
unification scale $M_{GUT}$.
Supersymmetry \cite{14} provides a solution of
this hierarchy problem. 
Once the fundamental parameters 
of the supersymmetric theory and 
its breaking mechanism are generated at the Terascale,
the quadratically
divergent contributions to the radiative
corrections of the scalar Higgs boson
mass are cancelled by the destructive
interference between bosonic
and fermionic loops in supersymmetric theories \cite{15}. 
The Minimal Supersymmetric extension of the Standard
Model (MSSM) provides an illustrative example 
for deriving a Terascale theory from a supersymmetric 
grand unified theory. A strong indication for the realization
of the basic components of this physical picture in nature 
is the excellent agreement between the value of
the electroweak mixing angle $\sin^2 \theta_W$
predicted by the unification of the gauge
couplings, and the experimentally measured value. If the
gauge couplings are unified in the
minimal supersymmetric theory at a scale
$M_{GUT} = {\cal O}(10^{16}~\mbox{GeV})$, 
the electroweak mixing angle is predicted
to be $\sin^2\theta_W = 0.2336 \pm 0.0017$
\cite{16} for a mass spectrum of the supersymmetric
particles of order $M_Z$ to 1 TeV.
This theoretical prediction is matched very well by 
 the experimental result
$\sin^2\theta_W^{exp} = 0.23153 \pm 0.00016$
\cite{8}; the difference between the two numbers
is less than 2 per-mille.

In the MSSM, the Higgs sector is built up 
by two Higgs doublets \cite{17}. The doubling is
necessary to generate masses for up- and
down-type fermions in a supersymmetric
theory and to render the theory anomaly-free. 
The Higgs particle spectrum consists
of a quintet of states: two $\mathcal{CP}$-even
scalar neutral ($h,H$), one $\mathcal{CP}$-odd pseudoscalar
neutral ($A$), and a pair of charged ($H^\pm$)
Higgs bosons \cite{19}. The masses of the heavy
Higgs bosons, $H,A,H^\pm$, are expected to be of order $v$, 
but they may extend up to the TeV range. By contrast,
since the quartic Higgs self-couplings are 
determined by the gauge couplings, the mass
of the lightest Higgs boson $h$ is constrained
stringently. At tree level, the mass
has been predicted to be smaller than the
$Z$ mass \cite{19}. Radiative corrections,
increasing as the fourth power of the
top mass, shift the upper limit to a value
between $\sim 100$ GeV and $\sim 140$
GeV, depending on the parameter $\tgb$,
the ratio of the vacuum expectation values
of the two neutral scalar Higgs fields.

Extensions beyond the minimal supersymmetric form
of the theory may be motivated by slight fine-tuning
problems in accommodating the experimentally 
observed $Z$-boson mass. The expansion of the
Higgs sector introduces new couplings which
grow with rising scale. In parallel with the
Standard Model, the lightest Higgs boson mass
is bounded \cite{EspinQuiros} to less than 
about 200 GeV, however,
if the fields remain weakly interacting  
up to scales close to the Planck scale.

A general lower bound of 91 GeV in $\mathcal{CP}$-invariant
theories has been experimentally established
for the Higgs particle $h$ at
LEP \cite{9}. 
The search for $h$
masses in excess of $\sim 100$ GeV
and the search for the heavy Higgs bosons
continues at the Tevatron, the LHC and $e^+e^-$
linear colliders. 

\vspace*{2mm}
\noindent
{\bf 4.$\,$} A light Higgs boson may also be generated 
as a (pseudo-)Goldstone boson by spontaneous 
breaking of global symmetries of new interactions
at multi-TeV scales, the mass kept small 
by collective symmetry breaking mechanisms.
Alternatively to supersymmetry, 
the quadratic divergencies of the Standard Model are
cancelled by new partners 
of the Standard Model particles that do not differ 
in the fermionic/bosonic character. 
Symmetry schemes constrain the couplings 
in such a way that the cancellations 
are achieved in a natural way. Such 
scenarios are realized in Little Higgs models 
\cite{2A} which predict a large 
ensemble of new SM-type particles 
in the mass range of a few TeV.

\vspace*{2mm}
\noindent
{\bf 5.$\,$} Elastic-scattering amplitudes 
of massive vector bosons grow indefinitely 
with energy if they are calculated in a
perturbative expansion in the weak coupling
of a non-Abelian gauge theory. As a
result, they violate unitarity beyond
a critical energy scale of 
$\sim 1.2$ TeV. Apart from
introducing a light Higgs boson, this problem can
also be solved by assuming the $W$ bosons to become 
strongly interacting at TeV energies,
thus damping the rise of the elastic-scattering 
amplitudes. Naturally, the strong
forces between the $W$ bosons may be traced
back to new fundamental interactions
characterized by a scale of order 1 TeV \cite{2}.
If the underlying theory is globally
chiral-invariant, this  symmetry may be broken
spontaneously. The Goldstone bosons 
associated with the spontaneous 
breaking of the symmetry can be absorbed by  gauge
bosons to generate their masses and to build
up the longitudinal components of their wave functions.

Since the longitudinally polarized $W$ bosons
are associated with the Goldstone modes
of chiral symmetry breaking, the scattering
amplitudes of the $W_L$
bosons can be predicted for high energies
by a systematic expansion in the energy.
The leading term is parameter-free, a
consequence of the chiral symmetry-breaking
mechanism {\it per se}, which is independent of
the particular structure of the dynamical theory. The 
higher-order terms in the chiral expansion however 
are defined by the detailed structure 
of the underlying theory. With rising
energy the chiral expansion is expected to diverge
and new resonances may be generated in
$WW$ scattering at mass scales between 1
and 3 TeV. This picture is analogous to
pion dynamics in QCD, where the threshold
amplitudes can be predicted in a chiral
expansion, while at higher energies vector
and scalar resonances are formed in $\pi \pi$
scattering.

Such a scenario can be studied in $WW$
scattering experiments, where the $W$ bosons
are radiated, as quasi-real particles \cite{22},
off high-energy quarks in the proton
beams of the LHC \cite{12}, \citer{23,23B} or off electrons
and positrons in TeV linear colliders \cite{13,24,24a}.

\vspace*{2mm}
\noindent
{\bf 6.$\,$} In theories formulated in extra space dimensions,
suitably chosen boundary conditions for fields 
in the compactified space can be exploited 
to break symmetries \cite{RI2}. In one class of models, the 
Higgs fields are identified with the zero-mass fifth components 
of vector boson fields, associated with broken gauge symmetries 
beyond the Standard Model,
while other massive fifth components are transformed 
to the longitudinal degrees of freedom for the 
vector bosons of the Standard Model. 
Alternatively, the electroweak symmetries can be broken 
by transforming all fifth components to longitudinal
components of the vector fields, ground state vectors as well
as Kaluza-Klein state vectors, 
so that higgsless theories emerge in such a scenario.
In any such theory of extra space dimensions, 
Kaluza-Klein towers are generated above the 
Standard Model states. The additional exchange 
of the Kaluza-Klein towers in $WW$ scattering 
damps the scattering amplitude of the Standard Model 
and allows in principle to extend the theory 
to energies beyond the 1.2 TeV unitarity bound of 
naive higgsless scenarios. 

\vspace*{2mm}
\noindent
{\bf 7.$\,$} This report is divided into three 
parts. A basic introduction and a summary of the
main theoretical and experimental results
will be presented in the next section on
the Higgs sector of the Standard Model.
Also the search for the Higgs particle
at hadron and future $e^+e^-$
colliders will be described. In
the same way, the Higgs spectrum of 
supersymmetric theories will be discussed
in the subsequent section. The main 
features of strong $W$ interactions and their
analysis in $WW$ scattering experiments will
be presented in the last section.

Only basic elements of electroweak symmetry
breaking and Higgs mechanism can be reviewed  
in this report which is an updated version 
of the reports Ref.$\,$\cite{XAL} and Refs.$\,$\cite{24A}. 
Other aspects 
may be traced back from Refs.$\,$\cite{Quigg}, 
the canon Ref.$\,$\cite{24b} and the
recent reports Refs.$\,$\cite{DJ}. 

\vspace*{6mm}
\section{The Higgs Sector of the Standard Model}

\vspace*{2mm}
\subsection{Physical Basis}

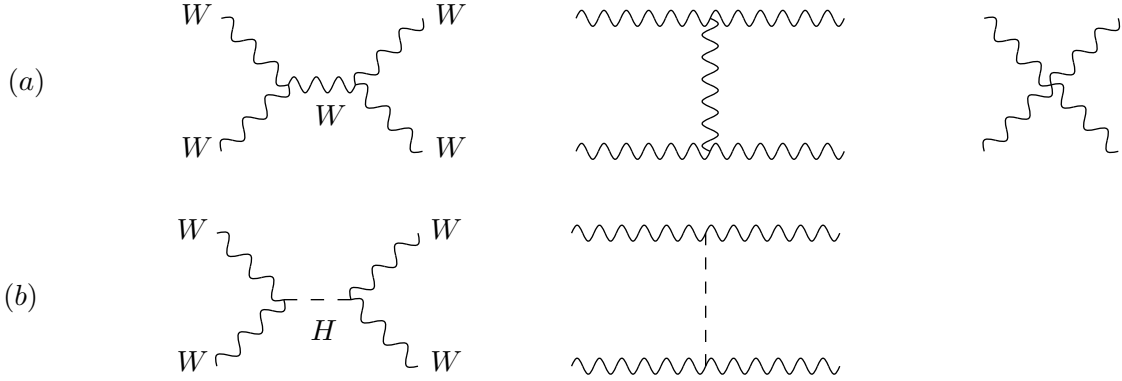
\begin{figure}[hbt]
\begin{center}
\begin{picture}(80,80)(50,-20)
\Photon(0,50)(25,25){3}{3}
\Photon(0,0)(25,25){3}{3}
\Photon(25,25)(50,25){3}{3}
\Photon(50,25)(75,50){3}{3}
\Photon(50,25)(75,0){3}{3}
\put(-80,23){$(a)$}
\put(-15,-2){$W$}
\put(-15,48){$W$}
\put(80,-2){$W$}
\put(80,48){$W$}
\put(35,10){$W$}
\end{picture}
\begin{picture}(60,80)(0,-20)
\Photon(0,50)(100,50){3}{12}
\Photon(0,0)(100,0){3}{12}
\Photon(50,50)(50,0){3}{6}
\end{picture}
\begin{picture}(60,80)(-90,-20)
\Photon(0,50)(25,25){3}{3}
\Photon(0,0)(25,25){3}{3}
\Photon(25,25)(50,50){3}{3}
\Photon(25,25)(50,0){3}{3}
\end{picture} \\
\begin{picture}(80,60)(83,0)
\Photon(0,50)(25,25){3}{3}
\Photon(0,0)(25,25){3}{3}
\DashLine(25,25)(50,25){6}
\Photon(50,25)(75,50){3}{3}
\Photon(50,25)(75,0){3}{3}
\put(-80,23){$(b)$}
\put(-15,-2){$W$}
\put(-15,48){$W$}
\put(80,-2){$W$}
\put(80,48){$W$}
\put(35,10){$H$}
\end{picture}
\begin{picture}(60,60)(33,0)
\Photon(0,50)(100,50){3}{12}
\Photon(0,0)(100,0){3}{12}
\DashLine(50,50)(50,0){5}
\end{picture}  \\
\end{center}
\caption[]{\it \label{fg:wwtoww} Generic diagrams of elastic $WW$ scattering:
(a) pure gauge-boson dynamics, and (b) Higgs-boson exchange.}
\end{figure}

\vspace*{2mm}
\noindent
{\bf 1.$\,$} At high energies, the amplitude for the 
elastic scattering of massive $W$ bosons, $WW \to WW$, 
grows indefinitely with energy for longitudinally 
polarized particles, Fig.~\ref{fg:wwtoww}a. This is a consequence
of the linear rise of the longitudinal $W_L$ wave function, 
$\epsilon_L = (p,0,0,E)/M_W$,
with the energy of the particle. Even though the term of 
the scattering amplitude rising as the fourth power in the energy
is cancelled by virtue of the non-Abelian 
gauge symmetry, the amplitude remains quadratically
divergent in the energy. On the other hand, 
unitarity requires elastic-scattering 
amplitudes of partial waves $J$ to be bounded by
$\Re e A_J \leq 1/2$.
Applied to the asymptotic $S$-wave amplitude
$A_0 = G_F s/8\pi\sqrt{2}$ of the isospin-zero channel
$2W_L^+W_L^- + Z_L Z_L$,  the bound  \cite{25}
\begin{equation}
s \leq 4\pi\sqrt{2}/G_F \sim (1.2~\mbox{TeV})^2
\end{equation}
on the c.m. energy $\sqrt{s}$ can be derived for 
the validity of a theory of weakly 
coupled massive gauge bosons.
\noindent

However, the quadratic rise in the energy can be damped by 
exchanging a new scalar particle, Fig.~\ref{fg:wwtoww}b. 
To achieve the 
cancellation, the size of the coupling must be given 
by the product of the gauge coupling with the gauge
boson mass. For high energies, the
amplitude $A'_0 = -G_F s/8\pi\sqrt{2}$
cancels exactly the quadratic divergence of the pure 
gauge-boson amplitude $A_0$.
Thus, unitarity can be restored by introducing a 
fundamental, weakly
coupled \underline{\it Higgs particle}.

In the same way the linear divergence of the amplitude 
$A(f\bar f\to W_L W_L)\sim gm_f\sqrt{s}$
for the annihilation of a fermion--antifermion pair to 
a pair of longitudinally polarized gauge bosons
can be damped by adding the Higgs exchange to 
the gauge-boson exchange. In this case
the Higgs particle must couple proportionally
to the mass $m_f$ of the fermion $f$.

These observations can be summarized in a rule:
{\it A theory of massive gauge bosons and fermions
which are weakly coupled up to asymptotic
energies, requires, by unitarity, the existence 
of a Higgs particle; the Higgs particle is a
scalar $0^+$ particle that couples to other particles 
proportionally to the masses of the particles.}

The assumption that the couplings of the 
fundamental particles are weak up to asymptotic
energies is qualitatively supported
by the perturbative renormalization of the
electroweak mixing angle $\sin^2\theta_W$
from the symmetry value 3/8 at the GUT scale
down to $\sim 0.2$ at the electroweak scale, 
which is close to the experimentally observed
value.

\begin{figure}[hbt]
\begin{center}
\begin{picture}(60,10)(80,40)
\Photon(0,25)(50,25){3}{6}
\LongArrow(65,25)(90,25)
\put(-15,21){$V$}
\put(-15,50){$(a)$}
\end{picture}
\begin{picture}(60,10)(40,40)
\Photon(0,25)(50,25){3}{6}
\put(60,23){$+$}
\end{picture}
\begin{picture}(60,10)(15,40)
\Photon(0,25)(50,25){3}{6}
\DashLine(25,25)(12,50){3}
\DashLine(25,25)(38,50){3}
\Line(9,53)(15,47)
\Line(9,47)(15,53)
\Line(35,53)(41,47)
\Line(35,47)(41,53)
\put(45,45){$H$}
\put(70,23){$+$}
\end{picture}
\begin{picture}(60,10)(-10,40)
\Photon(0,25)(75,25){3}{9}
\DashLine(20,25)(8,50){3}
\DashLine(20,25)(32,50){3}
\DashLine(55,25)(43,50){3}
\DashLine(55,25)(67,50){3}
\Line(5,53)(11,47)
\Line(5,47)(11,53)
\Line(29,53)(35,47)
\Line(29,47)(35,53)
\Line(40,53)(46,47)
\Line(40,47)(46,53)
\Line(64,53)(70,47)
\Line(64,47)(70,53)
\put(90,23){$+ \cdots$}
\end{picture} \\
\begin{picture}(60,80)(80,20)
\ArrowLine(0,25)(50,25)
\LongArrow(65,25)(90,25)
\put(-15,23){$f$}
\put(-15,50){$(b)$}
\end{picture}
\begin{picture}(60,80)(40,20)
\ArrowLine(0,25)(50,25)
\put(65,23){$+$}
\end{picture}
\begin{picture}(60,80)(15,20)
\ArrowLine(0,25)(25,25)
\ArrowLine(25,25)(50,25)
\DashLine(25,25)(25,50){3}
\Line(22,53)(28,47)
\Line(22,47)(28,53)
\put(35,45){$H$}
\put(65,23){$+$}
\end{picture}
\begin{picture}(60,80)(-10,20)
\ArrowLine(0,25)(25,25)
\ArrowLine(25,25)(50,25)
\ArrowLine(50,25)(75,25)
\DashLine(25,25)(25,50){3}
\DashLine(50,25)(50,50){3}
\Line(22,53)(28,47)
\Line(22,47)(28,53)
\Line(47,53)(53,47)
\Line(47,47)(53,53)
\put(90,23){$+ \cdots$}
\end{picture}  \\
\end{center}
\caption[]{\it \label{fg:massgen} Generating (a) gauge boson and (b)
fermion masses through interactions with the scalar background field.}
\end{figure}
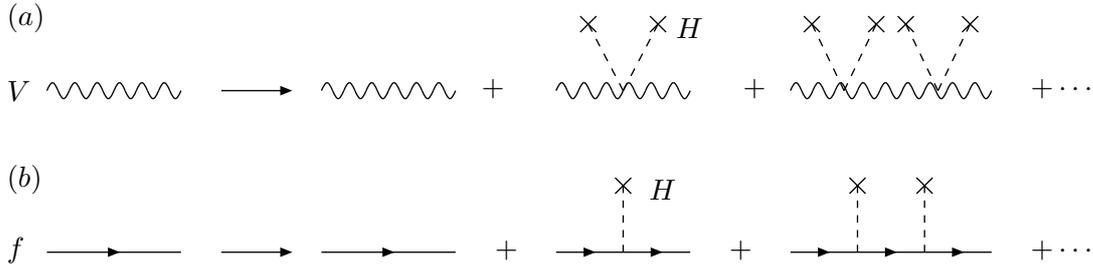

\vspace*{2mm}
\noindent
{\bf 2.$\,$} These ideas can be cast into an elegant
mathematical form by interpreting the electroweak
interactions as a gauge theory with spontaneous
symmetry breaking in the scalar sector\footnote{The mechanisms of spontaneous
symmetry breaking, including the Goldstone theorem as well as the Higgs 
mechanism, are exemplified for the illustrative $O(3)$ $\sigma$ model in 
Appendix A.}. 
Such a theory consists of fermion fields, gauge
fields and a scalar field coupled by the 
standard gauge interactions and Yukawa interactions
to the other fields. Moreover, a self-interaction
\begin{equation}
V = \frac{\lambda}{2} \left[ |\phi|^2 - \frac{v^2}{2} \right]^2
\label{eq:potential}
\end{equation}
is introduced in the scalar sector, which leads to 
a non-zero ground-state value $v/\sqrt{2}$
of the scalar field. By fixing the phase of the 
vacuum amplitude at an arbitrarily chosen value, 
say zero, the gauge symmetry
is broken spontaneously in the scalar sector. 
Interactions of the 
gauge fields with the scalar background field,
Fig.~\ref{fg:massgen}a, and Yukawa interactions of the 
fermion fields with the background field, Fig.~\ref{fg:massgen}b, 
shift the masses of these fields from 
zero to non-zero values:
\begin{equation}
\begin{array}{lrclclcl}
\displaystyle
(a) \hspace*{2.0cm} &
\displaystyle
\frac{1}{q^2} & \to & \displaystyle \frac{1}{q^2} + \sum_j \frac{1}{q^2}
\left[ \left( \frac{gv}{{2}} \right)^2 \frac{1}{q^2} \right]^j & = &
\displaystyle \frac{1}{q^2-M^2} & : & \displaystyle M^2 = g^2 \frac{v^2}{4}
\\ \\
(b) &
\displaystyle
\frac{1}{\not \! q} & \to & \displaystyle \frac{1}{\not \! q} +
\sum_j \frac{1}{\not \! q} \left[ \frac{g_fv}{\sqrt{2}} \frac{1}{\not
\! q} \right]^j & = & \displaystyle \frac{1}{\not \! q-m_f} & : &
\displaystyle m_f = g_f \frac{v}{\sqrt{2}} 
\end{array}
\end{equation}
Thus, in theories with gauge and Yukawa interactions,
in which the scalar field acquires a non-zero
ground-state value, the couplings are naturally
proportional to the masses. This ensures the
unitarity of the theory as discussed before.
These theories are renormalizable (as a result
of the gauge invariance, which is only disguised 
in the unitary formulation adopted here), and 
thus they describe a well-defined physical system. 

\vspace*{4mm}

\subsection{The Higgs Mechanism in the Standard Model}

\vspace*{2mm}
\noindent
Besides the Yang--Mills and the fermion parts, the
electroweak $SU_2 \times U_1$
Lagrangian includes a scalar isodoublet field
$\phi$, coupled to itself in the potential $V$,
cf. eq. (\ref{eq:potential}),
to the gauge fields through the covariant derivative
$iD = i\partial - g \vec{I} \vec{W} - g'YB$,
and to the up and down fermion fields $u,d$
by Yukawa interactions:
\begin{equation}
{\cal L}_0 = |D\phi|^2 - \frac{\lambda}{2} \left[ |\phi|^2
- \frac{v^2}{2} \right]^2 - g_d \bar d_L \phi d_R - g_u \bar u_L
\phi_c u_R + {\rm hc} ~.
\end{equation}
In the unitary gauge, the isodoublet $\phi$
is effectively replaced by the physical Higgs field
$H$, $\phi\to [0,(v+H)/\sqrt{2}]$,
which describes the fluctuation of the $I_3=-1/2$
component about the ground-state 
value $v/\sqrt{2}$. The scale $v$
of the electroweak symmetry breaking is fixed
by the weak gauge coupling and the $W$ mass, 
which in turn can be reexpressed by the
Fermi coupling:
\begin{equation}
  v = 1/\sqrt{\sqrt{2}G_F} \approx 246 \,{\rm GeV}.
\end{equation}
While the $W$ mass is related to $v$ by the gauge coupling,
the Yukawa couplings $g_f$ and
the quartic coupling $\lambda$
can likewise be reexpressed in terms of the 
Higgs mass $M_H$ and the fermion masses $m_f$: 
\begin{eqnarray}
M^2_W & = & g^2 v^2 / 4       \nonumber \\
m_f   & = & g_f v / \sqrt{2}  \nonumber \\
M_H^2 & = & \lambda v^2
\end{eqnarray}
respectively.

Since the couplings of the Higgs particle to gauge particles, 
to fermions and to itself are given by the gauge couplings
and the masses of the particles, the only unknown
parameter in the Higgs sector (apart from the CKM 
mixing matrix) is the Higgs mass. When this mass is fixed,
all properties of the Higgs particle can be predicted,
i.e. the lifetime and decay branching ratios, as 
well as the production mechanisms and the corresponding
cross sections.

\vspace*{4mm}
\subsubsection{The SM Higgs Mass}

\vspace*{2mm}
\noindent
Even though the mass of the Higgs boson 
cannot be predicted in the Standard Model, stringent 
upper and lower bounds can nevertheless be derived 
from internal consistency conditions and extrapolations
of the model to high energies.

\vspace*{2mm}
\noindent
{\bf 1.$\,$} The Higgs boson has been introduced as a fundamental
particle to render 2--2 scattering amplitudes involving
longitudinally polarized $W$
bosons compatible with unitarity. Based on the general
principle of time-energy uncertainty, particles must
decouple from a physical system if their mass grows
indefinitely.  
The mass of the Higgs particle must 
therefore be bounded to restore unitarity in the 
perturbative regime. From the asymptotic expansion of 
the elastic $W_L W_L$ $S$-wave scattering amplitude including 
$W$ and Higgs exchanges, $A(W_L W_L \to W_L W_L) \to -G_F M_H^2/4\sqrt{2}\pi$,
it follows \cite{25} that
\begin{equation}
M_H^2 \leq 2\sqrt{2}\pi/G_F \sim (850~\mbox{GeV})^2 ~.
\end{equation}
Within the canonical formulation of the Standard Model,
consistency conditions  therefore require a Higgs mass below 1 TeV.\\

\begin{figure}[hbt]
\vspace*{-0.5cm}

\begin{center}
\begin{picture}(90,80)(60,-10)
\DashLine(0,50)(25,25){3}
\DashLine(0,0)(25,25){3}
\DashLine(50,50)(25,25){3}
\DashLine(50,0)(25,25){3}
\put(-15,45){$H$}
\put(-15,-5){$H$}
\put(55,-5){$H$}
\put(55,45){$H$}
\end{picture}
\begin{picture}(90,80)(10,-10)
\DashLine(0,50)(25,25){3}
\DashLine(0,0)(25,25){3}
\DashLine(75,50)(50,25){3}
\DashLine(75,0)(50,25){3}
\DashCArc(37.5,25)(12.5,0,360){3}
\put(-15,45){$H$}
\put(-15,-5){$H$}
\put(35,40){$H$}
\put(80,-5){$H$}
\put(80,45){$H$}
\end{picture}
\begin{picture}(50,80)(-40,2.5)
\DashLine(0,0)(25,25){3}
\DashLine(0,75)(25,50){3}
\DashLine(50,50)(75,75){3}
\DashLine(50,25)(75,0){3}
\ArrowLine(25,25)(50,25)
\ArrowLine(50,25)(50,50)
\ArrowLine(50,50)(25,50)
\ArrowLine(25,50)(25,25)
\put(-15,70){$H$}
\put(-15,-5){$H$}
\put(35,55){$t$}
\put(80,-5){$H$}
\put(80,70){$H$}
\end{picture}  \\
\setlength{\unitlength}{1pt}
\caption[]{\label{fg:lambda} \it Diagrams generating the evolution of
the Higgs self-interaction $\lambda$.}
\end{center}
\end{figure}
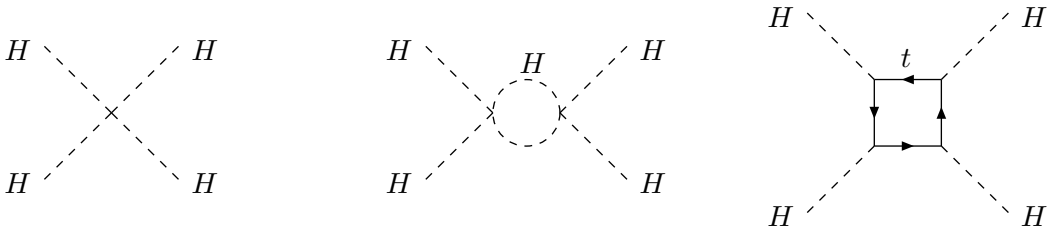

\begin{figure}[hbtp]

\vspace*{0.8cm}

\hspace*{3.0cm}
\epsfxsize=8.5cm \epsfbox{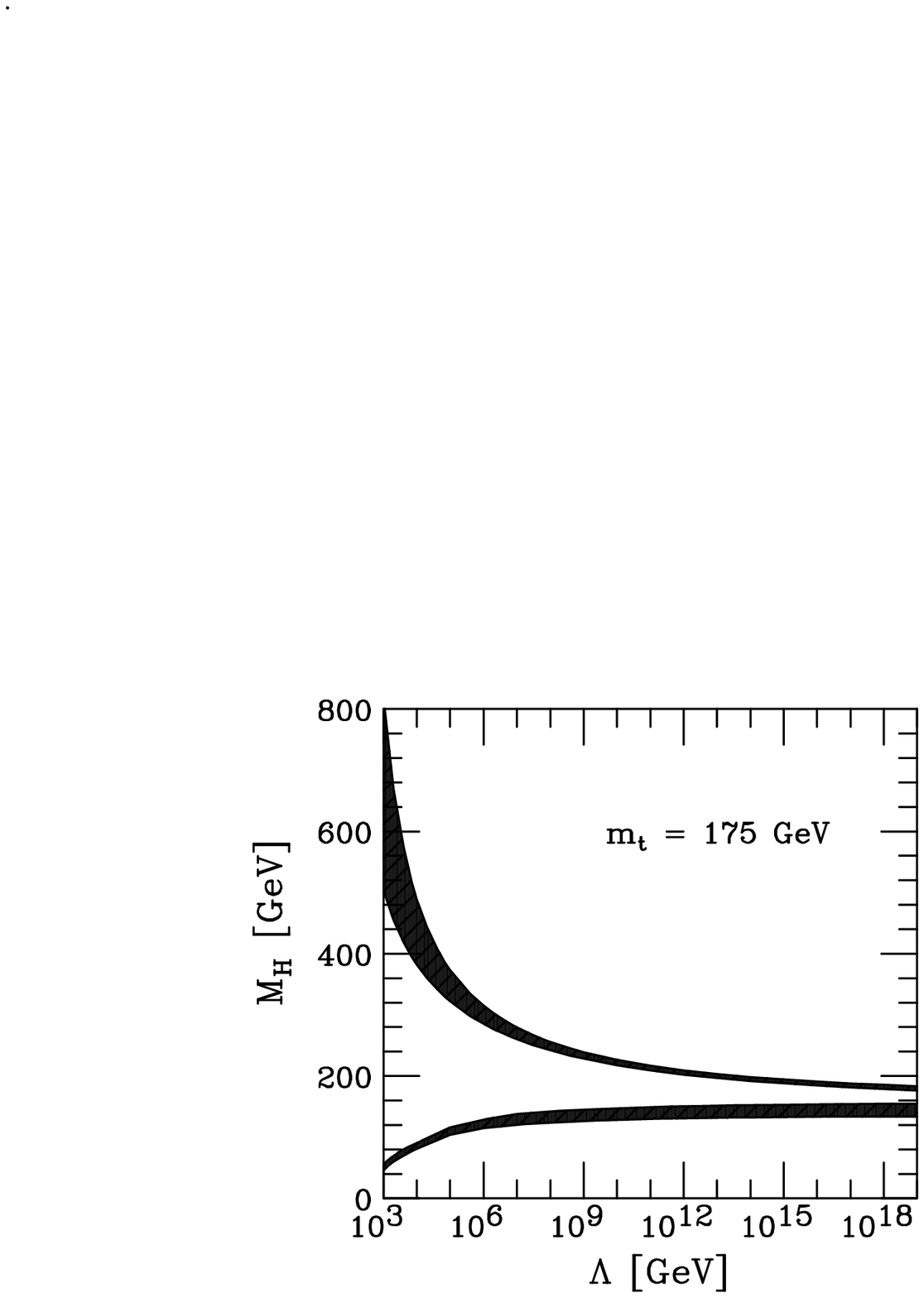}
\vspace*{-0.2cm}

\caption[]{\label{fg:triviality} \it Bounds on the mass of the Higgs boson in
the Standard Model. $\Lambda$ denotes the energy scale at which the SM Higgs boson 
with mass $M_H$ would become strongly interacting (upper bound); the lower bound follows
from the requirement of vacuum stability. Refs. \cite{4,5}.}
\vspace*{-0.3cm}
\end{figure}

\vspace*{2mm}
\noindent
{\bf 2.$\,$} Quite restrictive bounds on the value of the SM Higgs
mass follow from limits on the 
energy scale $\Lambda$
up to which the Standard Model can be extended before
new strong interaction phenomena emerge. The 
key to these bounds is the evolution of the quartic
coupling $\lambda$
with the energy 
due to quantum fluctuations \cite{4}. The basic
contributions are depicted in Fig.~\ref{fg:lambda}. The
Higgs loop itself gives rise to an indefinite increase
of the coupling while the fermionic top-quark
loop, with increasing top mass, drives the coupling
to smaller values, finally even to values below zero.
The variation of the quartic Higgs coupling
$\lambda$ and the top-Higgs Yukawa coupling $g_t$
with energy, parametrized by $t=\log \mu^2/v^2$,
may be written as \cite{4}
\begin{equation}
\begin{array}{rclcl}
\displaystyle \frac{d\lambda}{dt} & = & \displaystyle \frac{3}{8\pi^2}
\left[ \lambda^2 + \lambda g_t^2 - g_t^4 \right] & : & \displaystyle
\lambda(v^2) = M_H^2/v^2 \\ \\
\displaystyle \frac{d g_t}{dt} & = & \displaystyle \frac{1}{32\pi^2}
\left[ \frac{9}{2} g_t^3 - 8 g_t g_s^2 \right] & : & \displaystyle
g_t(v^2) = \sqrt{2}~m_f/v ~.
\end{array}
\end{equation}
Only the leading loop contributions from Higgs, top
and QCD [coupling $g_s$] are taken into account. \\

For moderate top masses, the quartic coupling $\lambda$
rises indefinitely, $d\lambda / dt \sim + \lambda^2$,
and the coupling becomes strong shortly before 
reaching the Landau pole:
\begin{equation}
\lambda (\mu^2) = \frac{\lambda(v^2)}{1- \frac{3\lambda(v^2)}{8\pi^2} \log
\frac{\mu^2}{v^2}} ~.
\end{equation}
Reexpressing the initial value of $\lambda (v^2)$
by the Higgs mass, the condition $\lambda (\Lambda^2) < \infty$,
can be translated to an \underline{upper bound} on the Higgs
mass: 
\begin{equation}
M_H^2 \leq \frac{8\pi^2 v^2}{3\log \Lambda^2/v^2} ~.
\end{equation}
This mass bound is related logarithmically to the energy $\Lambda$
up to which the Standard Model is assumed to be valid.
The maximal value of $M_H$ for the minimal cut-off $\Lambda \sim $ 1~TeV
is given by $\sim 750$ GeV. This bound is close to the estimate
of $\sim 700$ GeV
in lattice calculations for $\Lambda \sim 1$ TeV,
which allow  proper control of non-perturbative 
effects near the boundary \cite{6}.\\

\begin{table}[hbt]
\renewcommand{\arraystretch}{1.5}
\begin{center}
\begin{tabular}{|l||l|} \hline
$\Lambda$ & $M_H$ \\ \hline \hline
1 TeV & 60 GeV $\lessim M_H \lessim 700$ GeV \\
$10^{19}$ GeV & 130 GeV $\lessim M_H \lessim 190$ GeV \\ \hline
\end{tabular}
\renewcommand{\arraystretch}{1.2}
\caption[]{\label{tb:triviality}
\it Higgs mass bounds for two values of the cut-off $\Lambda$.}
\end{center}
\end{table}
A \underline{lower bound} on the Higgs mass can be derived from 
the requirement of vacuum stability \cite{4,5}. Since
top-loop corrections reduce $\lambda$,
driven finally to negative values,
the self-energy potential becomes 
unbounded negative and the ground state is no longer 
stable. To avoid the instability for cut-off values less than $\Lambda$, 
the Higgs mass must exceed a minimal value 
depending on the cut-off $\Lambda$.\\

The scales $\Lambda$ up to which the Standard Model can be extended
before new interactions must become effective, are displayed 
in Fig.~\ref{fg:triviality} as a function of the Higgs mass. 
The allowed Higgs mass values are
collected in Table~\ref{tb:triviality} for two specific cut-off values
$\Lambda$.  If the Standard Model 
is assumed to be valid up to the Planck scale, 
the Higgs mass is restricted to a narrow window
between 130 and 190~GeV. The window is widened to 200~ GeV 
for the cut-off near the grand unification scale. 
The observation of a Higgs mass above or below
this window would demand a new physics scale 
below the Planck/GUT scales.

\vspace*{2mm}
\noindent
{\bf 3.$\,$} Indirect evidence for a light Higgs
boson can be derived from the
high-precision measurements of electroweak
observables at LEP and elsewhere. Indeed, the
fact that the Standard Model is renormalizable only
after including the top and Higgs particles in
the loop corrections, indicates that the
electroweak observables are sensitive to the
masses of these particles.

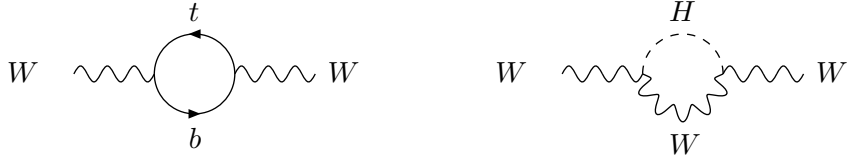
\begin{figure}[hbtp]
\begin{picture}(100,60)(-80,-0)
\Photon(0,30)(30,30){3}{3}
\Photon(60,30)(90,30){3}{3}
\ArrowArc(45,30)(15,360,180)
\ArrowArc(45,30)(15,180,0)
\put(-25,27){$W$}
\put(43,50){$t$}
\put(43,2){$b$}
\put(95,27){$W$}
\end{picture}
\begin{picture}(100,60)(-160,-0)
\Photon(0,30)(30,30){3}{3}
\Photon(60,30)(90,30){3}{3}
\PhotonArc(45,30)(15,180,360){-3}{5.5}
\DashCArc(45,30)(15,0,180){3}
\put(-25,27){$W$}
\put(40,50){$H$}
\put(40,0){$W$}
\put(95,27){$W$}
\end{picture}
\caption[]{\label{fg:WtH} \it Virtual t,b and W,Higgs radiative
corrections to the propagators of the electroweak gauge boson.} 
\end{figure}

The Fermi coupling can be rewritten in terms of
the weak coupling and the $W$ mass; to lowest order,
$G_F/\sqrt{2} = g^2/8M_W^2$.
After substituting the electromagnetic coupling $\alpha$,
the electroweak mixing angle and the $Z$
mass for the weak coupling and the $W$
mass, this relation can be rewritten as
\begin{equation}
\frac{G_F}{\sqrt{2}} = \frac{2\pi\alpha}{\sin^2 2\theta_W M_Z^2}
[1+\Delta r_\alpha + \Delta r_t + \Delta r_H ] ~.
\end{equation}
The terms $\Delta$ take account
of the radiative corrections, cf. Fig.~\ref{fg:WtH}:
$\Delta r_\alpha$ describes the shift in the electromagnetic coupling
$\alpha$ if evaluated at the scale $M_Z^2$ instead of zero-momentum;
$\Delta r_t$ denotes the top/bottom quark contributions to the
$W$ and $Z$ masses, which are quadratic in the top mass. 
Finally,
$\Delta r_H$ accounts for the virtual Higgs contribution to the masses;
this term depends only logarithmically \cite{7} on
the Higgs mass at leading order:
\begin{equation}
\Delta r_H = \frac{G_F M_Z^2 (1+9\sin^2\theta_W)}{24\sqrt{2}\pi^2} \,
\log \frac{M_H^2}{M^2_W} \, + \, ...  \hspace{1cm} (M_H^2 \gg M_W^2)
~.
\end{equation}
The screening effect reflects the role of the Higgs
field as a regulator that renders the electroweak theory
renormalizable.

Although the sensitivity on the Higgs mass is
only logarithmic, the increasing precision in the
measurement of the electroweak observables allows
us to derive interesting estimates and constraints
on the Higgs mass \cite{8}, cf. Fig.~\ref{fg:SMHiggs}:
\begin{eqnarray}
M_H & = & 76^{+33}_{-24}~\mbox{GeV} \\
    & \lessim \phantom{1}\!\! & 144 ~\mbox{GeV~~~(95\% CL)}  ~. \nonumber
\end{eqnarray}
With a value of 15.1\% the probability for the fit
is not overwhelmingly large but not forbiddingly small 
either. The 95\% confidence level is still significantly
above the direct search limit,
\begin{equation}
  {\hspace*{-18mm}} M_H \;\;\;\,\geq\;\;\;\, 114.1 \, {\rm GeV}
\end{equation}
derived from LEP2 analyses, Ref.$\,$\cite{9}, 
of the dominant Higgs-strahlung channel $e^+e^- \to ZH$.
 
It may be concluded from these numbers that the
canonical formulation of the Standard Model
including the existence of a light Higgs boson,
is compatible with the electroweak data. However,
alternative mechanisms cannot be ruled out if the
system is opened up to contributions from physics
areas beyond the Standard Model.

\vspace{4mm}
\subsubsection{Decays of the Higgs Particle}

\vspace*{2mm}
\noindent
The profile of the Higgs particle is uniquely
determined if the Higgs mass is fixed. The
strength of the coupling of the Higgs boson
to the electroweak gauge bosons 
$V=W,Z$ is fixed by their masses $M_V$ 
and the strength of the Yukawa 
couplings of the Higgs
boson to fermions is set by the fermion masses $m_f$;
the couplings may be defined uniformly as
\begin{eqnarray}
g_{HVV} & = & \left[2 \sqrt{2} G_F \right]^{1/2} M_V      \\
g_{Hff} & = & \left[2 \sqrt{2} G_F \right]^{1/2} m_f \, . \nonumber
\end{eqnarray}
The total decay width and lifetime, as well as the
 branching ratios for specific decay channels, are 
determined by these parameters. The measurement
of the decay characteristics can therefore
be exploited to establish, experimentally, 
that Higgs couplings grow with the masses
of the particles, a direct consequence of
the Higgs mechanism.

For Higgs particles in the intermediate mass range
${\cal O}(M_Z) \leq M_H \leq 2M_Z$, 
the main decay modes are decays into 
$b\bar b$ pairs and $WW,ZZ$
pairs,  one of the gauge bosons being virtual
below the respective thresholds. Above the
$WW,ZZ$ pair thresholds, the Higgs particles decay almost
exclusively into these two channels, with a small
admixture of top decays near the $t\bar t$ threshold.
Below 140 GeV, the decays $H\to \tau^+\tau^-, c\bar c$
and $gg$ are also important besides the dominating
$b\bar b$ channel; $\gamma\gamma$
decays, though suppressed in rate, nevertheless provide 
 a clear 2-body signature for the
formation of Higgs particles in this mass range.

\vspace*{3mm}
\noindent
{\it (a) \underline{Higgs decays to fermions}} 

\vspace*{1mm}
\noindent
The partial width of Higgs decays to lepton
and quark pairs is given by \cite{26}
\begin{equation}
\Gamma (H\to f\bar f) = {\cal N}_c \frac{G_F}{4\sqrt{2}\pi} m_f^2(M_H^2) M_H 
~,
\end{equation}
${\cal N}_c = 1$ or 3 being the color factor. [Near the threshold the partial
width is suppressed by the additional $P$-wave factor $\beta_f^3$, where 
$\beta_f$
is the fermion velocity.] Asymptotically, the
fermionic width grows only linearly with the
Higgs mass.
The bulk of QCD radiative corrections can be 
mapped into the scale dependence of the quark mass,
evaluated at the Higgs mass. For $M_H\sim 100$ GeV
the relevant parameters are $m_b (M_H^2) \simeq 3$ GeV and
$m_c (M_H^2) \simeq$ 0.6~GeV.
The reduction of the effective $c$-quark mass
overcompensates the color factor in the ratio 
between charm and $\tau$
decays of Higgs bosons. The residual QCD corrections,
$\sim 5.7 \times (\alpha_s/\pi)$, modify the widths only slightly. 

\vspace*{3mm}
\noindent
{\it (b) \underline{Higgs decays to $WW$ and $ZZ$ boson pairs}} 

\vspace*{1mm}
\noindent
Above the $WW$ and $ZZ$ decay thresholds, the partial widths for these 
channels may be written as \cite{27}
\begin{equation}
\Gamma (H\to VV) = \delta_V \frac{G_F}{16\sqrt{2}\pi} M_H^3 (1-4x+12x^2)
\beta_V ~,
\end{equation}
where $x=M_V^2/M_H^2$ and $\delta_V = 2$ and 1 for $V=W$ and $Z$, 
respectively. For large Higgs masses, the vector bosons
are longitudinally polarized. Since the wave functions
of these states are linear in the energy, the widths
grow as the third power of the Higgs mass. Below the
threshold for two real bosons, the Higgs particle can
decay into $VV^*$
pairs, one of the vector bosons being virtual. The partial 
width is given in this case \cite{28} by
\begin{equation}
\Gamma(H\to VV^*) = \frac{3G^2_F M_V^4}{16\pi^3}~M_H  R(x)~\delta'_V ~,
\end{equation}
where $\delta'_W = 1$, $\delta'_Z = 7/12 - 10\sin^2\theta_W/9 + 40
\sin^4\theta_W/27$ and
\begin{displaymath}
R(x) = \frac{3(1-8x+20x^2)}{(4x-1)^{1/2}}\arccos\left(\frac{3x-1}{2x^{3/2}}
\right) - \frac{1-x}{2x} (2-13x+47x^2)
- \frac{3}{2} (1-6x+4x^2) \log x ~.
\end{displaymath}
The $ZZ^*$ channel becomes relevant for Higgs masses beyond 
$\sim 140$ GeV.  Above the threshold, the 4-lepton channel 
$H\to ZZ \to 4 \ell^\pm$
provides a very clear signal for Higgs bosons. Despite of escaping 
neutrinos in leptonic $W$ decays, also the $WW$ decay channel proves useful 
if the on-shell $ZZ$ channel is still closed kinematically.

\vspace*{3mm}
\noindent
{\it (c) \underline{Higgs decays to $gg$ and $\gamma\gamma$ pairs}} 

\vspace*{1mm}
\noindent
In the Standard Model, gluonic Higgs decays are 
mediated by top- and bottom-quark loops, photonic decays in addition by 
$W$ loops. Since these decay modes are significant
only far below the top and $W$
thresholds, they are described by the approximate
expressions \cite{29,30}
\begin{eqnarray}
\Gamma (H\to gg) & = & \frac{G_F \alpha_s^2(M_H^2)}{36\sqrt{2}\pi^3}M_H^3
\left[ 1+ \left(\frac{95}{4} - \frac{7N_F}{6} \right) \frac{\alpha_s}{\pi}
\right] \label{eq:htogg} \\ \nonumber \\
\Gamma (H\to \gamma\gamma) & = & \frac{G_F \alpha^2}{128\sqrt{2}\pi^3}M_H^3
\left[  \frac{4}{3} {\cal N}_C e_t^2 - 7 \right]^2 ~,
\end{eqnarray}
which are valid in the limit $M_H^2 \ll 4M_W^2, 4M_t^2$.
The QCD radiative corrections, which include the $ggg$ and $gq\bar q$
final states in (\ref{eq:htogg}), are very important; they
increase the partial width by about 65\%. Even
though photonic Higgs decays are very rare, 
they nevertheless open an attractive resonance-type channel 
for the search of Higgs particles. 

\vspace*{3mm}
\noindent
{\it \underline{Summary}}

\begin{figure}[hbtp]

\vspace*{0.5cm}
\hspace*{1.0cm}
\begin{turn}{-90}%
\epsfxsize=8.5cm \epsfbox{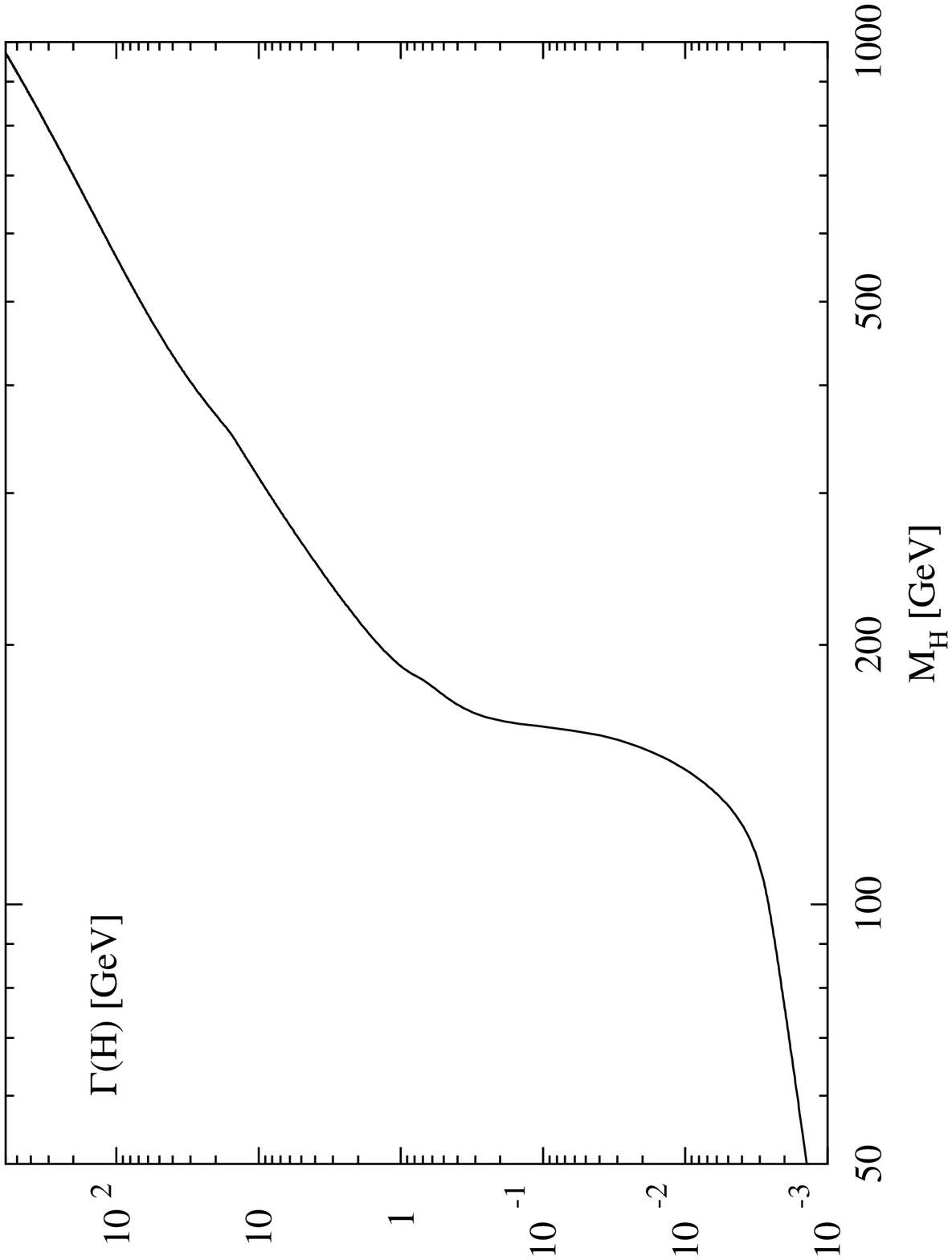}
\end{turn}

\vspace*{1.0cm}
\hspace*{1.0cm}
\begin{turn}{-90}%
\epsfxsize=8.5cm \epsfbox{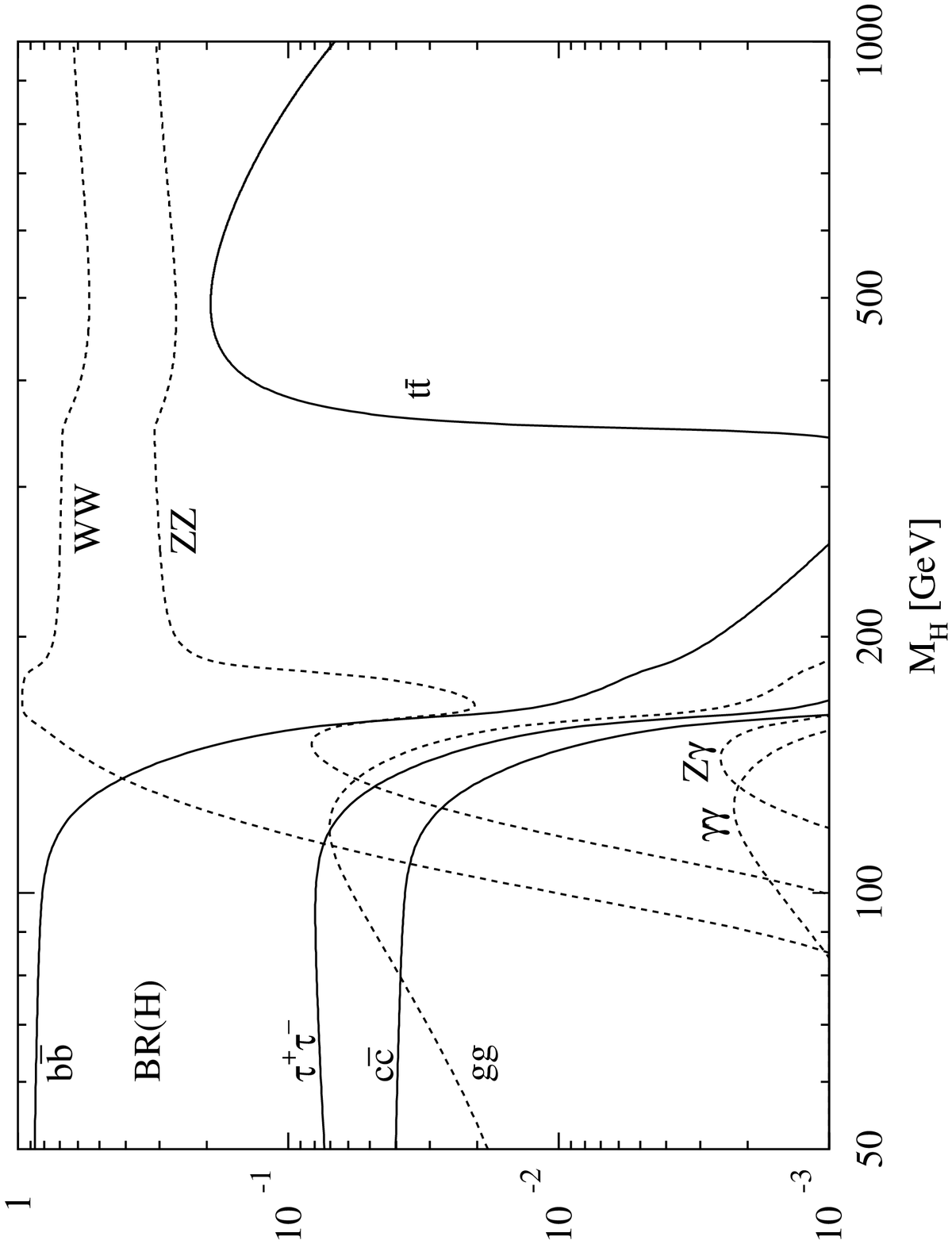}
\end{turn}
\vspace*{-0.0cm}

\caption[]{\label{fg:wtotbr} \it (a) Total decay width (in GeV) of the SM
Higgs boson as a function of its mass. (b) Branching ratios of the
dominant decay modes of the SM Higgs particle. All relevant higher-order
corrections are taken into account. Code: HDECAY, Ref.\cite{HDECAY}.}
\end{figure}

\vspace*{1mm}
\noindent
By adding up all possible decay channels, we obtain 
the total width shown in Fig.~\ref{fg:wtotbr}a. Up to masses of 
140 GeV, the Higgs particle is very narrow, $\Gamma(H) \leq 10$ MeV.
After the real and virtual gauge-boson
channels open up, the state rapidly becomes  wider,
reaching a width of $\sim 1$ GeV at the $ZZ$
threshold. The width cannot be measured directly
in the intermediate mass region at the LHC or $e^+ e^-$
colliders. However it can be determined indirectly; measuring, for
example, the
partial width $\Gamma (H\to WW)$ in the fusion process $WW\to H$, and
the branching fraction $BR(H\to WW)$ in the decay process $H\to WW$, the total 
width follows from the ratio of the two observables. 
Above a mass of $\sim 250$ GeV,
the state would become wide enough to be resolved experimentally.

The branching ratios of the main decay modes are
displayed in Fig.~\ref{fg:wtotbr}b. A large variety of channels
will be accessible for Higgs masses below 140 GeV. 
The dominant mode are $b\bar b$ decays, yet
$c\bar c, \tau^+\tau^-$ and $gg$ decays 
still occur at a level of several per-cent.
At $M_H=$ 120~GeV for instance, the branching ratios are 68\% for
$b\bar b$, 3.1\% for $c\bar c$, 6.9\% for $\tau^+\tau^-$ and 7\% for $gg$.
$\gamma\gamma$
decays occur at a level of 1~per-mille. Above this mass value, 
the Higgs boson decay into $W$'s
becomes dominant, overwhelming all other channels if 
the decay mode into two real $W$'s is kinematically possible.
For Higgs masses far above the thresholds, 
$ZZ$ and $WW$ decays occur at a ratio of 1:2, slightly modified only
just above the $t\bar t$ threshold. Since the decay widths
to vector-boson pairs grow as the third
power of the mass, the Higgs particle becomes very wide
asymptotically, $\Gamma(H) \sim
\frac{1}{2} M_H^3$ [TeV]. In fact, for $M_H\sim 1$ TeV,
the width reaches $\sim \frac{1}{2}$ TeV.

\vspace*{4mm}
\subsection{Higgs Production at Hadron Colliders}

\vspace*{2mm}
\noindent
Several processes can be exploited to produce 
Higgs particles in hadron colliders \cite{24A,32}: \\[0.5cm]
\begin{tabular}{llll}
\hspace*{21mm} & gluon fusion & :              & $gg\to H$ \\ \\
& $WW,ZZ$ fusion           & :    & $W^+ W^-, ZZ \to H$ \\ \\
& Higgs-strahlung off $W,Z$ & :   & $q\bar q \to W,Z \to W,Z + H$ \\ \\
& Higgs bremsstrahlung off top & : & $q\bar q, gg \to t\bar t + H$ \;.
\end{tabular} \\[0.5cm]
Gluon fusion plays the dominant role 
throughout the entire Higgs mass range of the
Standard Model. While the $WW/ZZ$
fusion process becomes increasingly important
with rising Higgs mass, it plays also an important role in the 
search for the Higgs boson and the study of its properties in the intermediate 
mass range. The last two radiation 
processes are of interest only for light Higgs masses.

The production cross sections at hadron colliders, at
the LHC in particular, are quite sizeable so that a 
large sample of SM Higgs particles can be produced
in this machine. Experimental difficulties 
arise from the huge number  of background events
that come along with the Higgs signal events.
This problem will be tackled by either
triggering on leptonic decays of $W,Z$ and $t$
in the radiation processes or by exploiting
the resonance character of the Higgs decays
$H\to \gamma\gamma$ and $H\to ZZ \to 4\ell^\pm$.
In this way, the Tevatron is expected to 
search for Higgs particles in the mass range
above LEP2 up to about 110 to 130 GeV \cite{11}. 
The LHC is expected to cover the entire canonical
Higgs mass range $M_H \lessim 700$ GeV
of the Standard Model \cite{12}.

\vspace*{4mm}
\noindent
{\it (a) \underline{Gluon fusion}}

\vspace*{1mm}
\noindent
The gluon-fusion mechanism \cite{29,32,39A,39B}
\begin{displaymath}
pp \to gg \to H
\end{displaymath}
provides the dominant production mechanism of Higgs 
bosons at the LHC in the entire relevant Higgs
mass range up to about 1 TeV. The gluon coupling
to the Higgs boson in the SM is mediated by
triangular loops of top and bottom quarks, 
cf. Fig.~\ref{fg:gghlodia}.
Since the Yukawa coupling of the Higgs particle
to heavy quarks grows with the quark mass, thus
balancing the decrease of the triangle amplitude, the 
form factor approaches a non-zero value for large 
loop-quark masses. [If the masses of heavy quarks
beyond the third generation were generated solely
by the Higgs mechanism, these particles
would add the same amount to the form factor as
the top quark in the asymptotic heavy-quark limit.]
\begin{figure}[hbt]
\vspace*{3mm}
\begin{center}
\setlength{\unitlength}{1pt}
\begin{picture}(180,90)(0,0)

\Gluon(0,20)(50,20){-3}{5}
\Gluon(0,80)(50,80){3}{5}
\ArrowLine(50,20)(50,80)
\ArrowLine(50,80)(100,50)
\ArrowLine(100,50)(50,20)
\DashLine(100,50)(150,50){5}
\put(155,46){$H$}
\put(25,46){$t,b$}
\put(-15,18){$g$}
\put(-15,78){$g$}

\end{picture}  \\
\setlength{\unitlength}{1pt}
\caption[ ]{\label{fg:gghlodia} \it Diagram contributing to the
formation of Higgs bosons in gluon-gluon collisions
at lowest order.}
\end{center}
\end{figure}
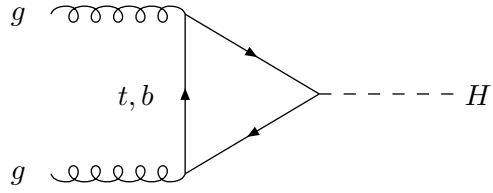\\

The partonic cross section, Fig.~\ref{fg:gghlodia}, can be expressed 
by the gluonic width of the Higgs boson at 
lowest order \cite{32}:
\begin{eqnarray}
\hat \sigma_{LO} (gg\to H) & = & \sigma_0 M_H^2 \times BW(\hat{s}) \\
\sigma_0 & = & \frac{\pi^2}{8M_H^2} \Gamma_{LO} (H\to gg) =
\frac{G_F\alpha_s^2}{288\sqrt{2}\pi} \left| \sum_Q A_Q^H (\tau_Q) \right|^2 ~,
\nonumber
\end{eqnarray}
where the scaling variable is defined as $\tau_Q = 4M_Q^2/M_H^2$ and $\hat s$
denotes the partonic c.m. energy squared. The
form factor can easily be evaluated:
\begin{eqnarray}
A_Q^H (\tau_Q) & = & \frac{3}{2} \tau_Q \left[ 1+(1-\tau_Q) f(\tau_Q)
\right] \label{eq:ftau} \\
f(\tau_Q) & = & \left\{ \begin{array}{ll}
\displaystyle \arcsin^2 \frac{1}{\sqrt{\tau_Q}} & \tau_Q \geq 1 \\
\displaystyle - \frac{1}{4} \left[ \log \frac{1+\sqrt{1-\tau_Q}}
{1-\sqrt{1-\tau_Q}} - i\pi \right]^2 & \tau_Q < 1
\;. \end{array} \right. \nonumber 
\end{eqnarray}
For small loop masses the form factor vanishes, $A_Q^H(\tau_Q) \sim -3/8 \tau_Q
[\log (\tau_Q/4)+i\pi]^2$,
while for large loop masses it approaches a non-zero value,
$A_Q^H (\tau_Q) \to 1$. The final term $BW$ is the normalized Breit-Wigner 
function
\beq
BW(\hat{s}) = \frac{M_H \Gamma_H/\pi}{[\hat{s}-M_H^2]^2 + M_H^2 \Gamma_H^2}
\eeq
approaching in the narrow-width approximation a $\delta$-function at 
$\hat{s}=M_H^2$.

In the narrow-width approximation, the hadronic
cross section can be cast into the form
\begin{equation}
\sigma_{LO} (pp\to H) = \sigma_0 \tau_H \frac{d{\cal L}^{gg}}{d\tau_H} ~,
\end{equation}
with $d{\cal L}^{gg}/d\tau_H$ denoting the $gg$ luminosity of the 
$pp$ collider, 
\begin{equation}
  d{\cal L}^{gg}/d\tau_H = \int_{\tau_H}^1 \frac{d\xi}{\xi} \,
                           g(\xi; \tau_H s) \, g(\tau_H / \xi; \tau_H s) \,,
\end{equation}
built up by the gluon densities $g$ and 
evaluated for the Drell--Yan variable
$\tau_H = M_H^2/s$, where $s$ is the total hadronic energy squared. \\

%
%
%
%

The QCD corrections to the gluon fusion 
process \cite{29,32,39B} are very important. They
stabilize the theoretical predictions for the 
cross section when the renormalization and
factorization scales are varied. Moreover,
they are large and positive, thus increasing the
production cross section for Higgs bosons. 
The QCD corrections consist of virtual 
corrections to the basic process $gg\to H$, 
and of real corrections due to the associated
production of the Higgs boson with massless
partons, $gg\to Hg$ and $gq\to Hq,\, q\bar q\to Hg$.
These subprocesses contribute to Higgs production
at ${\cal O}(\alpha_s^3)$.
The virtual corrections rescale the lowest-order
fusion cross section with a coefficient that depends
only on the ratios of the Higgs and quark masses. 
Gluon radiation leads to two-parton final states
with invariant energy $\hat s \geq M_H^2$ in the 
$gg, gq$ and $q\bar q$ channels.\\

\begin{figure}[hbt]

\vspace*{0.4cm}
\hspace*{2.0cm}
\begin{turn}{-90}%
\epsfxsize=7cm \epsfbox{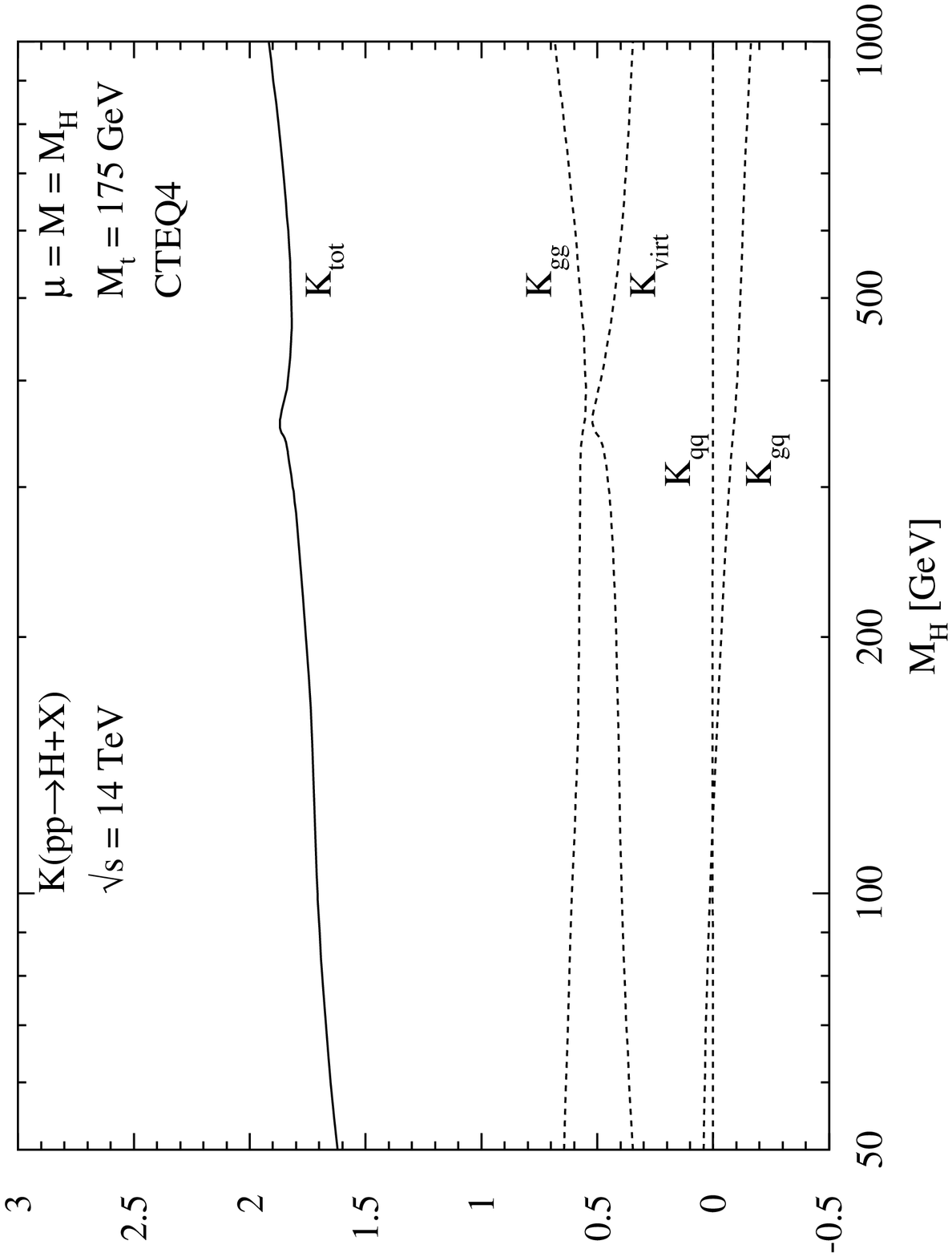}
\end{turn}
\vspace*{-0.2cm}

\caption[]{\label{fg:gghk} \it K factors of the QCD-corrected gluon-fusion
cross section $\sigma(pp \to H+X)$ at the LHC with c.m.~energy $\sqrt{s}=14$
TeV. The dashed lines show the individual contributions of 
the QCD corrections. The renormalization and
factorization scales have been identified with the Higgs mass,  
and  CTEQ4 parton densities have been adopted.}
\end{figure}
\begin{figure}[hbt]

\vspace*{-0.2cm}
\hspace*{2.7cm}
\epsfxsize=9.5cm \epsfbox{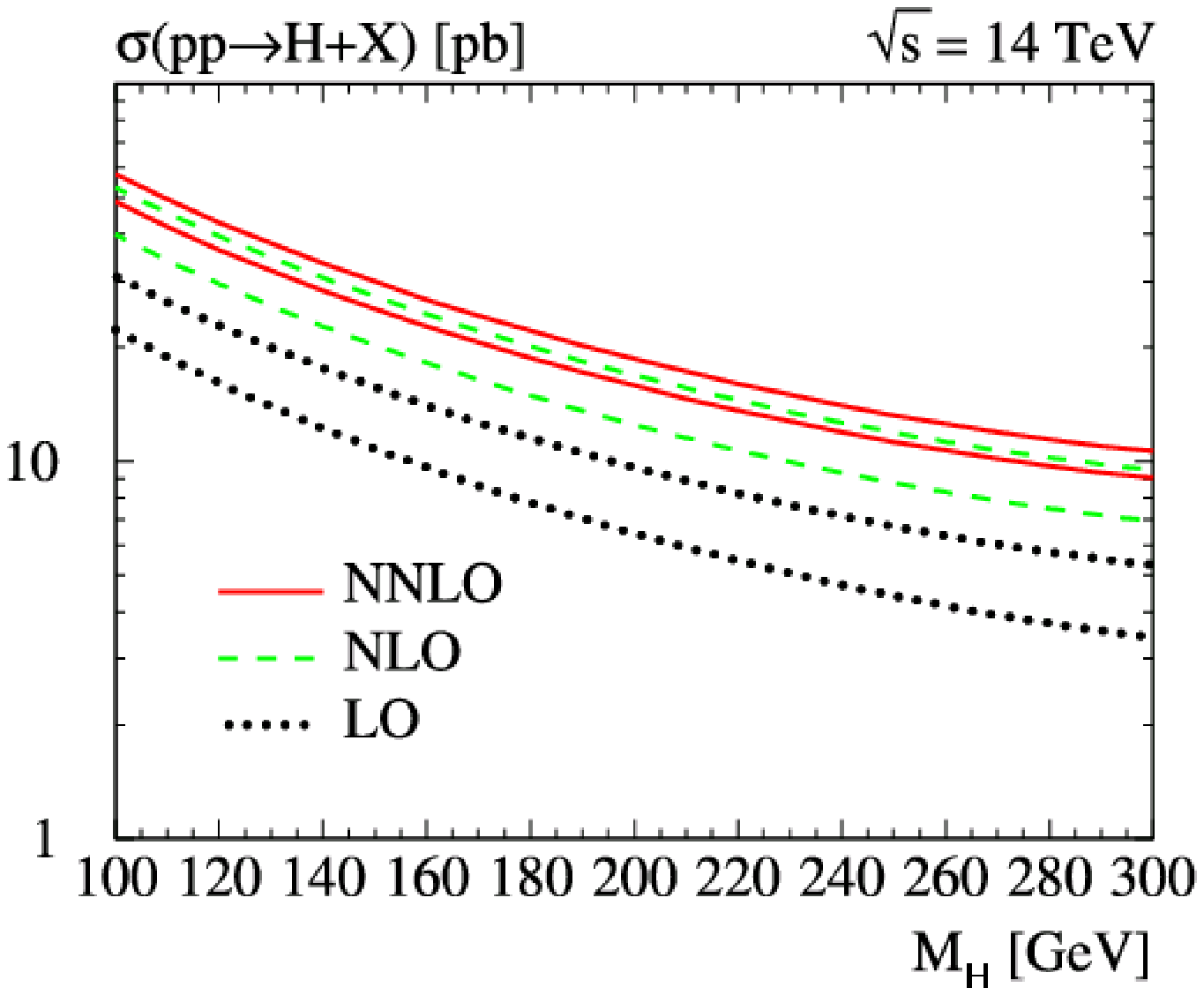}
\vspace*{-0.3cm}

\caption[]{\label{fg:MochVV(a)} \it Gluon-fusion cross section
$\sigma(pp \to H+X)$ at the LHC with c.m.~energy $\sqrt{s}=14$ TeV at
LO, NLO and NNLO. The size of the error bands is determined by the
variation of the renormalization and factorization scales between $M_H/2$
and $2M_H$. First reference in \cite{x}.}
\end{figure}
\begin{figure}[hbt]

\vspace*{-0.1cm}
\hspace*{3.5cm}
\epsfxsize=7.5cm \epsfbox{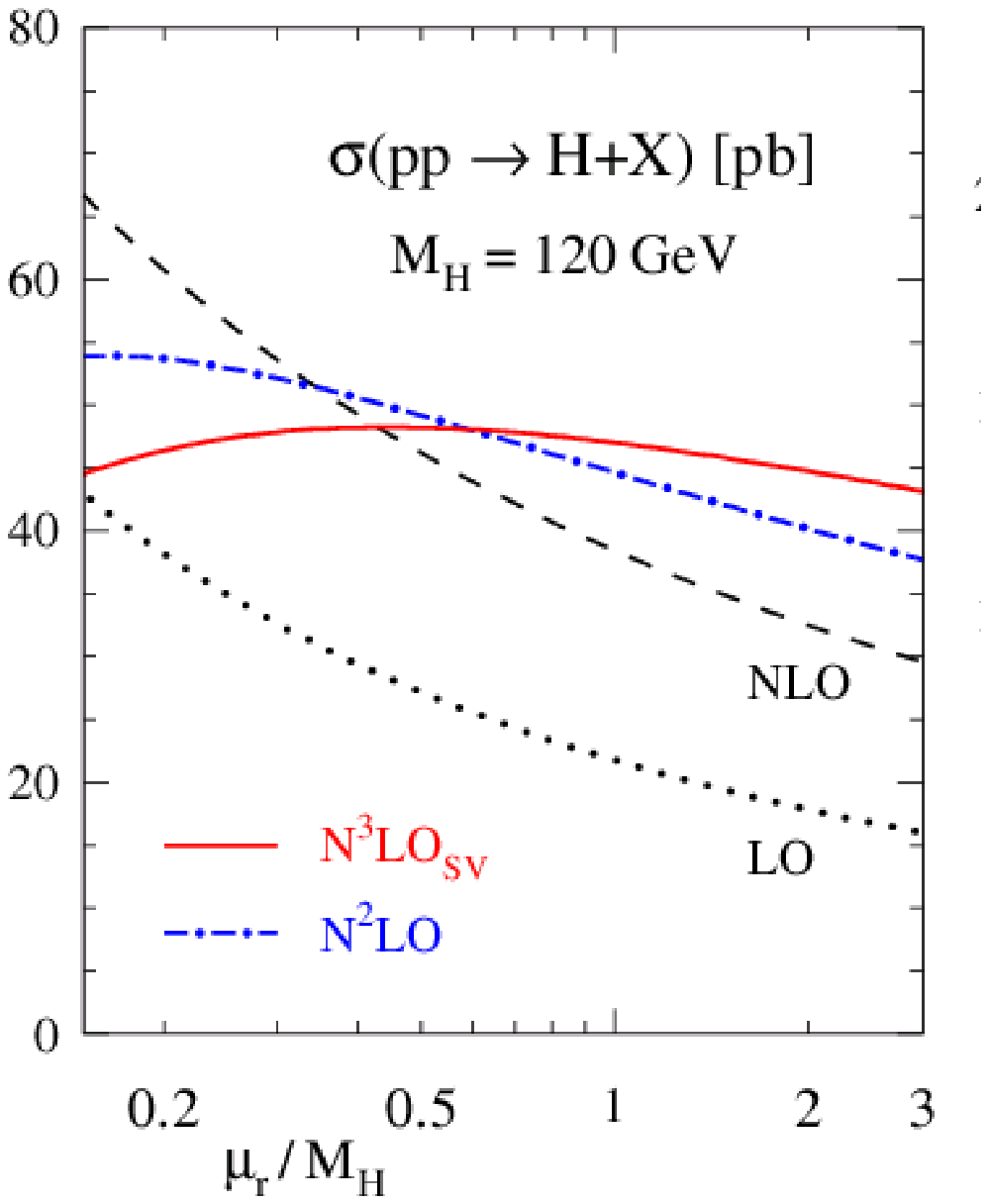}
\vspace*{-0.2cm}

\caption[]{\label{fg:MochVV(b)} \it Renormalization scale dependence of
the gluon-fusion cross section $\sigma(pp \to H+X)$ at the LHC with
c.m.~energy $\sqrt{s}=14$ TeV at LO, NLO, NNLO and the soft/virtual
approximation at N$^3$LO; from Ref.~\cite{MochVV}.}
\end{figure}

The size of the radiative corrections can be parametrized
by defining the $K$ factor as $K=\sigma_{NLO}/\sigma_{LO}$, in which
all quantities are evaluated in the  numerator and 
denominator in next-to-leading and leading order,
respectively. The results of this calculation are
shown in Fig.~\ref{fg:gghk}. The virtual corrections 
$K_{virt}$ and the real corrections $K_{gg}$ for the $gg$ collisions 
are apparently of the same size, and both are large
and positive; the corrections for $q\bar q$
collisions and the $gq$
inelastic Compton contributions are less important.
Depending only mildly on the Higgs bosons mass, the overall $K$ factor, 
$K_{tot}$, turns out to be close to 2 \cite{29,32,39B,R}.
The main contributions are 
generated by the virtual corrections and the 3-parton final states initiated by 
$gg$ initial states. Large NLO corrections are expected for these 
gluon processes as a result of the large color charges. However, by 
studying the next order of corrections in the large top-mass limit, the 
N$^{2}$LO corrections generate only a modest additional increase of the $K$ 
factor, $\delta_2 K_{tot} \lessim 0.2$ \cite{x},
and even less at N$^{3}$LO \cite{MochVV}. This proves the expansion to 
be convergent with the most important correction to be attributed to the 
next-to leading order contribution \cite{R}, cf. Fig.~\ref{fg:MochVV(a)}.
In addition, when the higher-order QCD corrections are included,
the dependence of the cross section on the renormalization
and factorization scales is significantly reduced, Fig.~\ref{fg:MochVV(b)}.

%
%

\begin{figure}[hbt]

\vspace*{0.5cm}
\hspace*{0.0cm}
\begin{turn}{-90}%
\epsfxsize=10cm \epsfbox{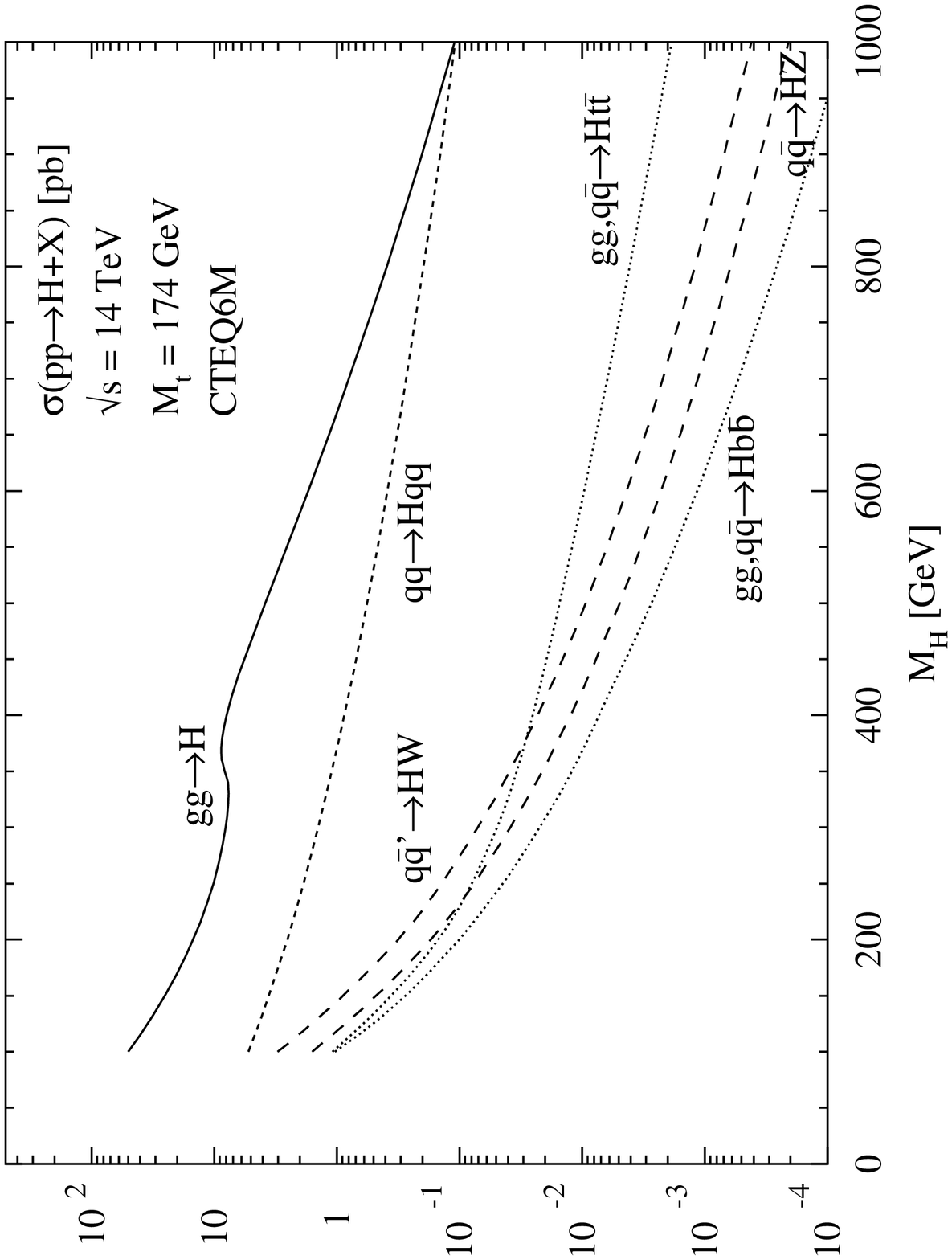}
\end{turn}
\vspace*{0.0cm}

\caption[]{\label{fg:lhcpro} \it Higgs production cross sections at the LHC
 for the various production mechanisms as a function of the
Higgs mass. The full QCD-corrected results for the gluon fusion $gg
\to H$, vector-boson fusion $qq\to VVqq \to Hqq$, vector-boson bremsstrahlung
$q\bar q \to V^* \to HV$ and associated production $gg,q\bar q \to Ht\bar t,
Hb\bar b$ are shown.}
\end{figure}

The theoretical prediction for the production cross 
section  of Higgs particles 
is presented in Fig.~\ref{fg:lhcpro} for the LHC as a 
function of the Higgs mass.
The cross section decreases with increasing Higgs mass.
This is, to a large extent, a consequence of the
sharply falling $gg$
luminosity for large invariant masses. The bump in 
the cross section is induced by the $t\bar t$
threshold
in the top triangle. The overall theoretical 
accuracy of this calculation is expected to be at
a level of 10 to 20\%.  

\vspace*{3mm}
\noindent
{\it (b) \underline{Vector-boson fusion}}

\vspace*{1mm}
\noindent
The second important channel for Higgs production at the
LHC is vector-boson fusion, $W^+W^- \to H$
\cite{23}. For large Higgs masses this mechanism becomes
competitive to gluon fusion; for intermediate
masses the cross section is smaller but very important nevertheless
for searching light Higgs bosons with a reduced background-to-signal
ratio and for exploring its properties \cite{Zep}. \\ 

For large Higgs masses, the two electroweak bosons $W,Z$
that form the Higgs boson are predominantly 
longitudinally polarized. At high energies, the 
equivalent particle spectra of the longitudinal $W,Z$
bosons in quark beams are given by
\begin{eqnarray}
f^W_L (x) & = & \frac{G_F M_W^2}{2\sqrt{2}\pi^2} \frac{1-x}{x} 
 \label{eq:xyz} \\ \non \\
f^Z_L (x) & = & \frac{G_F M_Z^2}{2\sqrt{2}\pi^2}
\left[(I_3^q - 2e_q \sin^2\theta_W)^2 + (I_3^q)^2\right] \frac{1-x}{x} ~, \non
\end{eqnarray}
where $x$ is the fraction of energy transferred from the quark
to the $W,Z$ boson in the splitting process
$q\to q +W/Z$. From these particle spectra, the $WW$ and $ZZ$
luminosities can easily be derived:
\begin{eqnarray}
\frac{d{\cal L}^{WW}}{d\tau_W} & = & \frac{G_F^2 M_W^4}{8\pi^4}
\left[ 2 - \frac{2}{\tau_W} -\frac{1+\tau_W}{\tau_W} \log \tau_W \right] \\
\non \\
\frac{d{\cal L}^{ZZ}}{d\tau_Z} & = & \frac{G_F^2 M_Z^4}{8\pi^4}
\left[(I_3^q - 2e_q \sin^2\theta_W)^2 + (I_3^q)^2\right]
\left[(I_3^{q'} - 2e_{q'} \sin^2\theta_W)^2 + (I_3^{q'})^2\right] \non \\
& & \hspace{1.5cm} \times \left[ 2 - \frac{2}{\tau_Z} -\frac{1+\tau_Z}{\tau_Z}
\log \tau_Z \right] \non
\end{eqnarray}
with the Drell--Yan variable defined as  $\tau_V = M_{VV}^2/s$.
The cross section for Higgs production in quark--quark
collisions is given by the convolution of the parton cross sections 
$WW,ZZ \to H$ with the luminosities:
\begin{equation}
\hat \sigma(qq\to qqH) = \frac{d{\cal L}^{VV}}{d\tau_V} \sqrt{2} \pi G_F ~.
\label{eq:vvhpart}
\end{equation}
The hadronic cross section is finally obtained by
summing the parton cross section (\ref{eq:vvhpart}) 
over the flux of all possible pairs
of quark--quark and antiquark combinations. \\

Since to lowest order the proton remnants are
color singlets in the $WW,ZZ$
fusion processes, no color will be exchanged between the
two quark lines from which the two vector bosons are
radiated. As a result, the leading QCD corrections to
these processes are already accounted for
by the corrections to the quark parton densities.\\

The $WW/ZZ$ fusion cross section for Higgs bosons at the LHC
is shown in Fig.~\ref{fg:lhcpro}. The process is apparently
very important for the search of the Higgs boson in the upper mass range, 
where the cross section approaches values close to gluon fusion. For 
intermediate masses, it comes close within an order of magnitude to the 
leading gluon-fusion cross section.

\vspace*{3mm}
\noindent
{\it (c) \underline{Higgs-strahlung off vector bosons}}

\vspace*{1mm}
\noindent
Higgs-strahlung $q\bar q \to V^* \to VH~(V=W,Z)$
is a very important mechanism (Fig.~\ref{fg:lhcpro}) for the 
search of light Higgs bosons at the hadron colliders
Tevatron and LHC. Though the cross section is 
smaller than for gluon fusion, leptonic decays
of the electroweak vector bosons are
useful to filter Higgs signal events
out of the huge background. Since the dynamical
mechanism is the same as for $e^+e^-$
colliders [{\it see later}], except for the  folding with
the quark--antiquark densities, 
intermediate steps of the
calculation will not be noted here, and merely  
the final values 
of the cross sections for the Tevatron and the 
LHC are recorded in Fig.~\ref{fg:lhcpro}.

\vspace*{3mm}
\noindent
{\it (d) \underline{Higgs bremsstrahlung off top quarks}}

\vspace*{1mm}
\noindent
Also the process $gg,q\bar q \to t\bar t H$
is relevant only for small Higgs masses, Fig.~\ref{fg:lhcpro}.
The analytical expression for the parton cross
section, even at lowest order, is quite involved, 
so that just the final results for the LHC
cross section are shown in Fig.~\ref{fg:lhcpro}.
Higher order corrections have been presented in Ref.~\cite{z}.
Separating the signal from the background is experimentally
very difficult for this channel.

Nevertheless, Higgs bremsstrahlung off top quarks may be an interesting
process for measurements of the fundamental $Htt$
Yukawa coupling in coherent LHC/LC analyses. 
The cross section $\sigma (pp\to t\bar t H)$
is directly proportional to the square of
this fundamental coupling.

\vspace*{3mm}
\noindent
{\it{\underline{Summary}}} 

\vspace*{1mm}
\noindent
A comprehensive overview of the production cross
sections for Higgs particles at the LHC
is presented in Fig.~\ref{fg:lhcpro}. Three classes
of channels can be distinguished. The gluon fusion of Higgs particles
is a universal process, dominant over the 
entire SM Higgs mass range. Higgs-strahlung
off electroweak $W,Z$
bosons or top quarks is prominent for light
Higgs bosons. The $WW/ZZ$
fusion channel, by contrast, becomes increasingly
important in the upper part of the SM Higgs
mass range, though it proves also very useful in the intermediate 
mass range.

The signatures for the search for Higgs particles are
dictated by the decay branching ratios. In the
lower part of the intermediate mass range, resonance
reconstruction in $\gamma\gamma$ final states and $b\bar b$
jets can be exploited. In the upper part of the
intermediate mass range, decays to $ZZ^*$ and $WW^*$
are important, with the two electroweak bosons 
decaying leptonically. In the mass range above
the on-shell $ZZ$ decay threshold, the charged-lepton decays
$H\to ZZ \to 4\ell^\pm$ provide a gold-plated signature. At the
upper end of the classical SM Higgs mass range,
decays to neutrinos and jets,  
generated in $W$ and $Z$ decays, complete the search techniques. \\

\begin{figure}[t]

\vspace*{-0.8cm}
\hspace*{3.0cm}
\epsfxsize=10cm \epsfbox{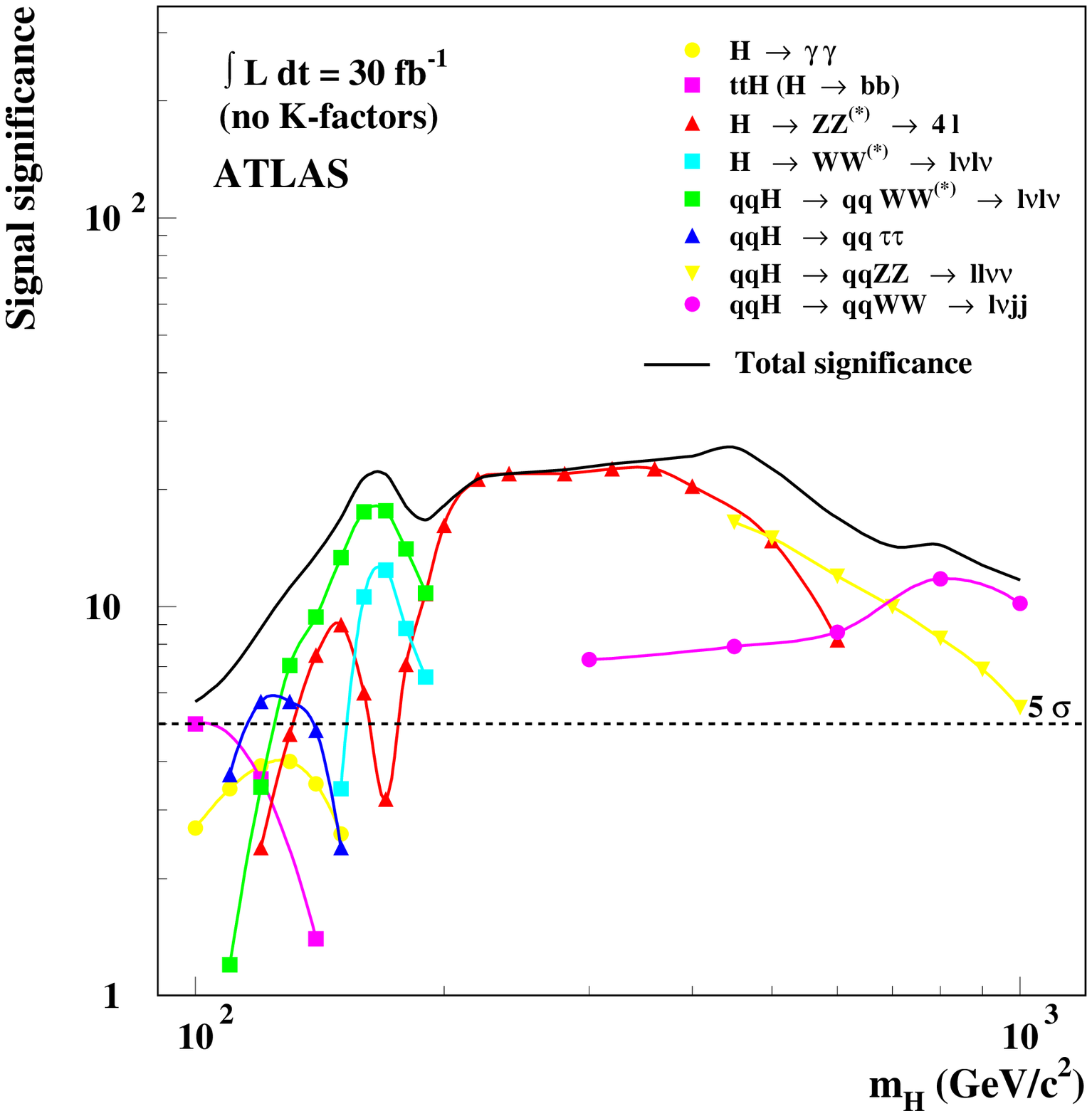}
\vspace*{-0.3cm}

\caption[]{\label{fg:lhcsig} \it Significances of various channels
in the search for the SM Higgs boson at the LHC as a function of
the mass for an integrated luminosity of 30 {\rm{fb}}$^{-1}$;
Ref.\cite{12,12A}.} 
\end{figure}

Experimental expectations at the LHC for the search of the Higgs particle
in the Standard Model are summarized in Fig.~\ref{fg:lhcsig}.
The significance of the Higgs signal is shown 
as a function of the Higgs mass for the integrated luminosity
of 30 fb$^{-1}$. The entire mass range can be covered 
for searching the SM Higgs boson at the LHC. \\
\begin{figure}[hbt]

\vspace*{-0.4cm}
\hspace*{1.6cm}
\epsfxsize=12cm \epsfbox{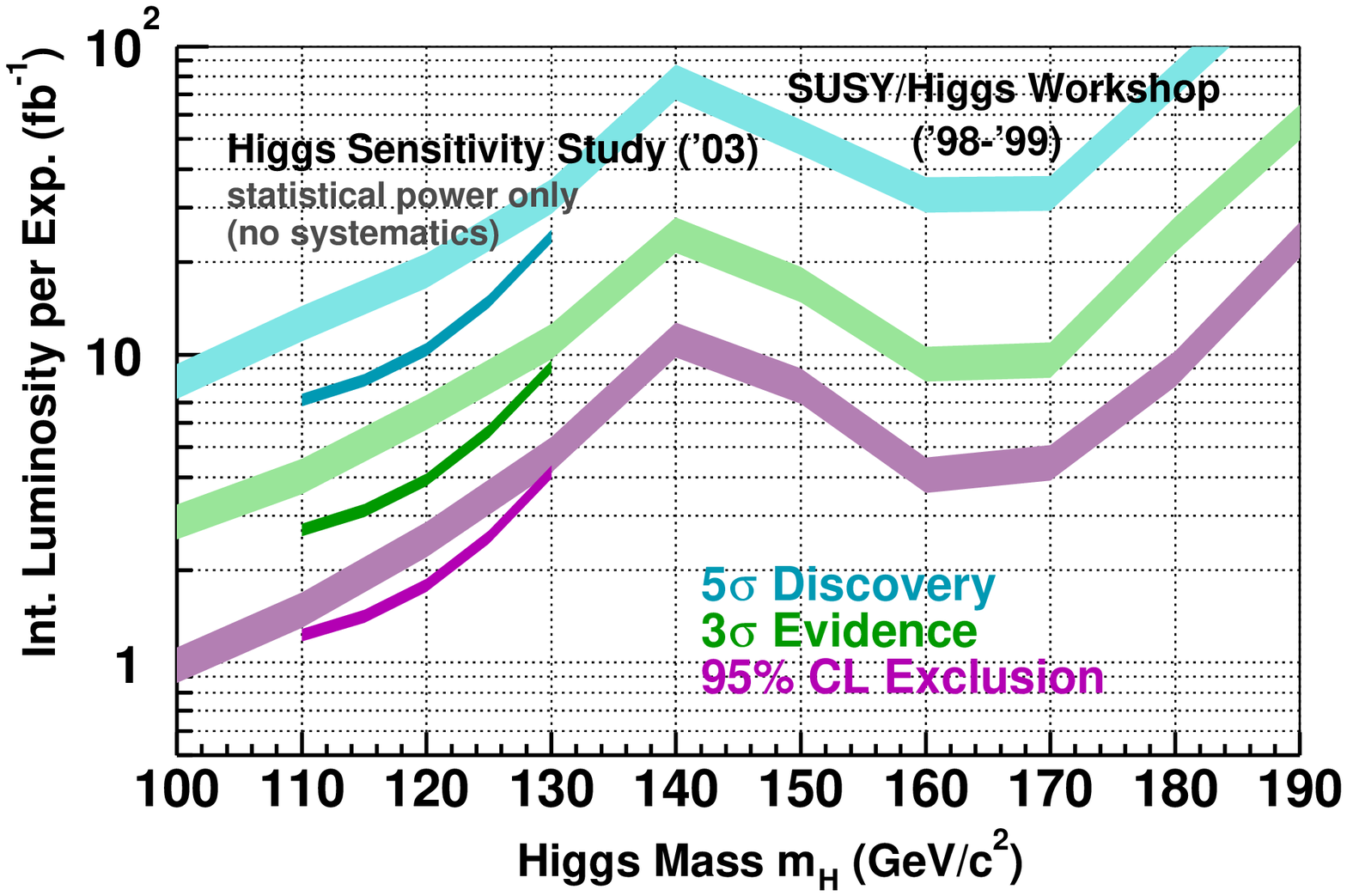}
\vspace*{-0.3cm}

\caption[]{\label{fg:Tevsig} \it Integrated luminosities
needed to exclude or to discover the SM Higgs boson at the Tevatron;
Ref.\cite{11}.}
\end{figure}

Expectations for the search of the SM Higgs boson at the
Tevatron are summarized in Fig.~\ref{fg:Tevsig}.
The SM Higgs boson may be excluded at the $2\sigma$ level
across the entire intermediate mass range. Discovering 
the particle, even in restricted mass intervals, is a
demanding task which requires the collection of a large
integrated luminosity. 

\vspace*{4mm}
\subsection{Higgs Production Channels at $e^+e^-$ Colliders}

\vspace*{2mm}
\noindent
The first process that was used to search
directly for Higgs bosons over a large mass range,
was the Bjorken process, $Z\to Z^* H, Z^* \to f\bar f$ \cite{34}.
By exploring this production channel,
Higgs bosons with masses less than 65.4 GeV
were excluded by the LEP1 experiments.
The search continued by reversing the
role of the real and virtual $Z$ bosons in the
$e^+e^-$ continuum at LEP2.

The  main production mechanisms for Higgs
bosons in $e^+e^-$ collisions are
\begin{eqnarray}
\mbox{Higgs-strahlung} & : & e^+e^- \to Z^* \to ZH \\
\mbox{$WW$ fusion}     & : & e^+e^- \to \bar \nu_e \nu_e (WW) \to \bar \nu_e
\nu_e H \; .
\label{eq:wwfusion}
\end{eqnarray}
In Higgs-strahlung \cite{30,34,35} the Higgs boson is emitted
from the $Z$-boson line, while $WW$ fusion is a formation
process of Higgs bosons in the collision of two quasi-real
$W$ bosons radiated off the electron and positron beams \cite{36}.

As evident from the subsequent analyses, LEP2 could cover
the SM Higgs mass range up to about 114 GeV
\cite{9}. The high-energy $e^+e^-$
linear colliders can cover the entire Higgs
mass range, the intermediate mass range already at a 500 GeV collider
\cite{13}, the upper mass range in the second phase of the machines in which
they are expected to reach a total energy of 3~TeV \cite{38A}.

\vspace*{3mm}
\noindent
{\it (a) \underline{Higgs-strahlung}}

\vspace*{1mm}
\noindent
The cross section for Higgs-strahlung can be
written in a compact form as
\begin{equation}
\sigma (e^+e^- \to ZH) = \frac{G_F^2 M_Z^4}{96\pi s} \left[ v_e^2 + a_e^2
\right] \lambda^{1/2} \frac{\lambda + 12 M_Z^2/s}{\left[ 1- M_Z^2/s \right]^2}
~,
\end{equation}
where $v_e = -1 + 4 \sin^2 \theta_W$ and $a_e=-1$
are the vector and axial-vector $Z$
charges of the electron and $\lambda = [1-(M_H+M_Z)^2/s] [1-(M_H-M_Z)^2/s]$
is the usual two-particle phase-space
function. The cross section is of the size $\sigma \sim \alpha_W^2/s$,
i.e. of second order in the weak coupling, and
it scales in the squared energy. Higher order contributions to the cross
sections are under theoretical control \cite{38B,38C}.

\begin{figure}[hbt]

\vspace*{-5.0cm}
\hspace*{-2.0cm}
\epsfxsize=20cm \epsfbox{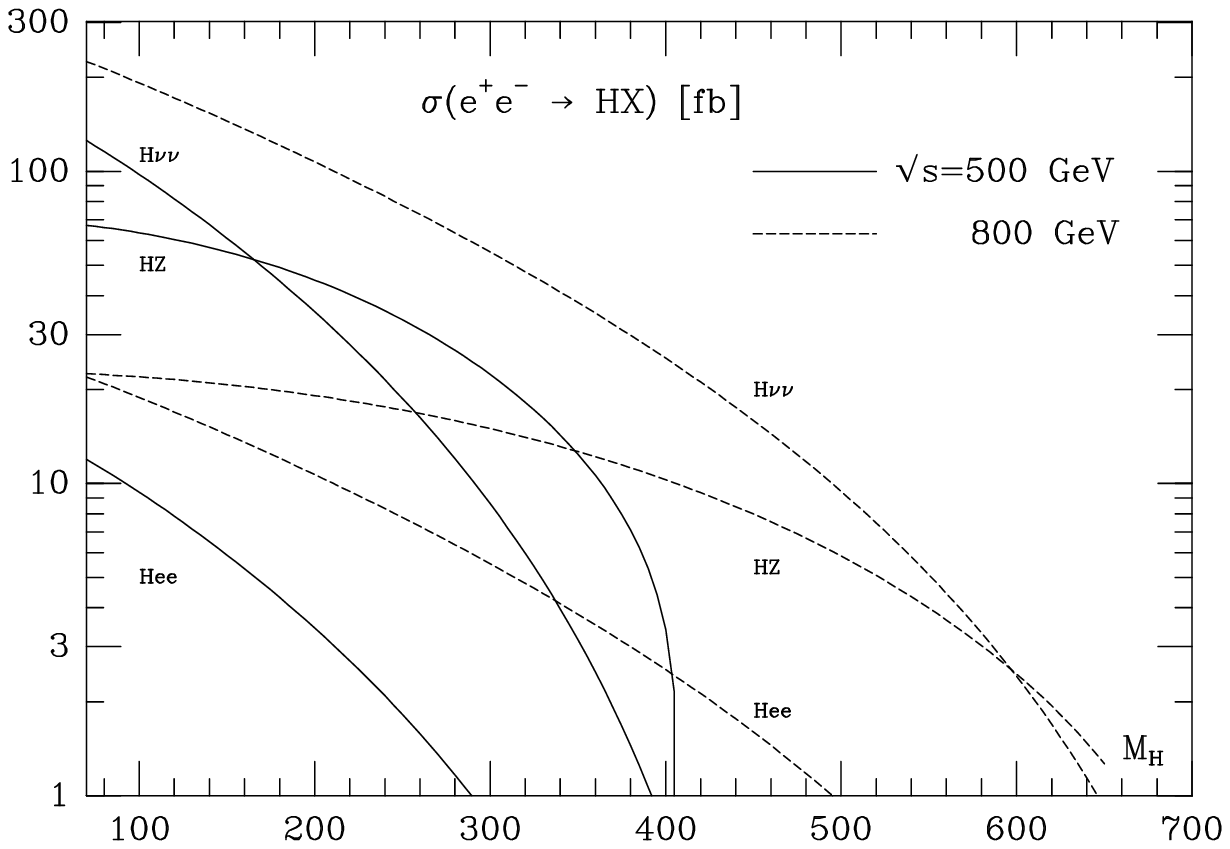}

\caption[]{\label{fg:eehx} \it The cross section for the production of SM
Higgs bosons in Higgs-strahlung $e^+e^-\to ZH$ and $WW/ZZ$ fusion $e^+e^- \to
\bar \nu_e \nu_e/e^+e^- H$; solid curves: $\sqrt{s}=500$ GeV, dashed curves:
$\sqrt{s}=800$ GeV.}
\end{figure}
\begin{figure}[hbt]
\begin{center}
\hspace*{-0.3cm}
\epsfig{figure=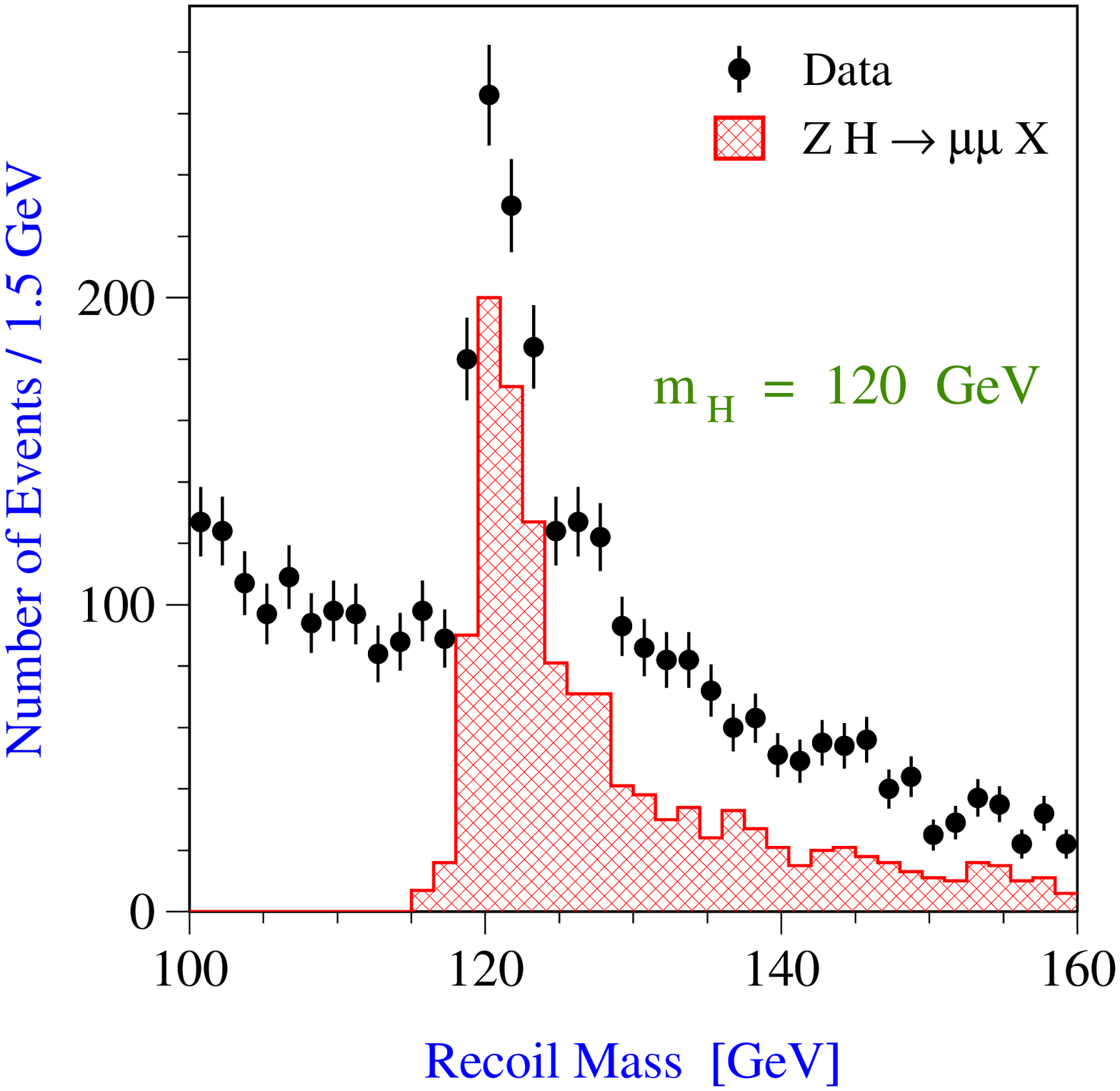,bbllx=0,bblly=0,bburx=567,bbury=561,width=10cm,clip=}
\end{center}
\vspace*{-0.4cm}

\caption[]{\label{fg:zrecoil} \it The $\mu^+\mu^-$ recoil mass distribution in
the process $e^+e^- \to H^0 Z\to X \mu^+\mu^-$ for $M_H=120$~GeV and
$\int {\cal L} = 500 fb^{-1}$ at $\sqrt{s}=$ 350~GeV. The dots with error
bars are Monte Carlo simulations of the Higgs signal and the background. The
shaded histogram represents the signal only. Ref. \cite{13}.}
\end{figure}

Since the cross section vanishes for asymptotic
energies, the Higgs-strahlung process is most
useful for studying Higgs bosons in the
range where the collider energy is of the
same order as the Higgs mass, $\sqrt{s} \gsim {\cal{O}}$$(M_H)$.
The size of the cross section is illustrated
in Fig.~\ref{fg:eehx} for the energy $\sqrt{s}=500$ GeV of
$e^+e^-$ linear colliders as a function of the Higgs mass.
Since the recoiling $Z$ mass in the two-body reaction
$e^+e^- \to ZH$
is mono-energetic, the mass of the Higgs boson
can be reconstructed from the energy of the
$Z$ boson, $M_H^2 = s -2\sqrt{s}E_Z + M_Z^2$,
without any need of analyzing the decay products
of the Higgs boson. For leptonic $Z$
decays, missing-mass techniques provide a
very clear signal, as demonstrated in Fig.~\ref{fg:zrecoil}.

\vspace*{3mm}
\noindent
{\it (b) \underline{$WW$ fusion}}

\vspace*{1mm}
\noindent
Also the cross section for the fusion process (\ref{eq:wwfusion})
can be cast implicitly into a compact form:
\begin{eqnarray}
\sigma (e^+e^-\to\bar \nu_e \nu_e H) & = & \frac{G_F^3 M_W^4}{4\sqrt{2}\pi^3}
\int_{\kappa_H}^1\int_x^1\frac{dx~dy}{[1+(y-x)/\kappa_W ]^2}f(x,y)
\\ \nonumber \\
f(x,y) & = & \left( \frac{2x}{y^3} - \frac{1+3x}{y^2} + \frac{2+x}{y} -1 \right)
\left[ \frac{z}{1+z} - \log (1+z) \right] + \frac{x}{y^3} \frac{z^2(1-y)}{1+z}
\nonumber
\end{eqnarray}
with $\kappa_H=M_H^2/s$, $\kappa_W=M_W^2/s$ and $z=y(x-\kappa_H)/(\kappa_Wx)$.

Since the fusion process is a
$t$-channel exchange process, the size is set by the
$W$ Compton wavelength, suppressed however with
respect to Higgs-strahlung by the third power
of the electroweak coupling, $\sigma \sim \alpha_W^3/M_W^2$.
As a result, $W$
fusion becomes the leading production process
for Higgs particles at high energies. At
asymptotic energies the cross section simplifies to
\begin{equation}
\sigma (e^+e^- \to \bar \nu_e \nu_e H) \to \frac{G_F^3 M_W^4}{4\sqrt{2}\pi^3}
\left[ \log\frac{s}{M_H^2} - 2 \right] ~.
\end{equation}
In this limit, $W$
fusion to Higgs bosons can be interpreted as a
two-step process: the $W$
bosons are radiated as quasi-real particles
from electrons and positrons, $e \to \nu W$,
with a lifetime of the split state of order $E_W / M^2_W$;
the Higgs bosons are formed subsequently in the
colliding $W$ beams. The electroweak higher order corrections are under
control \cite{38C}.

The size of the
fusion cross section is compared with Higgs-strahlung
in Fig.~\ref{fg:eehx}. At $\sqrt{s}=500$ GeV
the two cross sections are of the same order, yet the
fusion process becomes increasingly important with
rising energy.

\vspace*{4mm}
\subsection{The Profile of the SM Higgs Particle}

\vspace*{2mm}
\noindent
To establish the Higgs mechanism
experimentally, the nature of this particle
must be explored by measuring all its
characteristics, the mass and lifetime,
the external quantum numbers spin-parity,
the couplings to gauge bosons and fermions,
and last but not least, the Higgs self-couplings.
While part of this program
can be realized at the LHC \cite{12,xx}, the complete
profile of the particle can be reconstructed
across the entire mass range in $e^+ e^-$ colliders \cite{13}.

\vspace*{3mm}
\noindent
{\it (a) \underline{Mass}}

\vspace*{1mm}
\noindent
The mass of the Higgs particle can be
measured by collecting the decay products
of the particle at hadron and $e^+e^-$ colliders. 
Values at the level of 2 per-mille may be reached
by this method at LHC \cite{12,12A}.
Moreover, in
$e^+e^-$ collisions Higgs-strahlung can be exploited
to reconstruct the mass very precisely from
the $Z$ recoil energy in the two-body
process $e^+e^-\to ZH$.
An overall
accuracy of about $\delta M_H \sim 100$ MeV can be expected
\cite{13}.

\vspace*{3mm}
\noindent
{\it (b) \underline{Width / Lifetime}}

\vspace*{1mm}
\noindent
The width of the state, i.e. the lifetime of
the particle, can be measured directly
above the $ZZ$ decay threshold where the
width grows rapidly. In the lower part of
the intermediate mass range the width can be
measured indirectly \citer{12,13}, by combining the branching
ratio for $H\to WW$ with the measurement
of the partial $WW$ width, accessible through the cross 
section for $W$ boson fusion:
$\Gamma_{tot} = \Gamma_{WW} / BR_{WW}$.
Thus, the total width of the Higgs particle can be
determined throughout the entire  mass
range when the experimental results from the LHC and
$e^+e^-$ colliders can be combined. 

\vspace*{3mm}
\noindent
{\it (c) \underline{Spin-parity}}

\vspace*{1mm}
\noindent
{\bf 1.$\,$} The zero-spin of the
Higgs particle can be determined from the
isotropic distribution of the decay
products \cite{Muhll,vdB}. Moreover, the parity can be
measured by observing the spin correlations
of the decay products. According to the
equivalence theorem, the azimuthal angles
of the decay planes in $H\to ZZ\to (\mu^+\mu^-) (\mu^+\mu^-)$
are asymptotically uncorrelated, $d\Gamma^+/d\phi_* \to 0$,
for a $0^+$ particle; this is to be contrasted with
$d\Gamma^-/d\phi_* \; \to 1-\frac{1}{4} \cos 2\phi_*$
for the distribution of the azimuthal angle
between the planes for the decay of a $0^{-}$
particle. The difference between the angular distributions
is a consequence of the different polarization
states of the vector bosons in the two
cases. While they approach states of
longitudinal polarization for scalar Higgs
decays, they are transversely polarized
for pseudoscalar particle decays. 

In the low mass range in which Higgs decays 
to $Z$-boson pairs are suppressed, the azimuthal angular
distribution between the accompanying quark jets 
in $WW$ fusion can be exploited to measure the parity
\cite{Zeppaz}. While the jets are nearly uncorrelated
for Higgs boson production in the Standard Model, the
correlation is of markedly different oscillatory
character for the production of a pseudoscalar particle
[i.e. $\mathcal{CP}$-odd], the jets pointing preferentially into 
directions perpendicular to each other.

\begin{figure}[hbt]

\vspace*{0.5cm}
\hspace*{0.0cm}
\epsfxsize=8cm \epsfbox{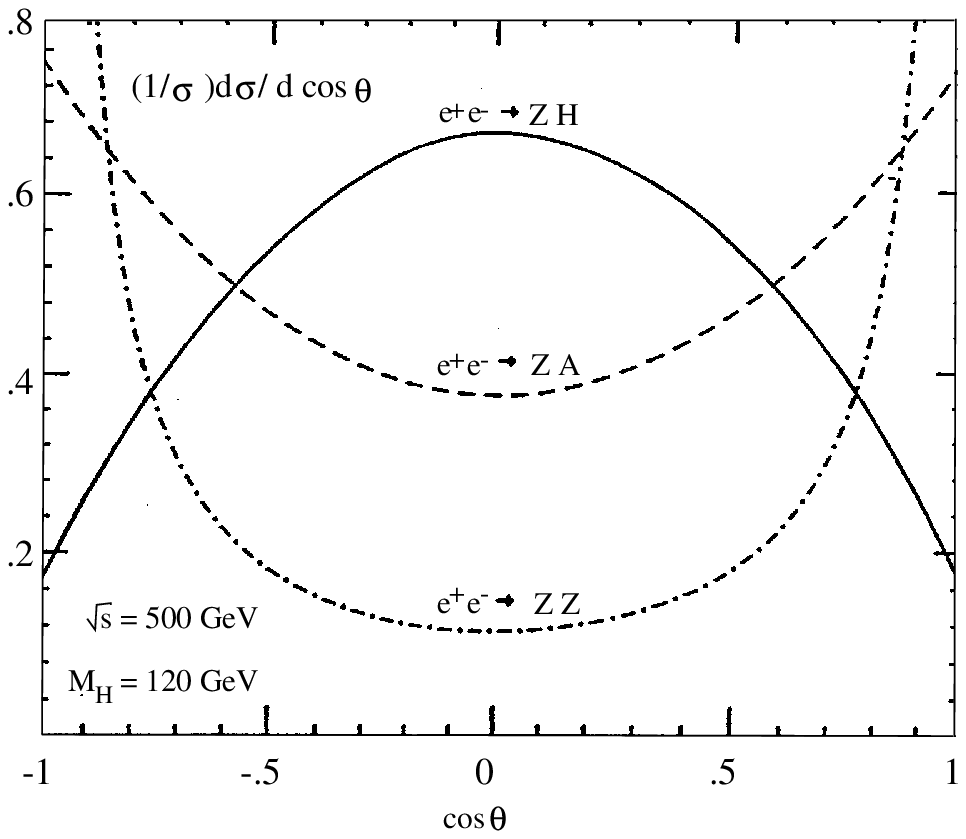}
\vspace*{0.0cm}

\vspace*{-6.3cm}
\hspace*{9.0cm}
\epsfig{figure=spinexp.eps,bbllx=0,bblly=0,bburx=560,bbury=539,width=6.5cm,clip=}
\vspace*{0.0cm}

\caption[]{\it \label{fg:spinpar} Left: Angular distribution of $Z/H$ bosons in
Higgs-strahlung, compared with the production of pseudoscalar
particles and the $ZZ$ background final states. Ref.~\cite{41}.
Right: Threshold excitation of Higgs-strahlung which discriminates spin=0 
from other assignments, Ref. \cite{MMM,exp}.}
\end{figure}

\vspace*{2mm}
\noindent
{\bf 2.$\,$} The angular distribution of the $Z/H$ bosons
in the Higgs-strahlung process is 
sensitive to the spin and parity of the
Higgs particle \cite{41}. Since the production
amplitude is given by ${\cal A}(0^+) \sim \vec{\epsilon}_{Z^*} \cdot
\vec{\epsilon}_Z$, the $Z$ boson is produced in a state of
longitudinal polarization at high
energies -- in accordance  with the equivalence
theorem. As a result, the angular distribution
\begin{equation}
\frac{d\sigma}{d\cos\theta} \sim \sin^2 \theta + \frac{8M_Z^2}{\lambda s}
\end{equation}
approaches the spin-zero $\sin^2\theta$
law asymptotically. This may be contrasted
with the distribution $\sim 1 + \cos^2\theta$
for negative parity states, which follows
from the transverse polarization amplitude
${\cal A}(0^-) \sim \vec{\epsilon}_{Z^*} \times \vec{\epsilon}_Z \cdot
\vec{k}_Z$. It is also characteristically different
from the distribution of the background
process $e^+e^- \to ZZ$, which, as a result of $t/u$-channel $e$ exchange,
is strongly peaked in the forward/backward
direction, Fig.~\ref{fg:spinpar}~left.

A different method to determine the spin of the Higgs boson is provided by 
scanning the onset of the excitation curve in Higgs-strahlung \cite{MMM} 
$e^+e^- \to ZH$. For Higgs spin $S_H = 0$ 
the excitation curve rises steeply at the 
threshold $\sim \sqrt{s - (M_H + M_Z)^2}$. This behavior is distinctly 
different from higher spin excitations which rise with a power $> 1$ of the 
threshold factor. An ambiguity for states with spin/parity 
$1^+$ and $2^+$ can be resolved by evaluating 
also the angular distribution of the Higgs and $Z$ boson in 
the Higgs-strahlung process. 
The experimental precision will be 
sufficient to discriminate the spin-0 assignment to the Higgs boson from 
other assignments as shown in Fig.~\ref{fg:spinpar}~right.

\vspace*{3mm}
\noindent
{\it (d) \underline{Higgs couplings}} 

\vspace*{1mm}
\noindent
Since fundamental particles acquire
mass by the interaction with the
Higgs field, the strength of the Higgs
couplings to fermions and gauge bosons
is set by the masses of the particles.
It will therefore be a crucial experimental
task to measure these couplings, which
are uniquely predicted by the very
nature of the Higgs mechanism.

\begin{figure}[hbt]
\hspace*{4.3cm}
\epsfig{figure=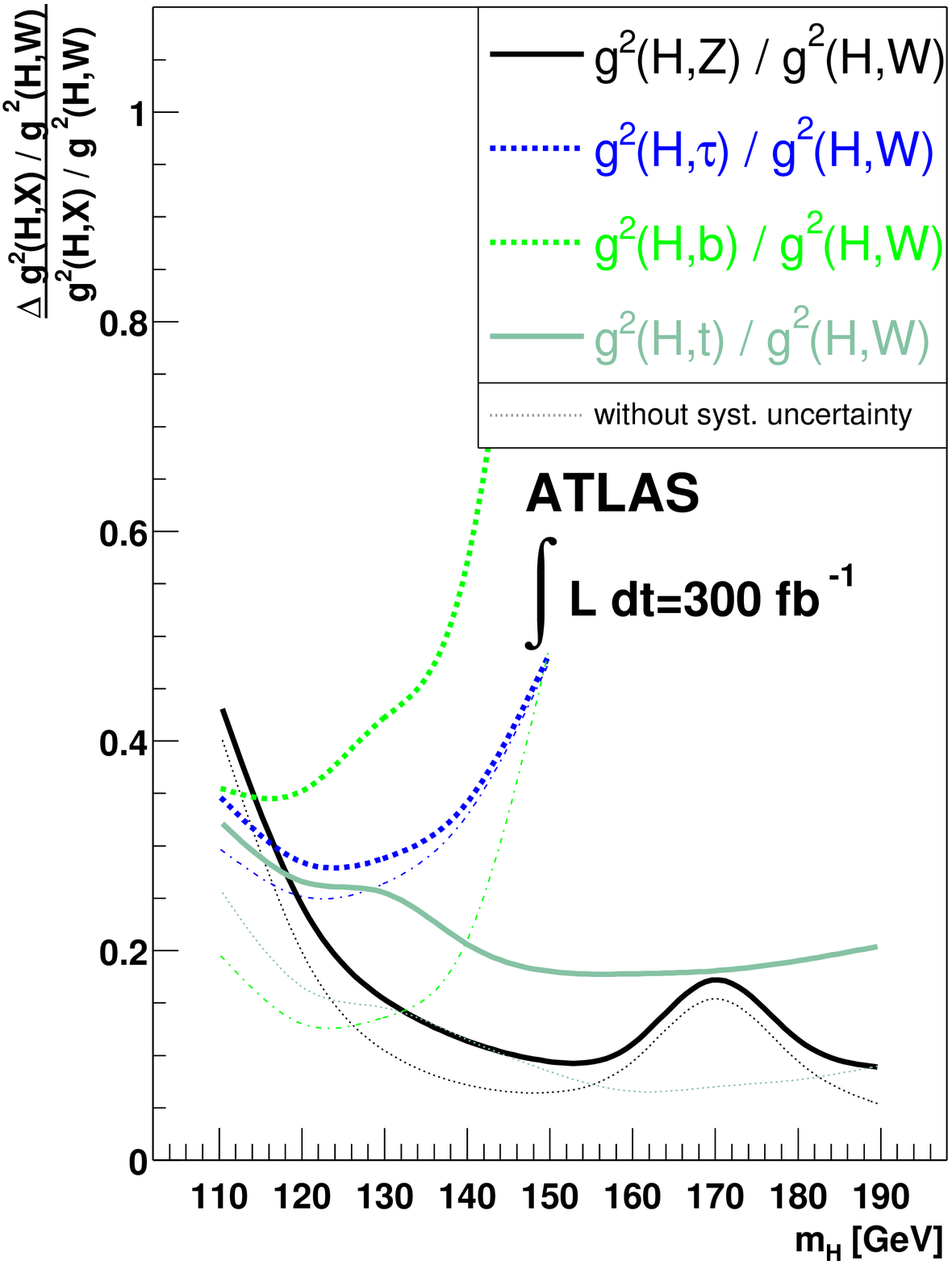,height=8.5cm,clip=}
\vspace*{-0.4cm}

\caption[]{\label{fg:Hrat} \it Expected accuracies in measurements 
of ratios of Higgs couplings at the LHC. Ref. \cite{Duhr}.}
\end{figure}

\vspace*{3mm}
\noindent
{\bf 1.$\,$} At the LHC only ratios 
of Higgs couplings can be determined 
in the intermediate mass range in a model-independent way. Since only
the product $\sigma_i \cdot BR_f \sim \Gamma_i \Gamma_f / \Gamma_{tot}$
can be measured, the partial widths $\Gamma_{i,f}$ may be rescaled and
the shifts balanced in $\Gamma_{tot}$ by unidentified decay channels.
The expected accuracy for the ratios of various decay channels 
is displayed in Fig.~\ref{fg:Hrat} \cite{Duhr}. 
Apparently
first insight can be obtained at the LHC into the fundamental rule
\begin{equation}
  g_i/g_j = m_i/m_j
\end{equation}
for various particles $i,j = W,Z,\tau, etc$.
\begin{figure}[hbt]
\begin{center}
\hspace*{-0.5cm}
\epsfig{figure=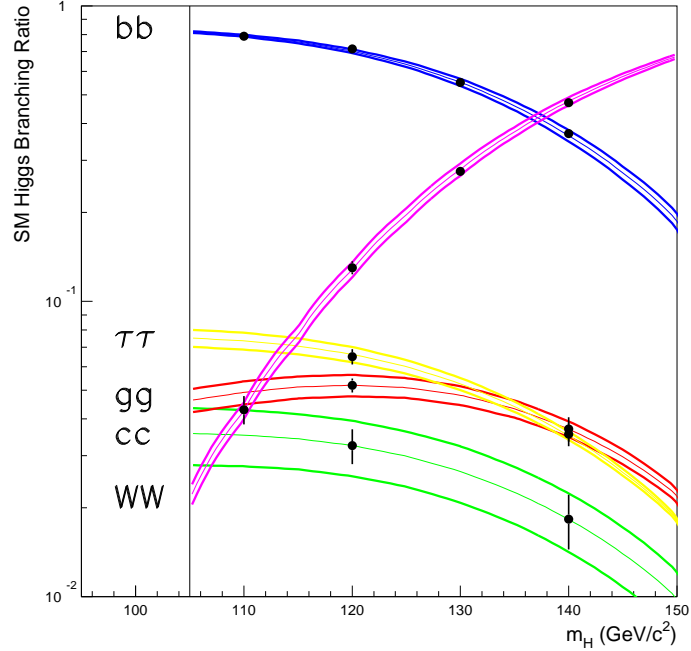,bbllx=0,bblly=18,bburx=521,bbury=517,width=9cm,clip=}
\end{center}
\vspace*{-0.5cm}

\caption[]{\it \label{fg:brmeas} The predicted SM Higgs boson branching 
ratios. Points with error bars show the expected experimental accuracy, while 
the lines show the estimated uncertainties on the SM predictions. 
Ref.~\cite{13}.}
\end{figure}

\vspace*{3mm}
\noindent 
{\bf 2.$\,$} At $e^+e^-$ colliders the absolute values 
of the Higgs couplings can be measured in a model-independent way
and with high presision.

The Higgs couplings to massive gauge
bosons can be determined from the
production cross sections in
Higgs-strahlung and $WW,ZZ$ fusion, with an
accuracy expected at the per-cent level.
For heavy enough Higgs bosons the decay
width can be exploited to determine the
couplings to the electroweak gauge bosons.
For Higgs couplings to fermions the
branching ratios $H\to b\bar b, c\bar c, \tau^+\tau^-$
can be used in the lower part of the
intermediate mass range, cf.~Fig.~\ref{fg:brmeas}; these observables
allow the direct measurement of the Higgs
Yukawa couplings.

A particularly interesting coupling
is the Higgs coupling to top quarks.
Since the top quark is by far the
heaviest fermion in the Standard Model,
irregularities in the standard picture
of electroweak symmetry breaking by
a fundamental Higgs field may become
apparent first in this coupling. Thus 
the $Htt$ Yukawa coupling may eventually provide
essential clues to the nature of the
mechanism breaking the electroweak
symmetries.

\begin{figure}[hbt]
\vspace*{-0.01cm}
\hspace*{4.0cm}
\epsfxsize=8.2cm \epsfbox{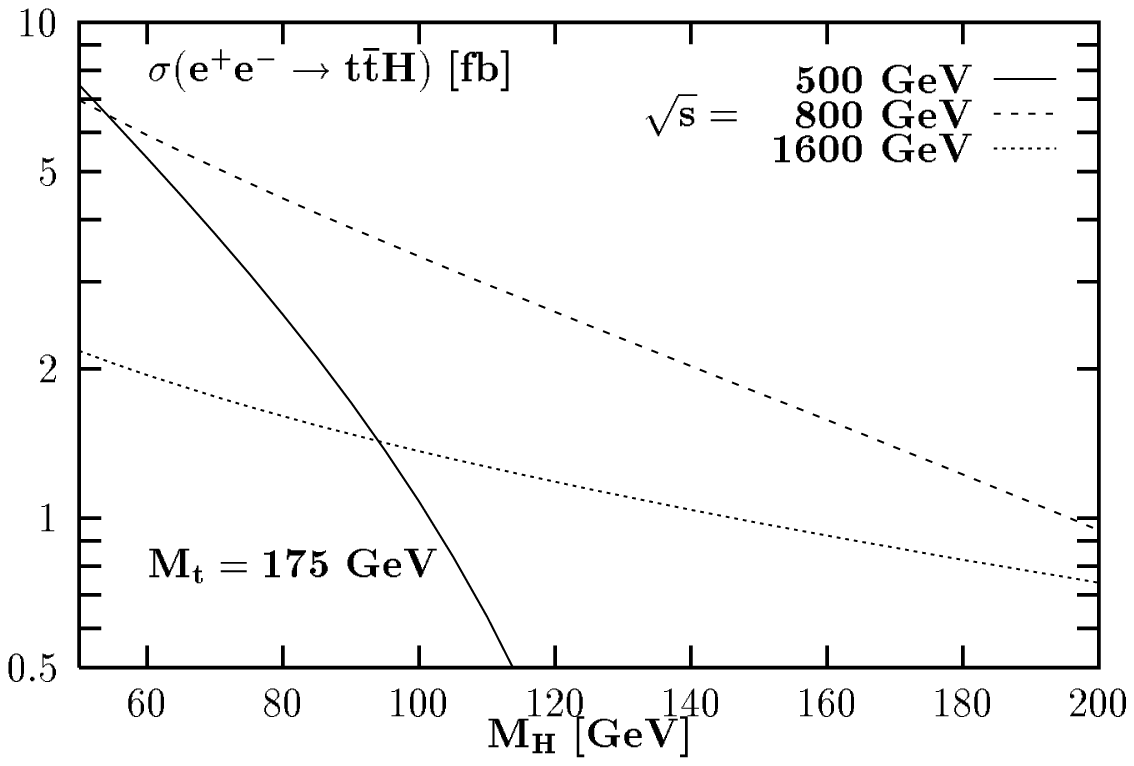}
\vspace*{-0.3cm}

\caption[]{\it \label{fg:eetth} The cross section for bremsstrahlung of SM
Higgs bosons off top quarks in the Yukawa
process $e^+e^-\to t\bar t H$.
[The amplitude for radiation off the
intermediate $Z$-boson line is small.] Ref. \cite{44}.}
\end{figure}
\begin{figure}[hbt]
\vspace*{-0.3cm}
\hspace*{2.7cm}
\epsfxsize=8.5cm \epsfbox{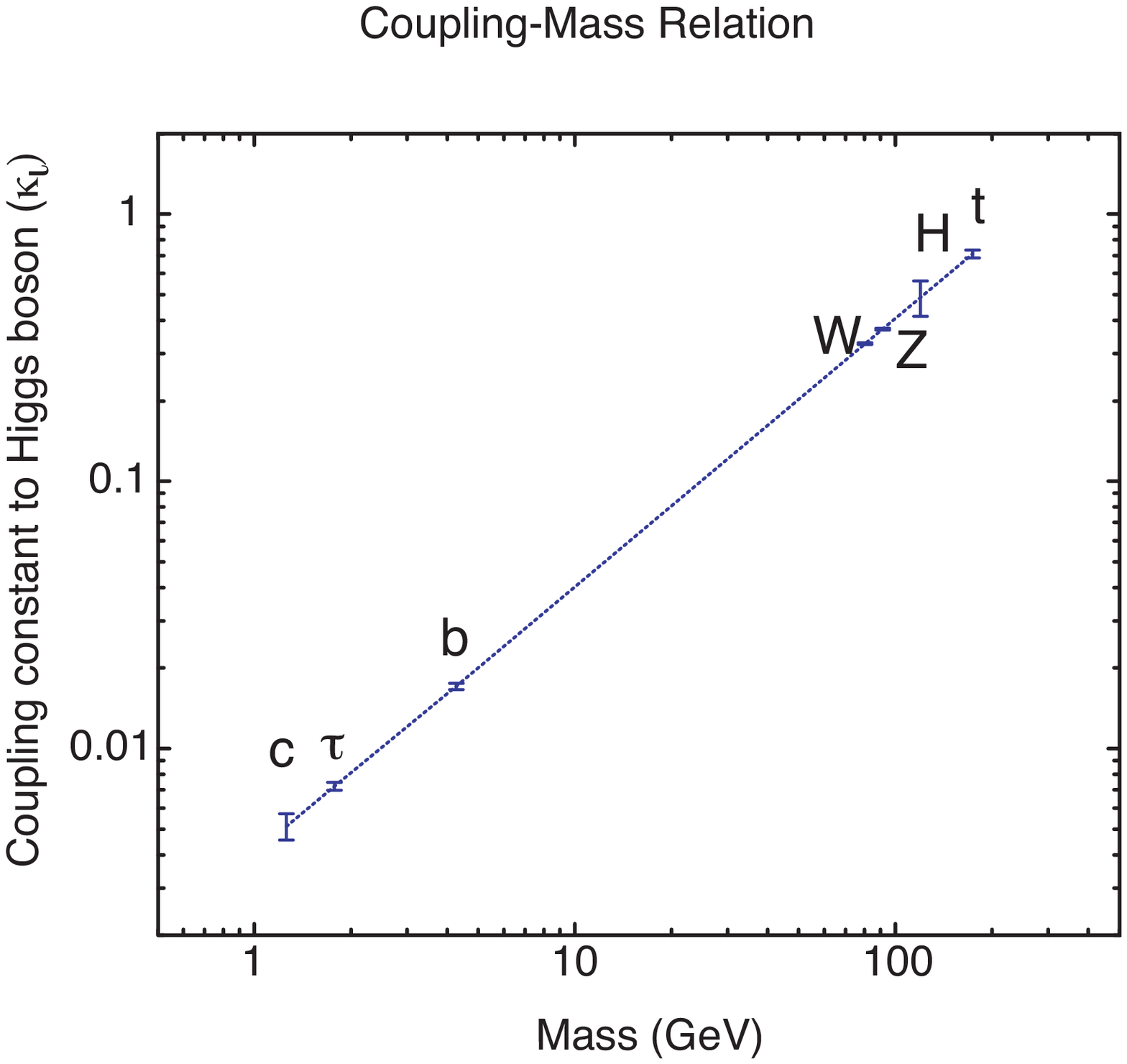}
\vspace*{-0.3cm}

\caption[]{\it \label{fg:Hcplg} The predicted normalized coupling
constants of the SM Higgs boson to the heavy SM fermions and vector bosons
as well as the Higgs boson itself as a function of the correponding SM
particle mass.  Points with error bars show the expected experimental
accuracy, while the lines show the linear rise of the couplings with the
corresponding masses.  Ref.~\cite{Hcplg}.}
\end{figure}
Top loops mediating the production
processes $gg\to H$ and $\gamma\gamma\to H$
(and the corresponding decay channels)
give rise to cross sections and partial
widths, which are proportional to the square of
the Higgs--top Yukawa coupling. This
Yukawa coupling can be measured directly,
for the lower part of the intermediate
mass range, in the bremsstrahlung
processes $pp\to t\bar t H$ and $e^+e^- \to t\bar t H$ \cite{44}.
The Higgs boson is radiated, in the first
process exclusively, in the second process
predominantly, from the heavy top quarks.
Even though these experiments are 
difficult because of  the small cross sections
[cf. Fig.~\ref{fg:eetth} for $e^+e^-$ collisions] 
and of the complex topology of
the $b\bar bb\bar bW^+W^-$ final state, this process 
is an important tool for exploring the
mechanism of electroweak symmetry breaking.
For large Higgs masses above the $t\bar t$
threshold, the decay channel $H\to t\bar t$
can be studied; in $e^+e^-$ collisions the cross section of
$e^+e^- \to t\bar t Z$ increases through the reaction
$e^+e^- \to ZH (\to t\bar t)$ \cite{45}. Higgs exchange between
$t\bar t$ quarks also affects the excitation curve
near the threshold at a level of a few per cent.

The expected accuracies for some of the couplings 
are collected in Table~\ref{tab:muehll}. The linear 
rise of the Higgs couplings with the masses of the 
particles is clearly visible in Fig.~\ref{fg:Hcplg}
in which the slope is uniquely predicted within 
the Standard Model. Mixing between the Higgs
boson and other scalar particles, like radions,
may change these couplings in a universal way. It 
is necessary therefore to scrutinize not only 
the mass dependence but also the absolute 
values of the Higgs couplings.
\begin{table}[h]
\begin{center}
{\small
\begin{tabular}{|lll|}
\hline
Coupling & $M_H=120$~GeV & 140~GeV \\
\hline
$g_{HWW}$ & $\pm 0.012$ & $\pm 0.020$ \\
$g_{HZZ}$ & $\pm 0.012$ & $\pm 0.013$ \\
\hline
$g_{Htt}$ & $\pm 0.030$ & $\pm 0.061$ \\
$g_{Hbb}$ & $\pm 0.022$ & $\pm 0.022$ \\
$g_{Hcc}$ & $\pm 0.037$ & $\pm 0.102$ \\
\hline
$g_{H\tau\tau}$ & $\pm 0.033$ & $\pm 0.048$ \\
\hline
\end{tabular}
}
\end{center}
\caption{Relative accuracy on the Higgs couplings assuming $\int\!{\cal L}=500$~fb$^{-1}$, $\sqrt{s}=500$~GeV ($\int\!{\cal L}=1$~ab$^{-1}$, $\sqrt{s}=800$~GeV for $g_{Htt}$).}
\label{tab:muehll}
\vspace*{-0.4cm}
\end{table}

\vspace*{3mm}
\noindent
{\it (e) \underline{Higgs self-couplings}}

\vspace*{1mm}
\noindent
The Higgs mechanism, based on a non-zero
value of the Higgs field in the vacuum, must
finally be made manifest experimentally by
reconstructing the interaction potential
that  generates the non-zero field in
the vacuum. This program can be carried out
by measuring the strength of the  trilinear
and quartic self-couplings of the Higgs
particles:
\begin{eqnarray}
g_{H^3} & = & 3 \sqrt{\sqrt{2} G_F} M_H^2 \\ \non \\
g_{H^4} & = & 3 \sqrt{2} G_F M_H^2 ~.
\end{eqnarray}

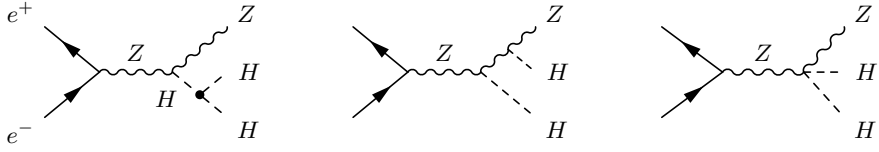
\begin{figure}[hbtp]
\begin{center}
\vspace*{0.25cm}
\begin{fmffile}{fd}
{\footnotesize
\unitlength1mm
\begin{fmfshrink}{0.7}
\begin{fmfgraph*}(24,12)
  \fmfstraight
  \fmfleftn{i}{3} \fmfrightn{o}{3}
  \fmf{fermion}{i1,v1,i3}
  \fmflabel{$e^-$}{i1} \fmflabel{$e^+$}{i3}
  \fmf{boson,lab=$Z$,lab.s=left,tens=3/2}{v1,v2}
  \fmf{boson}{v2,o3} \fmflabel{$Z$}{o3}
  \fmf{phantom}{v2,o1}
  \fmffreeze
  \fmf{dashes,lab=$H$,lab.s=right}{v2,v3} \fmf{dashes}{v3,o1}
  \fmffreeze
  \fmf{dashes}{v3,o2}
  \fmflabel{$H$}{o2} \fmflabel{$H$}{o1}
  \fmfdot{v3}
\end{fmfgraph*}
\hspace{15mm}
\begin{fmfgraph*}(24,12)
  \fmfstraight
  \fmfleftn{i}{3} \fmfrightn{o}{3}
  \fmf{fermion}{i1,v1,i3}
  \fmf{boson,lab=$Z$,lab.s=left,tens=3/2}{v1,v2}
  \fmf{dashes}{v2,o1} \fmflabel{$H$}{o1}
  \fmf{phantom}{v2,o3}
  \fmffreeze
  \fmf{boson}{v2,v3,o3} \fmflabel{$Z$}{o3}
  \fmffreeze
  \fmf{dashes}{v3,o2}
  \fmflabel{$H$}{o2} \fmflabel{$H$}{o1}
\end{fmfgraph*}
\hspace{15mm}
\begin{fmfgraph*}(24,12)
  \fmfstraight
  \fmfleftn{i}{3} \fmfrightn{o}{3}
  \fmf{fermion}{i1,v1,i3}
  \fmf{boson,lab=$Z$,lab.s=left,tens=3/2}{v1,v2}
  \fmf{dashes}{v2,o1} \fmflabel{$H$}{o1}
  \fmf{dashes}{v2,o2} \fmflabel{$H$}{o2}
  \fmf{boson}{v2,o3} \fmflabel{$Z$}{o3}
\end{fmfgraph*}
\end{fmfshrink}
\\[0.5cm]
}
\end{fmffile}
\end{center}
\caption[]{\label{fg:wwtohh} \it Generic diagrams contributing to the 
double Higgs-strahlung process $e^+e^- \to ZHH$.}
\end{figure}

This is a  difficult task since the
processes to be exploited are suppressed
by small couplings and phase space.
At the LHC it does not seem possible to determine the self-couplings,
leaving some hope though for the high-luminosity version VLHC.
However, this problem can be solved for the trilinear
coupling $g_{H^3}$
in the high-energy phase
of an $e^+e^-$ linear collider for sufficiently high
luminosities \cite{selfMMM}. A well-suited reaction
at $e^+e^-$ colliders for the measurement of the trilinear
coupling for Higgs masses in the theoretically
preferred mass range of ${\cal O}(100~\mbox{GeV})$, is the
double Higgs-strahlung process
\begin{equation}
e^+e^- \to ZH^{\ast} \to ZHH
\end{equation}
in which, among other mechanisms, the two-Higgs
final state is generated by the
exchange of a virtual Higgs particle so that this
process is sensitive to the trilinear $HHH$
coupling in the Higgs potential, Fig.~\ref{fg:wwtohh}. Since
the cross section is only a fraction of 1 fb,
an integrated luminosity of
$\sim$ 1 ab$^{-1}$ is needed to isolate the events at linear
colliders.
If combined with measurements of the double-Higgs fusion
process
\begin{equation}
 e^+e^- \to \bar{\nu} \nu H^\ast \to \bar{\nu} \nu H H
\end{equation}
experimental accuracies close to 12\% can be expected
\cite{Rexp}. The quartic coupling $H^4$
seems to be accessible only through loop
effects in the foreseeable future.\\

\vspace*{2mm}
\noindent
{\underline{To sum up:}} {\it The essential elements of the
Higgs mechanism can be established experimentally
at the LHC and TeV $e^+e^-$ linear colliders.}

\vspace*{4mm}
\section{Higgs Bosons in Supersymmetric Theories}

\vspace*{2mm}
\noindent
{\bf 1.$\,$} Arguments  deeply rooted in the Higgs sector 
play an eminent role in introducing
supersymmetry as a fundamental symmetry
of nature \cite{14}. This is the only symmetry
which correlates bosonic with fermionic
degrees of freedom, cf. Ref.\cite{SUSY}.

\vspace*{2mm}
{\bf (a)} The cancellation between bosonic and
fermionic contributions to the radiative
corrections of the light Higgs masses
in supersymmetric theories provides part of the
solution to the hierarchy problem in
the Standard Model. If the Standard Model
is embedded in a grand-unified theory,
the large gap between the high grand-unification
scale and the low scale of
electroweak symmetry breaking can be
stabilized in a natural way in boson--fermion symmetric
theories \cite{15,601}. Denoting
the bare Higgs mass by $M_{H,0}^2$,
the radiative corrections due to vector-boson
loops in the Standard Model by $\delta M_{H,V}^2$, 
and the contributions of supersymmetric
fermionic gaugino partners by $\delta M_{\tilde H,\tilde V}^2$, 
the physical Higgs mass is given by the sum
$M_H^2 = M_{H,0}^2 + \delta M_{H,V}^2 + \delta M_{\tilde H,\tilde V}^2$.
The vector-boson correction is quadratically
divergent, $\delta M_{H,V}^2 \sim \alpha [\Lambda^2 - M^2]$, 
so that, if present alone, for a cut-off scale $\Lambda \sim \Lambda_{GUT}$
extreme fine-tuning between  the
intrinsic bare mass and the radiative quantum fluctuations 
would be needed to
generate a Higgs mass of order $M_W$.
However, owing  to Pauli's principle, the
additional fermionic gaugino contributions
in supersymmetric theories are just
opposite in sign, $\delta M_{\tilde H,\tilde V}^2\sim -\alpha
[\Lambda^2-\tilde M^2]$,
so that the divergent terms cancel\footnote{Different statistics for bosons 
and fermions are sufficient for the cancellation of the divergencies. However, 
they are not necessary; symmetry relations among couplings, as realized in 
Little Higgs models, may also lead to cancellations individually between 
boson-boson or fermion-fermion amplitudes.}. Since
$\delta M_H^2\sim\alpha [\tilde M^2-M^2]$,
any fine-tuning is avoided for supersymmetric
particle masses $\tilde M \lessim {\cal O}$(1 TeV).
Thus, within this symmetry scheme the Higgs 
sector is stable in the low-energy range $M_H\sim M_W$
even in the context of high-energy GUT scales. This mechanism leads in a 
natural way to low-energy supersymmetry.

\vspace*{2mm}
{\bf (b)} The concept of supersymmetry is strongly
supported by the successful prediction
of the electroweak mixing angle in
the minimal version of this theory \cite{16}.
The extended particle spectrum of the theory 
drives the evolution of the
electroweak mixing angle from the
GUT value 3/8 down to $\sin^2\theta_W = 0.2336 \pm 0.0017$,
the error including unknown threshold
contributions at the low and the
high supersymmetric mass scales.
The prediction coincides with the
experimentally measured value $\sin^2\theta_W^{exp} = 0.23153 \pm 0.00016$
within the theoretical uncertainty
of less than 2 per-mille.

\vspace*{2mm}
{\bf (c)} Conceptually very interesting is 
the interpretation
of the Higgs mechanism in supersymmetric
theories as a quantum effect \cite{50A}. The
breaking of the electroweak symmetry $SU(2)_L \times U(1)_Y$
can be induced radiatively while
leaving the electromagnetic gauge
symmetry $U(1)_{EM}$
and the color gauge symmetry $SU(3)_C$
unbroken for top-quark masses
between 150 and 200 GeV. Starting
with a set of universal scalar
masses at the high GUT scale, one of the
squared mass parameters in the Higgs
sector evolves to negative values
at the low electroweak scale, while
the squared squark and slepton
masses remain positive.

\vspace*{2mm}
\noindent
{\bf 2.$\,$} The Higgs sector of supersymmetric
theories differs in several aspects
from the Standard Model \cite{17}. To preserve
supersymmetry and gauge invariance,
at least two iso-doublet fields must
be introduced, leaving us with a
spectrum of five or more physical
Higgs particles. In the minimal
supersymmetric extension of the
Standard Model the Higgs
self-interactions are generated
by the scalar-gauge field action, so that the
quartic couplings are related to
the gauge couplings in this scenario. 
After including radiative corrections
this leads to strong bounds \cite{19} of less than
about 140 GeV for the mass of 
the lightest Higgs boson. If the
system is assumed to remain
weakly interacting up to scales
of the order of the GUT or Planck
scale, the mass remains small, 
for reasons quite analogous to those described 
in the Standard Model,
even in more complex supersymmetric
theories involving additional Higgs fields and  
Yukawa interactions. The masses of the heavy 
Higgs bosons are expected to be
of the scale of electroweak symmetry
breaking up to order 1 TeV.

\vspace*{4mm}
\subsection{The Higgs Sector of the MSSM}

\vspace*{2mm}
\noindent
{\bf 1.$\,$} The particle spectrum of the MSSM \cite{14} consists
of leptons, quarks and their scalar 
supersymmetric partners, and gauge
particles, Higgs particles and their
spin-1/2 partners. The matter and force fields are coupled
in supersymmetric and gauge-invariant
actions:
\begin{equation}
\begin{array}{lrcll}
S = S_V + S_\phi + S_W: \hspace*{1cm}
& S_V    & = & \frac{1}{4} \int d^6 z \hat W_\alpha \hat W_\alpha
\hspace*{1cm} & \mbox{gauge action} , \\ \\
& S_\phi & = & \int d^8 z \hat \phi^* e^{gV} \hat \phi
& \mbox{matter action} , \\ \\
& S_W    & = & \int d^6 z W[\hat \phi]
& \mbox{superpotential} .
\end{array}
\end{equation}
Decomposing the superfields into fermionic
and bosonic components, and carrying out
the integration over the Grassmann
variables in $z\to x$,
the following Lagrangians are derived, which  
describe the interactions of the
gauge, matter and Higgs fields:
\begin{eqnarray*}
{\cal L}_V & = & -\frac{1}{4}F_{\mu\nu}F_{\mu\nu}+\ldots+\frac{1}{2}D^2 ~, \\ \\
{\cal L}_\phi & = & D_\mu \phi^* D_\mu \phi +\ldots+\frac{g}{2} D|\phi|^2  ~, \\ \\
{\cal L}_W & = & - \left| \frac{\partial W}{\partial \phi} \right|^2 ~. 
\end{eqnarray*}
The $D$ field is an auxiliary field that 
does not propagate in space-time and it
 can be eliminated by applying  
the equations of motion: $D=-\frac{g}{2} |\phi|^2$.
Reinserted into the Lagrangian, the
quartic self-coupling of the scalar Higgs
fields is generated,
\begin{equation}
{\cal L} [\phi^4] = -\frac{g^2}{8} |\phi^2|^2 ~,
\end{equation}
in theories, like MSSM, in which the superpotential does not 
generate a quartic term.
Thus, the quartic coupling of the Higgs 
fields is given, in the minimal
supersymmetric theory, by the square
of the gauge coupling. Unlike the Standard
Model case, the quartic coupling is not a free parameter. Moreover,
this coupling is weak.

\vspace*{2mm}
\noindent
{\bf 2.$\,$} Two independent Higgs doublet fields $H_1$ and $H_2$
must be introduced into the superpotential: 
\begin{equation}
W = -\mu \epsilon_{ij} \hat H_1^i \hat H_2^j + \epsilon_{ij} [f_1 \hat H_1^i
\hat L^j \hat R + f_2 \hat H_1^i \hat Q^j \hat D -
f_2' \hat H_2^i \hat Q^j \hat U]
\end{equation}
to provide the down-type particles ($H_1$)
and the up-type particles ($H_2$) with mass.
Unlike the Standard Model, the second Higgs
field cannot be identified with the
charge conjugate of the first Higgs field
since $W$ must be analytic to preserve
supersymmetry. In addition, the Higgsino
fields associated with a single Higgs
field would generate triangle anomalies;
they cancel if the two conjugate doublets
are added up, and the gauge
invariance of the classical interactions is not
destroyed at the quantum level. 
Integrating the superpotential over
the Grassmann coordinates generates
the supersymmetric Higgs self-energy
$V_0 = |\mu|^2 (|H_1|^2 + |H_2|^2)$.
The breaking of supersymmetry can be
incorporated in the Higgs sector by
introducing bilinear mass terms $\mu_{ij} H_i H_j$.
Added to the supersymmetric self-energy part $H^2$
and the quartic part $H^4$
generated by the gauge action, they
lead to the following Higgs potential
\begin{eqnarray}
V & = & m_1^2 H_1^{*i} H_1^i + m_2^2 H_2^{*i} H_2^i - m_{12}^2 (\epsilon_{ij}
H_1^i H_2^j + hc) \non \\ \non \\
& & + \frac{1}{8} (g^2 + g'^2) [H_1^{*i} H_1^i -
H_2^{*i} H_2^i]^2 + \frac{1}{2} g^2 \, |H_1^{*i} H_2^{i}|^2 ~. 
\end{eqnarray}
The Higgs potential includes three
bilinear mass terms, while the strength
of the quartic couplings is set by the
$SU(2)_L$ and $U(1)_Y$
gauge couplings squared. The three mass
terms are free parameters.

The potential develops a stable minimum
for $H_1 \to [v_1,0]$ and $H_2\to [0,v_2]$
if the following conditions are met:
$ m_1^2 +  m_2^2 >  2 | m^2_{12} |  \;\;{\rm and}\;\;
m_1^2    m_2^2  <  | m^2_{12} |^2 ~.$

\vspace*{2mm}
\noindent
Expanding the fields about the ground-state 
values $v_1$ and $v_2$,
\begin{equation}
\begin{array}{rclcl}
H_1^1 & = & v_1 &\!\! +\!\! & [H^0 \cos \alpha - h^0 \sin \alpha + i A^0 \sin \beta - i G^0
\cos \beta ]/\sqrt{2}                                   \\ \\
H_1^2 & = & &\!\!\!\! & H^- \sin \beta - G^- \cos \beta
\end{array}
\end{equation}
and
\begin{equation}
\begin{array}{rclcl}
H_2^1 & = & &\!\!\!\! & H^+ \cos \beta + G^+ \sin \beta   \\ \\
H_2^2 & = & v_2 &\!\!+\!\! & [H^0 \sin \alpha + h^0 \cos \alpha + i A^0 \cos \beta + i G^0
\sin \beta ]/\sqrt{2} \,,
\end{array}
\end{equation}
the mass eigenstates are given by the
neutral states $h^0,H^0$ and $A^0$,
which are even and odd under ${\cal CP}$
transformations, and by the charged states $H^\pm$;
the $G$ states correspond to the Goldstone
modes, which are absorbed by the gauge
fields to build up the longitudinal
components. After introducing the three
parameters
\begin{eqnarray}
M_Z^2 & = & \frac{1}{2} (g^2 + g'^2) (v_1^2 + v_2^2) \non \\ \non \\
M_A^2 & = & m_{12}^2 \,\frac{v_1^2 + v_2^2}{v_1v_2} \non \\ \non \\
\tgb  & = & \frac{v_2}{v_1} ~, 
\end{eqnarray}
the mass matrix can be decomposed into
three $2\times 2$ blocks, which are easy to
diagonalize: \\[-0.0mm]
\begin{displaymath}
\begin{array}{ll}
\mbox{\bf pseudoscalar mass:} & M_A^2 \\ \\
\mbox{\bf charged mass:} & M_\pm^2 = M_A^2+M_W^2 \\ \\
\mbox{\bf scalar mass:} &
M_{h,H}^2 = \frac{1}{2} \left[ M_A^2 + M_Z^2 \mp \sqrt{(M_A^2+M_Z^2)^2
- 4M_A^2M_Z^2 \cos^2 2\beta} \right] \\ \\
& \displaystyle \tg 2\alpha = \tg 2\beta \,\frac{M_A^2 + M_Z^2}{M_A^2 - M_Z^2}
\hspace*{0.5cm} \mbox{with} \hspace*{0.5cm} -\frac{\pi}{2} < \alpha < 0
\nonumber \;.
\end{array} 
\end{displaymath}

>From the mass formulae, two important
inequalities can readily be derived,
\begin{eqnarray}
M_h & \leq & M_Z, M_A \leq M_H \\ \nonumber \\
M_W & \leq & M_{H^\pm} ~, 
\end{eqnarray}
which, by construction, are valid in
the tree approximation. As a result,
the lightest of the scalar Higgs masses
is predicted to be bounded by the $Z$ mass,
{\it modulo} radiative corrections. These bounds
follow from the fact that the quartic
coupling of the Higgs fields is determined
in the MSSM by the size of the gauge
couplings squared. 

\vspace*{2mm}
\noindent
{\bf 3.$\,$} \underline{\it SUSY Radiative Corrections}:  
The tree-level relations between the
Higgs masses are strongly modified
by radiative corrections that involve
the supersymmetric particle spectrum
of the top sector \cite{50B}; cf.~Ref.~\cite{DJ,mhplot} for recent
summaries. These effects
are proportional to the fourth power
of the top mass and to the logarithm
of the stop mass. Their origin are
incomplete cancellations between virtual
top and stop loops, reflecting the
breaking of supersymmetry. Moreover,
the mass relations are affected by 
the potentially large mixing between
$\tilde t_L$ and $\tilde t_R$
due to the top Yukawa coupling.

\vspace*{2mm}
To leading order in $M_t^4$
the radiative corrections can be
summarized in the parameter
\begin{equation}
\epsilon = \frac{3G_F}{\sqrt{2}\pi^2}\frac{M_t^4}{\sin^2\beta}\log
\frac{\langle M^2_{\tilde{t}} \rangle}{M_t^2} 
\end{equation}
with $\langle M^2_{\tilde{t}} \rangle = M_{\tilde{t_1}} M_{\tilde{t_2}} \, .$
In this approximation the light Higgs mass $M_h$ 
can be expressed by $M_A$ and $\tgb$
in the following compact form:
\begin{eqnarray*}
M^2_h & = & \frac{1}{2} \left[ M_A^2 + M_Z^2 + \epsilon \right.
\non \\
& & \left. - \sqrt{(M_A^2+M_Z^2+\epsilon)^2
-4 M_A^2M_Z^2 \cos^2 2\beta
-4\epsilon (M_A^2 \sin^2\beta + M_Z^2 \cos^2\beta)} \right] \;.
\end{eqnarray*}
The heavy Higgs masses $M_H$ and $M_{H^\pm}$
follow from the sum rules
\begin{eqnarray*}
M_H^2 & = & M_A^2 + M_Z^2 - M_h^2 + \epsilon \non \\
M_{H^\pm}^2 & = & M_A^2 + M_W^2 ~.
\end{eqnarray*}
Finally, the mixing parameter $\alpha$,  
which diagonalizes the ${\cal CP}$-even mass
matrix, is given by the radiatively
improved relation:
\begin{equation}
\tg 2 \alpha = \tg 2\beta \frac{M_A^2 + M_Z^2}{M_A^2 - M_Z^2 +
\epsilon/\cos 2\beta} ~. 
\label{eq:mssmalpha}
\end{equation}

\begin{figure}[ht]
\begin{center}
\hspace*{-1.6cm}
\epsfig{figure=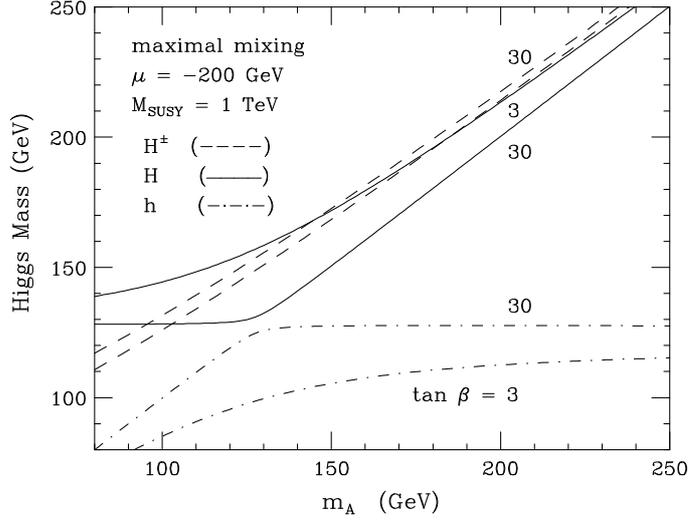,bbllx=80,bblly=205,bburx=556,bbury=570,width=9cm,clip=}
\end{center}
\vspace*{-0.6cm}
\caption[]{\label{kdfig} \it The $\mathcal{CP}$-even and charged MSSM Higgs boson 
masses as a function of $m_A$ for $\tan\beta=3$ and $30$, including radiative
corrections. Ref.~\cite{66a}.}
\end{figure}

For large $A$ mass, the
masses of the heavy Higgs particles
coincide approximately, $M_A\simeq M_H \simeq M_{H^\pm}$,
while the light Higgs mass approaches
a small asymptotic value. The spectrum for large
values of $\tgb$ is quite  regular: for small $M_A$ one finds
$\{ M_h\simeq M_A; M_H  \simeq \mbox{const} \}$ \cite{intense}; 
for large $M_A$ the opposite relationship
$\{ M_h\simeq \mbox{const}, M_H \simeq M_{H^\pm}\simeq M_A \}$,
cf.~Fig.~\ref{kdfig} which includes the radiative corrections.\\
\begin{figure}[hbt]
\begin{center}
\vspace*{7mm}
\vspace*{-2.0cm}
\hspace*{-0.3cm}
\epsfig{figure=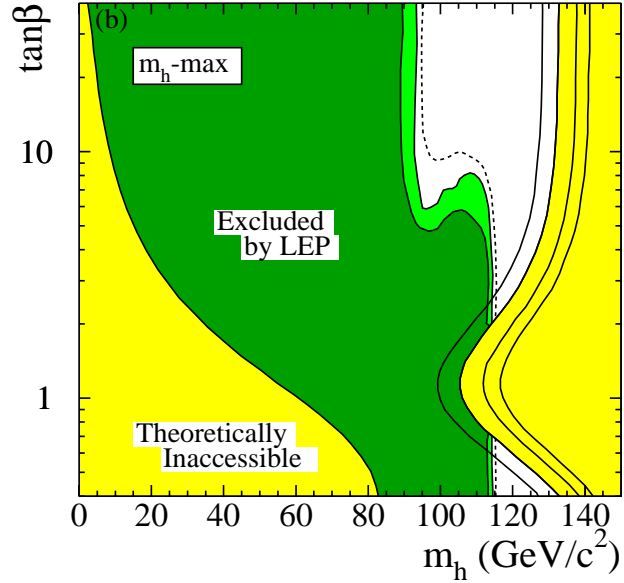,bbllx=48,bblly=42,bburx=567,bbury=557,width=9cm,clip=}
\end{center}
\vspace*{-0.0cm}
\caption[]{\label{fg:mssmhiggs} \it Upper bounds on the light Higgs boson mass
as a function of $\tg\beta$ for varying top mass, and the region excluded by the negative
searches at the LEP experiments. Ref. \cite{pdgplot}.}
\end{figure}

While the non-leading effects of
mixing on the Higgs mass relations are
quite involved, the impact on the upper
bound of the light Higgs mass $M_h$ 
can be summarized in a simple form:
\begin{equation}
M_h^2 \leq M_Z^2 +  \delta M_t^2 + \delta M_X^2 ~.
\end{equation}
The leading top contribution is related to
the parameter $\epsilon$ while
the second contribution
depends on the mixing parameter in the scalar top sector,
\begin{equation}
M_t X_t = M_t \left[A_t - \mu~\ctgb \right] ~, 
\end{equation}
which couples left- and right-chirality
states in the stop mass matrix:
\begin{equation}
M_h^2 \leq M_Z^2 + \frac{3G_F M_t^4}{2\sqrt{2}\pi^2} \, \left[  
                   \log \frac{\langle M_{\tilde{t}}^2 \rangle}{M_t^2} +  
                   \frac{X_t^2}{ \langle M_{\tilde{t}}^2 \rangle }
                   \left(1 - \frac{X_t^2}{\langle 12 M_{\tilde{t}}^2 \rangle}  \right) \right] \;.
\end{equation} 

Subdominant contributions can essentially
be reduced to higher-order QCD effects.
They can effectively be incorporated by
interpreting the top mass parameter
$M_t \to M_t(\mu_t)$ as the $\overline{\rm MS}$
top mass evaluated at the geometric mean
between top and stop masses, $\mu_t^2 = M_t M_{\tilde t}$.

\vspace*{2mm}
Upper bounds on the light Higgs mass are
shown in Fig.~\ref{fg:mssmhiggs} as a function of $\tg \beta$. The curves are 
the results of calculations taking into account the mixing effects. 
It turns out that the
general upper bound for maximal mixing is given by $M_h\lessim 140$ GeV, 
including large values of $\tgb$.  
Thus, the light Higgs sector could not entirely be
covered by the LEP2 experiments due to the increase of the mass limit with the 
top mass. 


\vspace*{4mm}
\subsection{SUSY Higgs Couplings to SM Particles}

\vspace*{2mm}
\noindent
The size of MSSM Higgs couplings to quarks,
leptons and gauge bosons is similar to
the Standard Model, yet modified by the
mixing angles $\alpha$ and $\beta$.
Normalized to the SM values, they are
listed in Table \ref{tb:hcoup}. The pseudoscalar Higgs
boson $A$ does not couple to gauge bosons
at the tree level, but the coupling, 
compatible with  ${\cal CP}$ symmetry, can be
generated by higher-order loops.
The charged Higgs bosons couple to up   
and down fermions with the left- and
right-chiral amplitudes $g_\pm = -
\left[ g_t (1 \mp \gamma_5) + g_b (1 \pm \gamma_5) \right]/\sqrt{2}$ where
$g_{t,b} = (\sqrt{2} G_F)^{1/2} m_{t,b}$.
\begin{table}[hbt]
\renewcommand{\arraystretch}{1.5}
\begin{center}
\begin{tabular}{|lc||ccc|} \hline
\multicolumn{2}{|c||}{$\Phi$} & $g^\Phi_u$ & $g^\Phi_d$ &  $g^\Phi_V$ \\
\hline \hline
SM~ & $H$ & 1 & 1 & 1 \\ \hline
MSSM~ & $h$ & $\cos\alpha/\sin\beta$ & $-\sin\alpha/\cos\beta$ &
$\sin(\beta-\alpha)$ \\
& $H$ & $\sin\alpha/\sin\beta$ & $\cos\alpha/\cos\beta$ &
$\cos(\beta-\alpha)$ \\
& $A$ & $ 1/\tg\beta$ & $\tg\beta$ & 0 \\ \hline
\end{tabular}
\renewcommand{\arraystretch}{1.2}
\caption[]{\label{tb:hcoup}
\it Higgs couplings in the MSSM to fermions and gauge bosons [$V=W,Z$]
relative to SM couplings.}
\end{center}
\end{table}

The modified couplings incorporate the
renormalization due to SUSY radiative
corrections, to leading order in $M_t$, 
if the mixing angle $\alpha$ is related to
$\beta$ and $M_A$
as given in the corrected formula Eq.~(\ref{eq:mssmalpha}).
For large $M_A$, in practice $M_A\gsim 200$ GeV, 
the couplings of the light Higgs boson
$h$ to the fermions and gauge bosons
approach the SM values asymptotically.
This is the essence of the \underline{decoupling theorem} in the 
Higgs sector \cite{66AA}: 
Particles with large masses
must decouple from the light-particle
system as a consequence of the
quantum-mechanical uncertainty principle.
In the same limit, the heavy Higgs boson $H$ 
decouples from vector bosons, and the coupling
to up-type fermions is suppressed by $1/{\tan\beta}$
while the coupling to down-type fermions is enhanced by
$\tan\beta$. Thus the couplings of the two degenerate
heavy Higgs bosons $A,H$ are isomorphic in the
decoupling limit.

\vspace*{4mm}
\subsection{Decays of Higgs Particles}

\vspace*{2mm}
\noindent
The lightest \underline{\it neutral Higgs boson} $h$ 
will decay mainly into fermion pairs
since the mass is smaller than $\sim 140$
GeV, Fig.~\ref{fg:mssmbr} (cf. \cite{613A} for a
comprehensive summary). This is, in general,
also the dominant decay mode of the
pseudoscalar boson $A$. For values of $\tgb$
larger than unity and for masses less than
$\sim 140$ GeV, the main decay modes of the neutral
Higgs bosons are decays into $b\bar b$ and $\tau^+\tau^-$
pairs; the branching ratios are of order $\sim 90\%$ and $8\%$,
respectively. The decays into $c\bar c$
pairs and gluons are suppressed, especially
for large $\tgb$.  
For large masses, the top decay channels
$H,A \to t\bar t$ open up; yet for large $\tgb$
this mode remains suppressed and the
neutral Higgs bosons decay almost 
exclusively into $b\bar b$ and $\tau^+\tau^-$
pairs. In contrast to the pseudoscalar
Higgs boson $A$, the
heavy ${\cal CP}$-even Higgs boson $H$
can in principle decay into weak gauge
bosons, $H\to WW,ZZ$, if the mass is large enough.
However, since the partial widths are proportional
to $\cos^2(\beta - \alpha)$,
they are strongly suppressed in general,
and the gold-plated $ZZ$ signal of the
heavy Higgs boson in the Standard Model
is lost in the supersymmetric extension.
As a result, the total widths of the Higgs
bosons are much smaller in supersymmetric
theories than in the Standard Model.
\begin{figure}[hbtp]

\vspace*{-2.5cm}
\hspace*{-4.5cm}
\begin{turn}{-90}%
\epsfxsize=16cm \epsfbox{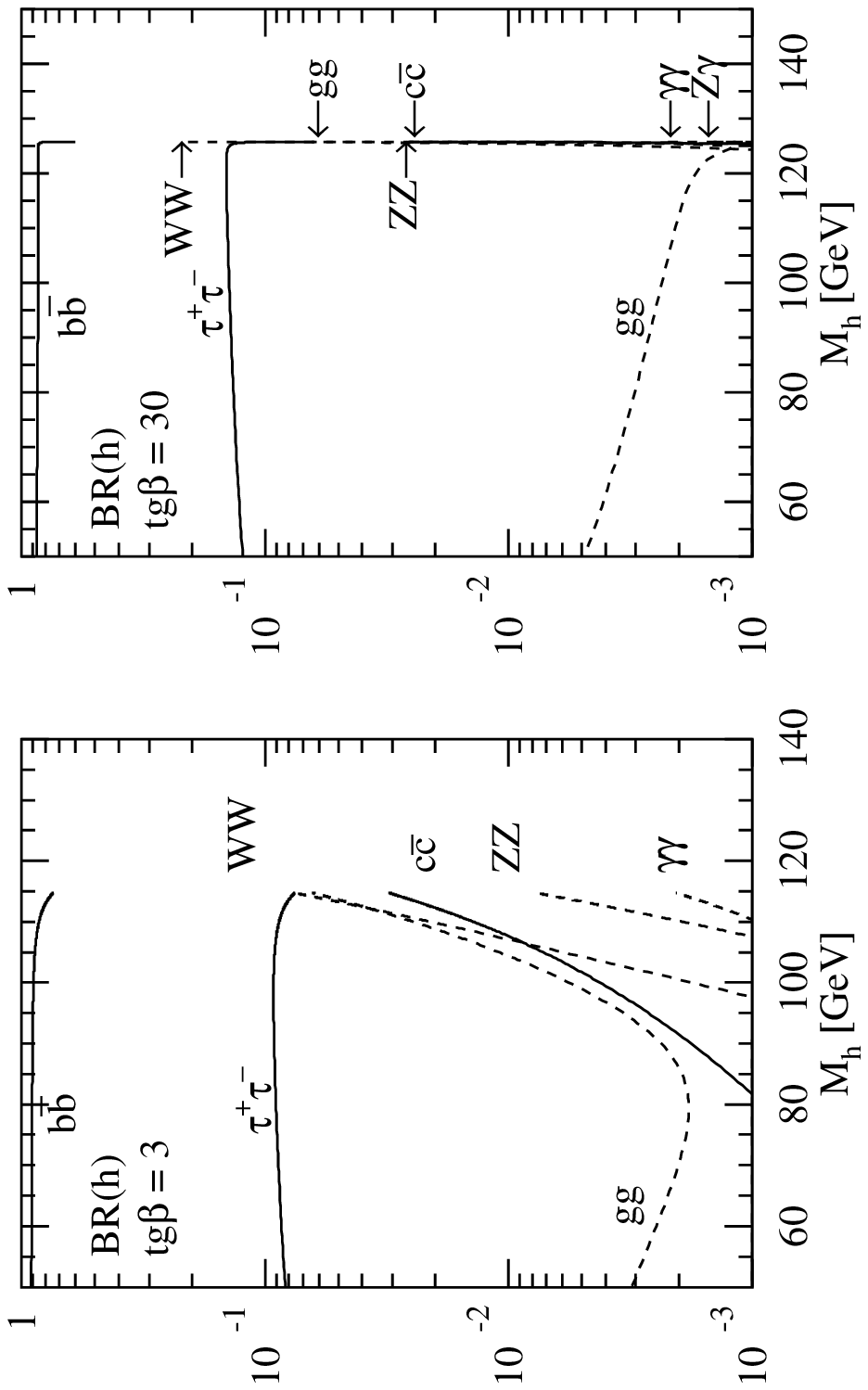}
\end{turn}
\vspace*{-4.2cm}

\centerline{\bf Fig.~\ref{fg:mssmbr}a}

\vspace*{-2.5cm}
\hspace*{-4.5cm}
\begin{turn}{-90}%
\epsfxsize=16cm \epsfbox{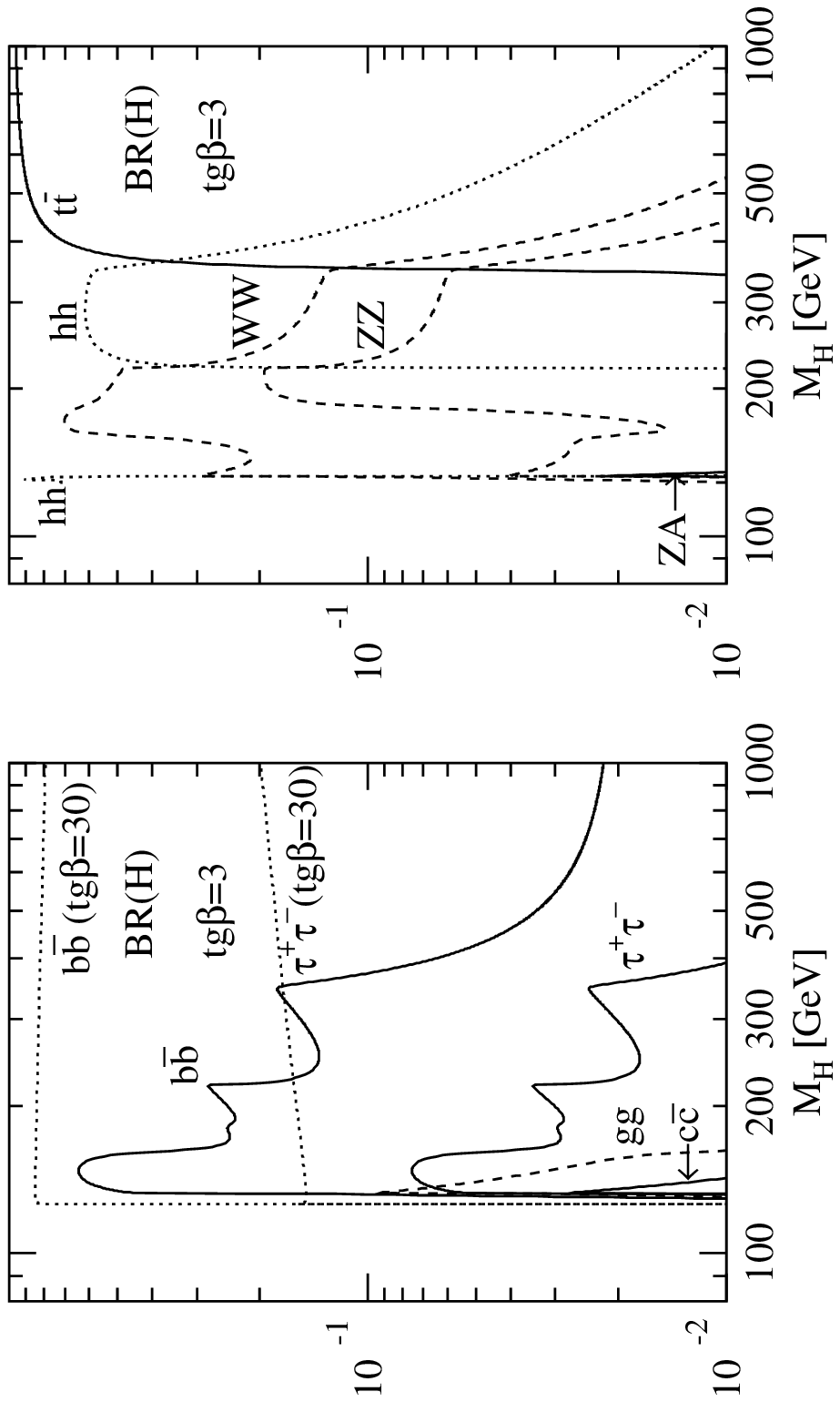}
\end{turn}
\vspace*{-4.2cm}

\centerline{\bf Fig.~\ref{fg:mssmbr}b}

\caption[]{\label{fg:mssmbr} \it Branching ratios of the MSSM Higgs bosons $h,
 H, A, H^\pm$ for non-SUSY decay modes as a function of the
masses for two values of $\tgb=3, 30$ and vanishing mixing. The common squark
mass has been chosen as $M_S=1$ TeV.}
\end{figure}
\addtocounter{figure}{-1}
\begin{figure}[hbtp]

\vspace*{-2.5cm}
\hspace*{-4.5cm}
\begin{turn}{-90}%
\epsfxsize=16cm \epsfbox{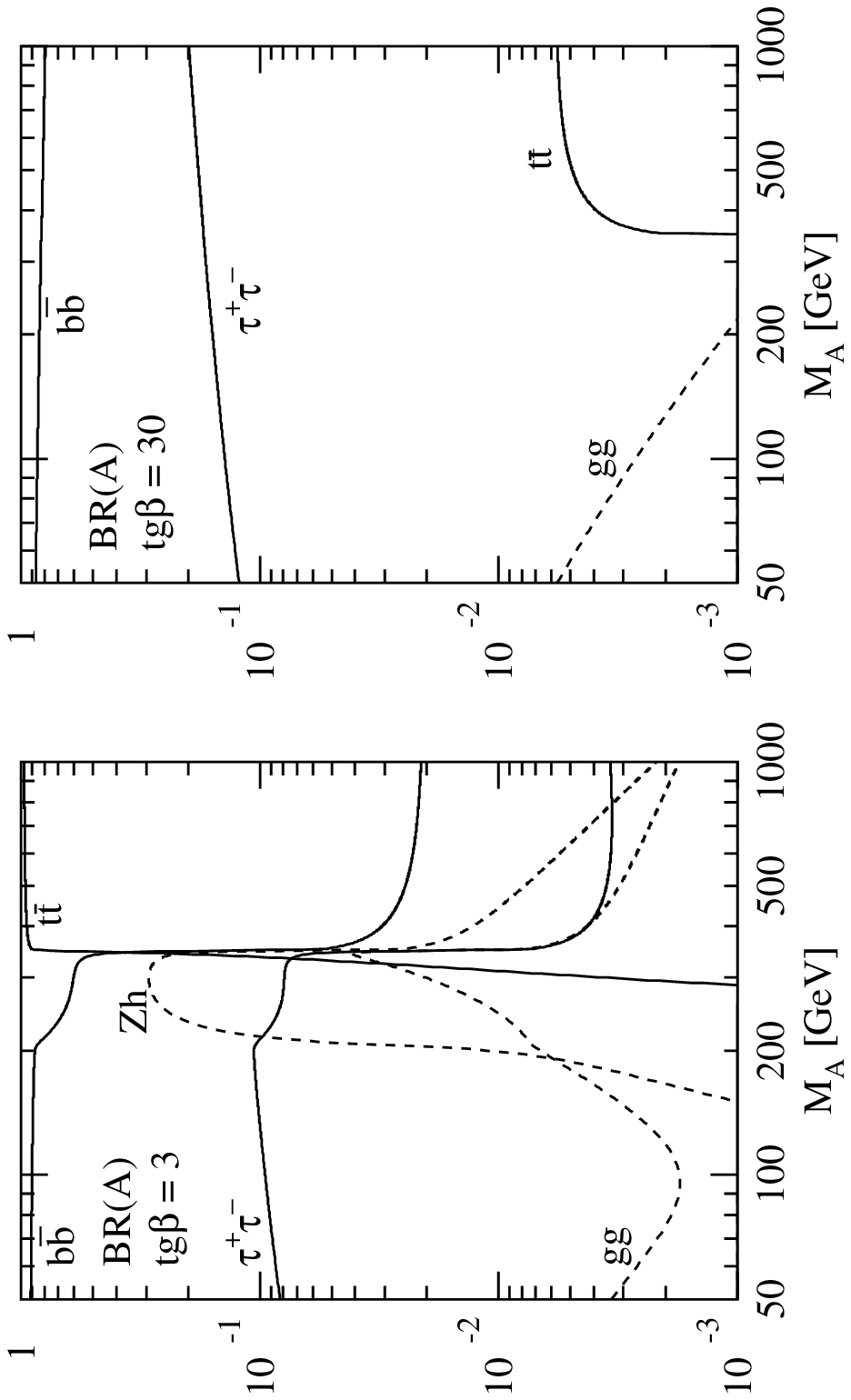}
\end{turn}
\vspace*{-4.2cm}

\centerline{\bf Fig.~\ref{fg:mssmbr}c}

\vspace*{-2.5cm}
\hspace*{-4.5cm}
\begin{turn}{-90}%
\epsfxsize=16cm \epsfbox{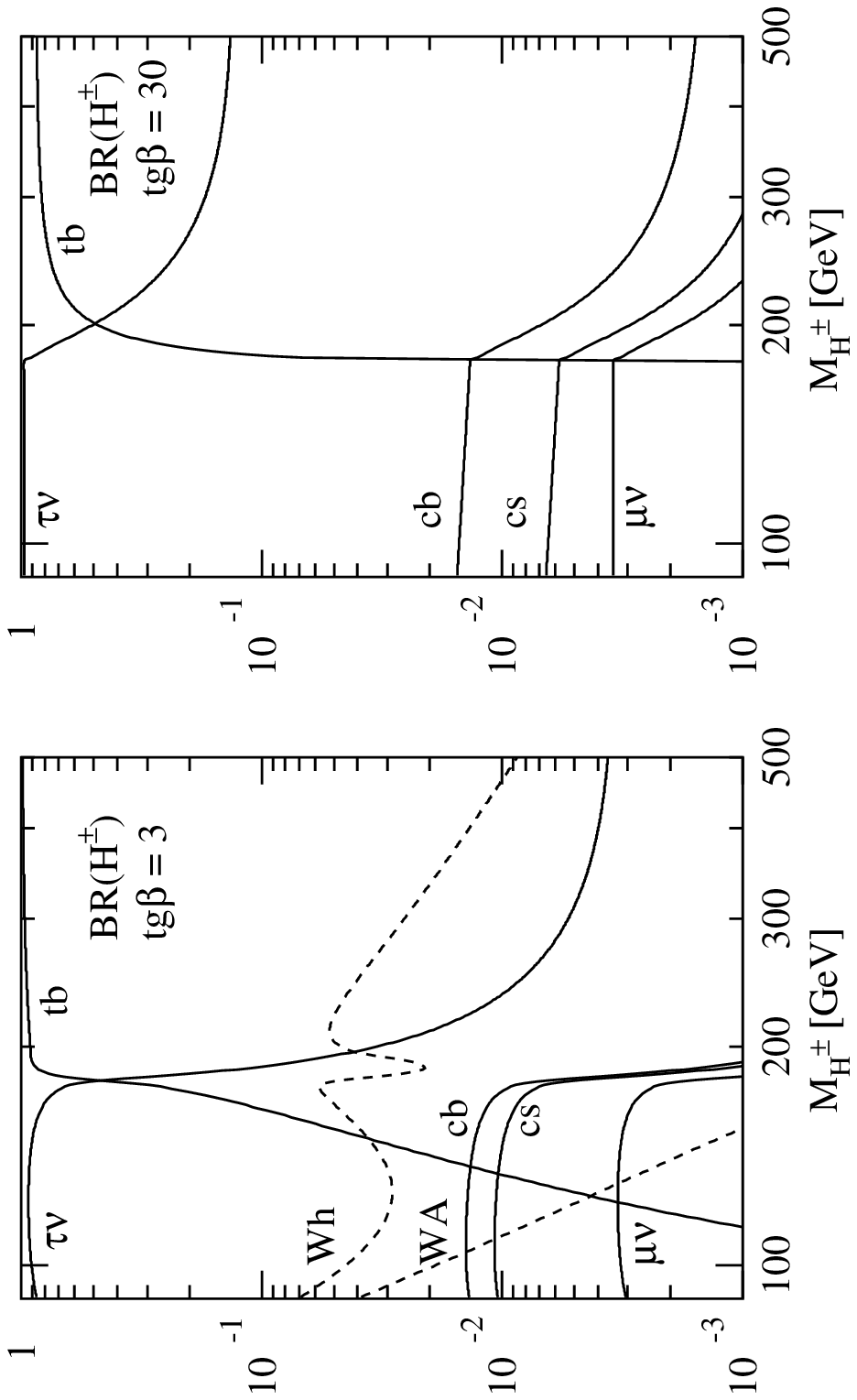}
\end{turn}
\vspace*{-4.2cm}

\centerline{\bf Fig.~\ref{fg:mssmbr}d}

\caption[]{\it Continued.}
\end{figure}

\begin{figure}[hbt]

\vspace*{-2.5cm}
\hspace*{-4.5cm}
\begin{turn}{-90}%
\epsfxsize=16cm \epsfbox{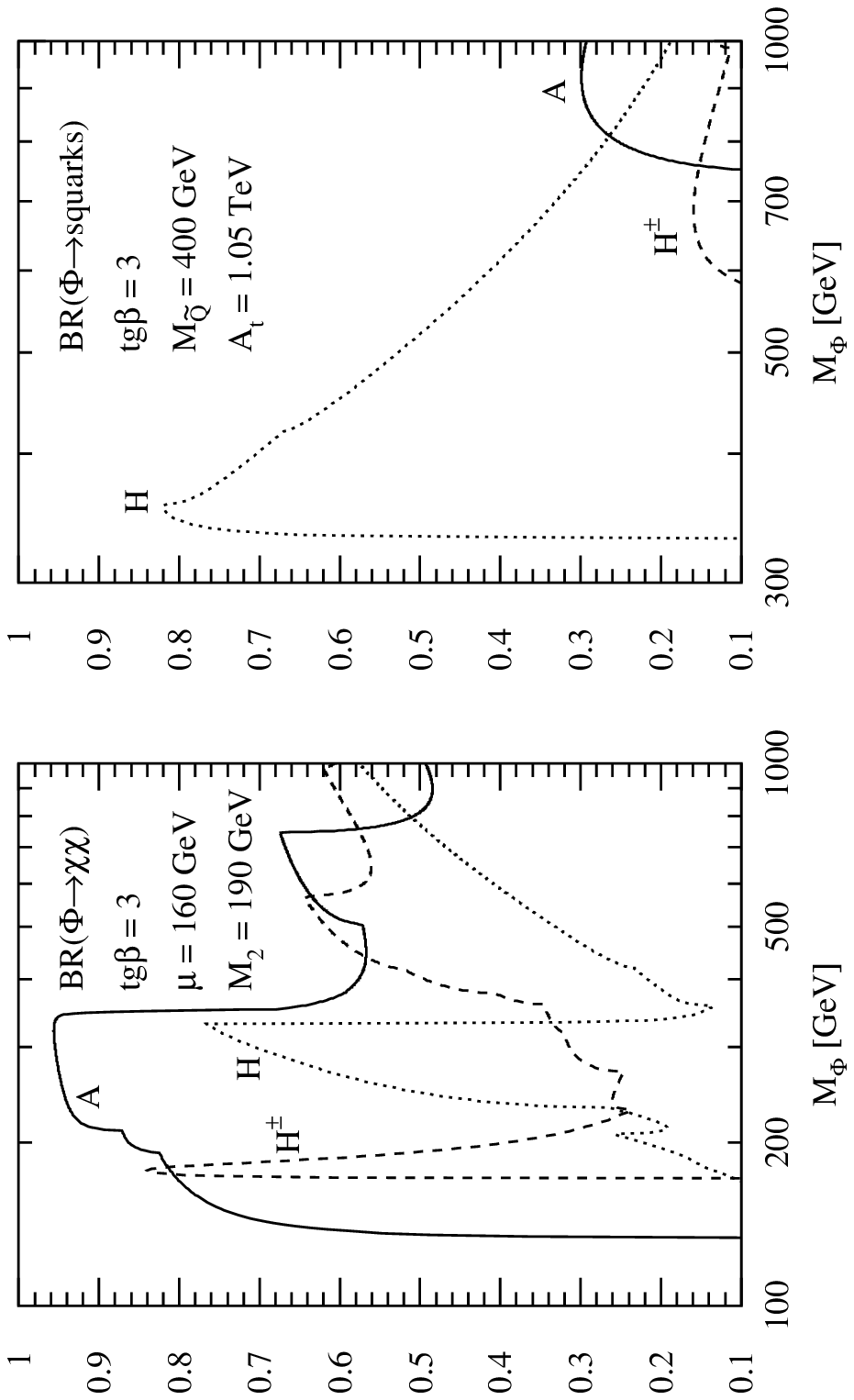}
\end{turn}
\vspace*{-4.2cm}

\caption[]{\label{fg:hcharneutsq} \it Branching ratios of the MSSM Higgs boson
$H,A,H^\pm$ decays into charginos/neutralinos and squarks as a function of their
masses for $\tgb=3$. The mixing parameters have been chosen as $\mu=160$ GeV,
$A_t=1.05$ TeV, $A_b=0$ and the squark masses of the first two generations as
$M_{\widetilde{Q}}=400$ GeV. The gaugino mass parameter has been set to
$M_2=190$ GeV.}
\end{figure}

The heavy neutral Higgs boson $H$
can also decay into two lighter Higgs bosons.
Other possible channels are Higgs cascade
decays and decays into supersymmetric
particles \citer{614,616}, Fig.~\ref{fg:hcharneutsq}. In addition to light
sfermions, Higgs boson decays into charginos
and neutralinos could eventually be important.
These new channels are kinematically accessible, 
at least for the heavy Higgs bosons $H,A$ and $H^\pm$;
in fact, the branching fractions can be very
large and they can even become dominant in some
regions of the MSSM parameter space. Decays of $h$
into the lightest neutralinos (LSP) are also
important, exceeding 50\% in some parts of
the parameter space. These decays
strongly affect experimental search techniques.

\noindent
The \underline{\it charged Higgs particles} decay into
fermions, but also, if allowed kinematically,
into the lightest neutral Higgs and a
$W$ boson. Below the $tb$ and $Wh$ thresholds,
the charged Higgs particles will 
decay mostly into $\tau \nu_\tau$ and $cs$
pairs, the former being dominant for $\tgb>1$.
For large $M_{H^\pm}$ values, the top--bottom decay mode
$H^+\to t\bar b$ becomes dominant. In some parts of
the SUSY parameter space, decays into
supersymmetric particles may exceed
50\%.

\vspace*{2mm}
Adding up the various decay modes,
the width of all five Higgs bosons
remains very narrow, being of order
10 GeV even for large masses.

\vspace*{4mm}
\subsection{The Production of SUSY Higgs Particles in Hadron Collisions}

\vspace*{2mm}
\noindent
The basic production processes of SUSY
Higgs particles at hadron colliders \cite{24A,32,620B}
are essentially the same as in the
Standard Model. Important differences
are  nevertheless generated by the
modified couplings, the extended particle
spectrum, and the negative parity of
the $A$ boson. For large $\tgb$
the coupling $hb\bar b$ is enhanced so that
the bottom-quark loop becomes competitive
to the top-quark loop in the effective
$hgg$ coupling. Moreover squark loops
will contribute to this coupling \cite{sqloop}.

\vspace*{2mm}
\noindent
{\bf 1.$\,$} The partonic cross section $\sigma(gg\to \Phi)$
for the gluon fusion of Higgs particles
can be expressed by couplings $g$, in units
of the corresponding SM couplings, and
form factors $A$; to lowest order \cite{32,sqloopqcd}:
\begin{eqnarray}
\hat\sigma^\Phi_{LO} (gg\to \Phi) & = & \sigma^\Phi_0 M_\Phi^2 \times
BW(\hat{s}) \\
\sigma^{h/H}_0 & = & \frac{G_{F}\alpha_{s}^{2}(\mu)}{128 \sqrt{2}\pi} \
\left| \sum_{Q} g_Q^{h/H} A_Q^{h/H} (\tau_{Q})
+ \sum_{\widetilde{Q}} g_{\widetilde{Q}}^{h/H} A_{\widetilde{Q}}^{h/H}
(\tau_{\widetilde{Q}}) \right|^{2} \nonumber \\
\sigma^A_0 & = & \frac{G_{F}\alpha_{s}^{2}(\mu)}{128 \sqrt{2}\pi} \
\left| \sum_{Q} g_Q^A A_Q^A (\tau_{Q}) \right|^{2} \nonumber \;.
\end{eqnarray}
While the quark couplings have been
defined in Table \ref{tb:hcoup}, the couplings of
the Higgs particles to squarks are given by
\begin{eqnarray}
g_{\tilde Q_{L,R}}^{h} & = & \frac{M_Q^2}{M_{\tilde Q}^2} g_Q^{h}
\mp \frac{M_Z^2}{M_{\tilde Q}^2} (I_3^Q - e_Q \sin^2 \theta_W)
\sin(\alpha + \beta) \nonumber \\ \nonumber \\
g_{\tilde Q_{L,R}}^{H} & = & \frac{M_Q^2}{M_{\tilde Q}^2} g_Q^{H}
\pm \frac{M_Z^2}{M_{\tilde Q}^2} (I_3^Q - e_Q \sin^2 \theta_W)
\cos(\alpha + \beta) \;.
\end{eqnarray}
Only ${\cal CP}$ non-invariance allows for non-zero
squark contributions to the pseudoscalar $A$
boson production.
The form factors can be expressed
in terms of the scaling function $f(\tau_i=4M_i^2/M_\Phi^2)$,
cf. Eq. (\ref{eq:ftau}):
\begin{eqnarray}
A_Q^{h/H} (\tau) & = & \tau [1+(1-\tau) f(\tau)] \nonumber \\
A_Q^A (\tau) & = & \tau f(\tau) \nonumber \\
A_{\tilde Q}^{h/H} (\tau) & = & -\frac{1}{2}\tau [1-\tau f(\tau)] ~.
\end{eqnarray}
For small $\tgb$ the contribution of the top loop is
dominant, while for large $\tgb$
the bottom loop is strongly enhanced.
The squark loops can be significant
for squark masses below $\sim 400$ GeV \cite{sqloopqcd,sqloopmass}.

\vspace*{2mm}
\noindent
{\bf 2.$\,$} Other production mechanisms for SUSY
Higgs bosons, vector boson fusion,
Higgs-strahlung off $W,Z$ bosons and
Higgs-bremsstrahlung off top quarks and bottom-quark
fusion, can be treated in analogy to
the corresponding SM processes.

Particularly important is the process of $b$-quark fusion
for large values of $\tan\beta$ when the Higgs couplings
to $b$ quarks are enhanced \cite{bH}. Since $b$ quarks 
are moderately
light, gluon splitting $g \to \bar{b} b$ gives rise to 
high-energy $b$-quark/anti-quark beams in fast moving 
protons/antiprotons. The fusion of $\bar{b} + b \to h,H,A$ 
is therefore a rich source of Higgs bosons at Tevatron and 
LHC for large $\tan\beta$.

\begin{figure}[hbtp]

\vspace*{0.3cm}
\hspace*{1.0cm}
\begin{turn}{-90}%
\epsfxsize=8.5cm \epsfbox{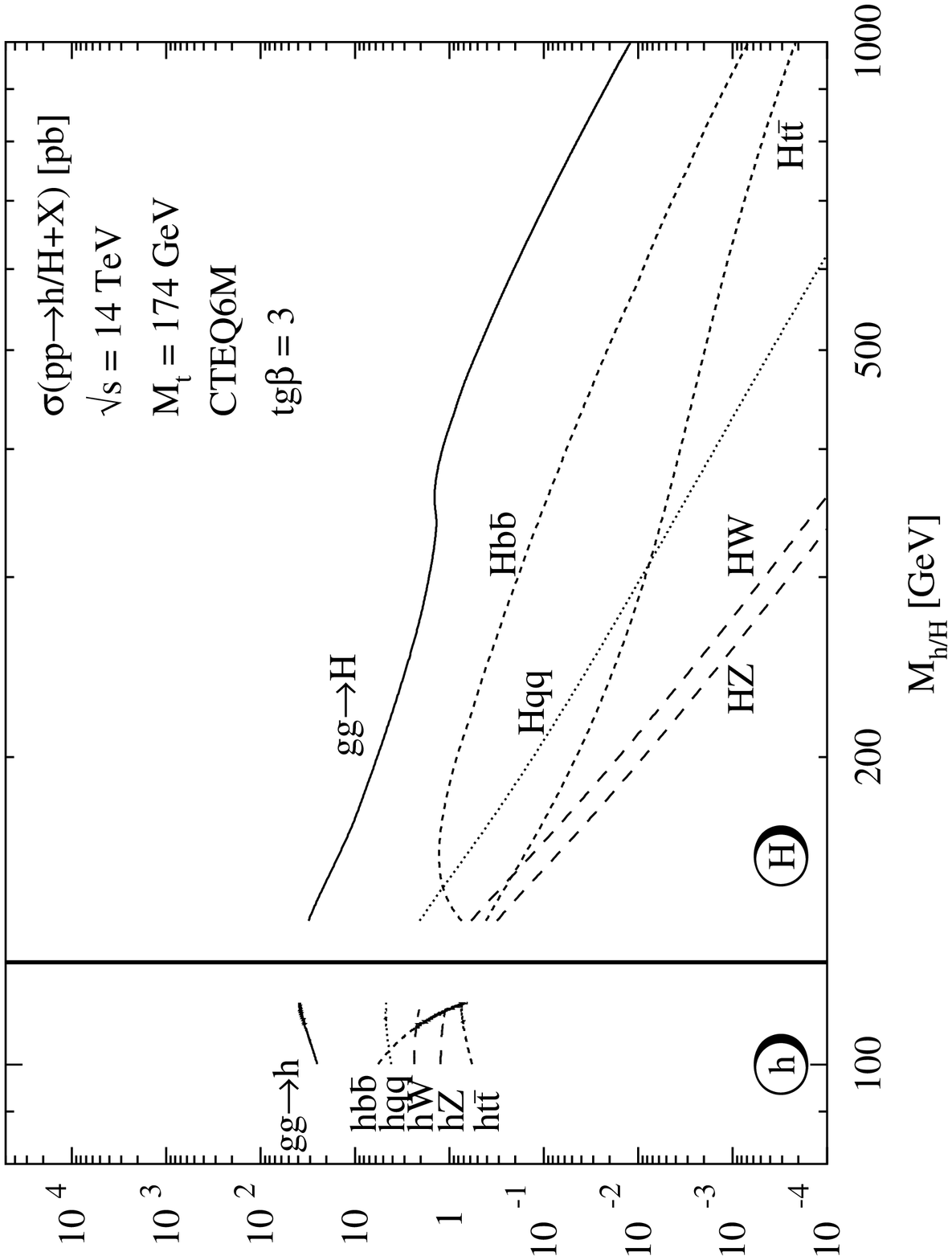}
\end{turn}
\vspace*{0.3cm}

\centerline{\bf Fig.~\ref{fg:mssmprohiggs}a}

\vspace*{0.2cm}
\hspace*{1.0cm}
\begin{turn}{-90}%
\epsfxsize=8.5cm \epsfbox{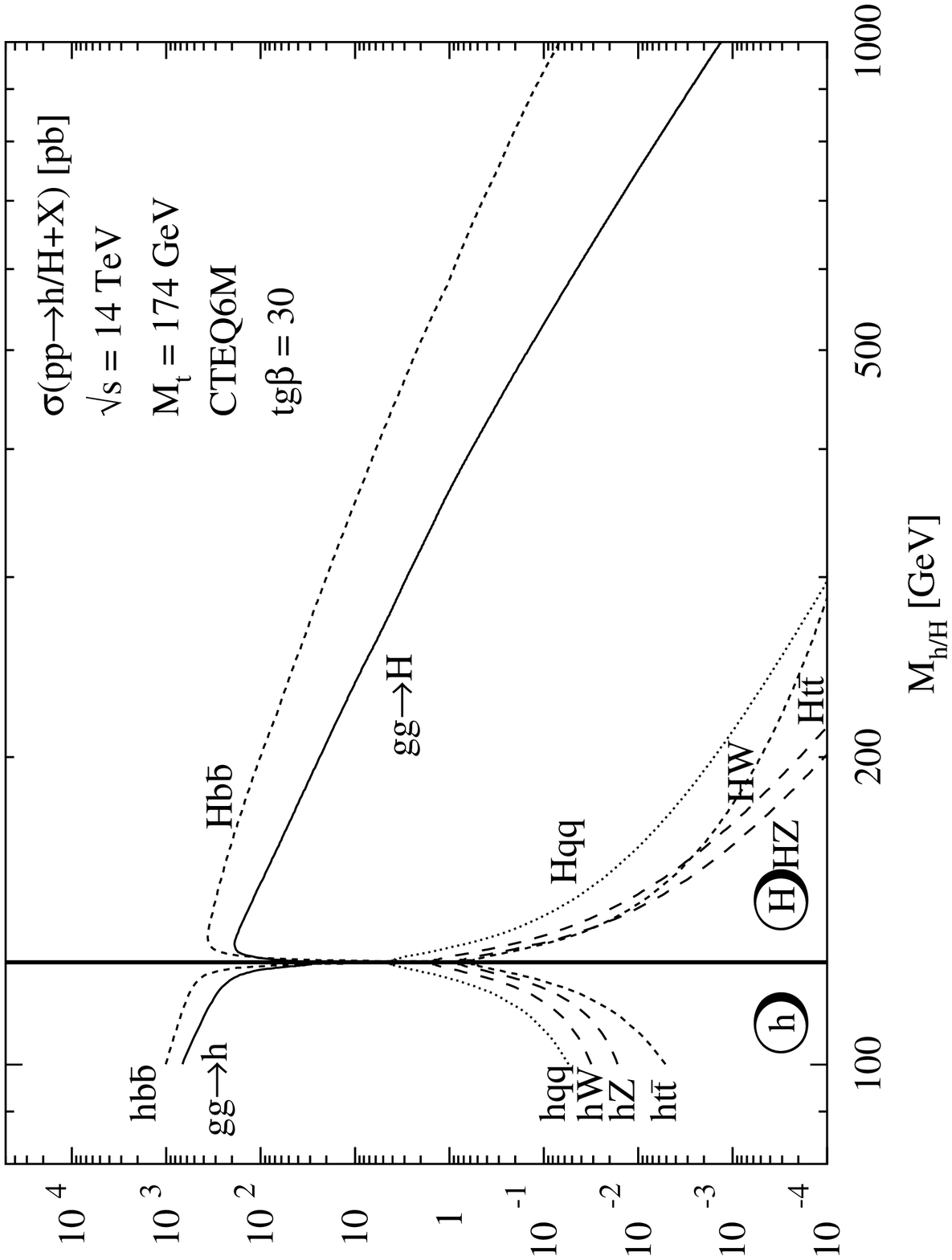}
\end{turn}
\vspace*{0.3cm}

\centerline{\bf Fig.~\ref{fg:mssmprohiggs}b}

\caption[]{\label{fg:mssmprohiggs} \it Neutral MSSM Higgs production cross
sections at the LHC  for gluon fusion $gg\to \Phi$,
vector-boson fusion $qq\to qqVV \to qqh/
qqH$, Higgs-strahlung $q\bar q\to V^* \to hV/HV$ and the associated
production $gg,q\bar q \to  b\bar b \Phi/ t\bar t \Phi$, including all known
QCD corrections. (a) $h,H$ production for $\tgb=3$, (b) $h,H$ production for
$\tgb=30$, (c) $A$ production for $\tgb=3$, (d) $A$ production for $\tgb=30$.}
\end{figure}
\addtocounter{figure}{-1}
\begin{figure}[hbtp]

\vspace*{0.3cm}
\hspace*{1.0cm}
\begin{turn}{-90}%
\epsfxsize=8.5cm \epsfbox{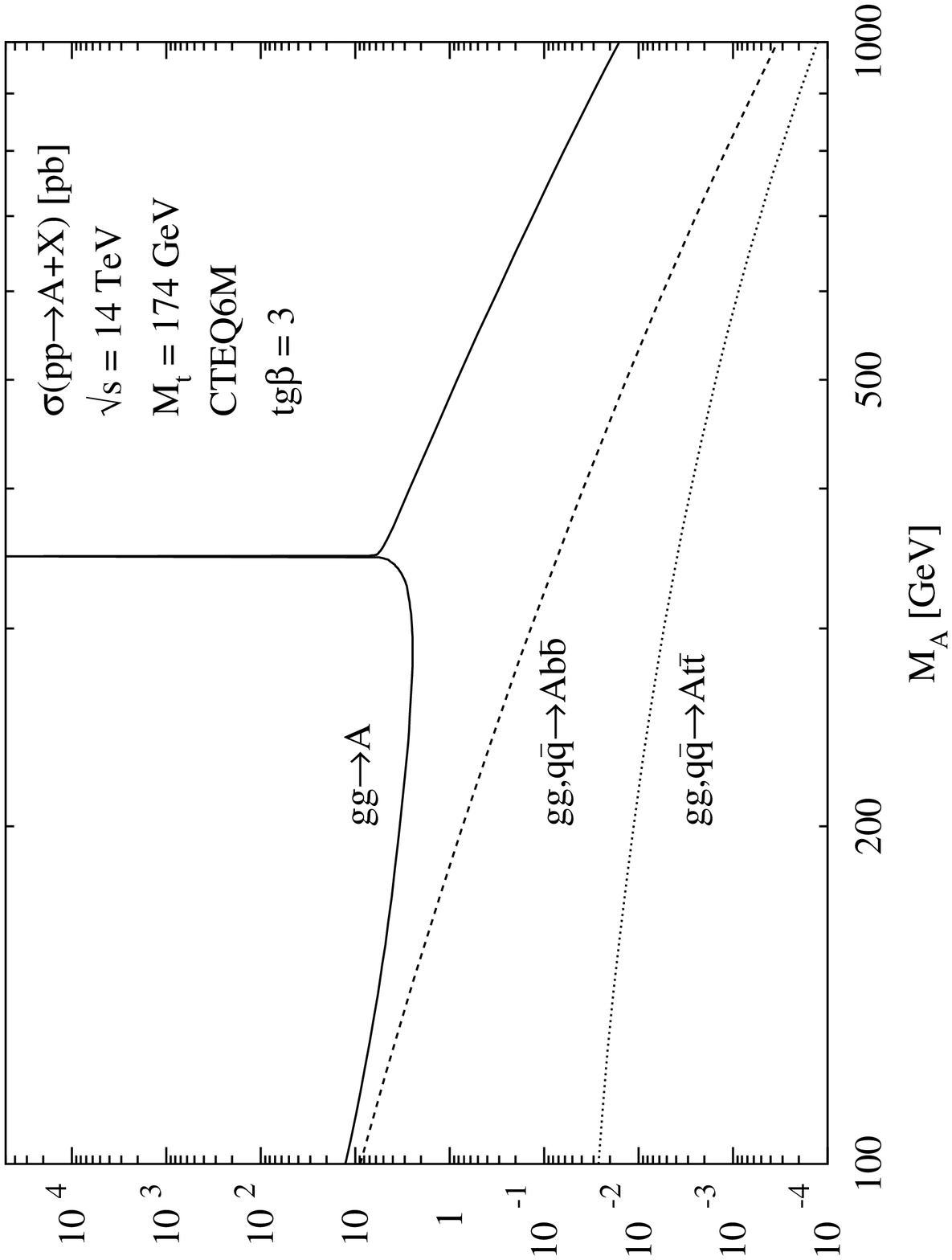}
\end{turn}
\vspace*{0.3cm}

\centerline{\bf Fig.~\ref{fg:mssmprohiggs}c}

\vspace*{0.2cm}
\hspace*{1.0cm}
\begin{turn}{-90}%
\epsfxsize=8.5cm \epsfbox{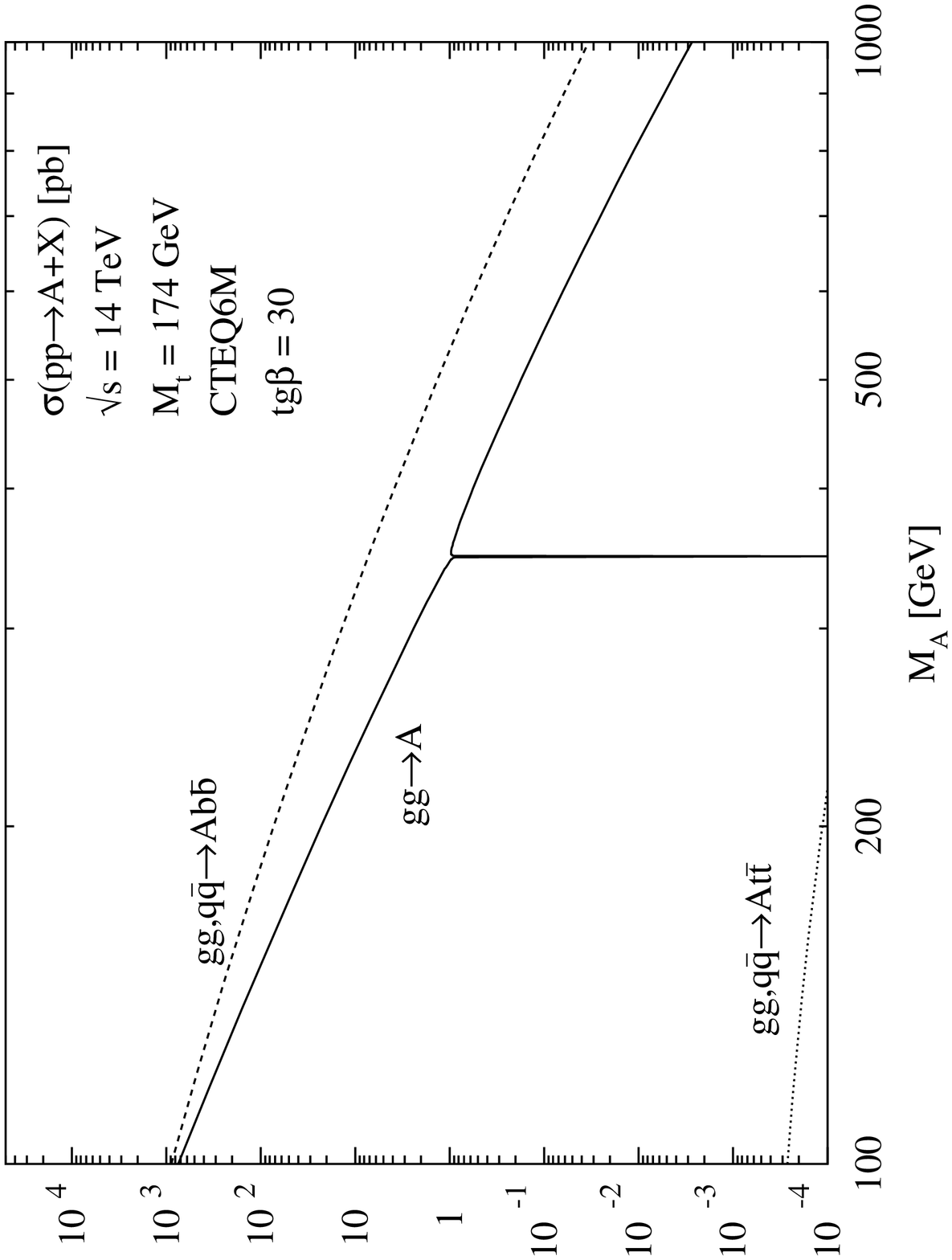}
\end{turn}
\vspace*{0.3cm}

\centerline{\bf Fig.~\ref{fg:mssmprohiggs}d}

\caption[]{\it Continued.}
\end{figure}

\vspace*{2mm}
\noindent
{\bf 3.$\,$} Data from the Tevatron in the channel 
$p \bar p \to b \bar b \tau^+ \tau^-$
have been exploited to exclude 
part of the supersymmetric Higgs
parameter space in the $[ M_A, \tgb]$ plane. In the
interesting range of $\tgb$ between 30 and 50,
pseudoscalar masses $M_A$ of up to 150 to 190 GeV appear
to be excluded.

\vspace*{2mm}
\noindent
{\bf 4.$\,$} The cross sections of the various 
MSSM Higgs production mechanisms at the LHC
are shown in Figs. \ref{fg:mssmprohiggs}a--d for two representative values of
$\tgb = 3$ and 30, as a function of the corresponding Higgs mass.
The CTEQ6M
parton densities have been adopted with $\alpha_s(M_Z)=0.118$; the top and
bottom masses have been set to $M_t=174$ GeV and $M_b=4.62$ GeV. For the
pseudoscalar Higgs bremsstrahlung off $t,b$ quarks, $pp \to Q\bar Q A +X$,  the
leading-order CTEQ6L1 parton densities have been used.
For small and moderate values of $\tgb\lessim 10$
the gluon-fusion cross section provides the dominant production cross section
for the entire Higgs mass region up to $M_\Phi\sim 1$ TeV. However, for large
$\tgb$, Higgs bremsstrahlung off bottom quarks, $pp\to b\bar b \Phi+X$,
dominates
over the gluon-fusion mechanism since  the  bottom Yukawa
couplings are strongly enhanced in this case.

\begin{figure}[hbtp]
\begin{center}
\epsfig{figure=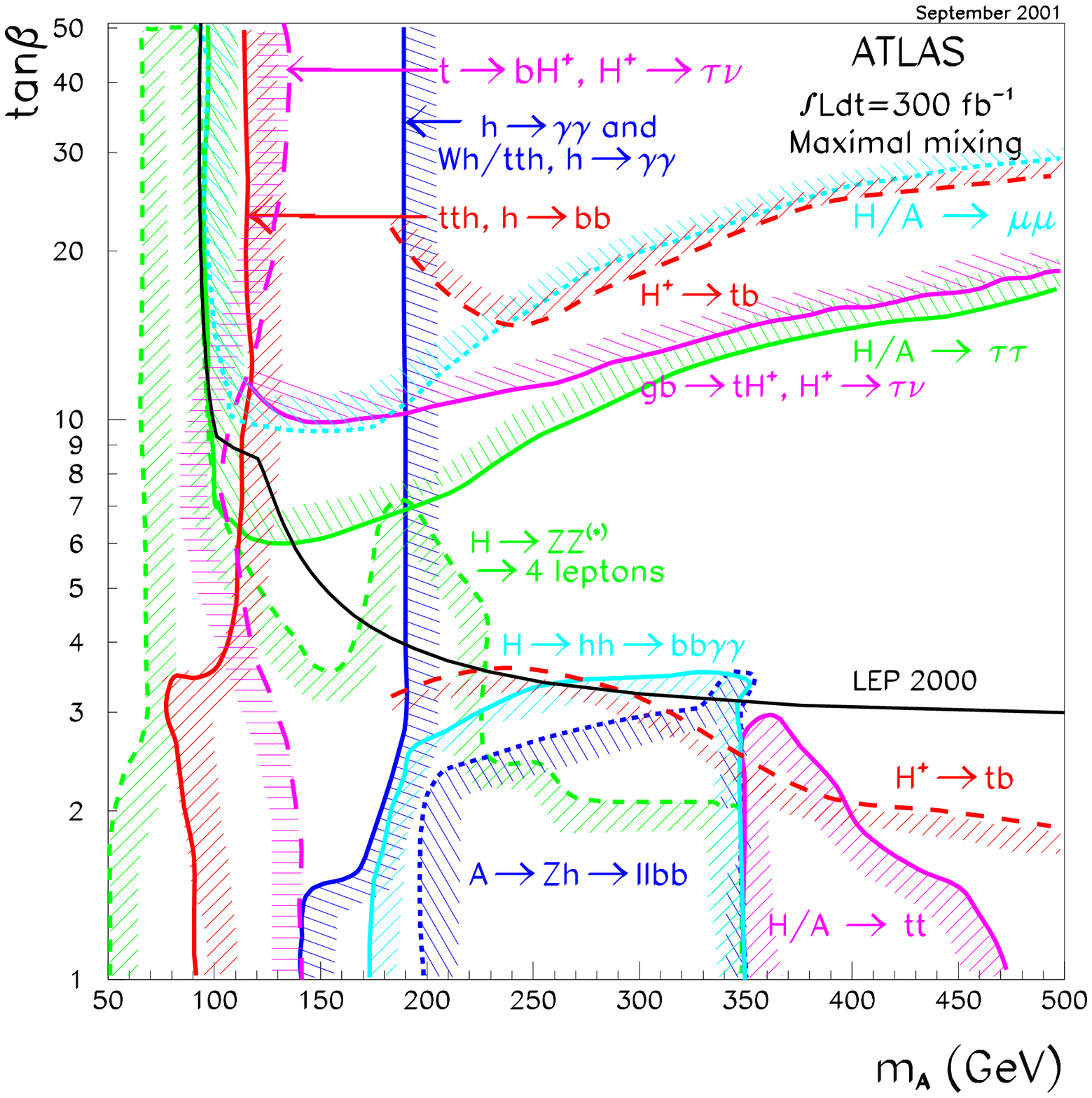,bbllx=0,bblly=0,bburx=523,bbury=504,width=10cm,clip=}
\end{center}
\caption[]{\label{fg:atlascms} \it
The ATLAS sensitivity for the discovery of the MSSM Higgs bosons in the
case of maximal mixing. The 5$\sigma$ discovery curves are shown in the
$(\tan\beta,m_A)$ plane for the individual channels and for an integrated
luminosity of 300 fb$^{-1}$. The corresponding LEP limit is also shown.
Ref.~\cite{richter}.}
\end{figure}

The MSSM Higgs search at the LHC will be more involved than the SM Higgs
search.
The final summary is presented in Fig.~\ref{fg:atlascms}. It exhibits a
difficult region for
the MSSM Higgs search at the
LHC. For $\tgb \sim 5$ and $M_A \sim 150$ GeV, the full
luminosity and the full data sample of both the ATLAS and CMS
experiments at the
LHC are needed to cover the problematic parameter region \cite{richter}.
On the other hand, if no excess of Higgs events
above the SM background processes beyond 2 standard deviations will be found,
the MSSM Higgs bosons can be excluded at 95\% C.L. 

Even though the
entire supersymmetric Higgs parameter
space of the MSSM is expected to be finally covered by the
LHC experiments, the entire ensemble of individual Higgs
bosons is accessible only in part of
the parameter space. Particularly in the blind wedge 
opening in the parameter space 
at about $M_A \sim$ 200 GeV and centered around $\tan\beta \sim 7$ 
only the lightest Higgs boson $h$ can be discovered, while the heavy
Higgs bosons $A,H,H^\pm$ cannot be found in non-supersymmetric decay
channels. Moreover, the search
for heavy $H,A$ Higgs particles is very
difficult, because of the $t\bar t$
continuum background for masses  $\gsim 500$ GeV.\\

\vspace*{4mm}
\subsection{The Production of SUSY Higgs Particles in $e^+e^-$ Collisions}

\vspace*{2mm}
\noindent
{\bf 1.$\,$} The search for the neutral SUSY Higgs bosons at \ee linear colliders will be
a straight\-forward ex\-tension of the search performed at LEP2, which
covered the mass range up to $\sim
100$~GeV for neutral Higgs bosons.  Higher
energies, $\sqrt{s}$ in excess of $250$~GeV, are required to sweep the
entire parameter space of the MSSM for moderate to large values
of $\tgb$. \\

\noindent
The main production mechanisms of \underline{\it neutral Higgs bosons} at
\ee colliders \cite{19, 615, 617} are the \Hs process and associated
pair production, as well as the fusion processes:
\begin{eqnarray}
{\hspace*{-0.7mm}}(a) \ \ \mbox{Higgs--strahlung{\hspace*{0.7mm}}:} \hspace{1.4cm} \epem &
\stackrel{Z}{\longrightarrow} & Z+h/H \hspace{5cm}
\nonumber  \\
\,(b) \ \ {\rm Pair \ production:}\! \hspace{13.6mm} \epem &
\stackrel{Z}{\longrightarrow} & A+h/H 
\nonumber \\
(c) \ \ {\rm Fusion \ processes:} \hspace{10.7mm} \ \epem &
\stackrel{WW}{\longrightarrow} & \overline{\nu}_e \ \nu_e \ + h/H 
\hspace{3.3cm} \nonumber  \\
\epem & 
\stackrel{ZZ}{\longrightarrow} &  \epem + h/H \;. \nonumber
\end{eqnarray}
The ${\cal CP}$-odd Higgs boson $A$ cannot be produced in fusion
processes to leading order.  The cross sections for the four \Hs and
pair production processes can be expressed as
\begin{eqnarray}
\sigma(\epem \ra Z + h/H) & =& \sin^2/\cos^2(\beta-\alpha) \ \sigma_{SM}
\nonumber \\
\sigma(\epem \ra A + h/H) & =& \cos^2/\sin^2(\beta-\alpha) \
\bar{\lambda} \  \sigma_{SM} ~, 
\end{eqnarray}
where $\sigma_{SM}$ is the SM cross section for \Hs and the coefficient
$\bar{\lambda} \sim \lambda^{3/2}_{Aj} / \lambda^{\demi}_{Zj}$ accounts 
for the suppression of the $P$-wave 
$Ah/H$ cross sections near the threshold.

\STS The cross sections for  Higgs-strahlung and for pair
production, much as those for the production of the
light and the heavy neutral Higgs bosons $h$ and $H$, are 
complementary, coming either with coefficients
$\sin^2(\beta-\alpha)$ or $\cos^2(\beta-\alpha)$.  As a result, since
$\sigma_{SM}$ is large, at least the lightest ${\cal CP}$-even Higgs
boson must be detected in $e^+e^-$ experiments.

\begin{figure}[hbtp]
\begin{center}
\vspace*{5mm}
\hspace*{5mm}
\epsfig{file=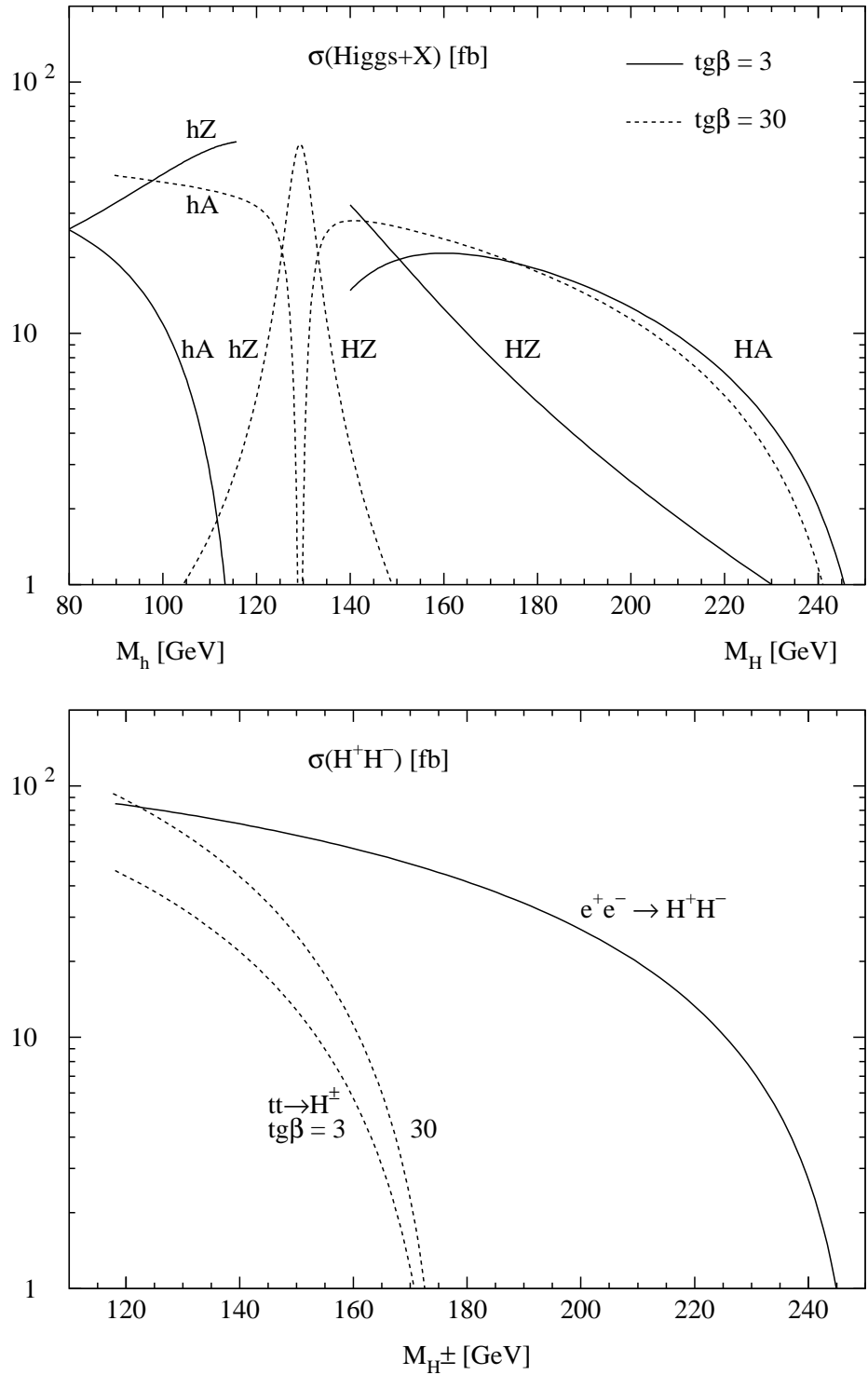,width=15cm}
\end{center}
\vspace{-2.5cm}
\caption[]{\it 
  Production cross sections of MSSM Higgs bosons at $\sqrt{s} =
  500$~GeV: Higgs-strahlung and pair production; upper part: neutral
  Higgs bosons, lower part: charged Higgs bosons.
  Ref. \protect\cite{613A}.  \protect\label{f604}\label{prodcs}}
\end{figure}
\STS Representative examples of cross sections for the production
mechanisms of the neutral Higgs bosons are exemplified in Fig.~\ref{f604}, 
as a function of the Higgs masses,  
 for $\tgb= 3$ and 30.  The cross
section for $hZ$ is large for $M_h$ near the maximum value allowed
for $\tgb$; it is of order 50~fb, corresponding to $\sim$ 2,500
events for an integrated luminosity of 50 fb$^{-1}$.  By contrast, the
cross section for $HZ$ is large if $M_h$ is sufficiently below the
maximum value  [implying small $M_H$].  For $h$
and for a low mass $H$, the signals consist of a $Z$ boson accompanied by
a $b\bar{b}$ or $\tau^+ \tau^-$ pair.  These signals are easy to separate
from the background,  which comes mainly from $ZZ$ production if the
Higgs mass is close to $M_Z$.  For the associated channels $\epem \to
Ah$ and $AH$, the situation is opposite to the previous case: the
cross section for $Ah$ is large for light $h$, whereas $AH$ pair
production is the dominant mechanism in the complementary region for
heavy $H$ and $A$ bosons.  The sum of the two cross sections
decreases from $\sim 50$ to 10~fb if $M_A$ increases from $\sim 50$ to
200~GeV at $\sqrt{s} = 500$~GeV.  In major parts of the parameter
space, the signals consist of four $b$ quarks in the final state,
requiring provisions for efficient $b$-quark tagging.  Mass
constraints will help to eliminate the backgrounds from QCD jets and
$ZZ$ final states.  For the $WW$ fusion mechanism, the cross sections
are larger than for Higgs-strahlung, if the Higgs mass is moderately small -- less
than 160~GeV at $\sqrt{s} = 500$ GeV.  However, since the final state
cannot be fully reconstructed, the signal is more difficult to
extract.  As in the case of the \Hs processes, the production of light
$h$ and heavy $H$ Higgs bosons complement each other in $WW$
fusion, too. \\

\noindent
The \underline{\it charged Higgs bosons}, if lighter than the top 
quark, can
be produced in top decays, $t \ra b + H^+$, with a branching ratio
varying between $2\%$ and $20\%$ in the kinematically allowed region.
Since the cross section for top-pair production is of order 0.5 pb at
$\sqrt{s} = 500$~GeV, this corresponds to 1,000 to 10,000 charged
Higgs bosons at a luminosity of 50~fb$^{-1}$.  Since, for $\tgb$ larger
than unity, the charged Higgs bosons will decay mainly into $\tau
\nu_\tau$, there is  a surplus of $\tau$ final states over $e,
\mu$ final states in $t$ decays, an apparent breaking of lepton
universality.  For large Higgs masses the dominant decay mode is the
top decay $H^+ \to t \overline{b}$.  In this case the charged Higgs
particles must be pair-produced in \ee colliders:
\[
              \epem \to H^+H^- ~.
\]
The cross section depends only on the charged Higgs mass.  It is of
order 100 fb for small Higgs masses at $\sqrt{s} = 500$~GeV, but it
drops very quickly due to the $P$-wave suppression $\sim \beta^3$
near the threshold.  For $M_{H^{\pm}} = 230$~GeV, the cross section
falls to a level of $\simeq 5\,$~fb. The \cs
is considerably larger for $\gamma \gamma$ collisions.

\vspace*{4mm}
\noindent
{\bf 2.$\,$} {\it \underline{Experimental Search Strategies}:}
Search strategies have been described for neutral and charged
Higgs bosons in Ref. \cite{13}. The overall experimental situation
can be summarized in the following two points:

\vspace*{1mm}
\noindent
{\hspace*{3mm}}{\bf (i)} The lightest ${\cal CP}$-even Higgs particle $h$ can be
detected in the entire range of the MSSM parameter space, either
via Higgs-strahlung  $\epem \to hZ$ or via pair
production $\epem \to hA$.  This conclusion holds true even at
a c.m. energy of 250 GeV, independently of the squark mass values; it
is also valid if decays to invisible neutralinos and other SUSY \ps
are realized in the Higgs sector.

\vspace*{1mm}
\noindent
{\hspace*{3mm}}{\bf (ii)} The area in the parameter space where {\it all SUSY Higgs
bosons} can be discovered at \ee colliders is characterized by $M_H,
M_A \lessim \frac{1}{2} \sqrt{s}$, independently of $\tgb$.  The $h,
H$ Higgs bosons can be produced either via \Hs or in $Ah, AH$
associated production; charged Higgs bosons will be produced in
$H^+H^-$ pairs. Thus the blind LHC wedge can be covered up to
$A,H,H^\pm$ Higgs masses of 500 GeV at the 1 TeV collider ILC,
and up to 1.5 TeV at the 3 TeV collider CLIC. If the $ee$ collider
is turned into a high-energy photon collider by Compton 
back-scattering of laser light \citer{novo,MMMPMZ},
the Higgs mass range in single formation experiments 
$\gamma \gamma \to A,H$ can be extended to 80\% of the total $e^+e^-$ energy,
i.e. 800 GeV and 2.4 TeV at ILC and CLIC, respectively,
cf. \cite{Muhlgamgam,Krawgamgam}.  
 
\vspace*{2mm}
The search for the lightest neutral SUSY
Higgs boson $h$ had been one of the most important experimental
tasks at LEP2. Mass values of the
pseudoscalar boson $A$ of less than about 90 GeV have 
been excluded, independently of $\tgb$, cf.~Fig.~\ref{fg:mssmhiggs}. \\

\vspace*{4mm}
\subsection{Measuring the Parity of Higgs Bosons}

\vspace*{2mm}
\noindent
Once the  Higgs bosons  are discovered, 
the properties of the particles must be established.
Besides the reconstruction of the supersymmetric Higgs potential \cite{66A,selfMMM},
which will be a very demanding task, the external quantum
numbers must be established, in particular the parity of the
heavy scalar and pseudoscalar Higgs particles $H$ and $A$ \cite{618}.

\vspace*{2mm}
\noindent
{\bf 1.$\,$} For large $H,A$ masses the decays $H,A\to t\bar t$ 
to top final states can be used to
discriminate between the different parity
assignments \cite{618}. For example, the $W^+$ and $W^-$
bosons in the $t$ and $\bar t$
decays tend to be emitted antiparallel
and parallel in the plane perpendicular
to the $t\bar t$ axis:
\begin{equation}
\frac{d\Gamma^\pm}{d\phi_*} \propto 1 \mp \left( \frac{\pi}{4} \right)^2
\cos \phi_*
\end{equation}
for $H$ and $A$ decays, respectively. 

Alternatively, the $\mathcal{CP}$ parity of Higgs bosons can be
measured by analyzing the magnitude of the total cross section and the
top-quark polarization in
associated top-Higgs production in $e^+e^-$ collisions \cite{bup}. \\[-0.1cm]

\vspace*{2mm}
\noindent
{\bf 2.$\,$} For light $H,A$ masses, $\gamma\gamma$
collisions appear to provide a viable
solution \cite{618}. The fusion of Higgs
particles in linearly polarized photon
beams depends on the angle between
the polarization vectors. For scalar $0^+$
particles the production amplitude
is non-zero for parallel polarization
vectors, while pseudoscalar $0^-$
particles require perpendicular
polarization vectors:
\begin{equation}
{\cal M}(H)^+  \sim  \vec{\epsilon}_1 \cdot \vec{\epsilon}_2  \hspace*{0.5cm}
\mbox{and} \hspace*{0.5cm}
{\cal M}(A)^-  \sim  \vec{\epsilon}_1 \times \vec{\epsilon}_2 ~.
\end{equation}
The experimental set-up for Compton
back-scattering of laser light can
be tuned in such a way that the
linear polarization of the hard-photon
beams approaches values close to 100\%.
 Depending on the $\pm$ parity 
of the resonance produced, the measured
asymmetry for photons of  parallel and perpendicular polarization, 
\begin{equation}
{\cal A} = \frac{\sigma_\parallel - \sigma_\perp}{\sigma_\parallel +
\sigma_\perp} ~, 
\end{equation}
is either positive or negative.

\vspace*{2mm}
Exciting observations in the Higgs sector
at photon colliders are predicted in $\mathcal{CP}$-violating extensions.
Particularly in the decoupling regime the near degeneracy of the
$\mathcal{CP}$-even and $\mathcal{CP}$-odd $H,A$ states gives rise potentially to
large mixing effects, rotating the current eigenstates $ H,A $
into the mass eigenstates $ H_2,H_3 $. In this configuration 
large asymmetries,
\begin{equation}
{\cal A} = \frac{\sigma_{++} - \sigma_{--}}{\sigma_{++} +
\sigma_{--}} ~,
\end{equation}
can be generated in single Higgs $\gamma \gamma$ formation  
between right- and left-circularly polarized $\gamma$ beams
\cite{CKZ}. 

\vspace*{4mm}
\subsection{Non-minimal Supersymmetric Extensions}

\vspace*{2mm}
\noindent
{\bf 1.$\,$} The minimal supersymmetric extension of the \SM may appear very
restrictive for supersymmetric theories in general, in particular in
the Higgs sector where the quartic couplings are identified with the
gauge couplings.  However, it turns out that the mass pattern of the
MSSM is quite typical if the theory is assumed to be valid up to the
GUT scale -- the motivation for supersymmetry {\it sui generis}.  This
general pattern has been studied thoroughly within the
next-to-minimal extension: the MSSM, incorporating two Higgs
isodoublets, is extended by introducing an additional isosinglet field $N$.
This extension leads to a model \citer{621,70A} that is generally
referred to as the NMSSM.

The additional Higgs singlet can solve the so-called
$\mu$-problem [i.e. $\mu \sim$ order $M_W$] by
eliminating the $\mu$ higgsino parameter from the potential and by 
replacing it by the vacuum expectation value of the $N$ field,
which can  naturally be related to the usual vacuum expectation values
of the Higgs isodoublet fields.  In this scenario the superpotential
involves the two trilinear couplings $H_1 H_2 N$ and $N^3$.  The
consequences of this extended Higgs sector will be outlined  in
the context of grand unification, including universal soft breaking
terms of the supersymmetry \cite{622,70A}.

\begin{figure}[hbt]
\begin{center}
\hspace*{-0.3cm}
\epsfig{figure=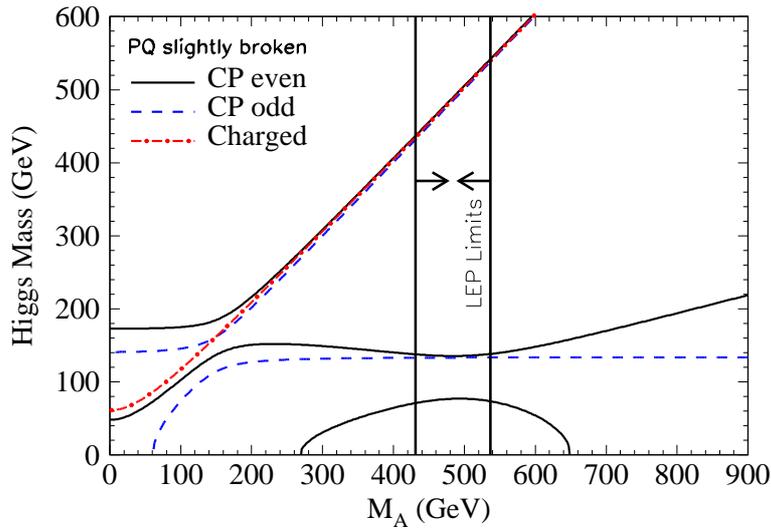,bbllx=0,bblly=14,bburx=439,bbury=320,width=10cm,clip=}
\end{center}
\vspace*{-0.4cm}
\caption[]{\label{fig:26} \it The one-loop Higgs boson masses as a 
function of $M_A$ for $\lambda=0.3$, $\kappa=0.1$, $v_s=3v$, $\tan\beta=3$
and $A_\kappa=-100$~GeV. The arrows denote the region allowed by LEP searches
with 95\% confidence. Ref. \cite{70A}.}
\end{figure}

\vspace*{2mm}
The Higgs spectrum of the NMSSM includes, besides the minimal
set of Higgs particles, one additional scalar and pseudoscalar Higgs
particle:  
\begin{eqnarray*}
  neutral\;CP=+              & &  H_1,\, H_2,\, H_3     \\ \non
  neutral\;CP=-              & &  A_1,\, A_2            \\ \non
  charged{\hspace*{14.5mm}}  & &  H^\pm                 \, .
\end{eqnarray*}
The neutral Higgs \ps are in general mixtures of 
isodoublets, which couple to $W, Z$ bosons and fermions, and
the isosinglet, decoupled from the non-Higgs sector.
The trilinear self-interactions contribute to the masses of the Higgs
particles; for the lightest Higgs boson of each species: 
\begin{eqnarray}
M^2 (H_1) & \leq & M^2_Z \cos^2 2\beta + \lambda^2 v^2 \sin^2 2 \beta \\
M^2 (A_1) & \leq & M^2 (A)   \nonumber \\
M^2 (H^{\pm}) & \leq & M^2 (W) + M^2 (A) - \lambda^2 v^2 \nonumber \;.
\end{eqnarray}
In contrast to the minimal model, the mass of the charged
Higgs \p could be smaller than the \W mass. An example of the mass spectrum
is shown in Fig.~\ref{fig:26}. Since the trilinear \cps
increase with energy, upper bounds on the mass of the lightest neutral
Higgs boson $H_1$ can be derived, in analogy to the Standard Model,
from the assumption that the theory be valid up to the GUT scale:
$m(H_1) \lessim 140 $~GeV.  Thus, despite the additional
interactions, the distinct pattern of the minimal extension remains
valid also in more complex \ssy scenarios.  
If $H_1$ is (nearly) pure isosinglet, it
decouples from the gauge boson and fermion system and its role is
taken by the next Higgs \p with a large isodoublet component, implying
the validity of the mass bound again.\\

For a primarily isosinglet Higgs \p $H_1$ the \cp
$ZZH_1$ is small and the \p cannot be produced by Higgs-strahlung.
However, in this case $H_2$ is generally light and couples with
sufficient strength to the $Z$ boson; if not, $H_3$ plays this role. 

\vspace*{5mm}
\noindent
{\bf 2.$\,$} A large variety of other extensions beyond the minimal supersymmetric
model has been analyzed theoretically. For example, if the gauge boson sector
is expanded by an additional $U(1)'$ Abelian symmetry at high energies \cite{CHKZ,BL}, 
the additional pseudoscalar Higgs field is absorbed to generate the mass of the
new $Z'$ boson while the scalar part of the Higgs field can be observed as a new
Higgs boson beyond the MSSM set. If generated by an extended symmetry like $E_6$,
the Higgs sector is expanded by an ensemble of new states \cite{E6} 
with quite unconventional properties.

\vspace*{2mm}
Quite generally, so long as the fields in supersymmetric theories remain
weakly interacting up to the canonical Planck scale, the mass of the
lightest Higgs bosons is bounded by about 200 GeV as the Yukawa couplings
are restricted to be small in the same way as the quartic coupling in
the standard Higgs potential.
Moreover, the mass bound of 140~GeV 
for the lightest Higgs particle is realized in almost all \ssy theories 
\cite{623}; 
cf. Ref.~\cite{EspinQuiros} for expansions of this limit.
Experiments at $e^+e^-$ colliders are in a `no-lose' situation 
\cite{L141A} for detecting the Higgs particles in general supersymmetric
theories, even for c.m. energies as low as $\sqrt{s} \sim 500$ GeV.

\vspace*{4mm}
\section{Dynamical Symmetry Breaking}

\vspace*{2mm}
\noindent
The Higgs mechanism is based on the
theoretical concept  of spontaneous symmetry
breaking \cite{1}. In the canonical
formulation, adopted in the Standard
Model, a four-component {\it fundamental}
scalar field is introduced, which is
endowed with a self-interation such
that the field acquires a non-zero
value in the ground state. The specific
direction in isospace, which is singled out by  the
ground-state solution, breaks the
isospin invariance of the interaction
spontaneously\footnote{We retain this language commonly used also in the
context of gauge theories, although the gauge symmetry is not broken,
in effect,  
by the Higgs mechanism.}. The interaction of the gauge fields with the
scalar field in the ground state
 generates the masses
of these fields. The longitudinal degrees
of freedom of the gauge fields are built
up by absorption of the Goldstone modes,
which are associated with the spontaneous
breaking of the electroweak symmetries
in the scalar field sector. Fermions
acquire masses through Yukawa interactions
with the ground-state field. While three
scalar components are absorbed by the
gauge fields, one degree of freedom
manifests itself as a physical particle,
the Higgs boson. The exchange of this
particle in scattering amplitudes, including
longitudinal gauge fields and massive
fermion fields, guarantees the unitarity
of the theory up to asymptotic energies. 

\vspace*{2mm}
Alternatively, to interpret the Higgs boson as a (pseudo-)Goldstone boson 
associated with the spontaneous breaking of new strong
interactions has been a very
attractive idea for a long time.

\vspace*{4mm}
\subsection{Little Higgs Models}

\vspace*{2mm}
The interest in this picture has been renewed 
within the Little Higgs scenarios \cite{2A} that have recently been developed 
to generate the electroweak symmetry breaking dynamically by new strong 
interactions. 
Little Higgs models are based on a complex system of symmetries and symmetry 
breaking mechanisms; for a recent review see Ref.~\cite{littlest}. 
Three points are central in realizing the idea:
\begin{itemize}
\item[(i)]  The Higgs field is a Goldstone field associated with the breaking 
of a global symmetry $G$. The strong interactions are characterized
by a scale $\Lambda \sim$ 10 to 30 TeV, while the dynamical Goldstone
scale is estimated to be $f \sim \Lambda / 4 \pi \sim$ 1 to 3 TeV;
\item[(ii)] In the same step, the gauge symmetry $G_0 \subset G$ is broken 
down to the gauge group $SU(2) \times U(1)$ of the Standard Model, generating 
masses for heavy vector bosons and fermions 
of the intermediate size $M \sim g f \sim 1$ TeV;
\item[(iii)] The Higgs boson acquires a mass finally by collective radiative 
symmetry breaking, i.e. to second order, at the standard electroweak scale  
$v \sim g^2 f / 4 \pi \sim$ 100 to 300 GeV. 
\end{itemize}

Thus three characteristic scales are encountered in these models: the strong 
interaction scale $\Lambda$, the new mass scale $M$ and the electroweak 
breaking scale $v$, ordered in the hierarchical chain $\Lambda \gg M \gg v$. 
The light Higgs boson mass is protected at small value by requiring the 
collective breaking of two symmetries. In contrast to the boson-fermion 
symmetry that cancels quadratic divergences in supersymmetry, 
the cancellation 
in Little Higgs models operates in the boson and fermion sectors individually, 
the cancellation ensured by the symmetries among the couplings of the SM 
fields and the new fields, $\mathcal{O}$$(M)$, to the Higgs field.

\vspace*{4mm}
\noindent
{\it{Example: \underline{Littlest Higgs Model}}}

\vspace*{2mm}
\noindent
An interesting example in which these concepts are realized, is provided by the 
``Littlest Higgs Model'' \cite{91A,91B}. The model is formulated as a 
non-linear sigma model with a global $SU(5)$ symmetry group. This group is 
broken down to $SO(5)$ by the non-zero vacuum expectation value
\beq
\Sigma_0 = crossdiag\,[{\scriptstyle '}\!\mathbb{I},1,
{\scriptstyle '}\!\mathbb{I}]
\eeq
of the $\Sigma$ field. Assuming the subgroup $[SU(2) \times U(1)]^2$ to be 
gauged, the global symmetry breaking leads also to the breaking of this gauge 
group down to $[SU(2) \times U(1)]$. The global symmetry breaking generates 
$24 - 10 = 14$ Goldstone bosons, four of which are absorbed by the gauge 
bosons associated with the broken gauge group. The remaining 10 Goldstone 
bosons, incorporated in the $\Sigma$ field
\beq
\Sigma = \exp[2i\Pi/f]: \quad \Pi = \left|\left| 
\begin{array}{ccc} 0 & h^\dagger/\sqrt{2} & 
\varphi^\dagger \\
h/\sqrt{2} & 0 & h^*/\sqrt{2} \\
\varphi & h^{\mathrm{T}}/\sqrt{2} & 0 
\end{array} \right|\right|
\eeq
are identified as an iso-doublet $h$ that becomes the light Higgs field 
of the Standard Model, and a Higgs triplet $\varphi$ that acquires a mass 
of order $M$.

\vspace*{2mm}
The main construction principles of the model should be illustrated by 
analyzing the gauge and the Higgs sector qualitatively. The top sector, 
extended by a new heavy $[T_L, T_R]$ doublet, can be treated in a similar way 
after introducing the appropriate top-Higgs interactions.

\vspace*{4mm}
\noindent
{\it Vector Boson Sector:}

\vspace*{2mm}
\noindent
Inserting the $[SU(2) \times U(1)]^2$ gauge fields into the sigma Lagrangian, 
\beq
{\cal L} = \frac{1}{2} \frac{f^2}{4} \mathrm{Tr} | {\cal D}_\mu \Sigma |^2
\eeq 
with 
\beq
{\cal D}_\mu \Sigma = \partial_\mu \Sigma - i \sum_{j=1}^2 
[ g_j (W_j \Sigma + \Sigma W_j^\mathrm{T}) + \{U(1)\} ] \,,
\eeq
the four vector bosons of the broken $[SU(2) \times U(1)]$ gauge symmetry 
acquire masses
\beq
M[W_H,Z_H,A_H] \sim g f
\eeq
where $W_H$ etc. denote the heavy electroweak gauge fields.

\vspace*{2mm}
Remarkably, the $W_H$ gauge bosons couple with the opposite sign to the square 
of the light Higgs boson compared with the standard $W$ bosons:
\beq
{\cal L} &=& + \frac{g^2}{4} W^2 \,\mathrm{Tr} h^\dagger h \nonumber \\
&& - \frac{g^2}{4} W_H^2 \,\mathrm{Tr} h^\dagger h + ... \; .
\eeq
The quadratic divergences of the two closed $W$ and $W_H$-loop diagrams 
attached to the 
light Higgs field, 
therefore cancel each other and, similarly to supersymmetric 
degrees of freedom, the new vector bosons should have masses not exceeding 1 
to 3 TeV to avoid excessive fine tuning.

\vspace*{2mm}
The Standard Model gauge bosons remain still massless at this point; they 
acquire non-zero masses after the standard electroweak breaking mechanism is in 
operation.

\vspace*{4mm}
\noindent
{\it Higgs Sector:}

\vspace*{2mm}
\noindent
Up to this level of the evolution of the theory, the global symmetries prevent 
a non-zero Higgs potential. Only if radiative corrections are switched on, the 
Coleman-Weinberg mechanism generates the Higgs potential that endows the Higgs 
bosons with masses and breaks the gauge symmetry of the Standard Model.

\vspace*{2mm}
Casting the Higgs potential into the form
\beq
V = m_\varphi^2 \,\mathrm{Tr} \varphi^\dagger \varphi - \mu^2 h h^\dagger + 
\lambda_4 (h h^\dagger)^2
\eeq
the first term provides a non-zero mass to the $\varphi$ Higgs boson while the 
next two terms are responsible for the symmetry breaking in the gauge sector 
of the Standard Model. 

\vspace*{2mm}
\noindent
-- Cutting-off the quadratically divergent contributions to the 
Coleman-Weinberg potential at $\Lambda$, the masses squared of the [now] 
pseudo-Goldstone bosons $\varphi$ are of the order
\beq
m_\varphi^2 \sim g^2 (\Lambda / 4 \pi )^2 \sim g^2 f^2 \;.
\eeq
Thus the heavy Higgs bosons acquire masses of the size of the heavy vector 
bosons.

\vspace*{2mm}
\noindent
-- The quartic coupling of the light Higgs boson is of order $g^2$. The 
coefficient $\mu^2$ however receives contributions only from one-loop 
logarithmically divergent and two-loop quadratically divergent parts in the 
Coleman-Weinberg potential:
\beq
\mu^2 = \mu_1^2 + \mu_2^2 : && \mu_1^2 \sim (\Lambda/4 \pi)^2 \log
\left(\Lambda^2/f^2\right)/16\pi^2 
\sim f^2 \log \left(\Lambda^2/f^2\right)/16\pi^2 \nonumber\\[0.3cm]
&& \mu_2^2 \sim \Lambda^2/(16\pi^2)^2 \sim f^2/16\pi^2 \;.
\eeq
Both contributions are naturally of the order $f/4\pi$, i.e. they are an order 
of magnitude smaller than the intermediate scale $M$ of the heavy Higgs and 
vector masses. \\

Thus, a light Higgs boson with mass of order 100 GeV can be 
generated in 
Little Higgs models as a pseudo-Goldstone boson associated with the
spontaneous breaking of new strong interactions. 
The light mass is protected against 
large radiative corrections individually in the boson and the fermion sectors. 

\vspace*{4mm}
\noindent
{\it{\underline{Phenomenology}}}

\vspace*{2mm}
\noindent
Such scenarios give rise to many predictions that can be checked 
experimentally.
Foremost, the spectrum of new heavy vector bosons and fermions should be 
observed with masses in the intermediate range of 1 to a few TeV at the LHC 
or TeV/multi-TeV $e^+e^-$ linear colliders.
Extensions beyond the minimal version may generate additional scalars with
a strong impact also on the spectrum of the light Higgs sector.

\begin{figure}[hbt]
\begin{center}
\epsfig{figure=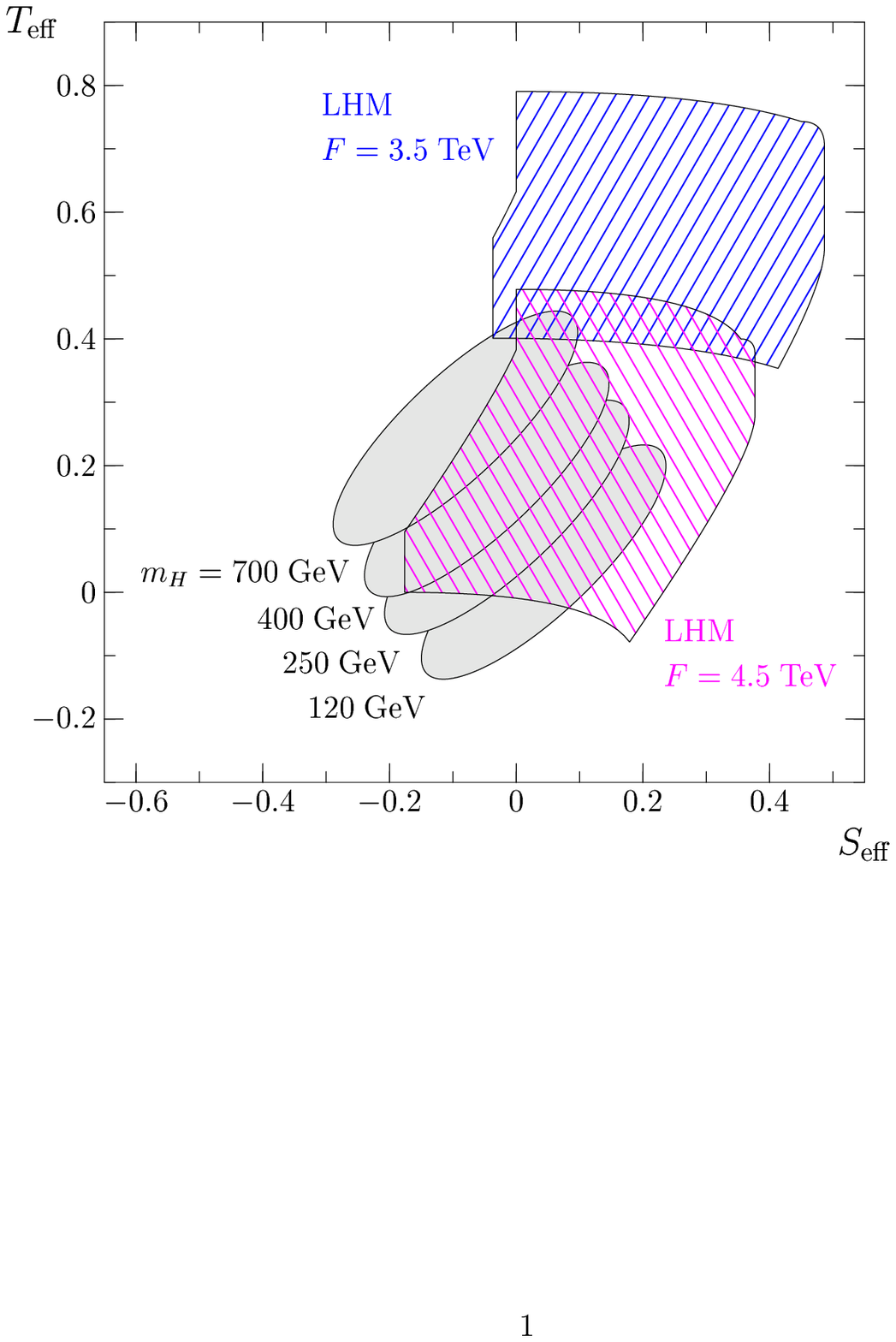,bbllx=114,bblly=296,bburx=451,bbury=620,width=10cm,clip=}
\end{center}
\vspace*{-0.4cm}

\caption[]{\label{fig:kilian} \it Predictions of the $S,T$ precision parameters
for the Littlest Higgs 
Model with standard $U(1)$ charge assigments. The shaded ellipses are the 
68 \% exclusion contours which follow from the electroweak precision data, 
assuming four different Higgs masses. The hatched areas are the allowed 
parameter ranges of the Littlest Higgs Model for two different values of the 
scale $F$. The limits from contact interactions have been taken into account.
Ref. \cite{kilfigure}.}
\end{figure}

\vspace*{2mm}
However, the model can already be checked by analyzing existing precision 
data from LEP and elsewhere. The impact of the new degrees of freedom on the 
Little Higgs models must be kept small enough not to spoil the success of the 
radiative corrections including just the light Higgs boson in the description 
of the data. This leads to a constraint of order 3 to 5 TeV on the parameter 
$f$, Fig.~\ref{fig:kilian}. Thus the theory is compatible with present 
precision data, but only marginally and the overlap is narrow. \\

Extensions of the system in which a new parity \cite{Tpar} is introduced, 
$T$-parity, reduce the radiative corrections to closed loops of the new
degrees of freedom so that lighter new particles are not excluded. 
However, potential anomalies \cite{HillHill} which break $T$-invariance
must be avoided in designing such scenarios.

\vspace*{4mm}
\subsection{Strongly Interacting $W$ Bosons}

\vspace*{2mm}
\noindent
In alternative scenarios \cite{2}
a system of novel 
fermions is introduced, which interact
strongly at a scale of order 1 TeV. In
the ground state of such a system a scalar
condensate of fermion--antifermion pairs
may form. Such a process is  generally
expected to be realized in any non-Abelian gauge theory
of the novel strong interactions [as in QCD]. 
Since the scalar condensate breaks the chiral
symmetry of the fermion system, Goldstone
fields will form, and they are absorbed
by the electroweak gauge fields to build
up the longitudinal components and the
masses of the gauge fields. Novel gauge
interactions must be introduced, which
couple the leptons and quarks of the
Standard Model to the new fermions in order
to generate lepton and quark masses
by interactions with the ground-state
fermion--antifermion condensate. In the
low-energy sector of the electroweak theory, 
the fundamental Higgs-field approach and
the dynamical alternative are effectively equivalent.
However, the two theories are quite
different at high energies. While the 
unitarity of the electroweak gauge theory
is guaranteed by the exchange of the scalar
Higgs particle in scattering processes, 
unitarity is restored in the dynamical
theory at high energies by the
non-perturbative strong interactions
between the particles. Since the longitudinal
gauge field components are equivalent to the
Goldstone fields associated with the microscopic
theory, their strong interactions at high
energies are transferred to the electroweak
gauge bosons. By unitarity, the $S$-wave scattering
amplitude of longitudinally polarized $W, Z$ bosons in the 
isoscalar channel $(2W^+W^- + ZZ) / \sqrt{3}$, 
$a^0_0 = \sqrt{2} G_F s/ 16 \pi$, is bounded by
1/2, so that the characteristic scale of the new strong interactions
must be close to 1.2 TeV. Thus near the critical energy of
1 TeV the $W, Z$ bosons interact strongly with each other.
Technicolor-type theories provide an elaborate form 
of such scenarios.

\vspace*{4mm}
\subsubsection{Theoretical Basis}

\vspace*{2mm}
\noindent
{\bf 1.$\,$} Physical scenarios of dynamical
symmetry breaking may be based on  new
strong interaction theories, which extend
the  spectrum of matter particles and of the interactions 
beyond the degrees of freedom realized in the
Standard Model. If the new strong interactions are 
invariant under transformations of a
chiral $SU(2) \times SU(2)$
group, the chiral invariance is generally  broken
spontaneously down to the diagonal custodial isospin group
$SU(2)$. This process is associated with the
formation of a chiral condensate in the
ground state and the existence of three
massless Goldstone bosons. \\[-1mm]

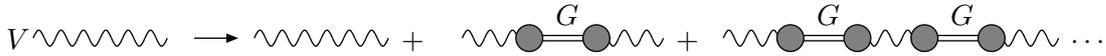
\begin{figure}[hbt]
\begin{center}
\begin{picture}(60,10)(90,30)
\Photon(0,25)(50,25){3}{6}
\LongArrow(60,25)(75,25)
\put(-10,21){$V$}
\end{picture}
\begin{picture}(60,10)(70,30)
\Photon(0,25)(50,25){3}{6}
\put(55,21){$+$}
\end{picture}
\begin{picture}(60,10)(55,30)
\Photon(0,25)(25,25){3}{3}
\Photon(50,25)(75,25){3}{3}
\Line(25,24)(50,24)
\Line(25,26)(50,26)
\GCirc(25,25){5}{0.5}
\GCirc(50,25){5}{0.5}
\put(80,21){$+$}
\put(35,30){$G$}
\end{picture}
\begin{picture}(60,10)(20,30)
\Photon(0,25)(25,25){3}{3}
\Photon(50,25)(75,25){3}{3}
\Photon(100,25)(125,25){3}{3}
\Line(25,24)(50,24)
\Line(25,26)(50,26)
\Line(75,24)(100,24)
\Line(75,26)(100,26)
\GCirc(25,25){5}{0.5}
\GCirc(50,25){5}{0.5}
\GCirc(75,25){5}{0.5}
\GCirc(100,25){5}{0.5}
\put(130,21){$\cdots$}
\put(35,30){$G$}
\put(85,30){$G$}
\end{picture}
\end{center}
\caption[]{\label{fg:gaugemass} \it Generating  gauge-boson masses (V)
 through the interaction with the Goldstone bosons (G).}
\end{figure}

\vspace*{1mm}
The Goldstone bosons can be absorbed by
the gauge fields, generating longitudinal
states and non-zero masses of the gauge bosons, as shown in
Fig.~\ref{fg:gaugemass}. Summing up the geometric series
of vector-boson--Goldstone-boson transitions
in the propagator leads to a shift of the  
mass pole:
\begin{eqnarray}
\frac{1}{q^2} & \to & \frac{1}{q^2} + \frac{1}{q^2} q_\mu \frac{g^2 F^2/2}{q^2}
q_\mu \frac{1}{q^2} + \frac{1}{q^2} \left[ \frac{g^2 F^2}{2} \frac{1}{q^2}
\right]^2 + \cdots \nonumber \\
& \to & \frac{1}{q^2-M^2} \;.
\end{eqnarray}
The coupling between gauge fields and
Goldstone bosons has been defined as $ig F/\sqrt{2} q_\mu$.
The mass generated for the gauge field is related
to this coupling by
\begin{equation}
M^2 = \frac{1}{2} g^2 F^2 ~.
\end{equation}
The numerical value of the coupling $F$ must coincide
with $v/\sqrt{2}=174$ GeV.

\vspace*{2mm}
The remaining custodial $SU(2)$ symmetry
guarantees that the $\rho$
parameter, the relative strength (squared) of the
neutral and charged current couplings, is one. Denoting the $W/B$ mass
matrix elements by
\begin{equation}
\begin{array}{rclcrcl}
\langle W^i | {\cal M}^2 | W^j \rangle & = & \displaystyle \frac{1}{2} g^2
F^2 \delta_{ij}
& \hspace*{1cm} & \langle W^3 | {\cal M}^2 | B \rangle & = & \langle B |
{\cal M}^2 | W^3 \rangle \\ \\
\langle B | {\cal M}^2 | B \rangle & = & \displaystyle \frac{1}{2} g'^2 F^2 &
& & = & \displaystyle \frac{1}{2} gg' F^2
\end{array}
\end{equation}
the universality of the coupling $F$ leads
to the ratio  $M_W^2/M_Z^2 = g^2/(g^2+g'^2) =
\cos^2\theta_W$ of the mass eigenvalues, equivalent to $\rho=1$.

\vspace*{2mm}
\noindent
{\bf 2.$\,$} Since the wave functions of longitudinally
polarized vector bosons grow with the
energy, the longitudinal field components are
the dominant degrees of freedom at high
energies. These states can, however, 
for asymptotic energies be identified with the
absorbed Goldstone bosons. This equivalence \cite{75} is apparent
in the 't Hooft--Feynman gauge where, for
asymptotic energies, 
\begin{equation}
\epsilon_\mu^L W_\mu \to k_\mu W_\mu \sim M^2 \Phi ~. 
\end{equation}
The dynamics of gauge bosons can therefore be
identified at high energies with the
dynamics of scalar Goldstone fields. An
elegant representation of the Goldstone
fields $\vec{G}$ in this context is provided by
the exponentiated form
\begin{equation}
U = \exp [-i \vec{G} \vec{\tau}/v ] ~, 
\end{equation}
which corresponds to an $SU(2)$ matrix field.

\vspace*{2mm}
The Lagrangian of a system of strongly interacting
 bosons  consists in such a scenario 
of the Yang--Mills part ${\cal L}_{YM}$
and the interactions ${\cal L}_G$
of the Goldstone fields, 
\begin{equation}
{\cal L}={\cal L}_{YM}+{\cal L}_G  \;.
\end{equation}
The Yang--Mills part is written in the
usual form ${\cal L}_{YM} = -\frac{1}{4} {\rm Tr} [W_{\mu\nu} W_{\mu\nu} +
B_{\mu\nu} B_{\mu\nu} ]$.  
The interaction of the Goldstone fields can be systematically expanded
in chiral theories 
in the derivatives of the
fields, corresponding to expansions in
powers of the energy for scattering
amplitudes \cite{76}:
\begin{equation}
{\cal L}_G = {\cal L}_0 + \sum_{dim=4} {\cal L}_i + \cdots \;.
\end{equation}
Denoting the SM covariant derivative of
the Goldstone fields by
\begin{equation}
D_\mu U = \partial_\mu U - i g W_\mu U + i g' B_\mu U  
\end{equation}
the leading term ${\cal L}_0$, which is  
of dimension = 2, is given by
\begin{equation}
{\cal L}_0 = \frac{v^2}{4} {\rm Tr} [ D_\mu U^+ D_\mu U ] \;.
\end{equation}
This term generates the masses of the $W,Z$
gauge bosons: $M_W^2 = \frac{1}{4} g^2 v^2$ and
$M_Z^2 = \frac{1}{4} (g^2+g'^2) v^2$.
The only parameter in this part of the
interaction is $v$, which however is fixed
uniquely by the experimental value of the
$W$ mass; thus the amplitudes predicted by
the leading term in the chiral expansion
can effectively be considered as parameter-free.

\vspace*{2mm}
The next-to-leading component in the expansion with
dimension = 4 consists of ten individual terms. If the
custodial $SU(2)$ symmetry is imposed, only two
terms are left, which do not affect propagators
and 3-boson vertices but only 4-boson vertices.
Introducing the vector field $V_\mu$ by 
\begin{equation}
V_\mu = U^+ D_\mu U
\end{equation}
these two terms are given by the interaction
densities
\begin{equation}
{\cal L}_4  =  \alpha_4 \left[Tr V_\mu V_\nu \right]^2 \hspace*{0.5cm}
\mbox{and} \hspace*{0.5cm}
{\cal L}_5  =  \alpha_5 \left[Tr V_\mu V_\mu \right]^2 \;.
\end{equation}

The two coefficients $\alpha_4,\alpha_5$
are parameters which characterize the 
underlying microscopic theory. In phenomenological
approaches they must be adjusted
experimentally from $WW$ scattering data.

Higher orders in the chiral expansion give
rise to an energy expansion of the scattering
amplitudes of the form ${\cal A} = \sum c_n (s/v^2)^n$.
This series  will diverge at energies for which
the resonances of the new strong interaction
theory can be formed in $WW$ collisions: $0^+$ `Higgs-like', 
$1^-$ `$\rho$-like' resonances, etc. The masses of these resonance
states are expected in the range $M_R \sim 4\pi v$ 
where chiral loop expansions diverge,
i.e. between about 1 and 3 TeV.

\vspace*{4mm}
\subsubsection{An Example: Technicolor-Type Theories}

\vspace*{2mm}
A simple example for such scenarios is provided by technicolor-type theories, 
see e.g. Ref.~\cite{94A}. They 
are built on a pattern similar to QCD but characterized by a scale 
$\Lambda_{TC}$ in the TeV range so that the interaction becomes strong
already at short distances of order $10^{-17}$~cm. 

\vspace*{2mm}
The basic degrees of freedom in the simplest version are a chiral set 
$[(U,D)_L;U_R,D_R]$ of massless fermions that interact with technicolor gauge 
fields. The chiral $SU(2)_L\times SU(2)_R$ symmetry of this theory is broken 
down to the diagonal $SU(2)_{L+R}$ vector symmetry by the formation of 
$\langle\bar{U} U\rangle =\langle\bar{D} D\rangle = {\cal O}(\Lambda^3_{TC})$ 
vacuum condensates. 
The breaking of the chiral symmetry generates three massless Goldstone bosons
$\sim \bar{Q} i \gamma_5 \stackrel{\to}{\tau} Q$, that can be 
absorbed by the gauge fields of the Standard Model to build the massive 
states with $M_W \sim 100$~GeV. From the chain
\beq
M_W = g F /\sqrt{2} \quad \mathrm{and} \quad F \sim \Lambda_{TC} / 4 \pi
\eeq 
the parameter $F$ is estimated to be 
below 1 TeV while $\Lambda_{TC}$ should be 
in the TeV range. 

\vspace*{2mm}
While the electroweak gauge sector can be formulated consistently in this 
picture, generating fermion masses leads to severe difficulties. Since gauge 
interactions couple only left-left and right-right field components, a 
helicity-flip left-right mass operator $\bar{f}_L f_R$ 
is not generated
for the fermions of 
the Standard Model. To solve this problem, new gauge 
interactions between the SM and TC fermions must be introduced 
[Extended Technicolor] so that the helicity can flip through the ETC 
condensate in the vacuum. The SM masses predicted this way are of order 
$m_f \sim g^2_E \Lambda^3_{ETC}/M_E^2$ with $g_E$ being the coupling in the 
extended technicolor gauge theory and $M_E$ the mass of the ETC gauge fields.
However, estimates of $M_E$ lead to a clash if one tries to reconcile the size
of the scale needed for generating the top mass, order TeV, with the 
suppression of flavor-changing  processes, like $K\bar{K}$ oscillations, 
which require a size of order PeV. 

\vspace*{2mm}
Thus, the simplest realization of the technicolor theories suffers from 
internal conflicts in the fermion sector. More involved theoretical models
are needed to reconcile these conflicting estimates \cite{94A}.
Nevertheless, the idea of generating 
electroweak symmetry breaking dynamically, is a theoretically attractive and 
interesting scenario. 

\vspace*{4mm}
\subsection{$WW$ Scattering at High-Energy Colliders}

\vspace*{2mm}
\noindent
{\bf 1.$\,$} Independently of specific realizations of dynamical symmetry breaking, 
theoretical tools have been developed which can serve to investigate these 
scenarios quite generally. The (quasi-) elastic 2--2 $WW$ scattering
amplitudes can be expressed at high
energies by a master amplitude
$A(s,t,u)$, which depends on the three
Mandelstam variables of the scattering processes:
\begin{eqnarray}
A(W^+ W^- \to ZZ) & = & A(s,t,u) \\
A(W^+ W^- \to W^+ W^-) & = & A(s,t,u) + A(t,s,u) \nonumber \\
A(ZZ \to ZZ) & = & A(s,t,u) + A(t,s,u) + A(u,s,t) \nonumber \\
A(W^- W^- \to W^- W^-) & = & A(t,s,u) + A(u,s,t)  \nonumber \;.
\end{eqnarray} 

\vspace*{2mm}
To lowest order in the chiral expansion, ${\cal L} \to {\cal L}_{YM} +
{\cal L}_0$, the master amplitude is given, in a
parameter-free form, by the energy squared $s$:
\begin{equation}
A(s,t,u) \to \frac{s}{v^2} ~.
\end{equation}
This representation is valid for energies $s \gg M_W^2$
but below the new resonance region, i.e. in
practice at energies $\sqrt{s}={\cal O}(1~\mbox{TeV})$.
Denoting the scattering length for the
channel carrying isospin $I$ and angular
momentum $J$ by $a_{IJ}$,
the only non-zero scattering channels
predicted by the leading term of the
chiral expansion correspond to
\begin{eqnarray}
a_{00} & = & + \frac{s}{16\pi v^2} \\
a_{11}   & = & + \frac{s}{96\pi v^2} \nonumber \\
a_{20}   & = & - \frac{s}{32\pi v^2} ~.
\end{eqnarray}
While the exotic $I=2$ channel is repulsive,
the $I=J=0$ and $I=J=1$ channels are attractive,
indicating the formation of non-fundamental
Higgs-type and $\rho$-type resonances.

Taking into account the next-to-leading terms
in the chiral expansion, the master amplitude
turns out to be \cite{24}
\begin{equation}
A(s,t,u) = \frac{s}{v^2} + \alpha_4 \frac{4(t^2+u^2)}{v^4}
+ \alpha_5 \frac{8s^2}{v^4} + \cdots ~,
\end{equation}
including the two parameters $\alpha_4$ and $\alpha_5$

\vspace*{2mm}
Increasing the energy, the
amplitudes will approach the resonance area.
There, the chiral character of the
theory does not provide any more guiding principle
for constructing  the scattering
amplitudes. Instead, {\it ad-hoc} hypotheses must
be introduced to define the nature of the
resonances; see e.g. Ref. \cite{24a}. \\[2mm]
 
\begin{figure}[hbt]
\begin{center}
\begin{picture}(60,50)(140,0)
\ArrowLine(0,50)(25,50)
\ArrowLine(25,50)(75,50)
\ArrowLine(0,0)(25,0)
\ArrowLine(25,0)(75,0)
\Photon(25,50)(45,30){-3}{3}
\Photon(25,0)(45,20){3}{3}
\Photon(55,20)(75,10){-3}{3}
\Photon(55,30)(75,40){3}{3}
\GBox(45,20)(55,30){0.5}
\put(-20,48){$q/e$}
\put(-20,-2){$q/e$}
\put(20,8){$W$}
\put(20,30){$W$}
\put(80,35){$W$}
\put(80,5){$W$}
\put(110,35){$I=0,2~~J~\mbox{even}$}
\put(110,15){$I=1~~~~~J~\mbox{odd}$}
\end{picture}
\begin{picture}(60,50)(-20,0)
\ArrowLine(25,25)(0,50)
\ArrowLine(0,0)(25,25)
\Photon(25,25)(75,25){3}{5}
\Photon(75,25)(100,50){3}{4}
\Photon(75,25)(100,0){3}{4}
\Photon(125,0)(175,0){3}{5}
\Photon(125,50)(175,50){3}{5}
\GBox(100,-5)(125,55){0.5}
\put(-15,48){$e^+$}
\put(-15,-2){$e^-$}
\put(70,2){$W^-$}
\put(70,40){$W^+$}
\put(180,48){$W^+$}
\put(180,-2){$W^-$}
\put(145,22){$I=J=1$}
\end{picture}
\end{center}
\caption[]{\label{fg:qqtoqqww} \it $WW$ scattering and rescattering 
at high energies at the LHC
and TeV $e^+e^-$ linear colliders.}
\end{figure}
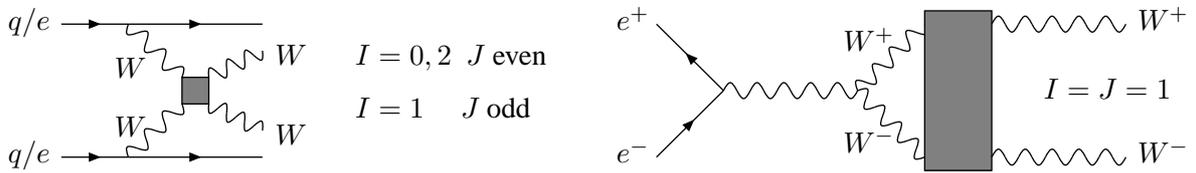

\begin{figure}[hbtp]
\begin{center}
\vspace*{-9.5cm}

\hspace*{-3.5cm}
\epsfig{file=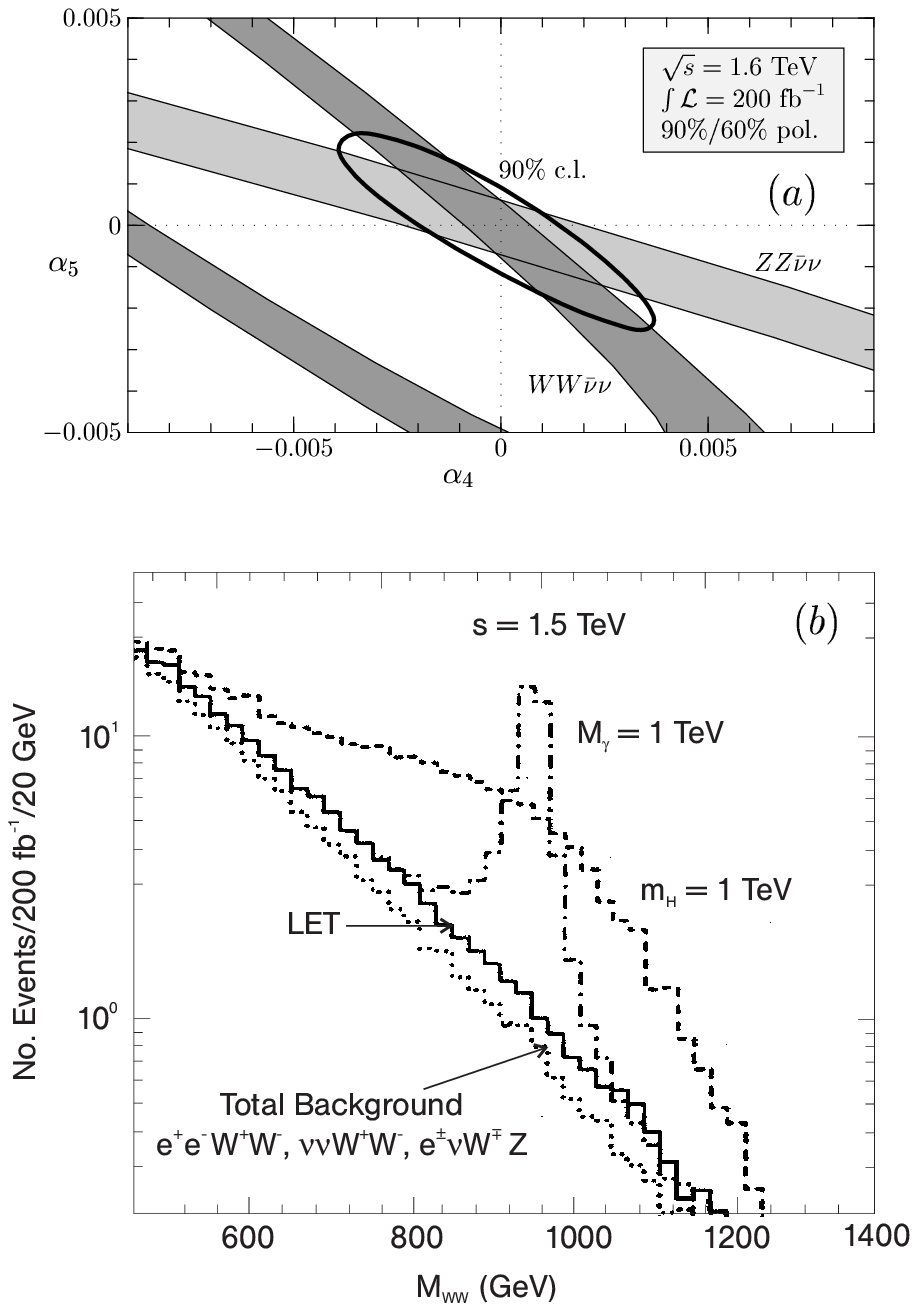,width=22.0cm,angle=0}
\vspace*{-8.5cm}

\end{center}
\caption[]{\it
  Upper part: Sensitivity to the expansion
  parameters in chiral electroweak models of $WW \to WW$ and $WW \to
  ZZ$ scattering at the strong-interaction threshold;
  Ref. \protect\cite{24}.
Lower part: The distribution of the $WW$ invariant energy in $e^+e^-
  \to \overline{\nu} \nu WW$ for scalar and vector resonance
  models [$M_H, M_V$ = 1 TeV];
  Ref. \protect\cite{24a}. 
\protect\label{17tt}\label{PKB}
}
\end{figure}

\vspace*{1mm}
\noindent
{\bf 2.$\,$} $WW$ scattering can be studied at the LHC
and at TeV $e^+e^-$ linear colliders. At high energies, 
equivalent $W$ beams accompany the quark
and electron/positron beams (Fig.~\ref{fg:qqtoqqww})
in the fragmentation processes $pp\to qq \to qqWW \, etc$ and
$ee\to \nu\nu WW \, etc$; the spectra of the longitudinally
polarized $W$ bosons have been given in Eq.~(\ref{eq:xyz}). 
In the hadronic LHC  environment the final-state 
$WW \; etc$ bosons can only be observed in 
leptonic and mixed hadronic/leptonic decays.
The clean environment of
$e^+e^-$ colliders will allow the reconstruction 
of resonances from copious $W$ decays to jet pairs.
The results of three experimental 
simulations are displayed in Fig.~\ref{PKB}.
In Fig.~\ref{PKB}a the sensitivity to the parameters
$\alpha_4,\alpha_5$ of the chiral expansion is shown for $WW$
scattering in $e^+e^-$ colliders \cite{24}. The results of this analysis
can be reinterpreted as sensitivity to the
parameter-free prediction of the chiral
expansion, corresponding to an error of
about 10\% in the first term of the master
amplitude $s/v^2$.  These experiments test the basic concept
of dynamical symmetry breaking through
spontaneous symmetry breaking. The production
of a vector-boson resonance of mass $M_V=1$ TeV
is exemplified in Fig.~\ref{PKB}b \cite{24a}. \\

LHC allows the observation of $WW$ resonances in these channels
up to a mass range of about 1.5~TeV \cite{12}. \\

A second powerful method measures the elastic 
$W^+W^- \to W^+W^-$ scattering in the $I=1, J=1$ channel. The
rescattering of $W^+W^-$ bosons produced in $e^+e^-$
annihilation, cf. Fig.~\ref{fg:qqtoqqww}, depends at high
energies on the $WW$ scattering phase $\delta_{11}$ 
\cite{78}. The production amplitude $F = F_{LO} \times R$
is the product of the lowest-order
perturbative diagram with the Mushkelishvili--Omn\`es rescattering amplitude
${\cal R}_{11}$,
\begin{equation}
{\cal R}_{11} = \exp \frac{s}{\pi} \int \frac{ds'}{s'}
\frac{\delta_{11}(s')}{s'-s-i\epsilon} ~,
\end{equation}
which is determined by the $I = J = 1$ $WW$ phase shift $\delta_{11}$.
The power of this method derives from the 
fact that 
the entire $e^+e^-$
collider energy is transferred to the $WW$ system
[while a major fraction of the energy
is lost in the fragmentation of $e \to \nu W$
if the $WW$ scattering is studied in the
process $ee\to \nu\nu WW$]. Detailed simulations \cite{78}
have shown that this process is sensitive
to vector-boson masses up to about $M_V \lessim 6$ TeV in technicolor-type 
theories. 

\vspace*{2mm}
The experimental analysis of the $\alpha$ parameters at the $e^+e^-$ linear 
collider in the first phase with energy up to $\sim 1$~TeV can be 
reinterpreted in the following way. Associating the parameters $\alpha$ 
with new 
strong interaction scales, $\Lambda_\star \sim v/\sqrt{\alpha}$, 
upper bounds on $\Lambda_\star$ 
of $\sim 3$~TeV can be probed
in $WW$ scattering. Thus this 
instrument allows to cover the entire threshold region $\lessim 4\pi v 
\sim 3$~TeV of the new strong interactions. In the $W^+W^-$ production channel 
of $e^+e^-$ collisions a range even up to order 10~TeV can be probed
indirectly. If a new scale $\Lambda_\star$ would be discovered below
$\sim 3$~TeV, novel $WW$ resonances could be searched for at the LHC while 
CLIC could investigate new resonance states with masses up to 3 TeV, and
virtual states even far beyond.

\vspace*{4mm}
\section{Summary}

\vspace*{2mm}
\noindent
The mechanism of electroweak symmetry
breaking can be established in the present
or the next generation of hadron and lepton
colliders:
\begin{itemize}
\item[$\star$] Whether there exists a light fundamental Higgs boson;
\item[$\star$] The profile of the Higgs particle can be
  reconstructed, which reveals the physical
  nature of the underlying mechanism of
  electroweak symmetry breaking;
\item[$\star$] Analyses of strong WW scattering can be
  performed if the symmetry breaking is 
  generated dynamically by novel
  strong interactions.
\end{itemize}
Moreover, depending on the experimental answer to these questions,
the electroweak sector will provide the
platform for extrapolations into physical
areas beyond the Standard Model: either
to the low-energy  supersymmetry sector; or
to a new strong interaction theory at a characteristic 
scale of order 1 TeV and beyond; or, alternatively,
to extra space dimensions.  
 
\vspace*{4mm}
\section*{Acknowledgments}

  P.M.Zerwas is very thankful to the organisers for the invitation 
  to the 4th CERN Latin American School on High Energy Physics,
  Vina del Mar, Chile, 2007. The writing of the report was
  supported in part by the grant PAPIIT-IN115207.

\newpage

\appendix

\vspace*{4mm}
\section{The O(3) $\sigma$ Model}

\vspace*{2mm}
\noindent
A transparent but, at the same time, sufficiently complex model to study all 
the aspects of electroweak symmetry breaking is the O(3) $\sigma$ model. By 
starting from the standard version, in a number of variants it allows to 
develop the idea of spontaneous symmetry breaking and the Goldstone theorem 
while gauging the theory leads to the Higgs phenomenon. This evolution will be
described step by step in the next three subsections. 

\vspace*{2mm}
The O(3) $\sigma$ model includes a triplet of field components:
\beq
\sigma = (\sigma_1,\sigma_2,\sigma_3) \;.
\eeq
If the self-interaction potential of the field depends only on the overall 
field-strength, the theory, described by the Lagrangian
\beq
{\cal L} = \frac{1}{2} (\partial \sigma)^2 - V(\sigma^2)
\eeq
is O(3) rotationally invariant. These iso-rotations are generated by the 
transformation
\beq
\sigma \to e^{i\alpha t}\sigma \quad \mathrm{with} \quad
(t^i)_{jk} = i \epsilon_{ijk}
\eeq
with rotation parameters $\alpha =
(\alpha_1,\alpha_2,\alpha_3)$. Choosing a quartic interaction for the 
potential, the theory is renormalizable and thus well-defined.

\vspace*{4mm}
\subsection{Normative Theory:}

\begin{figure}[hbt]
\begin{center}
\epsfig{figure=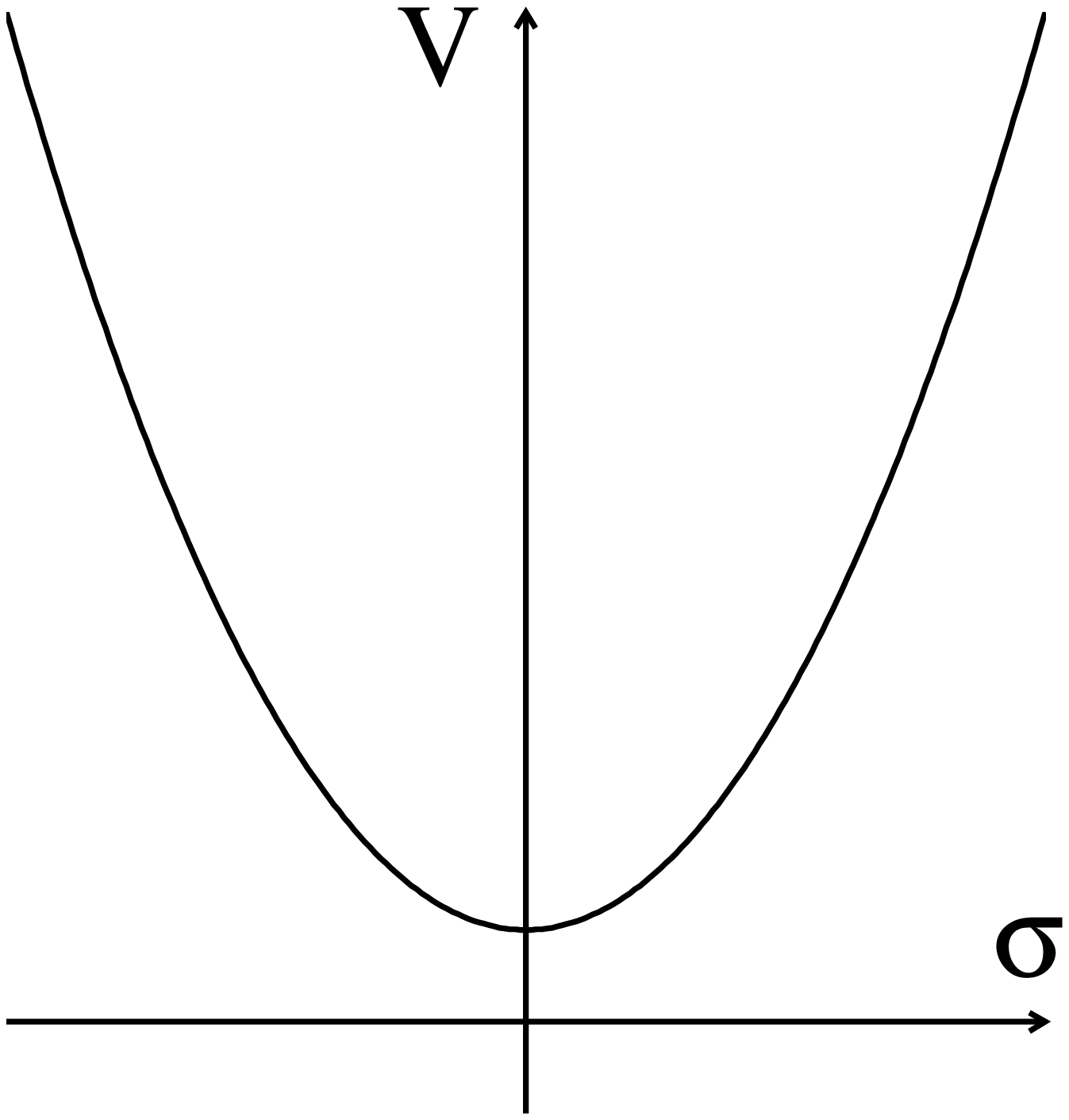,width=4cm}
\end{center}
\vspace*{-0.4cm}
\caption{\it \label{fig:pot} }
\end{figure}

\vspace*{2mm}
\noindent
If the quartic potential $V$ is chosen to be, cf.~Fig.~\ref{fig:pot},
\beq
V(\sigma^2) = \lambda^2 (\sigma^2 + \mu^2)^2
\eeq
the spectrum of particles and the interactions can easily be derived from the 
form
\beq
V(\sigma^2) = 2\lambda^2\mu^2\sigma^2 + \lambda^2 \sigma^4 + \mathrm{const.}
\;.
\eeq

\noindent
The bilinear field-term describes three degenerate masses
\beq
m(\sigma_1) = m(\sigma_2) = m(\sigma_3) = 2\lambda\mu
\eeq
corresponding to three physical particle degrees of freedom. The fields 
interact through the second quartic term. The ground state of the system is 
reached for zero field-strength:
\beq
\sigma^0 = (0,0,0) \;.
\eeq
This theory describes a standard particle system in which the ground state 
preserves the rotational invariance of the Lagrangian. Thus the Lagrangian 
and the solution of the field equation obey the same degree of symmetry.

\vspace*{4mm}
\subsection{Spontaneous Symmetry Breaking and Goldstone Theorem:}

\begin{figure}[hbt]
\begin{center}
\epsfig{figure=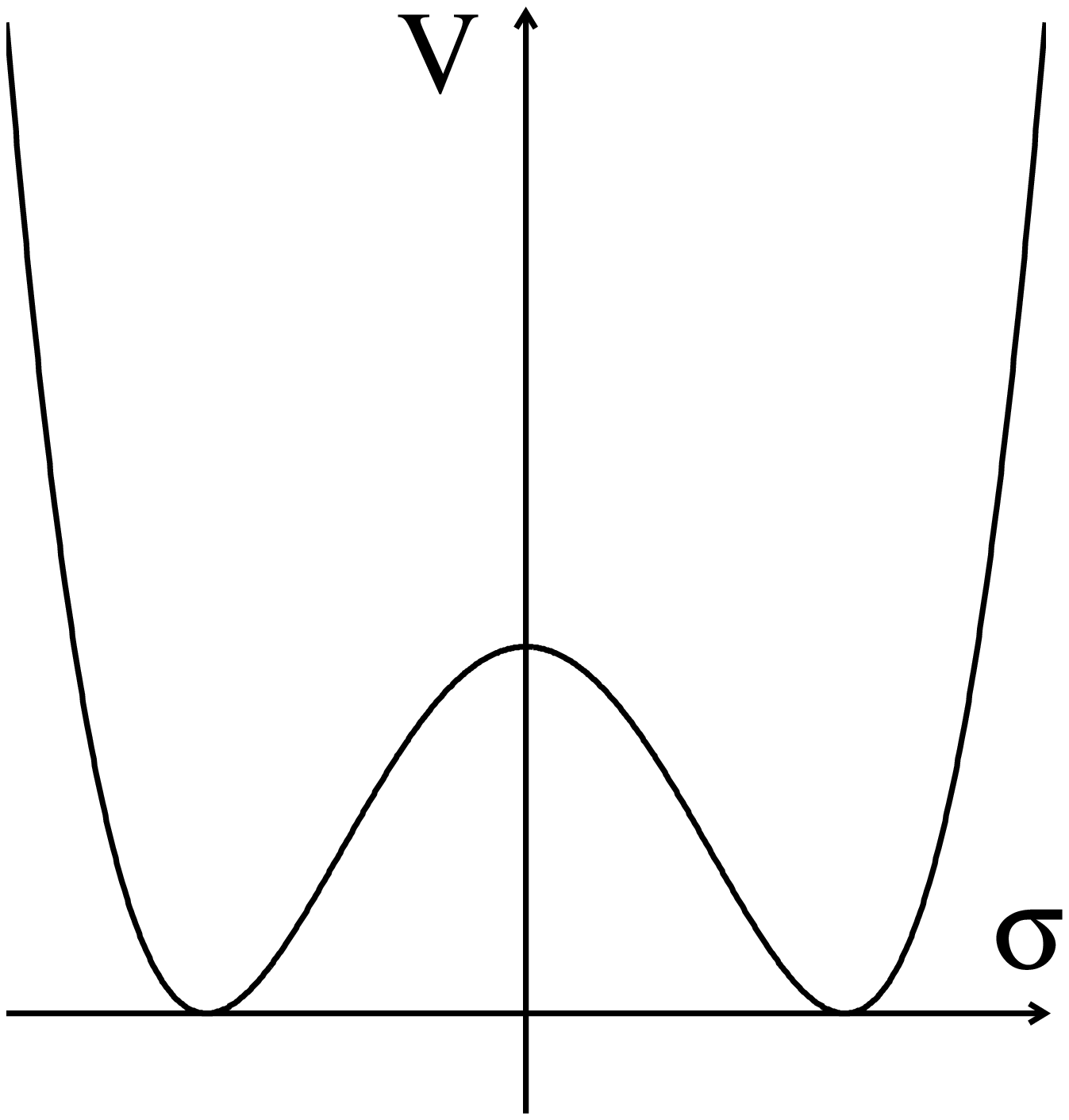,width=4cm}
\end{center}
\vspace*{-0.4cm}
\caption{\it \label{fig:pothiggs} }
\end{figure}

\vspace*{2mm}
\noindent
However, if the sign in the mass parameter in the potential flips to negative 
values,
\beq
V(\sigma^2) = \lambda^2(\sigma^2-\mu^2)^2
\eeq
the ground state is a state of non-zero field strength, 
cf.~Fig.~\ref{fig:pothiggs}. Fixing the axis of the ground state such that
\beq
\sigma^0 = (0,0,v) \quad \mathrm{with} \quad v=\mu
\eeq
the original O(3) rotational invariance of the Lagrangian is not obeyed any 
more by the ground-state solution which singles out a specific direction in 
iso-space. However, no principle determines the arbitrary direction of the 
ground state vector 
in iso-space. Such a phenomenon in which solutions of the field 
equations do not respect the symmetry of the Lagrangian, is generally termed 
``spontaneous symmetry breaking''.\\

Expanding the $\sigma$ field about the ground state,
\beq
\sigma = (\sigma_1', \sigma_2', v+\sigma_3')
\eeq
an effective theory emerges for the new dynamical degrees of freedom 
$\sigma_1', \sigma_2'$ and $\sigma_3'$. Evaluating the potential for the 
new fields,
\beq
V = 4v^2\lambda^2\sigma_3^{'2} + 4 v \lambda^2 \sigma_3' (\sigma_1^{'2} +
\sigma_2^{'2} + \sigma_3^{'2}) + \lambda^2 (\sigma_1^{'2} + \sigma_2^{'2} 
+ \sigma_3^{'2})^2 
\eeq
two massless particles plus one massive particle correspond to the bilinear 
field terms:
\beq
m(\sigma_1')=m(\sigma_2')=0 \nonumber\\
m(\sigma_3')=2\sqrt{2}\lambda v \neq 0 \;.
\eeq
The two massless particles are called Goldstone bosons, Ref.~\cite{appcite}.

\vspace*{2mm}
In addition to the standard quartic terms,
the Goldstone bosons and the massive particle interact with each other through 
trilinear terms in the effective potential.

\vspace*{2mm}
The symmetry of the effective theory is reduced from the original O(3) 
rotational invariance to O(2) invariance restricted to rotations about the 
ground-state axis. 

\vspace*{4mm}
\noindent
This $\sigma$ model is only a simple example of the general 

\vspace*{2mm}
{\it \noindent\underline{Goldstone theorem:}\\[1mm]
\noindent
If N is the dimension of the symmetry group of the basic Lagrangian, but the 
symmetry of the ground-state solution is reduced to M, then the theory 
includes (N-M) massless scalar Goldstone bosons.}

\vspace*{2mm}
For each destroyed symmetry degree of freedom, a massless particle appears in 
the spectrum. A most famous example of this theorem are the three 
nearly massless pions which emerge from spontaneously broken chiral isospin 
symmetry in QCD.

\vspace*{4mm}
\subsection{The Higgs mechanism}

\vspace*{2mm}
\noindent
The Higgs mechanism Ref.~\cite{1} 
provides the vector bosons in gauge theories with masses 
without destroying the renormalizability of the theory. Would masses be 
introduced by hand, the gauge invariance which ensures the renormalizability 
would be destroyed by the {\it ad-hoc} mass terms in the Lagrangian.

\vspace*{2mm}
The global isospin symmetry of the O(3) $\sigma$ model can be extended to a 
local symmetry by introducing an iso-triplet $W$ of gauge fields coupled 
minimally to the $\sigma$ field. Introducing the covariant derivative
\beq
\partial_\mu \sigma \to \partial_\mu \sigma + igtW_\mu \sigma
\eeq
into the Lagrangian,
\beq
{\cal L} = \frac{1}{2} [ (\partial + igtW)\sigma]^2 - V(\sigma^2) +
{\cal L}_{kin}(W)
\eeq
the theory is invariant under the local gauge transformation
\beq
\sigma \to e^{i\alpha t} \sigma \quad \mathrm{with} \quad
\alpha = \alpha(x)
\eeq
with the matter transformation complemented by the usual transformation of 
the non-abelian gauge field. The gauged Lagrangian includes the gauge kinetic 
part, the $\sigma$ kinetic part and the $\sigma$-gauge interaction, as well 
as the potential. 

\begin{itemize}
\item[--] If the $\sigma$ potential is just the standard potential, 
$V = \lambda^2 (\sigma^2 + \mu^2)^2$, the theory is a 
non-abelian Yang-Mills gauge theory with a mass-degenerate triplet of $\sigma$ particles 
interacting in the standard way with the massless $W$ gauge triplet fields.\\

\item[--] However, if the potential is chosen of the Mexican type, 
$V=\lambda^2 (\sigma^2-\mu^2)^2$, which leads in the $\sigma$ model to spontaneous symmetry 
breaking, the physical field/particle content of the theory changes 
dramatically [a phenomenon similar to the non-gauged theory]:
\end{itemize}

Parametrizing the $\sigma$ triplet-field by a rotation of the field about 
the ground-state axis,
\beq
\sigma = e^{i\Theta t/v} (\sigma^0 + \eta)
\eeq
with
\beq
\sigma^0 = (0,0,v)\,;\quad \eta = (0,0,\eta)\,; \quad
\Theta = (\Theta_1,\Theta_2,0)
\eeq
the $\Theta$ components of $\sigma$ perpendicular to the ground-state axis 
can be removed 
by the gauge transformation $\sigma \to exp[-i\Theta t/v] \sigma$ supplemented 
by the corresponding transformation of the gauge field. Keeping the original 
notation for the gauge-transformed fields, the new Lagrangian for the physical 
degrees of freedom is given by 
\beq
{\cal L} = \frac{1}{2} [(\partial + igWt)(\sigma^0 +\eta)]^2 - V([\sigma^0 +
\eta]^2) + {\cal L}_{kin}(W) \;.
\eeq
After writing the resulting Lagrangian of the effective theory as
\beq
{\cal L} = {\cal L}_{kin}(W) + \frac{1}{4} g^2 v^2 ( W_1^2 + W_2^2) + 
\frac{1}{2} (\partial\eta)^2 - V + {\cal L}_{int} (\eta, W)
\eeq
the physical particle/field content becomes manifest:

\begin{itemize}
\item[--] a massless vector field $W_3$ corresponding to the residual
rotational invariance about the ground-state 3-axis;\\

\item[--] two massive $W$ fields $W_1$ and $W_2$ perpendicular to the 
ground-state 
axis with masses determined by the ground-state $\sigma$ field-strength $v$ 
and the gauge coupling $g$. These two massive fields correspond to the 
symmetry degrees of freedom that were broken spontaneously in the non-gauged 
$\sigma$ model;\\

\item[--] the Goldstone bosons have disappeared from the spectrum, absorbed to 
build up the longitudinal degrees of the massive gauge bosons;\\

\item[--] a real scalar Higgs boson $\eta$. 
\end{itemize}

\vspace*{4mm}
\noindent
This example can easily be extended, in parallel to the Goldstone theorem, to 
formulate the general

\vspace*{2mm}
{\it \noindent \underline{Higgs mechanism:} \\[1mm]
\noindent
If N is the dimension of the symmetry group of the original Lagrangian, M the 
dimension of the symmetry group leaving invariant the ground state of the n 
scalar fields, then the physical theory consists of M massless vector fields, 
(N-M) massive vector fields, and n-(N-M) scalar Higgs fields.}

\newpage


\selectlanguage{spanish}
\renewcommand{\tablename}{Tabla}
\decimalpoint

\renewcommand\thefigure{\arabic{figure}}
\setcounter{figure}{0}
\setcounter{section}{0}
\setcounter{equation}{0}
\setcounter{footnote}{0}

\setcounter{table}{0}

\renewcommand{\thesection}{\arabic{section}}
\renewcommand{\thetable}{\arabic{table}}
\renewcommand{\thesubsection}{\thesection.\arabic{subsection}}
\renewcommand{\thesubsubsection}{\thesubsection.\arabic{subsubsection}}
\renewcommand{\theequation}{\arabic{equation}}
\numberwithin{equation}{section}
\numberwithin{table}{section}
\numberwithin{figure}{section}

\selectlanguage{spanish}
\selectlanguage{spanish}
\renewcommand{\tablename}{Tabla}
\renewcommand{\textfraction}{0.13}
\decimalpoint

\thispagestyle{empty}

{\Large\raggedright\noindent \bf\par
Conceptos del Rompimiento de la Simetr\'{\i}a Electrod\'{e}bil  \\[2mm]
y la  F\'{\i}sica del Higgs}\\[2ex]
\begin{enumerate}
      \item[]\normalsize\raggedright
{ \bf M. Gomez-Bock$^1$, M. Mondrag\'on$^2$, M. M\"uhlleitner$^{3,4}$,
M. Spira$^5$, P.M. Zerwas$^{6,7,8}$}
\end{enumerate}
\begin{enumerate}
\item[]\rm
{ $^1$ Inst. de F\'\i sica ``LRT'', Ben\'emerita Univ. Auton. de
Puebla,
72570 Puebla, Pue, M\'exico \\
$^2$ Inst. de F\'\i sica, Universidad Nacional Aut\'onoma de M\'exico,
01000 M\'exico D.F., M\'exico \\
$^3$ Laboratoire d'Annecy-Le-Vieux de Physique Th\'eorique, LAPTH,
Annecy-Le-Vieux, France \\
$^4$ CERN TH Division, CERN, Geneva, Switzerland \\
$^5$ Paul Scherrer Institut, CH-5232 Villigen PSI, Switzerland \\
$^6$ Deutsches Elektronen-Synchrotron DESY, D-22603 Hamburg, Germany \\
$^7$ Inst. Theor. Physik E, RWTH Aachen, D-52074 Aachen, Germany}    \\
$^8$ Laboratoire de Physique Th\'eorique, U. Paris-Sud, F-91405
Orsay, France
\end{enumerate}

\begin{abstract}
Presentamos una introducci\'on a los conceptos b\'asicos del
rompimiento de la simetr\'\i a electrod\'ebil y la f\'\i sica del
Higgs dentro del Modelo Est\'andar y sus extensiones
supersim\'etricas. Se presenta tambi\'en una breve perspectiva
ge\-ne\-ral de mecanismos alternativos del rompimiento de la
simetr\'\i a electrod\'ebil. Adem\'as de las bases te\'oricas, se
discute el estado actual de la f\'\i sica experimental del Higgs y
sus implicaciones para futuros experimentos en el Tevatron, el LHC y
en colisionadores lineales $\epem$.
\end{abstract}

\vspace*{4mm}
\section{Introducci\'on}

\vspace*{2mm}
{\bf 1.$\,$} Revelar el mecanismo f\'{\i}sico
responsable del rompimiento de las simetr\'{\i}as electrod\'ebiles,
es uno de los problemas principales en la F\'{\i}sica de
Part\'{\i}culas. Si las part\'{\i}culas fundamentales del Modelo
Est\'andar - leptones, quarks y bosones de norma (gauge)- siguen
interactuando d\'ebilmente a altas eneg\'{\i}as, potencialmente
cercanas a la escala de Planck, el sector en el cual la
simetr\'{\i}a electrod\'ebil es rota debe contener uno o m\'as
bosones escalares fundamentales de Higgs con masas ligeras del orden
de la escala del rompimiento de la simetr\'{\i}a $v\sim 246$ GeV. Las
masas de las part\'{\i}culas fundamentales son generadas por la
interacci\'on con el campo escalar de Higgs, el cual es diferente de
cero en su estado base \cite{1}.  De manera alternativa, el
rompimiento de la simetr\'\i a podr\'\i a ser generado
din\'amicamente por nuevas fuerzas fuertes caracterizadas por una
escala de interacci\'on $\Lambda \sim 1$ TeV o m\'as alta \cite{2}.
Si las simetr\'\i as globales de las interacciones fuertes son
rotas espont\'aneamente, los bosones de Goldstone asociados pueden
ser absorbidos por los campos de norma (gauge fields), generando las
masas de las part\'\i culas de norma. Las masas de los leptones y
quarks pueden ser generadas a trav\'es de interacciones con un
condensado de fermiones de la teor\'\i a de interacciones fuertes
nuevas. En otros escenarios de interacci\'on fuerte, $\Lambda >
\mathcal{O}$(10 TeV), el espectro de baja energ\'\i a incluye campos
escalares de Higgs \cite{2A} los cuales adquieren masas ligeras como
bosones de pseudo-Goldstone s\'olo por un rompimiento colectivo de
la simetr\'\i a. Otros mecanismos de rompimiento de las simetr\'{\i}as
electrod\'ebiles est\'an asociados con la din\'amica en el espacio
de dimensiones extra a bajas energ\'{\i}as \cite{RI2}. El campo de
Higgs podr\'a ser identificado con la quinta componente de un campo
vectorial en $D=5$ dimensiones, o no se tendr\'a un campo de Higgs
ligero en cuatro
dimensiones.

\vspace*{2mm}
\noindent {\bf 2.$\,$} Un mecanismo simple para
describir el rompimiento de la simetr\'{\i}a electrod\'ebil est\'a
incorporado en el Modelo Est\'andar (SM) \cite{3}. Se introduce un
campo escalar complejo iso-doblete el cual adquiere un valor esperado
del vac\'\i o no nulo a trav\'es de sus auto-interacciones,
rompiendo espont\'aneamente la simetr\'\i a electrod\'ebil SU(2)$_I
\times$ U(1)$_Y$ hasta la simetr\'{\i}a electomagn\'etica
U(1)$_{EM}$.  Las interacciones de los bosones de norma y los
fermiones con el campo de fondo generan las masas de estas
part\'{\i}culas. Una componente del campo escalar no se absorbe en
este proceso, manifiest\'andose como la part\'\i cula f\'\i sica
Higgs $H$.

La masa del bos\'on de Higgs es el \'unico par\'ametro desconocido
en el sector de rompimiento de simetr\'{\i}a del Modelo Est\'andar,
mientras que todos los acoplamientos est\'an fijos por las masas de
las part\'\i culas, una consecuencia del mecanismo de Higgs {\it sui
generis}. Sin embargo, la masa del bos\'on de Higgs est\'a acotada
de dos maneras. Dado que el auto-acoplamiento cu\'artico del campo
de Higgs crece indefinidamente conforme la energ\'{\i}a aumenta, un
l\'{\i}mite superior a la masa del Higgs puede ser derivado
requiriendo que las part\'{\i}culas del SM permanezcan d\'ebilmente
interactuantes a escalas de $\Lambda$ \cite{4}. Por otro lado,
l\'{\i}mites inferiores estrictos a la masa del Higgs provienen de
pedir que el vac\'{\i}o del campo electrod\'ebil sea estable
\cite{5}. Si el Modelo Est\'andar es v\'alido hasta la escala de
Planck, la masa del Higgs del SM est\'a restringido a un intervalo
estrecho entre los $130$ y los $190$ GeV. Para las masas del Higgs
que est\'en fuera de este intervalo, se esperar\'{\i}a que
ocurrieran nuevos fen\'omenos f\'\i sicos a escalas $\Lambda$ entre
$\sim 1$ TeV y la escala de Planck. Para masas del Higgs del orden
de $1$ TeV, la escala de las nuevas interacciones fuertes ser\'{\i}a
tan baja como $\sim 1$ TeV \cite{4,6}.

\begin{figure}[hbt]
\begin{center}
\hspace*{-1.1cm}
\epsfig{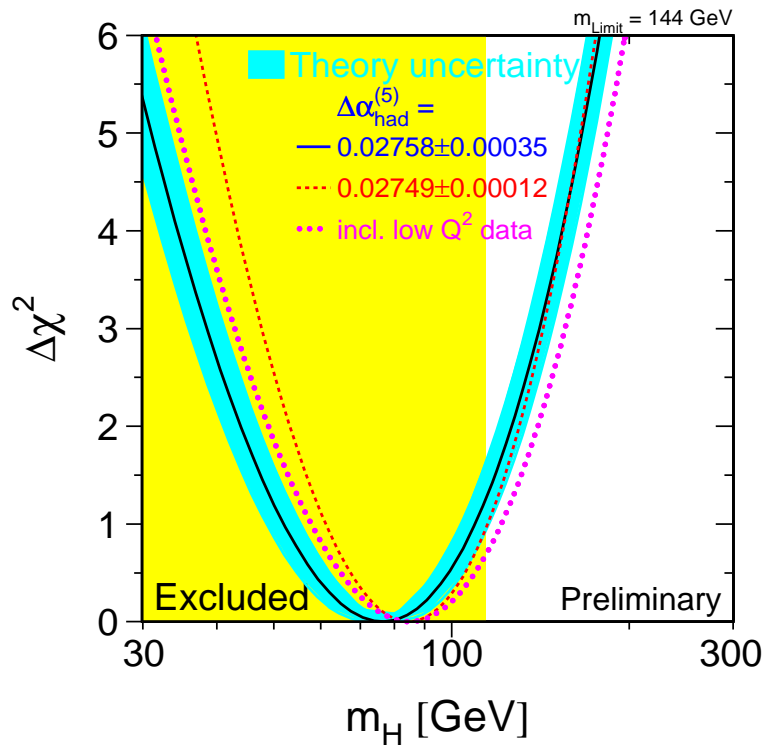}
\end{center}
\vspace*{-0.4cm}

\caption[]{\label{esp-fg:SMHiggs}\it La curva $\Delta\chi^2$
derivada de mediciones electrod\'ebiles de precisi\'on $Q^2$-alta,
realizadas en el LEP, SLD y Tevatron, como funci\'on de la masa del
bos\'on de Higgs en el Modelo Est\'andar.}
\end{figure}

Las observables electrod\'ebiles est\'an afectadas por la masa del
Higgs a trav\'es de correcciones radiativas \cite{7}. A pesar de la
dependencia logar\'{\i}tmica d\'ebil a primer orden, la alta
precisi\'on de los datos electrod\'ebiles, {\it c.f.}
Fig.~\ref{esp-fg:SMHiggs}, indican una preferencia por masas ligeras del
Higgs cercanas a $\sim 100$ GeV \cite{8}. A un 95\% del nivel de
confianza (CL), estos datos requieren un valor para la masa de Higgs
menor que $\sim 144$ GeV. Mediante la b\'usqueda directa de la
part\'\i cula Higgs del SM, los experimentos del LEP han fijado un
l\'{\i}mite inferior de $M_H\gsim 114$ GeV sobre la masa del Higgs
\cite{9}. Dado que el bos\'on de Higgs no ha sido encontrado en
LEP2, la b\'usqueda continuar\'a en el Tevatron, que puede alcanzar
masas de hasta $\sim 140$ GeV \cite{11}. El colisionador de protones
LHC abarca el intervalo can\'onico completo de la masa del Higgs del
Modelo Est\'andar \cite{12,12A}. Mientras en el LHC se dan los
primeros pasos en el an\'alisis de las propiedades de la
part\'\i cula de Higgs, un panorama extenso y de alta
resoluci\'on del mecanismo de Higgs puede establecerse
experimentalmente al llevar a cabo an\'alisis muy precisos en
colisionadores lineales $e^+ e^-$ \cite{13}.

\vspace*{2mm} \noindent {\bf 3.$\,$} Si el Modelo Est\'andar puede
encastrarse en una Teor\'{\i}a de Gran Unificaci\'on (GUT) a altas
energ\'{\i}as, la escala natural del rompimiento de la simetr\'{\i}a
electrod\'ebil se esperar\'{\i}a cercana a la escala de
unificaci\'on $M_{GUT}$. La supersimetr\'{\i}a \cite{14} ofrece una
soluci\'on a este problema de jerarqu\'{\i}as. Una vez que los
par\'ametros fundamentales de la teor\'ia supersim\'etrica y de su
mecanismo de rompimiento son generados a la Tera-escala, las
contribuciones cuadr\'aticamente divergentes a las correcciones
radiativas a la masa del bos\'on escalar de Higgs son canceladas por
la interferencia destructiva entre los lazos (loops) fermi\'onicos y
bos\'onicos en las teor\'{\i}as supersim\'etricas \cite{15}. La
Extensi\'on M\'{\i}nima supersim\'etrica del Modelo Est\'andar
(MSSM) prove\'e un ejemplo ilustrativo para derivar una teor\'{\i}a
a la Tera-escala a partir de una teor\'{\i}a de gran unificaci\'on
supersim\'etrica. Una fuerte indicaci\'on para la realizaci\'on de
las componentes b\'asicas de este esquema f\'\i sico en la
naturaleza es el excelente acuerdo entre el valor del \'angulo de
mezcla electrod\'ebil $\sin^2 \theta_W$ predicho por la
unificaci\'on de los acoplamientos de norma y el valor medido
experimentalmente.  Si los acoplamientos de norma est\'an unificados
en la Teor\'{\i}a M\'{\i}nima Supersim\'etrica a una escala $M_{GUT}
= {\cal O}(10^{16}~\mbox{GeV})$, el valor predicho del \'angulo de
mezcla electrod\'ebil es $\sin^2\theta_W = 0.2336 \pm 0.0017$
\cite{16} para el espectro de masas de las part\'{\i}culas
supersim\'etricas del orden de $M_Z$ a $1$ TeV.  Esta predicci\'on
te\'orica concuerda muy bien con el resultado experimental
$\sin^2\theta_W^{exp} = 0.23153 \pm 0.00016$ \cite{8}; la diferencia
entre los dos n\'umeros es menor que el 2 por mil.

En el MSSM, el sector de Higgs se establece mediante dos dobletes de
Higgs \cite{17}. Es necesario que sean dos para poder generar las
masas de los fermiones de tipo up y down en una teor\'{\i}a
supersim\'etrica y convertirla en una teor\'{\i}a sin anomal\'{\i}as. El
espectro de part\'{\i}culas de Higgs consiste en un quinteto de
estados: dos escalares neutrales $\mathcal{CP}$-pares ($h,H$), un
neutral pseudoescalar $\mathcal{CP}$-impar ($A$), y un par de bosones
de Higgs cargados ($H^{\pm}$) \cite{19}.  Se espera que la masa de los
bosones pesados de Higgs $H,A,H^\pm$ sean del orden de $v$, pero se
podr\'\i an extender hasta el rango de TeV. Por contraste, dado que los
acoplamientos propios cu\'articos de Higgs est\'an determinados por los
acoplamientos de norma, la masa del bos\'on de Higgs m\'as ligero $h$ est\'a
restringida muy rigurosamente.  A nivel \'arbol, la masa ha sido
predicha a ser menor a la masa del $Z$ \cite{19}. Las correcciones
radiativas, que aumentan como la cuarta potencia de la masa del top,
recorren el l\'\i mite superior a un valor entre $\sim 100$ GeV y $\sim 140$
GeV, dependiendo del par\'ametro $\tgb$, que es la raz\'on de los valores
esperados de los dos campos escalares neutrales de Higgs.

Las extensiones supersim\'etricas de la teor\'{\i}a  m\'as all\'a de
la forma m\'{\i}nima pueden ser motivadas por peque\~nos problemas
de precisi\'on fina al acomodar los datos experimentales observados
para la masa del bos\'on $Z$. La expansi\'on del sector de Higgs
introduce nuevos acoplamientos los cuales aumentan con la
escala de energ\'{\i}a. En paralelo con el Modelo Est\'andar, la
masa del bos\'on de Higgs m\'as ligero est\'a
acotada \cite{EspinQuiros} a un valor menor que alrededor de 200 GeV,
sin embargo, s\'olo si los campos se mantienen d\'ebilmente
interactuantes hasta escalas cercanas a la escala de Planck.

Un l\'{\i}mite bajo general de 91 GeV en teor\'{\i}as
$\mathcal{CP}$-invariantes ha sido establecido experimentalmente
para la part\'{\i}cula de Higgs $h$ por el LEP \cite{9}. La
b\'usqueda de la masa de $h$ en exceso de $\sim 100$ GeV y la
b\'usqueda para el bos\'on pesado de Higgs continuan en el Tevatron,
el LHC y el colisionador lineal $e^+e^-$.

\vspace*{2mm} \noindent {\bf 4.$\,$} Un bos\'on ligero de Higgs
tambi\'en puede ser generado como un (pseudo-)bos\'on de Goldstone
me\-dian\-te el rompimiento espont\'aneo de simetr\'\i as globales
de nuevas interacciones a escalas de multi-TeV, manteni\'endose
peque\~na la masa por mecanismos de rompimiento de simetr\'{\i}a
colectivos. Alternativamente a la supersimetr\'{\i}a, las
divergencias cuadr\'aticas podr\'\i an ser canceladas mediante los
nuevos compa\~neros de las part\'{\i}culas del Modelo Est\'andar que
no difieren en su caracter fermi\'onico/bos\'onico. Los esquemas de
simetr\'{\i}a restringen a los acoplamientos de tal forma que las
cancelaciones se logran de una manera natural. Tales escenarios se
realizan en Modelos de Higgs Peque\~nos (Little Higgs Models)
\cite{2A} los cuales predicen un gran conjunto de nuevas
part\'{\i}culas de tipo SM dentro del rango de masas de unos cuantos
TeV's.

\vspace*{2mm}
\noindent {\bf 5.$\,$} Las amplitudes de la
dispersi\'on el\'astica de bosones vectoriales masivos crecen
indefinidamente con la energ\'{\i}a si son calculadas en una
expansi\'on perturbativa del acoplamiento d\'ebil de una teor\'{\i}a
de norma no-Abeliana. Como resultado, se viola la unitaridad m\'as
all\'a de una escala cr\'{\i}tica de energ\'{\i}a $\sim 1.2$ TeV.
Aparte de introducir un bos\'on ligero de Higgs, este problema se
puede tambi\'en resolver suponiendo que el bos\'on $W$ se vuelve
fuertemente interactuante a energ\'{\i}as de TeV, y por lo tanto
amortiguando el aumento de las amplitudes de la dispersi\'on
el\'astica.  Naturalmente, las fuerzas fuertes entre los bosones $W$
pueden atribuirse a nuevas interacciones fundamentales
caracterizadas por un escala del orden de 1 TeV \cite{2}. Si la
teor\'{\i}a fundamental es invariante quiral globalmente, esta
simetr\'{\i}a puede romperse espont\'aneamente. Los bosones de
Goldstone asociados con el rompimiento espont\'aneo de la
simetr\'{\i}a pueden ser absorbidos por bosones de norma para
generar sus masas y para construir las componentes longitudinales de
sus funciones de onda.

Dado que los bosones $W$ longitudinalmente polarizados est\'an asociados
con modos de Goldstone del rompimiento de la simetr\'{\i}a quiral, las
amplitudes de dispersi\'on para el bos\'on $W_L$ pueden ser predichas para
altas energ\'{\i}as, mediante una expansi\'on sistem\'atica de la
energ\'{\i}a. El t\'ermino principal est\'a libre de par\'ametros, una
consecuencia del mecanismo de rompimiento de la simetr\'{\i}a quiral
{\it per se}, el cual es independiente de la estructura particular de
la teor\'{\i}a din\'amica. Los t\'erminos de orden superior en la
expansi\'on quiral sin embargo est\'an definidos por la estructura
detallada de la teor\'{\i}a fundamental.  Con el aumento de
energ\'{\i}a se espera que la expansi\'on quiral diverja y se podr\'\i an
generar nuevas resonancias en la dispersi\'on $WW$ a escalas de masa
entre 1 y 3 TeV's. Este esquema es an\'alogo a la din\'amica del pi\'on en
QCD, donde las amplitudes de umbral pueden ser predichas en una
expansi\'on quiral, mientras que a altas energ\'{\i}as las resonancias
vectoriales y escalares se forman en la dispersi\'on $\pi \pi$.  Este
escenario puede ser estudiado en los experimentos de dispersi\'on $WW$,
donde los bosones $W$ son radiados, como part\'{\i}culas cuasi-reales
\cite{22}, emitidas por quarks de alta energ\'\i a en el haz de protones
en el LHC \cite{12}, \citer{23,23B} o emitidas por electrones y
positrones en Colisionadores Lineales TeV \cite{13,24,24a}.

\vspace*{2mm}
\noindent {\bf 6.$\,$} En teor\'{\i}as formuladas en espacios con
dimensiones extra, se pueden escoger adecuadamente las condiciones a
la frontera de los campos en el espacio compactificado y explotar esta
propiedad para romper las simetr\'\i as \cite{RI2}. En una clase de
modelos los campos de Higgs son identificados con las quintas
componentes de los campos de los bosones vectoriales de masa cero,
asociadas con simetr\'{\i}as de norma rotas m\'as all\'a del Modelo
Est\'andar, mientras otras quintas componentes masivas son transformadas
en grados de libertad longitudinales para los bosones vectoriales del
Modelo Est\'andar. Alternativamente, las simetr\'{\i}as electrod\'ebiles
pueden romperse por la trasformaci\'on de todas las quintas componentes
a componentes longitudinales de los campos vectoriales, vectores de
estados base as\'{\i} como vectores de estado de Kaluza-Kein, de modo
que en \'este escenario emergen las teor\'{\i}as sin Higgs. En
cualquiera de estas teor\'{\i}as de dimensiones extra las torres
masivas de Kaluza-Klein son generadas arriba de los estados del Modelo
Est\'andar. El intercambio adicional de estas torres en la dispersi\'on
$WW$ amortigua la amplitud de dispersi\'on del Modelo Est\'andar y permite
en principio extender la teor\'{\i}a a energ\'{\i}as mayores al l\'\i mite
de unitaridad de 1.2 TeV de los escenarios aproximados sin Higgs.


\vspace*{2mm}
\noindent {\bf 7.$\,$}Este reporte est\'a dividido en
tres partes. Una introducci\'on b\'asica y un resumen de los
principales resultados te\'oricos y experimentales del sector de
Higgs en el Modelo Est\'andar se presentan en la siguiente
secci\'on. Tambi\'en describiremos el futuro de la b\'usqueda del
Higgs en los colisionadores hadr\'onicos y futuros $e^+e^-$. De la
misma forma en la secci\'on que le sigue, se discutir\'a el espectro
del Higgs de las teor\'{\i}as supersim\'etricas. Las principales
caracter\'{\i}sticas de las interacciones fuertes $WW$ y su
an\'alisis en los experimentos de dispersi\'on $WW$ se presentar\'an
en la secci\'on final.

S\'olo los elementos b\'asicos del rompimiento de la simetr\'ia
electrod\'ebil y el mecanismo de Higgs son examinados en este
reporte, el cual es una versi\'on actualizada de Ref.$\,$\cite{XAL}
y Refs.$\,$\cite{24A}. Otros aspectos pueden ser encontrados en la
referencia $\,$\cite{Quigg}, la referencia canon $\,$\cite{24b} y
los reportes recientes Refs.$\,$\cite{DJ}.

\vspace*{6mm}
\section{El Sector de Higgs del Modelo Est\'andar}

\vspace*{2mm}
\subsection{Bases f\'{\i}sicas}

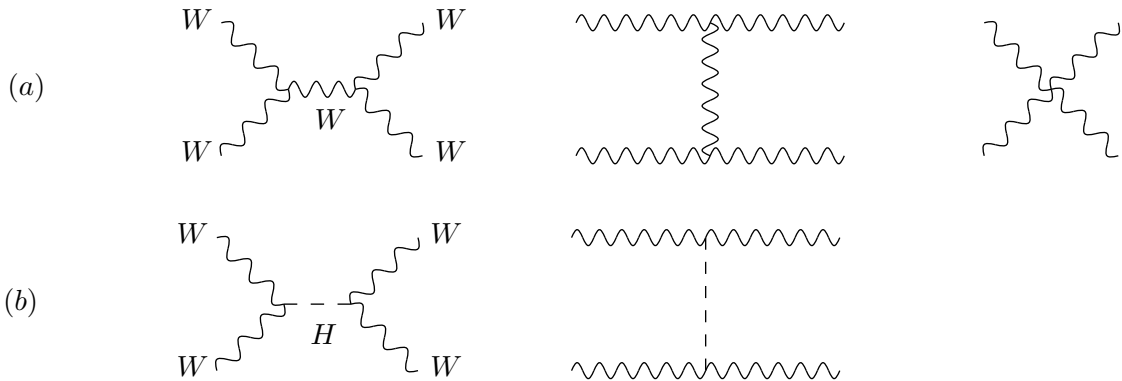
\begin{figure}[hbt]
\begin{center}
\begin{picture}(80,80)(50,-20)
\Photon(0,50)(25,25){3}{3} \Photon(0,0)(25,25){3}{3}
\Photon(25,25)(50,25){3}{3} \Photon(50,25)(75,50){3}{3}
\Photon(50,25)(75,0){3}{3} \put(-80,23){$(a)$} \put(-15,-2){$W$}
\put(-15,48){$W$} \put(80,-2){$W$} \put(80,48){$W$} \put(35,10){$W$}
\end{picture}
\begin{picture}(60,80)(0,-20)
\Photon(0,50)(100,50){3}{12} \Photon(0,0)(100,0){3}{12}
\Photon(50,50)(50,0){3}{6}
\end{picture}
\begin{picture}(60,80)(-90,-20)
\Photon(0,50)(25,25){3}{3} \Photon(0,0)(25,25){3}{3}
\Photon(25,25)(50,50){3}{3} \Photon(25,25)(50,0){3}{3}
\end{picture} \\
\begin{picture}(80,60)(83,0)
\Photon(0,50)(25,25){3}{3} \Photon(0,0)(25,25){3}{3}
\DashLine(25,25)(50,25){6} \Photon(50,25)(75,50){3}{3}
\Photon(50,25)(75,0){3}{3} \put(-80,23){$(b)$} \put(-15,-2){$W$}
\put(-15,48){$W$} \put(80,-2){$W$} \put(80,48){$W$} \put(35,10){$H$}
\end{picture}
\begin{picture}(60,60)(33,0)
\Photon(0,50)(100,50){3}{12} \Photon(0,0)(100,0){3}{12}
\DashLine(50,50)(50,0){5}
\end{picture}  \\
\end{center}
\caption[]{\it \label{esp-fg:wwtoww} Diagramas gen\'ericos para la
dispersi\'on el\'astica $WW$: (a) din\'amica norma-bos\'on pura, y
(b) el intercambio Higgs-bos\'on.}
\end{figure}

\vspace*{2mm}
\noindent {\bf 1.$\,$} A energ\'\i as altas, la
amplitud para la dispersi\'on el\'astica de bosones $W$ masivos, $WW
\to WW$, crece indefinidamente con la energ\'{\i}a para
part\'{\i}culas linealmente polarizadas longitudinalmente,
Fig.~\ref{esp-fg:wwtoww}a.  Esta es una consecuencia del crecimiento
lineal de la funci\'on de onda longitudinal $W_L$, $\epsilon_L =
(p,0,0,E)/M_W$, con la energ\'{\i}a de la part\'{\i}cula. A pesar de
que el t\'ermino de la amplitud de dispersi\'on que aumenta con la
cuarta potencia de la energ\'{\i}a se cancela debido a la
simetr\'{\i}a de norma no-Abeliana, la amplitud permanece
cuadr\'aticamente divergente en la energ\'{\i}a. Por otro lado, la
unitaridad requiere que las amplitudes de dispersi\'on el\'astica de
las ondas parciales $J$ est\'en acotadas por $\Re e A_J \leq 1/2$.
Aplicado a la amplitud asint\'otica de la onda $S$, $A_0 = G_F
s/8\pi\sqrt{2}$, del canal de isospin-cero $2W_L^+W_L^- + Z_L Z_L$,
la cota \cite{25}
\begin{equation}
s \leq 4\pi\sqrt{2}/G_F \sim (1.2~\mbox{TeV})^2
\end{equation}
a la energ\'{\i}a del c.m. $\sqrt{s}$ puede ser derivada para la validez
de una  teor\'ia de bosones masivos de norma d\'ebilmente acoplados.
\noindent

Sin embargo, el aumento cuadr\'atico de la energ\'{\i}a puede ser
amortiguado si se intercambia una nueva part\'icula escalar,
Fig.~\ref{esp-fg:wwtoww}b.  Para lograr la cancelaci\'on, el tama\~no
del acoplamiento debe estar dado por el producto del acoplamiento
de norma con la masa del bos\'on de norma.  Para altas energ\'{\i}as,
la amplitud $A'_0 = -G_F s/8\pi\sqrt{2}$ cancela exactamente la
divergencia cuadr\'atica de la amplitud  pura del bos\'on de norma
$A_0$. Por lo tanto, la unitaridad puede restaurarse
introduciendo una  \underline{\it part\'{\i}cula de Higgs }
d\'ebilmente acoplada.

De la misma forma, la divergencia lineal de
la amplitud  $A(f\bar f\to W_L W_L)\sim gm_f\sqrt{s}$ para la
aniquilaci\'on de un par fermi\'on--antifermi\'on a un par de
bosones de norma longitudinalmente polarizados, puede ser aminorada
a\~nadiendo el intercambio del Higgs al intercambio de bos\'on de norma.
En este caso la part\'{\i}cula de Higgs debe acoplarse
proporcionalmente a la masa $m_f$ del fermi\'on $f$.

Estas observaciones pueden ser resumidas en una regla:
 {\it Una teor\'{\i}a de bosones de norma
 y  fer\-mio\-nes masivos que est\'an d\'ebilmente acoplados a muy altas
energ\'{\i}as, requiere, por unitaridad, la existencia de una
part\'{\i}cula de Higgs; la part\'{\i}cula de Higgs es una
part\'{\i}cula  escalar $0^+$ que se acopla a otras part\'{\i}culas
proporcionalmente a las masas de las part\'{\i}culas.}

La suposici\'on de que los acoplamientos de las part\'{\i}culas
fundamentales son d\'ebiles hasta energ\'{\i}as altas se apoya
cualitativamente en la renormalizaci\'on perturbativa del \'angulo
de mezcla electrod\'ebil $\sin^2\theta_W$ a partir del valor de
simetr\'\i a 3/8 a la escala de GUT hasta $\sim 0.2$ a la escala
electrod\'ebil, el cual es
cercano al valor observado experimentalmente.

\begin{figure}[hbt]
\begin{center}
\begin{picture}(60,10)(80,40)
\Photon(0,25)(50,25){3}{6} \LongArrow(65,25)(90,25) \put(-15,21){$V$}
\put(-15,50){$(a)$}
\end{picture}
\begin{picture}(60,10)(40,40)
\Photon(0,25)(50,25){3}{6} \put(60,23){$+$}
\end{picture}
\begin{picture}(60,10)(15,40)
\Photon(0,25)(50,25){3}{6} \DashLine(25,25)(12,50){3}
\DashLine(25,25)(38,50){3} \Line(9,53)(15,47) \Line(9,47)(15,53)
\Line(35,53)(41,47) \Line(35,47)(41,53) \put(45,45){$H$}
\put(70,23){$+$}
\end{picture}
\begin{picture}(60,10)(-10,40)
\Photon(0,25)(75,25){3}{9} \DashLine(20,25)(8,50){3}
\DashLine(20,25)(32,50){3} \DashLine(55,25)(43,50){3}
\DashLine(55,25)(67,50){3} \Line(5,53)(11,47) \Line(5,47)(11,53)
\Line(29,53)(35,47) \Line(29,47)(35,53) \Line(40,53)(46,47)
\Line(40,47)(46,53) \Line(64,53)(70,47) \Line(64,47)(70,53)
\put(90,23){$+ \cdots$}
\end{picture} \\
\begin{picture}(60,80)(80,20)
\ArrowLine(0,25)(50,25)
\LongArrow(65,25)(90,25)
\put(-15,23){$f$}
\put(-15,50){$(b)$}
\end{picture}
\begin{picture}(60,80)(40,20)
\ArrowLine(0,25)(50,25)
\put(65,23){$+$}
\end{picture}
\begin{picture}(60,80)(15,20)
\ArrowLine(0,25)(25,25)
\ArrowLine(25,25)(50,25)
\DashLine(25,25)(25,50){3}
\Line(22,53)(28,47)
\Line(22,47)(28,53)
\put(35,45){$H$}
\put(65,23){$+$}
\end{picture}
\begin{picture}(60,80)(-10,20)
\ArrowLine(0,25)(25,25)
\ArrowLine(25,25)(50,25)
\ArrowLine(50,25)(75,25)
\DashLine(25,25)(25,50){3}
\DashLine(50,25)(50,50){3}
\Line(22,53)(28,47)
\Line(22,47)(28,53)
\Line(47,53)(53,47)
\Line(47,47)(53,53)
\put(90,23){$+ \cdots$}
\end{picture}  \\
\end{center}
\caption[]{\it \label{esp-fg:massgen} A trav\'es de interacciones
con el campo escalar de fondo se generan (a)  bosones de norma y (b)
masas de fermiones.}
\end{figure}
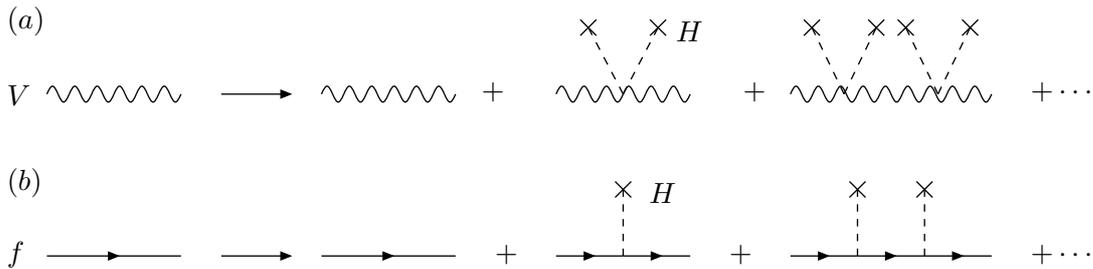

\vspace*{2mm}
\noindent {\bf 2.$\,$} Estas ideas pueden expresarse
en una forma matem\'atica elegante interpretando a las interacciones
electrod\'ebiles como una teor\'{\i}a de norma con rompimiento
espont\'aneo de la simetr\'{\i}a en el sector escalar\footnote{El
mecanismo del rompimiento espont\'aneo de la simetr\'{\i}a,
incluyendo el teorema de Goldstone as\'{\i} como el mecanismo de
Higgs, son ejemplificados por el ilustrativo modelo $O(3)$ $\sigma$
en el Ap\'endice A.}.  Dicha teor\'{\i}a consiste en campos
fermi\'onicos, campos de norma y campos escalares acoplados por las
interacciones estandares de norma y las interacciones de Yukawa a
los otros campos. Adem\'as, una auto-interacci\'on
\begin{equation}
V = \frac{\lambda}{2} \left[ |\phi|^2 - \frac{v^2}{2} \right]^2
\label{esp-eq:potential}
\end{equation}
es introducida en el sector escalar, la cual conduce a un valor
diferente de cero $v/\sqrt{2}$ para el estado base  del campo
escalar. Fijando la fase de la amplitud del vac\'{\i}o en un valor
arbitrario, digamos cero, la simetr\'{\i}a de norma se rompe
espont\'aneamente en el sector escalar. Las interacciones del campo
de norma con el campo escalar de fondo, Fig.~\ref{esp-fg:massgen}a, y
las interacciones de Yukawa de los campo fermi\'onicos  con el campo
de fondo, Fig.~\ref{esp-fg:massgen}b, recorren las masas de estos campos
de un valor cero a uno diferente de cero:
\begin{equation}
\begin{array}{lrclclcl}
\displaystyle
(a) \hspace*{2.0cm} & \displaystyle \frac{1}{q^2} & \to
& \displaystyle \frac{1}{q^2} + \sum_j \frac{1}{q^2} \left[ \left(
\frac{gv}{\sqrt{2}} \right)^2 \frac{1}{q^2} \right]^j & = &
\displaystyle \frac{1}{q^2-M^2} & : & \displaystyle M^2 = g^2
\frac{v^2}{2} \\ \\
(b) & \displaystyle \frac{1}{\not \! q} & \to &
\displaystyle \frac{1}{\not \! q} + \sum_j \frac{1}{\not \! q} \left[
\frac{g_fv}{\sqrt{2}} \frac{1}{\not \! q} \right]^j & = &
\displaystyle \frac{1}{\not \! q-m_f} & : & \displaystyle m_f = g_f
\frac{v}{\sqrt{2}}
\end{array}
\end{equation}
Por lo tanto, en teor\'{\i}as con interacciones de norma y Yukawa,
en las cuales el campo escalar adquiere un valor diferente de cero
para el estado base, los acoplamientos son naturalmente
proporcionales a las masas. Esto garantiza la unitaridad de la
teor\'\i a como se discuti\'o anteriormente.  Estas teor\'{\i}as son
renormalizables (como resultado de la invariancia de norma, la cual
est\'a solamente disfrazada en la formulaci\'on unitaria que se ha
adoptado aqu\'{\i}), y por tanto describen un sistema f\'{\i}sico
bien definido.\\

\subsection{El mecanismo de Higgs en el Modelo Est\'andar}

\phantom{h} Adem\'as de las partes de Yang--Mills y fermi\'onicas, el
Lagrangiano electrod\'ebil $SU_2 \times U_1$ incluye un campo escalar
iso-doblete, $\phi$, acoplado a s\'\i{} mismo en el potencial $V$, cf. eq.
(\ref{esp-eq:potential}), a los campos de norma a trav\'es de la derivada
covariante $iD = i\partial - g \vec{I} \vec{W} - g'YB$, y a los campos
fermi\'onicos up y down $u,d$ mediante las interacciones de Yukawa:
\begin{equation}
{\cal L}_0 = |D\phi|^2 - \frac{\lambda}{2} \left[ |\phi|^2 -
  \frac{v^2}{2} \right]^2 - g_d \bar d_L \phi d_R - g_u \bar u_L
\phi_c u_R + {\rm hc} ~.
\end{equation}
En la norma unitaria, el iso-doblete $\phi$ es reemplazado por el
campo f\'{\i}sico de Higgs $H$, $\phi\to [0,(v+H)/\sqrt{2}]$, el
cual describe las fluctuaciones de la componente $I_3=-1/2$
alrededor del valor del estado base $v/\sqrt{2}$. La escala $v$ del
rompimiento de la simetr\'{\i}a electrd\'ebil es fijada por el
acoplamiento de norma d\'ebil y la masa del $W$, los cuales a su vez
pueden re-expresarse por medio del acoplamiento de Fermi:
\begin{equation}
  v = 1/\sqrt{\sqrt{2}G_F} \approx 246 \,{\rm GeV}.
\end{equation}
mientras la masa del $W$ est\'a relacionada con $v$ por el
acoplamiento de norma, el acoplamiento de Yukawa $g_{f}$ y el
acoplamiento cu\'artico $\lambda$ pueden de la misma forma ser
re-expresados en t\'erminos de la masa f\'isica del Higgs $M_H$ y
las masas de los fermiones $m_f$:

\begin{eqnarray}
M^2_W & = & g^2 v^2 / 4       \nonumber \\
m_f   & = & g_f v / \sqrt{2}  \nonumber \\
M_H^2 & = & \lambda v^2
\end{eqnarray}
respectivamente.

Dado que los acoplamientos de la part\'{\i}cula de Higgs con las
part\'\i culas de norma, con los fer\-mio\-nes y con ellas mismas
est\'an dados por los acoplamientos de norma y las masas de las
part\'{\i}culas, el \'unico par\'ametro desconocido en el sector de
Higgs (aparte de la matriz de mezcla CKM) es la masa del Higgs.
Cuando esta masa se fija, todas las propiedades de la part\'{\i}cula
del Higgs pueden ser predichas, i.e. el tiempo de vida y las razones
de desintegraci\'on (branching ratios), as\'{\i} como los mecanismos
de producci\'on y las respectivas secciones eficaces.\\

\vspace*{4mm}
\subsubsection{La Masa del Higgs del Modelo Est\'andar}
\vspace*{2mm} \noindent
A\'un cuando la masa del bos\'on de Higgs no
puede ser predicha en el Modelo Est\'andar, l\'{\i}mites superior e
inferior estrictos pueden ser derivados de las condiciones de
consistencia interna y extrapolaciones del modelo a altas
energ\'{\i}as.

\vspace*{2mm}
\noindent {\bf 1.$\,$} El bos\'on de Higgs ha sido
introducido como una part\'{\i}cula fundamental para hacer a las
amplitudes de dispersi\'on 2--2 que involucran bosones $W$
polarizados longitudinalmente compatibles con la unitaridad.
Basados en el principio general de incertidumbre de
tiempo-energ\'{\i}a, las part\'{\i}culas deben desacoplarse de un
sistema f\'{\i}sico si su energ\'{\i}a crece indefinidamente. La
masa de la part\'{\i}cula de Higgs debe por tanto estar obligada a
restablecer la unitaridad en el r\'egimen perturbativo. De la
expansi\'on asint\'otica de la amplitud de dispersi\'on el\'astica
$W_L W_L$ onda $S$, incluyendo $W$ y los intercambios de Higgs,
$A(W_L W_L \to W_L W_L) \to -G_F M_H^2/4\sqrt{2}\pi$, se sigue que
\cite{25}
\begin{equation}
M_H^2 \leq 2\sqrt{2}\pi/G_F \sim (850~\mbox{GeV})^2 ~.
\end{equation}
Dentro de la formulaci\'on can\'onica del Modelo Est\'andar, las condiciones
de consistencia requieren por tanto una masa del Higgs por abajo de 1
TeV.\\

\begin{figure}[hbt]
\vspace*{-0.5cm}

\begin{center}
\begin{picture}(90,80)(60,-10)
\DashLine(0,50)(25,25){3}
\DashLine(0,0)(25,25){3}
\DashLine(50,50)(25,25){3}
\DashLine(50,0)(25,25){3}
\put(-15,45){$H$}
\put(-15,-5){$H$}
\put(55,-5){$H$}
\put(55,45){$H$}
\end{picture}
\begin{picture}(90,80)(10,-10)
\DashLine(0,50)(25,25){3}
\DashLine(0,0)(25,25){3}
\DashLine(75,50)(50,25){3}
\DashLine(75,0)(50,25){3}
\DashCArc(37.5,25)(12.5,0,360){3}
\put(-15,45){$H$}
\put(-15,-5){$H$}
\put(35,40){$H$}
\put(80,-5){$H$}
\put(80,45){$H$}
\end{picture}
\begin{picture}(50,80)(-40,2.5)
\DashLine(0,0)(25,25){3}
\DashLine(0,75)(25,50){3}
\DashLine(50,50)(75,75){3}
\DashLine(50,25)(75,0){3}
\ArrowLine(25,25)(50,25)
\ArrowLine(50,25)(50,50)
\ArrowLine(50,50)(25,50)
\ArrowLine(25,50)(25,25)
\put(-15,70){$H$}
\put(-15,-5){$H$}
\put(35,55){$t$}
\put(80,-5){$H$}
\put(80,70){$H$}
\end{picture}  \\
\setlength{\unitlength}{1pt}
\caption[]{\label{esp-fg:lambda} \it Diagramas generando la evoluci\'on de
la auto-interacci\'on del Higgs $\lambda$.}
\end{center}
\end{figure}
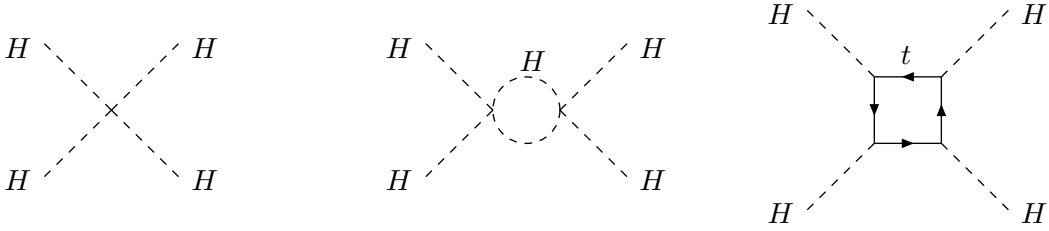

\begin{figure}[hbtp]

\vspace*{0.8cm}

\hspace*{3.0cm}
\epsfxsize=8.5cm \epsfbox{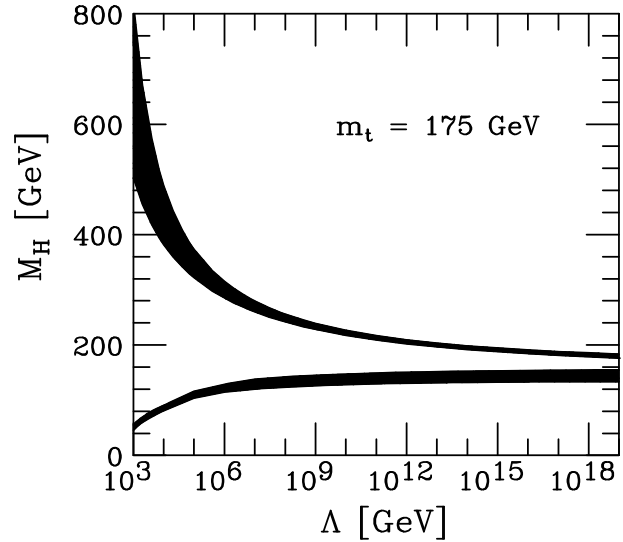}
\vspace*{-0.2cm}

\caption[]{\label{esp-fg:triviality} \it L\'{\i}mites a la masa del
bos\'on Higgs en el SM. $\Lambda$ denota la escala de energ\'{\i}a a
la cual el sistema del bos\'on de Higgs del SM con masa $M_H$ se
volver\'\i a fuertemente interactuante (l\'{\i}mite superior); el
l\'{\i}mite inferior proviene de pedir la estabilidad del
vac\'{\i}o. Refs. \cite{4,5}.}
\vspace*{-0.3cm}
\end{figure}

\vspace*{2mm}
\noindent {\bf 2.$\,$} L\'{\i}mites muy restrictivos
para el valor de la masa del Higgs del Modelo Est\'andar provienen
de los l\'{\i}mites en la escala de energ\'{\i}a $\Lambda$ hasta la
cual el Modelo Est\'andar puede extenderse antes de que nuevos
fen\'omenos de interacci\'on aparezcan. La clave para estos
l\'{\i}mites es la evoluci\'on del acoplamiento cu\'artico $\lambda$
con la energ\'{\i}a debida a fluctuaciones cu\'anticas \cite{4}. Las
contribuciones b\'asicas se presentan en la Fig.~\ref{esp-fg:lambda}. El
lazo del Higgs por s\'\i{} mismo da pie a un aumento indefinido del
acoplamiento, mientras que el lazo fermi\'onico-top quark, conforme
la masa del top aumenta, lleva al acoplamiento a valores peque\~nos,
finalmente incluso a valores abajo de cero. La variaci\'on del
acoplamiento cu\'artico del Higgs $\lambda$ y el acoplamiento de
Yukawa top-Higgs $g_t$ con la energ\'{\i}a, parametrizada por
$t=\log \mu^2/v^2$, pueden ser escritos como sigue \cite{4}
\begin{equation}
\begin{array}{rclcl}
\displaystyle \frac{d\lambda}{dt} & = & \displaystyle \frac{3}{8\pi^2}
\left[ \lambda^2 + \lambda g_t^2 - g_t^4 \right] & : & \displaystyle
\lambda(v^2) = M_H^2/v^2 \\ \\ \displaystyle \frac{d g_t}{dt} & = &
\displaystyle \frac{1}{32\pi^2} \left[ \frac{9}{2} g_t^3 - 8 g_t g_s^2
\right] & : & \displaystyle g_t(v^2) = \sqrt{2}~m_f/v ~.
\end{array}
\end{equation}

S\'olo las contribuciones principales de lazos de Higgs, top y
QCD  [acoplamiento $g_s$] son tomadas en cuenta. \\

Para masas del top moderadas, el acoplamiento cu\'artico $\lambda$
aumenta indefinidamente, $d\lambda / dt \sim + \lambda^2$, y el
acoplamiento se vuelve fuerte poco antes de alcanzar el polo de
Landau:
\begin{equation}
\lambda (\mu^2) = \frac{\lambda(v^2)}{1- \frac{3\lambda(v^2)}{8\pi^2}
\log \frac{\mu^2}{v^2}} ~.
\end{equation}
Reexpresando el valor inicial de $\lambda$ por la masa del Higgs, la
condici\'on $\lambda (\Lambda^2) < \infty$, puede ser traducida a un
\underline{l\'{\i}mite superior} para la masa del Higgs:
\begin{equation}
M_H^2 \leq \frac{8\pi^2 v^2}{3\log \Lambda^2/v^2} ~.
\end{equation}

Este l\'{\i}mite para la masa est\'a relacionado
logar\'{\i}tmicamente con la energ\'{\i}a $\Lambda$ hasta la cual el
Mo\-de\-lo Est\'andar se supone v\'alido.  El valor m\'aximo de
$M_H$ para el corte m\'\i nimo $\Lambda \sim $ 1~TeV est\'a dado por
$\sim 750$ GeV. Esta cota es cercana a la estimada $\sim 700$ GeV en
los c\'alculos hechos con lattices para $\Lambda \sim 1$ TeV, los
que permiten un control apropiado de
efectos no preturbativos cerca del l\'\i mite \cite{6}.\\

\begin{table}[hbt]
\renewcommand{\arraystretch}{1.5}
\begin{center}
\begin{tabular}{|l||l|} \hline
$\Lambda$ & $M_H$ \\ \hline \hline 1 TeV & 60 GeV $\lessim M_H \lessim
700$ GeV \\ $10^{19}$ GeV & 130 GeV $\lessim M_H \lessim 190$ GeV \\
\hline
\end{tabular}
\renewcommand{\arraystretch}{1.2}
\caption[]{\label{esp-tb:triviality} \it L\'{\i}mites de la masa del Higgs
para dos valores del corte $\Lambda$.}
\end{center}
\end{table}
Un \underline{l\'{\i}mite inferior} en la masa del Higgs puede ser
obtenido a partir de requerir la estabilidad del vac\'{\i}o
\cite{4,5}. Dado que las correcciones top-lazo reducen $\lambda$
llev\'andalo finalmente a valores negativos, el potencial de la
auto-energ\'{\i}a se volver\'\i a ilimitadamente negativo y el
estado base ya no ser\'\i a estable. Para evitar la inestabilidad,
para valores de corte menores a $\Lambda$, la masa del Higgs debe de
exceder un valor m\'{\i}nimo para un valor dado de la masa del top.
Este l\'{\i}mite inferior
depende del valor de corte $\Lambda$.\\

Las escalas de $\Lambda$ hasta las cuales el Modelo Est\'andar puede
ser extendido antes de que nuevas interacciones se vuelvan efectivas
se muestran en Fig.~\ref{esp-fg:triviality} como funci\'on de la masa
del Higgs. Los valores permitidos de la masa del Higgs  est\'an
reunidos en la Tabla \ref{esp-tb:triviality}, para dos valores espec\'\i
ficos del corte $\Lambda$.  Si el Modelo Est\'andar se supone
v\'alido hasta la escala de Planck, la masa del Higgs est\'a
restringida a un intervalo estrecho entre 130 y 190~GeV. Esta
ventana puede ensancharse a 200~ GeV para un corte cercano a la
escala de gran unificaci\'on. La observaci\'on de una masa del Higgs
por arriba o por abajo de estos valores requerir\'\i a una escala de
nueva f\'{\i}sica por debajo de la escala de Planck/GUT.

\vspace*{2mm}
\noindent {\bf 3.$\,$} Se puede derivar evidencia
indirecta del bos\'on ligero de Higgs de las mediciones de alta
precisi\'on de las observables electrod\'ebiles en LEP y otros
lugares. En efecto, el hecho de que el Modelo Est\'andar sea
renormalizable s\'olo despu\'es de incluir las part\'\i culas top y
Higgs en las correcciones de lazo, indica que las observables
electrod\'ebiles son sensibles a las masas de estas part\'{\i}culas.

\begin{figure}[hbtp]
\begin{picture}(100,60)(-80,-0)
\Photon(0,30)(30,30){3}{3} \Photon(60,30)(90,30){3}{3}
\ArrowArc(45,30)(15,360,180) \ArrowArc(45,30)(15,180,0)
\put(-25,27){$W$} \put(43,50){$t$} \put(43,2){$b$} \put(95,27){$W$}
\end{picture}
\begin{picture}(100,60)(-160,-0)
\Photon(0,30)(30,30){3}{3} \Photon(60,30)(90,30){3}{3}
\PhotonArc(45,30)(15,180,360){-3}{5.5} \DashCArc(45,30)(15,0,180){3}
\put(-25,27){$W$} \put(40,50){$H$} \put(40,0){$W$} \put(95,27){$W$}
\end{picture}
\caption[]{\label{esp-fg:WtH} \it Correcciones raditivas con t,b y
W,Higgs virtuales en los propagadores del bos\'on de norma
electrod\'ebil.}
\end{figure}
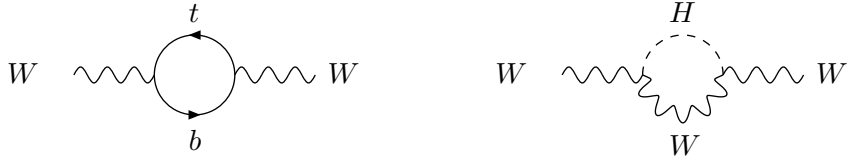

El acoplamiento de Fermi puede ser reescrito en t\'erminos del
acoplamiento d\'ebil y la masa del $W$; al orden m\'as bajo,
$G_F/\sqrt{2} = g^2/8M_W^2$. Despu\'es de sustituir el acoplamiento
electromagn\'etico $\alpha$, el \'angulo de mezcla electrod\'ebil y
la masa del $Z$ para el acoplamiento d\'ebil, y la masa del $W$,
esta relaci\'on puede reescribirse como
\begin{equation}
\frac{G_F}{\sqrt{2}} = \frac{2\pi\alpha}{\sin^2 2\theta_W M_Z^2}
[1+\Delta r_\alpha + \Delta r_t + \Delta r_H ] ~.
\end{equation}
Los t\'erminos $\Delta$ toman en cuenta las correcciones radiativas,
cf. Fig.~\ref{esp-fg:WtH}: $\Delta r_\alpha$ describe el cambio en el
acoplamiento electromagn\'etico $\alpha$ si se eval\'ua en la escala
$M_Z^2$ en lugar de en el momento cero; $\Delta r_t$ denota las
contribuciones del quark top (y bottom) a las masas de los bosones
$W$ y $Z$, las cuales son cuadr\'aticas en la masa del top.
Finalmente, $\Delta r_H$ toma en cuenta las contribuciones del Higgs
virtual a las masas; este t\'ermino depende s\'olo
logar\'{\i}tmicamente \cite{7} en la masa del Higgs al orden
dominante:
\begin{equation}
\Delta r_H = \frac{G_F M_Z^2 (1+9\sin^2\theta_W)}{24\sqrt{2}\pi^2}
\, \log \frac{M_H^2}{M^2_W} \, + \, ...  \hspace{1cm} (M_H^2 \gg
M_W^2) ~.
\end{equation}
El efecto de apantallamiento refleja el papel del campo de Higgs
como un regulador que hace a la teor\'{\i}a electrod\'ebil
renormalizable.

Aunque la sensibilidad a la masa del Higgs es \'unicamente
logar\'{\i}tmica, la creciente precisi\'on de las medidas de
las observables electrod\'ebiles nos permite derivar estimaciones
interesantes y restricciones a la masa del Higgs \cite{8}, cf.
Fig.~\ref{esp-fg:SMHiggs}:
\begin{eqnarray}
M_H & = & 76^{+33}_{-24}~\mbox{GeV} \\
    & \lessim \phantom{1}\!\! & 144 ~\mbox{GeV~~~(95\% CL)}  ~. \nonumber
\end{eqnarray}

Con un valor de $15.1 \ \% $ la probabilidad para el ajuste no es
exageradamente grande, pero tampoco prohibitivamente peque\~na. El
$95 \ \% $ del nivel de confiabilidad  todav\'ia est\'a
significativamente
por arriba del l\'imite de b\'usqueda directa,
\begin{equation}
  {\hspace*{-18mm}} M_H \;\;\;\,\geq\;\;\;\, 114.1 \, {\rm GeV}
\end{equation}
derivado de los an\'alisis del LEP2, Ref.$\,$\cite{9}, del canal
dominante de Higgs-strahlung $e^+e^- \to ZH$.\\

Se puede concluir de estos n\'umeros que la formulaci\'on can\'onica
del Modelo Est\'andar incluyendo la existencia de un bos\'on de
Higgs ligero, es compatible con los datos electrod\'ebiles. Sin
embargo, mecanismos alternativos no se pueden descartar si el
sistema est\'a abierto a contribuciones de \'areas de f\'{\i}sica
m\'as all\'a del Modelo Est\'andar.

\vspace{4mm}
\subsubsection{Decaimientos de la part\'{\i}cula de Higgs}

\vspace*{2mm}
\noindent El perfil de la part\'{\i}cula de Higgs
est\'a determinado un\'{\i}vocamente si se fija la masa del Higgs.
La intensidad de los acoplamientos del bos\'on de Higgs con los
bosones de norma electrod\'ebiles $V=W,\, Z$ est\'a determinado por
sus masas $M_V$ y la intensidad del acoplamiento de Yukawa del
bos\'on de Higgs a fermiones est\'a dado por las masas de los
fermiones $m_f$; los acoplamientos se podr\'an definir uniformemente
como
\begin{eqnarray}
g_{HVV} & = & \left[2 \sqrt{2} G_F \right]^{1/2} M_V      \\
g_{Hff} & = & \left[2 \sqrt{2} G_F \right]^{1/2} m_f \, . \nonumber
\end{eqnarray}

La anchura total de decaimiento y tiempo de vida, as\'{\i} como las
razones de desintegraci\'on para canales de decaimiento
espec\'{\i}ficos, est\'an determinados por estos par\'ametros. La
medici\'on de las caracter\'{\i}sticas de decaimiento puede, por lo
tanto, ser explotada para establecer, experimentalmente, que los
acoplamientos del Higgs crecen con las masas de las part\'\i culas,
una consecuencia directa del mecanismo de Higgs.

Para las part\'{\i}culas de Higgs en el rango intermedio de masas
${\cal O}(M_Z) \leq M_H \leq 2M_Z$, los modos principales de
decaimiento son en pares $b\bar b$ y pares $WW,ZZ$, siendo virtual
uno de los bosones de norma por abajo del umbral respectivo. Arriba
de los umbrales de los pares $WW,ZZ$, las part\'{\i}culas de Higgs
decaen casi exclusivamente en estos dos canales, con una peque\~na
contribuci\'on de los decaimientos del top cerca del umbral del
$t\bar t$.  Abajo de los 140 GeV, los decaimientos $H\to
\tau^+\tau^-, c\bar c$ y $gg$ son tambi\'en importantes adem\'as del
canal dominante $b\bar b$; los decaimientos a $\gamma\gamma$, aunque
de tasa suprimida, aportan no obstante una se\~nal clara de
2-cuerpos para la formaci\'on de la part\'{\i}cula de Higgs en este
rango de masa.\\

\vspace*{3mm}
\noindent {\it (a) \underline{Decaimientos de Higgs a
fermiones}}

\vspace*{1mm}
\noindent

La anchura parcial de los decaimientos de Higgs a pares de leptones y
quarks est\'a dado por \cite{26}
\begin{equation}
\Gamma (H\to f\bar f) = {\cal N}_c \frac{G_F}{4\sqrt{2}\pi}
m_f^2(M_H^2) M_H ~,
\end{equation}
siendo ${\cal N}_c = 1$ o 3 el factor de color. [Cerca del umbral el
anchura parcial est\'a suprimido por el factor de onda-P, $\beta_f^3$
adicional, donde $\beta_f$ es la velocidad del fermi\'on.]
Asint\'oticamente, el ancho fermi\'onico crece s\'olo linealmente
con la masa del Higgs. La mayor parte de las correcciones radiativas
de QCD se pueden mapear a la dependencia de escala de la masa del
quark, evaluada en la masa del Higgs.  Para $M_H\sim 100$ GeV los
par\'ametros relevantes son $m_b (M_H^2) \simeq 3$ GeV y $m_c
(M_H^2) \simeq$ 0.6~GeV. La reducci\'on de la masa efectiva del
quark-$c$ sobrecompensa el factor de color en la raz\'on entre los
decaimientos a charm y a $\tau$ del bos\'on de Higgs. Las
correcciones residuales de QCD, $\sim 5.7 \times (\alpha_s/\pi)$,
modifican los anchos s\'olo ligeramente.

\vspace*{3mm}
\noindent {\it (b) \underline{Decaimientos del Higgs a
pares de bosones $WW$ y $ZZ$}}

\vspace*{1mm}
\noindent
Por arriba de los umbrales de los
decaimientos $WW$ y $ZZ$, los anchos parciales para estos canales
pueden ser escritos como \cite{27}
\begin{equation}
\Gamma (H\to VV) = \delta_V \frac{G_F}{16\sqrt{2}\pi} M_H^3
(1-4x+12x^2) \beta_V ~,
\end{equation}
donde $x=M_V^2/M_H^2$ y $\delta_V = 2$ y 1 para $V=W$ y $Z$,
respectivamente. Para masas grandes del Higgs, los bosones vectoriales
est\'an polarizados longitudinalmente. Dado que las funciones de onda de
estos estados son lineales en la energ\'{\i}a, los anchos crecen como
la tercera potencia de la masa de Higgs. Por debajo del umbral de
dos bosones reales, la part\'{\i}cula de Higgs puede decaer en pares
$VV^*$, con uno de los bosones vectoriales siendo virtual. El ancho de
decaimiento est\'a dado en este caso \cite{28} por

\begin{equation}
\Gamma(H\to VV^*) = \frac{3G^2_F M_V^4}{16\pi^3}~M_H
R(x)~\delta'_V ~,
\end{equation}
donde $\delta'_W = 1$, $\delta'_Z = 7/12 - 10\sin^2\theta_W/9 + 40
\sin^4\theta_W/27$ y
\begin{displaymath}
R(x) =
\frac{3(1-8x+20x^2)}{(4x-1)^{1/2}}\arccos\left(\frac{3x-1}{2x^{3/2}}
\right) - \frac{1-x}{2x} (2-13x+47x^2) - \frac{3}{2} (1-6x+4x^2)
\log x ~.
\end{displaymath}
El canal $ZZ^*$ se vuelve relevante para masas del Higgs m\'as all\'a de
$\sim 140$ GeV. Por arriba del umbral, el canal de 4-leptones $H\to ZZ \to
4 \ell^\pm$ proporciona una se\~nal muy clara para los bosones de Higgs.
Tambi\'en el canal de decaimiento $WW$ demuestra su utilidad, a pesar
del escape de los neutrinos en decaimientos lept\'onicos del $W$, si el
canal $ZZ$ dentro de la capa de masa (on-shell) est\'a cerrado
cinem\'aticamente.

\vspace*{3mm}
\noindent {\it (c) \underline{Decaimiento del Higgs a
pares de $gg$ y $\gamma\gamma$}}

\vspace*{1mm}
\noindent

En el Modelo Est\'andar, los decaimientos glu\'onicos del Higgs son
mediados por lazos (loops) de quarks top y bottom, para los
decaimientos fot\'onicos, adem\'as son mediados por lazos del $W$.  Como
estos decaimientos son significantivos s\'olo muy por debajo de los
umbrales del top y del $W$,  son descritos por las expresiones
aproximadas \cite{29,30}
\begin{eqnarray}
\Gamma (H\to gg) & = & \frac{G_F
\alpha_s^2(M_H^2)}{36\sqrt{2}\pi^3}M_H^3 \left[ 1+
\left(\frac{95}{4} - \frac{7N_F}{6} \right) \frac{\alpha_s}{\pi}
\right] \label{esp-eq:htogg} \\ \nonumber \\
\Gamma (H\to \gamma\gamma) & = & \frac{G_F
\alpha^2}{128\sqrt{2}\pi^3}M_H^3 \left[  \frac{4}{3} {\cal N}_C
e_t^2 - 7 \right]^2 ~,
\end{eqnarray}
las cuales son v\'alidas en el l\'{\i}mite $M_H^2 \ll 4M_W^2,
4M_t^2$. Las correcciones radiativas de QCD, las cuales incluyen los
estados finales $ggg$ y $gq\bar q$ en (\ref{esp-eq:htogg}), son muy
importantes; estas correcciones incrementan el ancho parcial en
alrededor de un 65\%. Aunque los decaimientos fot\'onicos del Higgs
son muy raros, ofrecen, no obstante, un canal atractivo tipo
resonancia  para la b\'usqueda de las part\'{\i}culas de
Higgs.


%
\vspace*{3mm}
\noindent {\it \underline{Resumen}}

\begin{figure}[hbtp]

\vspace*{0.5cm}
\hspace*{1.0cm}
\begin{turn}{-90}%
\epsfxsize=8.5cm \epsfbox{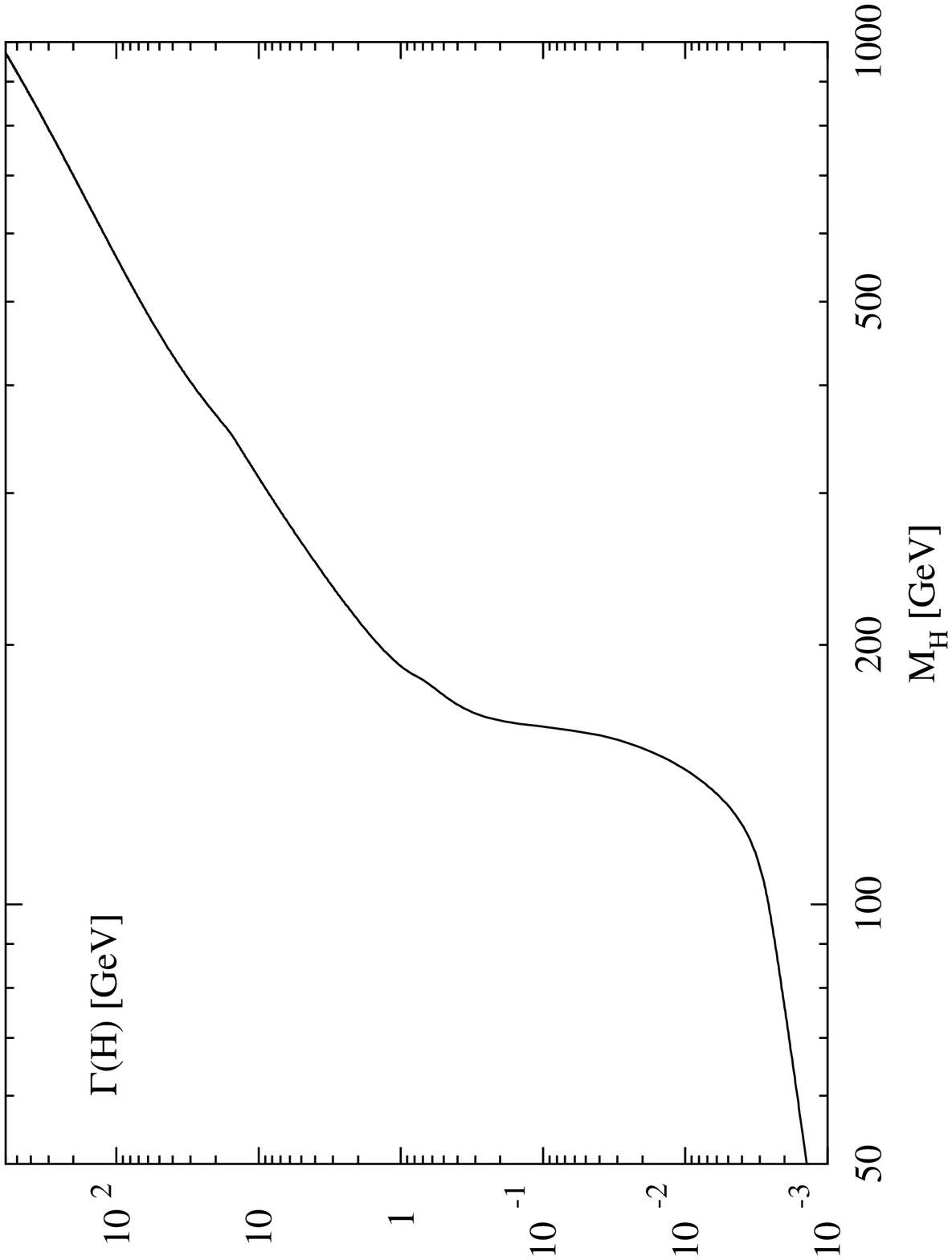}
\end{turn}

\vspace*{0.5cm}
\hspace*{1.0cm}
\begin{turn}{-90}%
\epsfxsize=8.5cm \epsfbox{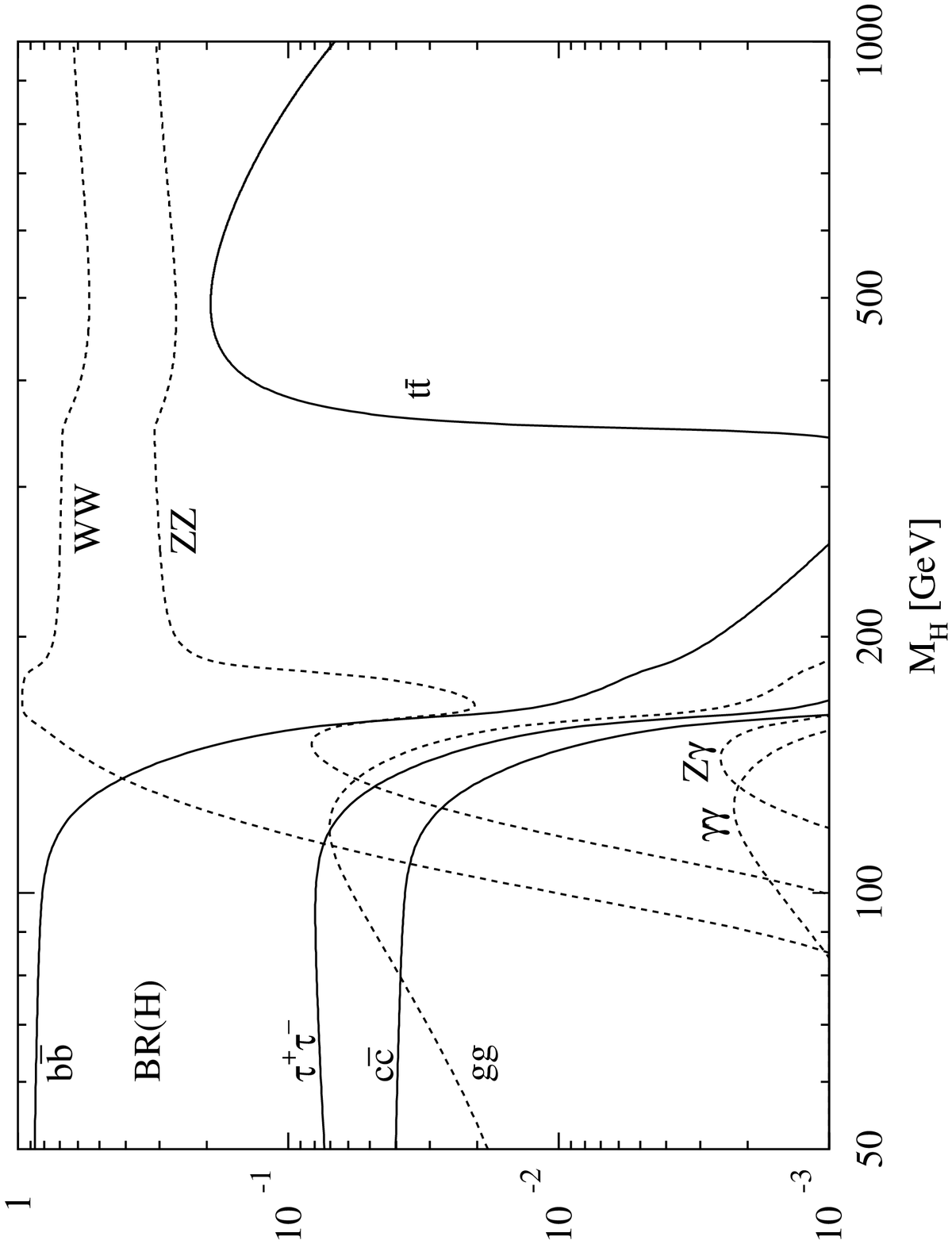}
\end{turn}
\vspace*{-0.0cm} \caption[]{\label{esp-fg:wtotbr} \it (a) Ancho total de
  decaimiento (en GeV) bos\'on de Higgs del SM como funci\'on de su masa.
  (b) Razones de desintegraci\'on (branching ratios) de los modos
  dominantes de decaimiento de
  la part\'{\i}cula de Higgs del SM. Se toman en consideraci\'on todas
  las correcciones relevantes de orden m\'as alto. Code: HDECAY, Ref.\cite{HDECAY}.}
\end{figure}

\vspace*{1mm}
\noindent Al sumar todos los posibles canales de
decaimiento, obtenemos el ancho total mostrado en la
Fig.~\ref{esp-fg:wtotbr}a. Hasta masas de 140 GeV, la part\'{\i}cula del
Higgs es muy estrecha, $\Gamma(H) \leq 10$ MeV. Despu\'es de que los
canales real y virtual de los bosones de norma se abren, el estado
r\'apidamente se ensancha, alcanzando un ancho de $\sim 1$ GeV en el
umbral de $ZZ$. El ancho no se puede medir directamente en la
regi\'on intermedia de masas en los colisionadores LHC o $e^+ e^-$.
Sin embargo, puede determinarse indirectamente midiendo, por
ejemplo, el ancho parcial $\Gamma (H\to WW)$ en el proceso de
fusi\'on $WW\to H$, y la fracci\'on de desintegraci\'on $BR(H\to
WW)$ en el proceso de decaimiento $H\to WW$, el ancho total se sigue
de la raz\'on de las dos observables. Por arriba de una masa de
$\sim 250$ GeV, el estado se ensancha lo suficiente para ser
resuelto experimentalmente.

Las razones de desintegraci\'on de los modos de decaimiento
principales se muestran en la Fig.~\ref{esp-fg:wtotbr}b. Una gran
variedad de canales ser\'a accesible para masas del Higgs por debajo
de 140 GeV. El modo dominante es el decaimiento a $b\bar b$, aunque
los decaimientos a $c\bar c, \tau^+\tau^-$ y $gg$ ocurren tambi\'en
a un nivel de porcentaje considerable. A $M_H=$ 120~GeV, por
ejemplo, las razones de desintegraci\'on son de 68\% para $b\bar b$,
3.1\% para $c\bar c$, 6.9\% para $\tau^+\tau^-$ y 7\% para $gg$.
Decaimientos a $\gamma\gamma$ ocurren a un nivel de 1 por mil.
Arriba de este valor de masa, el decaimiento del bos\'on de Higgs a
$W$'s se vuelve dominante, sobrepasando todos los dem\'as canales si
el modo de decaimiento a dos $W$'s reales es cinem\'aticamente
posible. Para masas del Higgs muy por arriba de los umbrales, los
decaimientos $ZZ$ y $WW$ se dan a un raz\'on de 1:2, ligeramente
modificados s\'olo justo por arriba del umbral de $t\bar t$. Como el
ancho de decaimiento a pares de bosones vectoriales crece como la
tercera potencia de la masa, la part\'{\i}cula de Higgs se vuelve
muy ancha asint\'oticamente, $\Gamma(H) \sim \frac{1}{2} M_H^3$
[TeV]. De hecho, para $M_H\sim 1$ TeV, el ancho alcanza $\sim
\frac{1}{2}$ TeV.

\vspace*{4mm}

\subsection{La Producci\'on de Higgs en Colisionadores Hadr\'onicos}
\vspace*{2mm}
\noindent Varios procesos pueden ser explotados para
producir
part\'\i culas de Higgs en colisionadores hadr\'onicos \cite{24A,32}: \\[0.5cm]
\begin{tabular}{llll}
\hspace*{21mm} & fusi\'on de gluones & :              & $gg\to H$ \\ \\
& fusi\'on de $WW,ZZ$            & :    & $W^+ W^-, ZZ \to H$ \\ \\
& Higgs-strahlung emitido por $W,Z$ & :   & $q\bar q \to W,Z \to W,Z + H$ \\ \\
& Higgs bremsstrahlung emitido por top & : & $q\bar q, gg \to t\bar t + H$
\end{tabular} \\[0.5cm]
La fusi\'on del gluones juega un papel dominante en todo el rango de
masa del Higgs  del Modelo Est\'andar. Mientras que el proceso de
fusi\'on de $WW/ZZ$ se vuelve cada vez m\'as importante al aumentar
la masa del Higgs, tambi\'en juega un papel importante en la
b\'usqueda del bos\'on de Higgs el estudio de sus propiedades en el
rango intermedio de masas. Los \'ultimos dos procesos de radiaci\'on
son relevantes s\'olo para masas ligeras del Higgs.

Las secciones eficaces de producci\'on en colisionadores
hadr\'onicos, en particular en el LHC, son bastante considerables  de
modo que en esta m\'aquina se puede producir una muestra grande de
part\'{\i}culas de Higgs del SM. Las dificultades experimentales
surgen del enorme n\'umero de eventos de fondo (background) que
aparecen junto con los eventos de se\~nales de Higgs. Este
problema ser\'a abordado ya sea en el desencadenamiento de los decaimientos
lept\'onicos de $W,Z$ y $t$ en los procesos de radiaci\'on o
explotando el car\'acter de resonancia de los decaimientos del
Higgs $H\to \gamma\gamma$ y $H\to ZZ \to 4\ell^\pm$. De esta
manera, se espera que el Tevatron realice la b\'usqueda  de las
part\'{\i}culas de Higgs en el rango de masas que est\'a por arriba
del alcanzado en el LEP2 y hasta alrededor de 110 a 130 GeV
\cite{11}. Se espera que el LHC cubra el rango can\'onico completo
de la masa del Higgs $M_H \lessim 700$ GeV del Modelo Est\'andar
\cite{12}.

\vspace*{4mm}
\noindent {\it (a) \underline{Fusi\'on de gluones}}

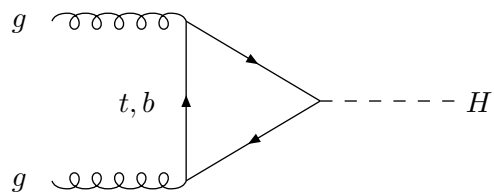
\begin{figure}[hbt]
\begin{center}
\setlength{\unitlength}{1pt}
\begin{picture}(180,90)(0,0)

\Gluon(0,20)(50,20){-3}{5}
\Gluon(0,80)(50,80){3}{5}
\ArrowLine(50,20)(50,80)
\ArrowLine(50,80)(100,50)
\ArrowLine(100,50)(50,20)
\DashLine(100,50)(150,50){5}
\put(155,46){$H$}
\put(25,46){$t,b$}
\put(-15,18){$g$}
\put(-15,78){$g$}

\end{picture}  \\
\setlength{\unitlength}{1pt} \caption[ ]{\label{esp-fg:gghlodia} \it
Diagrama que contribuye a la formaci\'on de los bosones de Higgs en
las colisiones glu\'on-glu\'on al orden m\'as bajo.}
\end{center}
\end{figure}

\vspace*{1mm}
\noindent El mecanismo de fusi\'on de gluones
\cite{29,32,39A,39B}
\begin{displaymath}
pp \to gg \to H
\end{displaymath}
proporciona el mecanismo de producci\'on dominante de los bosones
de Higgs en el LHC en el rango completo de masas relevantes del
Higgs de hasta alredeor de 1 TeV. El acoplamiento del glu\'on al
bos\'on de Higgs en el SM est\'a mediado por lazos tiangulares de
los quarks top y bottom, cf. Fig.~\ref{esp-fg:gghlodia}. Como el
acoplamiento de Yukawa de las part\'{\i}culas de Higgs a quarks
pesados crece con la masa del quark, equilibrando as\'{\i} la
disminuci\'on de la amplitud triangular, el factor de forma se
aproxima a un valor distinto de cero para masas grandes del lazo
del quark. [Si las masas de los quarks pesados m\'as all\'a de la
tercera generaci\'on fueran generadas s\'olamente por el mecanismo
de Higgs, estas part\'{\i}culas a\~nadir\'{\i}an al factor de
forma el mismo monto que el quark top en el l\'{\i}mite
asint\'otico de quark pesado.]
\\

La secci\'on eficaz part\'onica, Fig.~\ref{esp-fg:gghlodia}, puede ser
expresada por el ancho glu\'onico del bos\'on
de Higgs a m\'as bajo orden \cite{32}:
\begin{eqnarray}
\hat \sigma_{LO} (gg\to H) & = & \sigma_0 M_H^2 \times BW(\hat{s}) \\
\sigma_0 & = & \frac{\pi^2}{8M_H^2} \Gamma_{LO} (H\to gg) =
\frac{G_F\alpha_s^2}{288\sqrt{2}\pi} \left| \sum_Q A_Q^H (\tau_Q) \right|^2 ~,
\nonumber
\end{eqnarray}
donde la variable de escalamiento est\'a definida como $\tau_Q =
4M_Q^2/M_H^2$ y $\hat s$ denota la energ\'{\i}a del c.m.
part\'onica al cuadrado. El factor de forma puede ser evaluado
f\'acilmente:
\begin{eqnarray}
A_Q^H (\tau_Q) & = & \frac{3}{2} \tau_Q \left[ 1+(1-\tau_Q) f(\tau_Q)
\right] \label{esp-eq:ftau} \\
f(\tau_Q) & = & \left\{ \begin{array}{ll}
\displaystyle \arcsin^2 \frac{1}{\sqrt{\tau_Q}} & \tau_Q \geq 1 \\
\displaystyle - \frac{1}{4} \left[ \log \frac{1+\sqrt{1-\tau_Q}}
{1-\sqrt{1-\tau_Q}} - i\pi \right]^2 & \tau_Q < 1
\end{array} \right. \nonumber
\end{eqnarray}
Para masas peque\~nas de los lazos el factor de forma se hace
cero, $A_Q^H(\tau_Q) \sim -3/8 \tau_Q [\log (\tau_Q/4)+i\pi]^2$,
mientras que para masas grandes de los lazos se aproxima a un
valor distinto de cero, $A_Q^H (\tau_Q) \to 1$. El t\'ermino
final $BW$ es la funci\'on de Breit-Wigner normalizada
\beq
BW(\hat{s}) = \frac{M_H \Gamma_H/\pi}{[\hat{s}-M_H^2]^2 + M_H^2
\Gamma_H^2}
\eeq
acerc\'andose, en la aproximaci\'on de
anchura-estrecha (narrow-width) a una funci\'on $\delta$ en $\hat{s}=M_H^2$.

En la aproximaci\'on de anchura-estrecha, la secci\'on eficaz
hadr\'onica puede ponerse en la forma
\begin{equation}
\sigma_{LO} (pp\to H) = \sigma_0 \tau_H \frac{d{\cal L}^{gg}}{d\tau_H} ~,
\end{equation}
donde $d{\cal L}^{gg}/d\tau_H$ denota la luminosidad $gg$ del
colisionador $pp$,
\begin{equation}
  d{\cal L}^{gg}/d\tau_H = \int_{\tau_H}^1 \frac{d\xi}{\xi} \,
                           g(\xi; \tau_H s) \, g(\tau_H / \xi; \tau_H s) \,,
\end{equation}
establecido para densidades de gluones $g$ y evaluado para la
variable de Drell--Yan
$\tau_H = M_H^2/s$, donde $s$ es la energ\'{\i}a hadr\'onica total al cuadrado. \\

Las correcciones de QCD al proceso de fusi\'on de gluones
\cite{29,32,39B} son muy importantes. Estas estabilizan las
predicciones te\'oricas para la secci\'on eficaz cuando se
var\'{\i}an las escalas de re\-nor\-ma\-li\-za\-ci\'on y de
factorizaci\'on. M\'as a\'un, estas son grandes y positivas,
incrementando as\'{\i} la secci\'on eficaz de producci\'on para los
bosones de Higgs. Las correcciones de QCD consisten en correcciones
virtuales al proceso b\'asicos $gg\to H$, y a correcciones reales
debidas a la producci\'on asociada del bos\'on de Higgs con partones
sin masa, $gg\to Hg$ y $gq\to Hq,\, q\bar q\to Hg$. Estos
subprocesos contribuyen a la producci\'on del Higgs a orden de
${\cal O}(\alpha_s^3)$. Las correcciones virtuales reescalan la
secci\'on eficaz de fusi\'on al orden m\'as bajo con un coeficiente
que depende s\'olo en las razones de las masas del Higgs y de los
quarks.  La radiaci\'on del glu\'on conduce a estados finales de dos
partones con energ\'{\i}a invariante $\hat s \geq M_H^2$ en los
canales
$gg, gq$ y $q\bar q$.\\

\begin{figure}[hbt]

\vspace*{0.4cm}
\hspace*{2.0cm}
\begin{turn}{-90}%
\epsfxsize=7cm \epsfbox{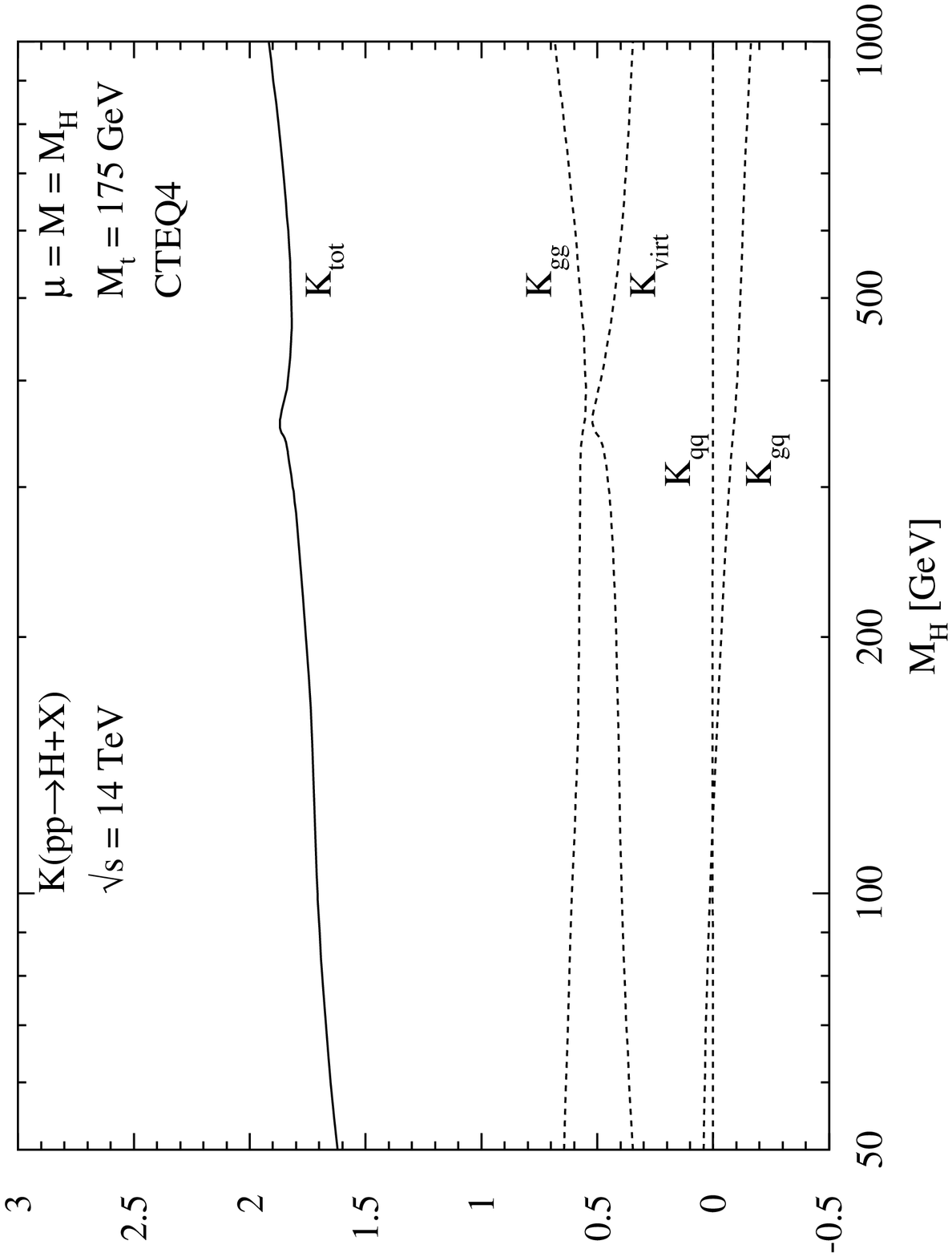}
\end{turn}
\vspace*{-0.2cm}

\caption[]{\label{esp-fg:gghk} \it Los factores K de la secci\'on
eficaz de la fusi\'on del glu\'on corregida por QCD $\sigma(pp \to
H+X)$ en el LHC con una energ\'{\i}a del c.m de $\sqrt{s}=14$ TeV.
Ls l\'{\i}neas punteadas muestran las contribuciones individuales
de las correcciones de QCD. Las escalas de renormalizaci\'on y
factorizaci\'on han sido identificadas con la masa del Higgs, y se
han adoptado densidades CTEQ4 del part\'on .}
\end{figure}

\begin{figure}[hbt]

\vspace*{-0.2cm}
\hspace*{2.7cm}
\epsfxsize=9.5cm
\epsfbox{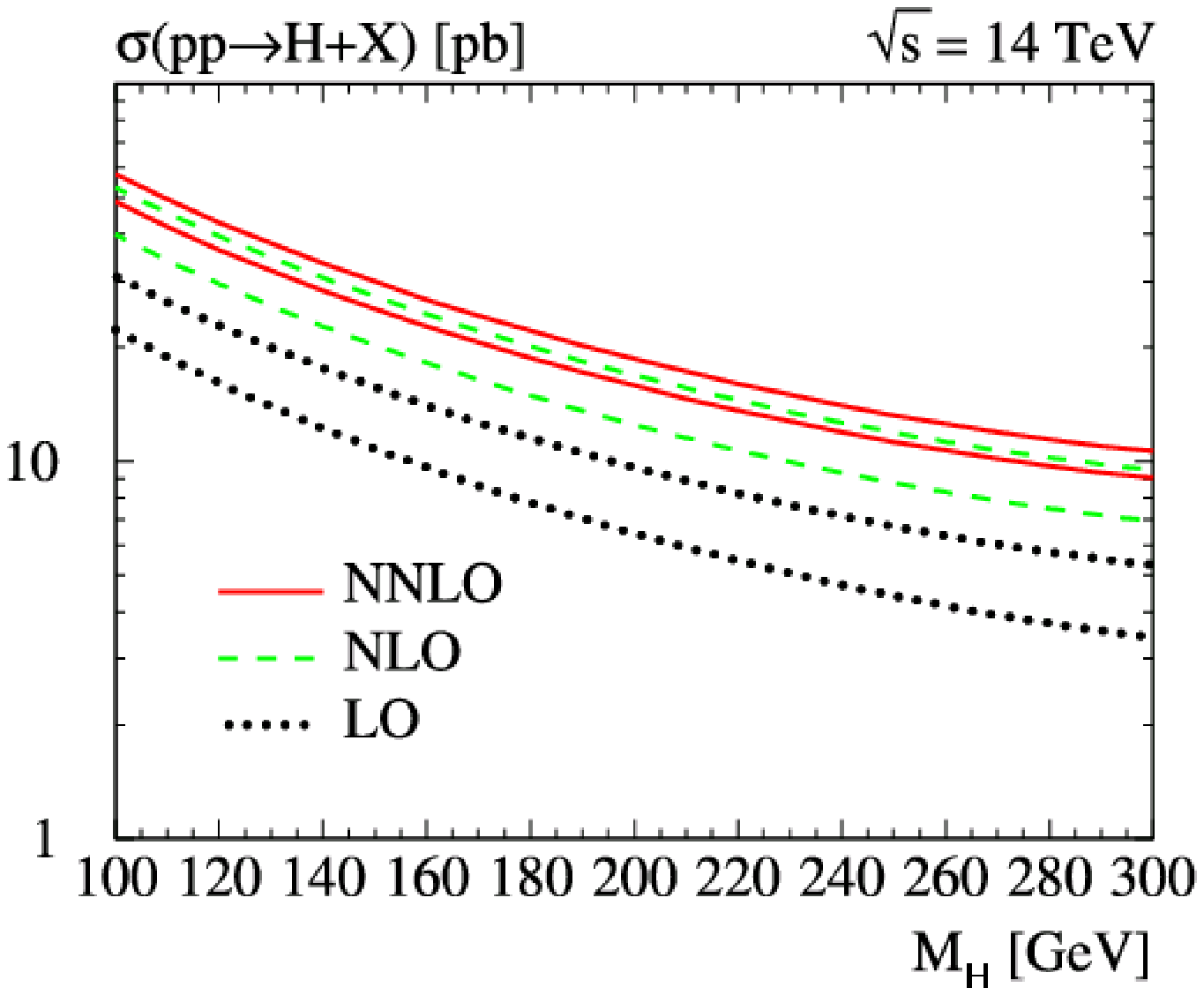} \vspace*{-0.3cm}

\caption[]{\label{esp-fg:MochVV(a)} \it La secci\'on eficaz de la
fusi\'on de gluones $\sigma(pp \to H+X)$ en el LHC con energ\'{\i}a
del c.m.~ $\sqrt{s}=14$ TeV a LO, NLO y NNLO. El tama\~no de las
bandas de error est\'a determinado por la variaci\'on de las escalas
de factorizaci\'on y renormalizaci\'on entre $M_H/2$ y $2M_H$. La
primera referencia en \cite{x}.}
\end{figure}

\begin{figure}[hbt]

\vspace*{-0.5cm} \hspace*{3.5cm} \epsfxsize=7.5cm
\epsfbox{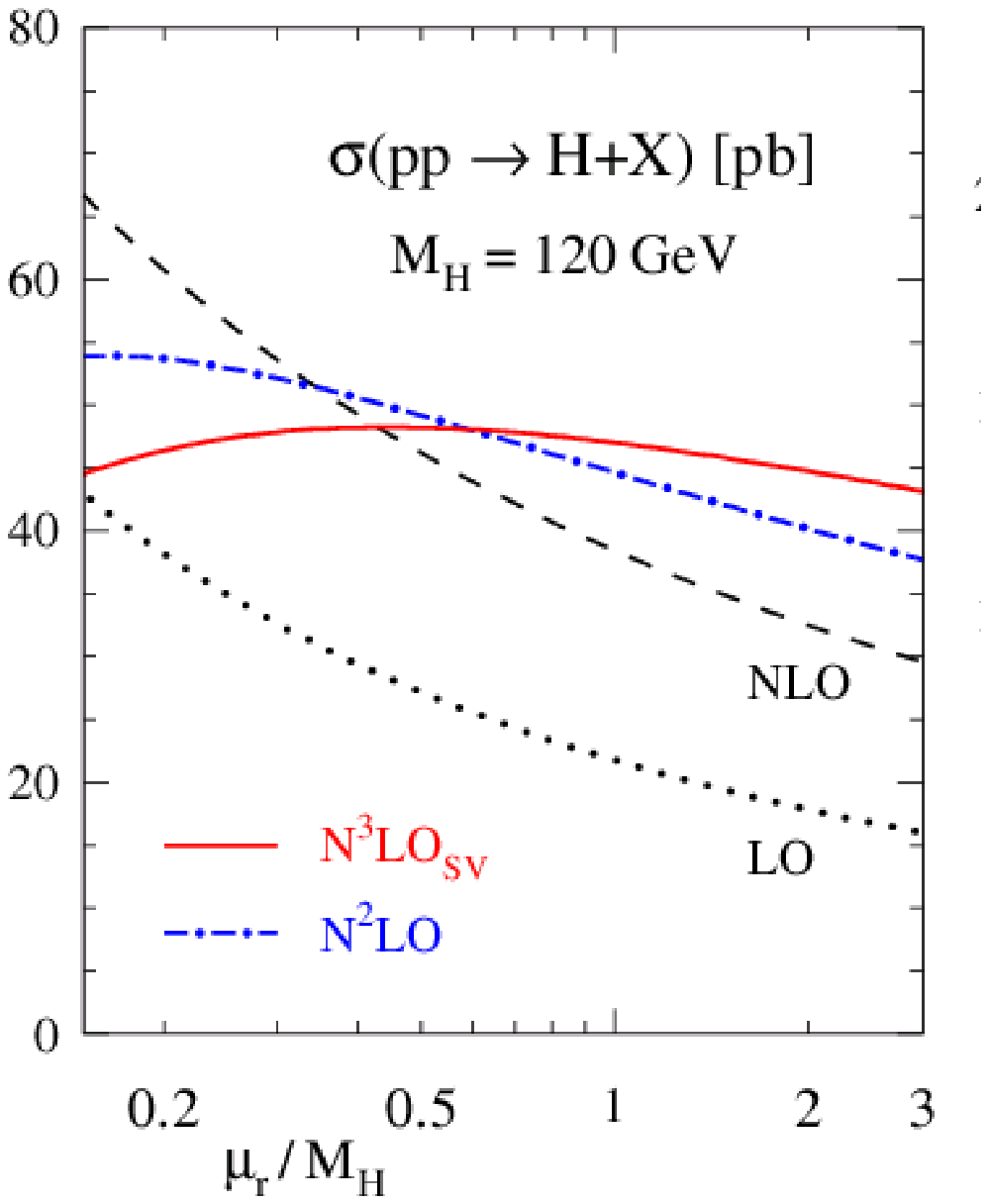} \vspace*{-0.2cm}

\caption[]{\label{esp-fg:MochVV(b)} \it La dependencia en la escala de
renormalizaci\'on de la secci\'on eficaz de la fusi\'on del glu\'on
 $\sigma(pp \to H+X)$ en el LHC
con energ\'{\i}a del c.m.~ $\sqrt{s}=14$ TeV a LO, NLO, NNLO y la
aproximaci\'on suave/virtual a N$^3$LO; de Ref.~\cite{MochVV}.}
\end{figure}

El tama\~no de las correcciones radiativas puede ser parametrizado
definiendo el factor $K$ como $K=\sigma_{NLO}/\sigma_{LO}$, en el
cual todas las cantidades son evaluadas en el numerador y
denominador, en el siguiente al primer orden (next-to-leading, NLO)
y en el primer orden respectivamente.  Los resultados de estos
c\'alculos se muestran en la Fig.~\ref{esp-fg:gghk}. Las correciones
virtuales $K_{virt}$ y las correcciones reales $K_{gg}$ para las
colisiones $gg$ son, aparentemente del mismo tama\~no, y ambas son
grandes y positivas; las correcciones para las colisiones $q\bar q$
y las contribuciones inel\'asticas de Compton $gq$ son menos
importantes. Dependiendo s\'olo suavemente de la masa del bos\'on de
Higgs, el factor total $K$, $K_{tot}$, resulta estar cerca de 2
\cite{29,32,39B,R}. Las contribuciones principales son generadas por
las correcciones virtuales y estados finales de 3-partones iniciados
por estados iniciales $gg$. Se esperan grandes correcciones NLO para
estos procesos de gluones como resultado de las grandes cargas de
color. Sin embargo, al estudiar los ordenes siguientes en la
correcci\'on en el l\'{\i}mite de masa grande del top, las
correcciones N$^{2}$LO generan \'unicamente un modesto incremento
adicional del factor $K$, $\delta_2 K_{tot} \lessim 0.2$ \cite{x}, e
incluso menor al N$^{3}$LO \cite{MochVV}. Esto prueba que la
expansi\'on es convergente, con las correcciones m\'as importantes
atribuidas a la contribuci\'on del orden siguiente al primero NLO
\cite{R}, cf. Fig.~\ref{esp-fg:MochVV(a)}. Adicionalmente, cuando se
incluyen las correcciones de QCD de orden m\'as alto, la dependencia
de la secci\'on eficaz sobre las escalas de factorizaci\'on y
renormalizaci\'on se reduce significativamente,
Fig.~\ref{esp-fg:MochVV(b)}.



\begin{figure}[hbt]

\vspace*{0.5cm} \hspace*{0.0cm}
\begin{turn}{-90}%
\epsfxsize=10cm \epsfbox{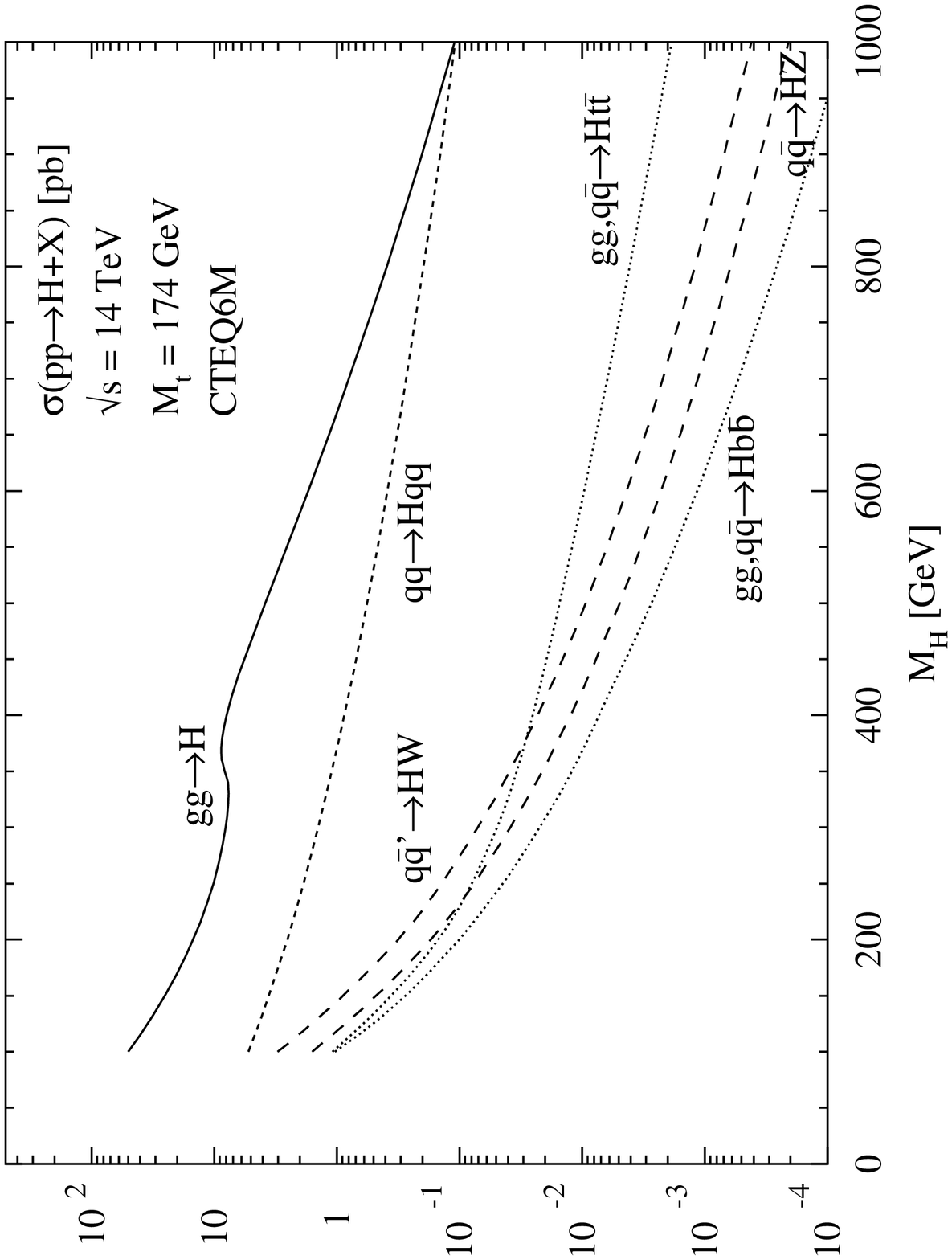}
\end{turn}
\vspace*{0.0cm}

\caption[]{\label{esp-fg:lhcpro} \it Las secciones eficaces de la
producci\'on de Higgs en el LHC para los varios mecanismos de
producci\'on como funciones de la masa del Higgs. Se muestran los
resultados con correcciones completas de QCD para la fusi\'on de
gluones $gg \to H$, para la fusi\'on de bosones vectoriales $qq\to
VVqq \to Hqq$, el bremsstrahlung del bos\'on vectorial $q\bar q \to
V^* \to HV$ y la producci\'on asociada $gg,q\bar q \to Ht\bar t,
Hb\bar b$.}
\end{figure}

La predicci\'on te\'orica para la secci\'on eficaz de producci\'on
de las part\'{\i}culas de Higgs es presentada en la
Fig.~\ref{esp-fg:lhcpro} para el LHC como funci\'on de la masa del
Higgs. La secci\'on eficaz disminuye conforme la masa del Higgs
aumenta. Esto es, en gran medida, una consecuancia de la ca\'{\i}da
abrubta de la luminosidad $gg$ para masas invariantes grandes.  La
protuberancia en la secci\'on eficaz es inducida por el umbral de
$t\bar t$
en el tri\'angulo del top. La precisi\'on te\'orica total de este
c\'alculo se espera al nivel de 10 al 20\%.

\vspace*{3mm}
\noindent {\it (b) \underline{Fusi\'on de bosones
vectoriales}}

\vspace*{1mm} \noindent El segundo canal importante para la
producci\'on de Higgs en el LHC es la fusi\'on de bosones
vectoriales, $W^+W^- \to H$ \cite{23}. Para masas grandes del Higgs,
este mecanismo se vuelve competitivo a la fusi\'on de gluones; para
masas intermedias la secci\'on eficaz es menor pero sin embargo muy
importante para la b\'usqueda de los bosones ligeros de Higgs con
una raz\'on reducida de se\~nales de fondo
 y para explorar tambi\'en sus propiedades \cite{Zep}.\\

Para masas grandes, los dos bosones electrod\'ebiles $W,Z$ que
forman el bos\'on de Higgs son predominantemente longitudinalmente
polarizados. A energ\'{\i}as altas, el espectro de part\'{\i}culas
equivalente de los bosones longitudinales $W,Z$ en haces de quark
est\'an dados por
\begin{eqnarray}
f^W_L (x) & = & \frac{G_F M_W^2}{2\sqrt{2}\pi^2} \frac{1-x}{x}
 \label{esp-eq:xyz} \\ \non \\
f^Z_L (x) & = & \frac{G_F M_Z^2}{2\sqrt{2}\pi^2}
\left[(I_3^q - 2e_q \sin^2\theta_W)^2 + (I_3^q)^2\right] \frac{1-x}{x} ~, \non
\end{eqnarray}
donde $x$ es la fracci\'on de la energ\'{\i}a transferida del
quark al bos\'on $W,Z$ en el proceso de escisi\'on (splitting) $q\to q +W/Z$.
De este espectro de part\'{\i}culas, las luminosidades de $WW$ y
$ZZ$ pueden f\'acilmente ser derivadas:
\begin{eqnarray}
\frac{d{\cal L}^{WW}}{d\tau_W} & = & \frac{G_F^2 M_W^4}{8\pi^4}
\left[ 2 - \frac{2}{\tau_W} -\frac{1+\tau_W}{\tau_W} \log \tau_W \right] \\
\non \\
\frac{d{\cal L}^{ZZ}}{d\tau_Z} & = & \frac{G_F^2 M_Z^4}{8\pi^4}
\left[(I_3^q - 2e_q \sin^2\theta_W)^2 + (I_3^q)^2\right]
\left[(I_3^{q'} - 2e_{q'} \sin^2\theta_W)^2 + (I_3^{q'})^2\right] \non \\
& & \hspace{1.5cm} \times \left[ 2 - \frac{2}{\tau_Z} -\frac{1+\tau_Z}{\tau_Z}
\log \tau_Z \right] \non
\end{eqnarray}
con la variable de Drell--Yan definida como $\tau_V = M_{VV}^2/s$.
La secci\'on eficaz para la producci\'on de Higgs en colisiones de
quark--quark est\'a dada por la convoluci\'on de las secciones
eficaces de partones $WW,ZZ \to H$ con luminosidades:
\begin{equation}
\hat \sigma(qq\to qqH) = \frac{d{\cal L}^{VV}}{d\tau_V} \sqrt{2} \pi G_F ~.
\label{esp-eq:vvhpart}
\end{equation}
La secci\'on eficaz hadr\'onica se obtiene finalmente al sumar la
seccion eficaz del part\'on (\ref{esp-eq:vvhpart}) sobre el flujo de
todos las posibles pares de combinaciones de quark--quark y antiquark. \\

Ya que al orden m\'as bajo los remanentes del prot\'on son singletes de
color en los procesos de fusi\'on de $WW,ZZ$, no habr\'a intercambio de
color entre las dos l\'{\i}neas de quarks de las cuales los dos
bosones vectoriales son radiados. Como resultado, las correcciones
dominantes de QCD a estos procesos ya est\'an tomadas en cuenta por las
correcciones a las densidades part\'onicas de los quarks.\\

La secci\'on eficaz de la fusi\'on de $WW/ZZ$ para el bos\'on de
Higgs en el LHC se muestra en la Fig.~\ref{esp-fg:lhcpro}. El proceso es
aparentemente muy importante para la b\'usqueda del bos\'on de Higgs
en el rango superior de la masa, donde la secci\'on eficaz se
aproxima a valores cercanos a la fusi\'on de gluones. Para masas
intermedias, se acerca m\'as, dentro de un orden de magnitud, a la
secci\'on eficaz dominante de fusi\'on--glu\'on.

\vspace*{3mm}
\noindent {\it (c) \underline{Emisi\'on radiativa de
Higgs por bosones vectoriales (Higgs-strahlung off vector bosons)}}

\vspace*{1mm} \noindent El proceso de radiaci\'on de Higgs
(Higgs-strahlung) $q\bar q \to V^* \to VH~(V=W,Z)$ es un
me\-ca\-nis\-mo muy importante (Fig.~\ref{esp-fg:lhcpro}) para la
b\'usqueda de bosones de Higgs ligeros en los colisionadores
hadr\'onicos Tevatron y LHC. Aunque la secci\'on eficaz es menor que
la de fusi\'on del glu\'on, los decaimientos lept\'onicos de los
bosones vectoriales electrod\'ebiles son \'utiles para filtrar
eventos de se\~nales de Higgs de el enorme fondo. Como el mecanismo
din\'amico es el mismo que para los colisionadores $e^+e^-$ [{\it
ver m\'as adelante}], excepto para el doblamiento con las densidades
quark--antiquark, los pasos intermedios del c\'alculo no necesitan
ser mencionados aqu\'\i, y simplemente se registran en la
Fig.~\ref{esp-fg:lhcpro} los valores finales de las secciones
eficaces para el Tevatron y el LHC.

\vspace*{3mm}
\noindent {\it (d) \underline{Higgs bremsstrahlung
emitido por quarks top}}

\vspace*{1mm}
\noindent

Tambi\'en el proceso $gg,q\bar q \to t\bar t H$ es relevante s\'olo
para masas peque\~nas del Higgs, Fig.~\ref{esp-fg:lhcpro}. La
expresi\'on anal\'{\i}tica para la secci\'on eficaz del part\'on,
incluso al orden m\'as bajo, es bastante complicada, de manera que
s\'olo se muestran los resultados finales para las secciones
eficaces del LHC en la Fig.~\ref{esp-fg:lhcpro}. Correcciones a m\'as
alto orden han sido presentadas en la Ref.~\cite{z}. Separar las
se\~nales de este canal de las se\~nales de fondo es
experimentalmente muy dif\'{\i}cil.

Sin embargo, el Higgs-bremsstrahlung emitido por quarks top
podr\'{\i}a ser tambi\'en un proceso interesante para las mediciones del
acoplamiento fundamental de Yukawa $Htt$ en an\'alisis coherentes de
LHC/LC. La secci\'on eficaz $\sigma (pp\to t\bar t H)$ es
directamente proporcional al cuadrado de este acoplamiento
fundamental.

\vspace*{3mm}
\noindent {\it{\underline{Resumen}}}

\vspace*{1mm}
\noindent
 Una visi\'on general de las secciones
eficaces de producci\'on para las part\'{\i}culas de Higgs en el LHC
es presentada en la Fig.~\ref{esp-fg:lhcpro}. Tres tipos de canales
pueden distinguirse. La fusi\'on de glu\'on de las part\'{\i}culas
de Higgs es un proceso universal, dominante sobre el rango entero de
masas de Higgs del SM. La emisi\'on radiativa de Higgs por bosones
electrod\'ebiles $W,Z$ o quarks top es prominente para bosones de
Higgs ligeros. El canal de fusi\'on de $WW/ZZ$, en contraste, se
vuelve cada vez m\'as importante en la parte superior del rango de
masas del Higgs del SM, aunque tambi\'en prueba ser \'util en el
rango intermedio de masa.

Las se\~nales para la b\'usqueda de part\'{\i}culas de Higgs son
dictadas por las razones de desintegraci\'on de decaimiento (decay
branching ratios). El la parte inferior del rango de masas
intermedio, la re\-cons\-truc\-ci\'on de resonancias en estados
finales $\gamma\gamma$ y $b\bar b$ de jets puede ser explotada. En
la parte superior del rango intermedio de masas, los decaimientos a
$ZZ^*$ y $WW^*$ son importantes, con los dos bosones
electrod\'ebiles decayendo lept\'onicamente. En el rango de masas
por arriba del umbral del decaimiento dentro de la capa de masa del
$ZZ$, los decaimientos a leptones cargados $H\to ZZ \to 4\ell^\pm$
proporcionan se\~nales muy valiosas.  En el l\'\i mite superior del
rango cl\'asico de masas de Higgs del SM, decaimientos a neutrinos y
jets, generados en los decaimientos de $W$ y $Z$, completan las
t\'ecnicas de b\'usqueda.

\begin{figure}[t]

\vspace*{-0.8cm}\hspace*{3.0cm} \epsfxsize=10cm \epsfbox{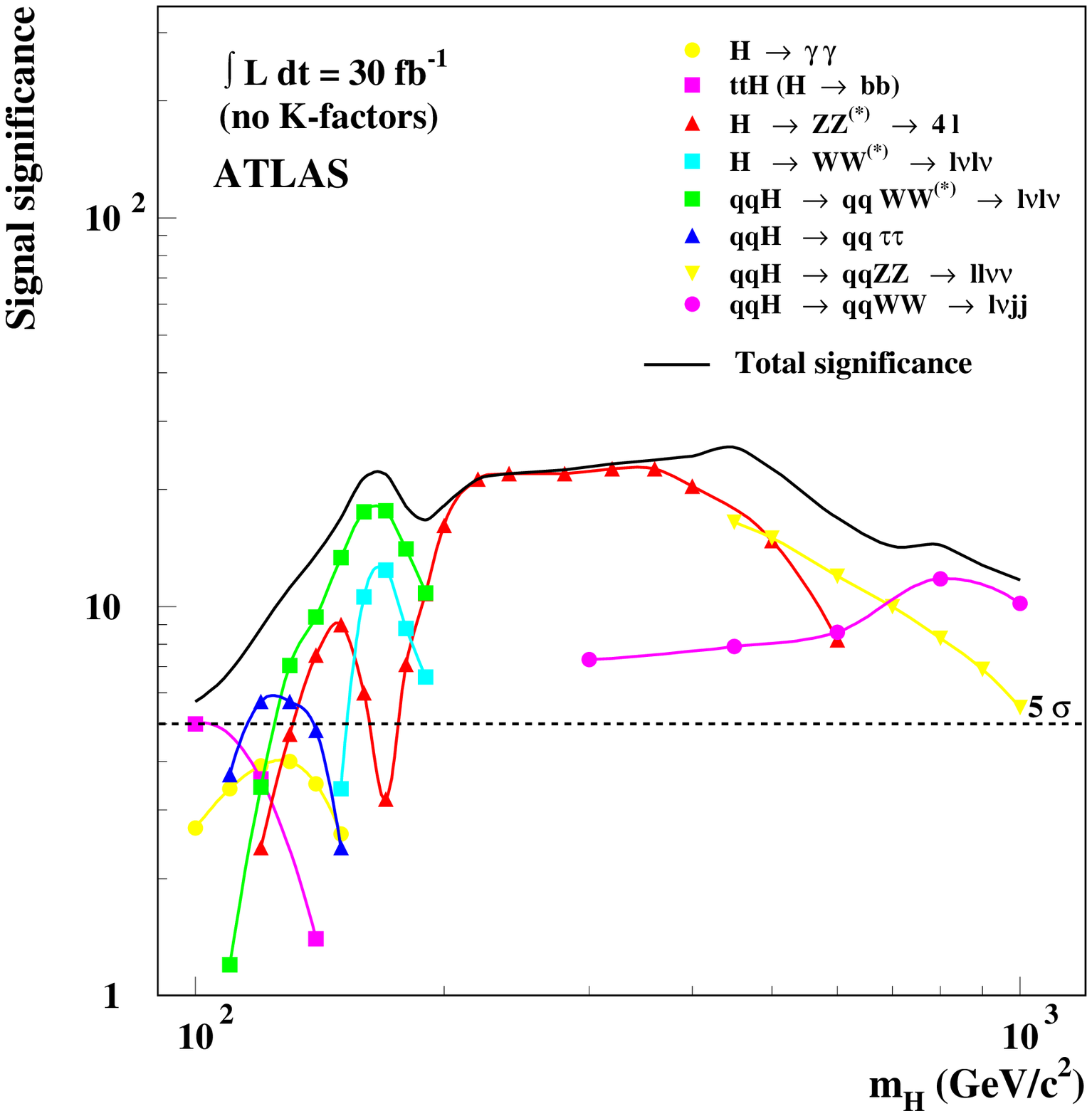}
\vspace*{-0.3cm}

\caption[]{\label{esp-fg:lhcsig} \it Relevancia significativa
 devarios canales en la b\'usqueda del bos\'on de Higgs del SM en el LHC
 como funci\'on de la masa  para una luminosidad integrada de 30 {\rm{fb}}$^{-1}$; Ref.\cite{12,12A}.}
\end{figure}

Las expectativas experimentales en el LHC para la b\'usqueda de la
part\'icula de Higgs del Modelo Est\'andar se resumen en la
Fig.~\ref{esp-fg:lhcsig}. La relevancia de la se\~nal del Higgs se
muestra como funci\'on de la masa del Higgs para una luminosidad
integrada de 30 fb$^{-1}$. El intervalo completo de masas para la
b\'usqueda del bos\'on de Higgs del ME puede ser cubierto en el
LHC.\\
\begin{figure}[hbt]

\vspace*{-0.4cm} \hspace*{1.6cm} \epsfxsize=12cm \epsfbox{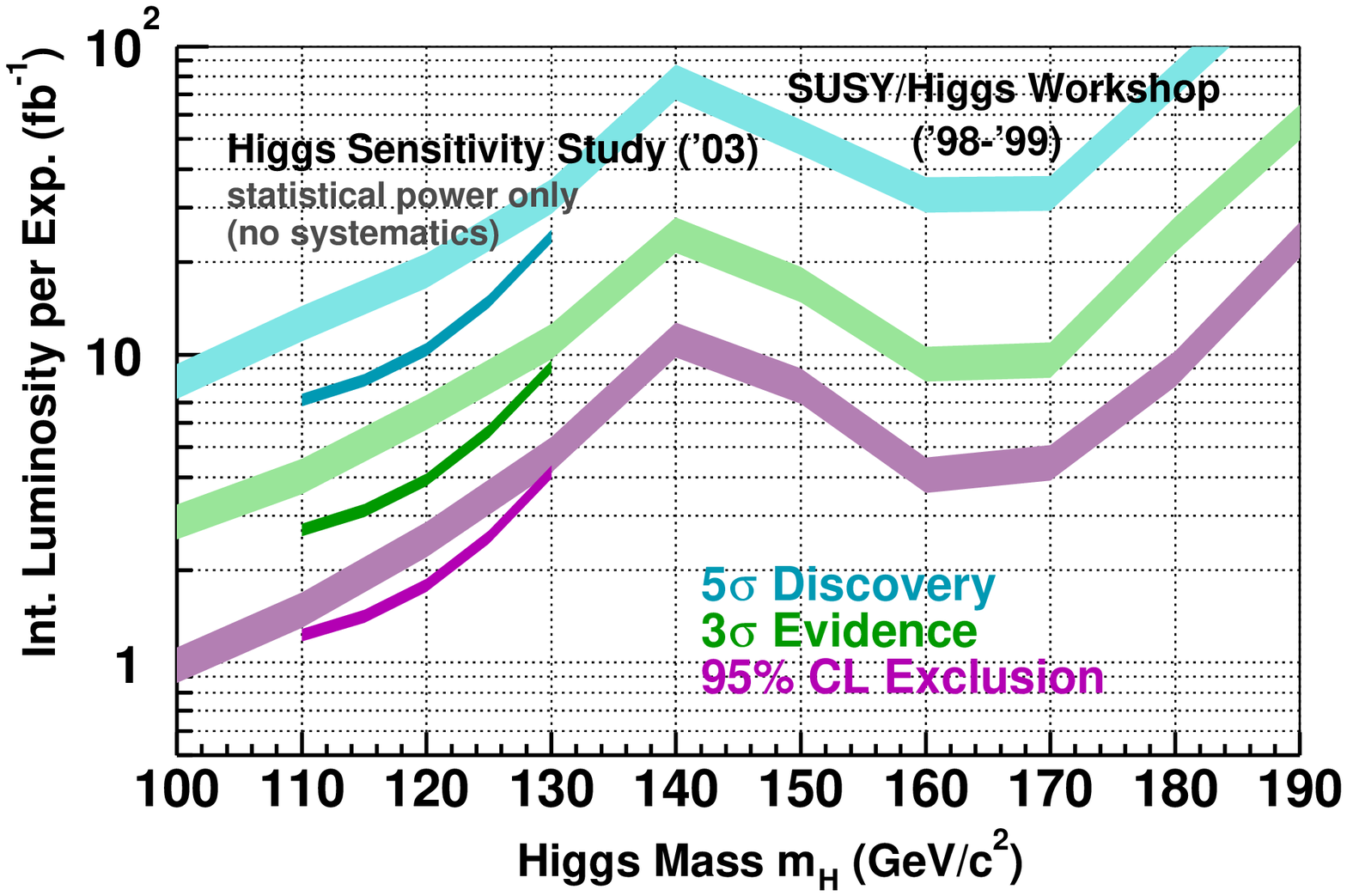}
\vspace*{-0.3cm}

\caption[]{\label{esp-fg:Tevsig} \it  Luminosidades integradas
necesarias para excluir o para decubrir el bos\'on de Higgs del SM
en el Tevatron; Ref.\cite{11}.}
\end{figure}

Las expectativas para la b\'usqueda del bos\'on de Higgs del ME en
el Tevatron est\'an resumidas en la Fig. ~\ref{esp-fg:Tevsig}. El
bos\'on de Higgs del ME puede quedar excluido a un nivel de
$2\sigma$ en todo el intervalo intermedio de masas. Descubrir la
part\'icula, incluso en intervalos restringidos de masa, es una
tarea muy exigente
que requiere de la acumulaci\'on de una luminosidad integrada grande.\\

\vspace*{4mm}
\subsection{Canales de Producci\'on de Higgs en colisionadores $e^+e^-$}

\vspace*{2mm}
\noindent
El primer proceso que fue usado para buscar
directamente los bosones de Higgs sobre un rango de masas grande fue
el proceso de Bjorken, $Z\to Z^* H, Z^* \to f\bar f$ \cite{34}. Al
explorar este canal de producci\'on se excluyeron valores para la
masa del bos\'on de Higgs menores a 65.4 GeV por los experimentos en
el LEP1. La b\'usqueda continu\'o al revertir los papeles de los
bosones $Z$ real y virtual en el continuo $e^+e^-$ en el LEP2.

EL principal mecanismo de producci\'on de los bosones de Higgs en
colisiones $e^+e^-$ son
\begin{eqnarray}
\mbox{Higgs-strahlung} & : & e^+e^- \to Z^* \to ZH \\
\mbox{fusi\'on $WW$}     & : & e^+e^- \to \bar \nu_e \nu_e (WW) \to
\bar \nu_e \nu_e H \label{esp-eq:wwfusion}
\end{eqnarray}
En Higgs-strahlung \cite{30,34,35} el bos\'on de Higgs es emitido de
la l\'{\i}nea de bos\'on $Z$, mientras que la fusi\'on $WW$ es un
proceso de formaci\'on de los bosones de Higgs en la colisi\'on de
dos bosones $W$ cuasi-reales radiados de los haces del electr\'on y
del positr\'on \cite{36}.

Como es evidente del an\'alisis subsecuente, el LEP2 pudo cubrir el
rango de masa del Higgs del SM hata alrededor de 114 GeV \cite{9}.
Los colisionadores lineales $e^+e^-$ de alta energ\'{\i}a pueden
cubrir el rango completo de la masa del Higgs, el rango de masa
intermedio lo cubre ya un colisionador de 500 GeV \cite{13}, el
rango superior de masa se cubrir\'a en la segunda fase de las
m\'aquinas en las cuales se espera alcanzar una energ\'{\i}a total
de al menos 3~TeV \cite{38A}.

\vspace*{3mm}
\noindent
{\it (a) \underline{Higgs-strahlung}}

\vspace*{1mm}
\noindent La secci\'on eficaz para el Higgs-strahlung
puede ser escrita en una forma compacta como
\begin{equation}
\sigma (e^+e^- \to ZH) = \frac{G_F^2 M_Z^4}{96\pi s} \left[ v_e^2 +
a_e^2 \right] \lambda^{1/2} \frac{\lambda + 12 M_Z^2/s}{\left[ 1-
M_Z^2/s \right]^2} ~,
\end{equation}
donde $v_e = -1 + 4 \sin^2 \theta_W$ y $a_e=-1$ son las cargas $Z$
vector y vector-axial del electr\'on, y $\lambda = [1-(M_H+M_Z)^2/s]
[1-(M_H-M_Z)^2/s]$ es la funci\'on usual de espacio fase de dos
part\'{\i}culas. La secci\'on eficaz es del tama\~no de $\sigma \sim
\alpha_W^2/s$, i.e. de segundo orden en el acoplamiento d\'ebil, y
se escala en el cuadrado de la energ\'{\i}a. Contribuciones a m\'as
alto orden a las secciones eficaces est\'an bajo control te\'orico
\cite{38B,38C}.

\begin{figure}[hbt]

\vspace*{-5.0cm} \hspace*{-2.0cm} \epsfxsize=20cm \epsfbox{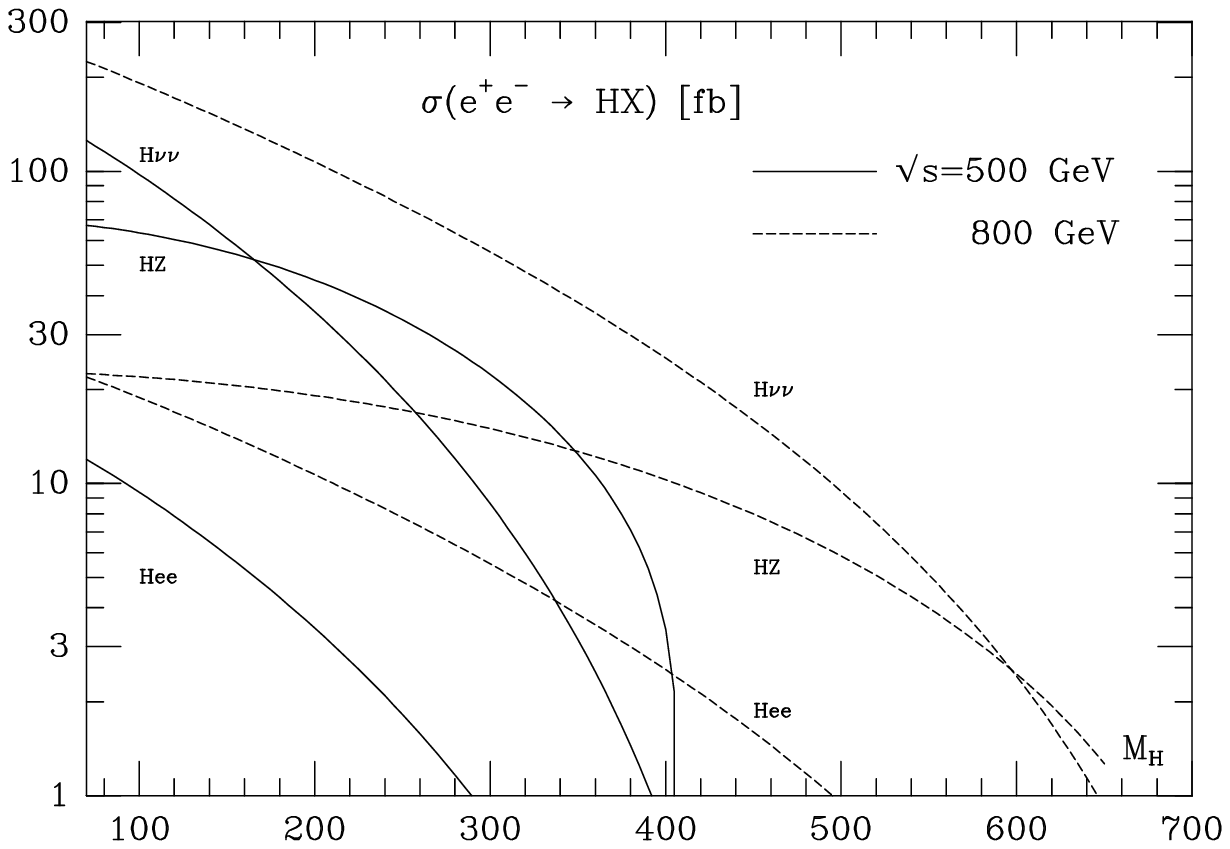}

\caption[]{\label{esp-fg:eehx} \it La secci\'on eficaz para la
producci\'on de bosones de Higgs del SM en el Higgs-strahlung
$e^+e^-\to ZH$, $WW/ZZ$ y fusi\'on $e^+e^- \to \bar \nu_e
\nu_e/e^+e^- H$; las curvas s\'olidas: $\sqrt{s}=500$ GeV, curvas
punteadas: $\sqrt{s}=800$ GeV.}
\end{figure}

\begin{figure}[hbt]
\begin{center}
\hspace*{-0.3cm}
\epsfig{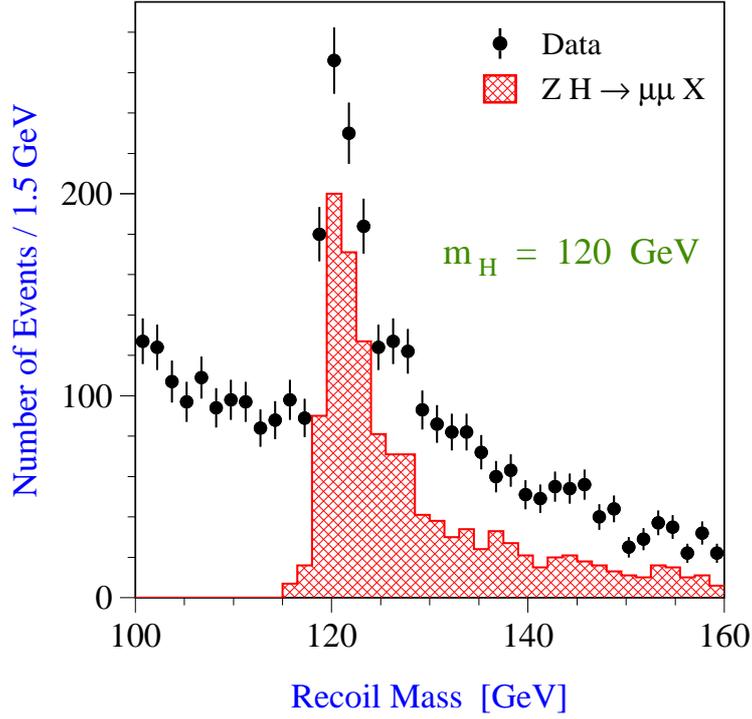}
\end{center}
\vspace*{-0.4cm}

\caption[]{\label{esp-fg:zrecoil} \it La distribuci\'on de masa de
retroceso $\mu^+\mu^-$ en el proceso $e^+e^- \to H^0 Z\to X
\mu^+\mu^-$ para $M_H=120$~GeV y $\int {\cal L} = 500 fb^{-1}$ a
$\sqrt{s}=$ 350~GeV. Los puntos con barra de error son simulaciones
de Monte Carlo de la se\~nal de Higgs y se\~nales de fondo. EL
histograma sombreado representa s\'olo la se\~nal. Ref. \cite{13}.}
\end{figure}

Como la secci\'on eficaz se hace cero para energ\'{\i}as
asint\'oticas, el proceso de Higgs-strahlung es m\'as \'util para la
b\'usqueda de bosones de Higgs en el rango donde la energ\'{\i}a del
colisionador es del mismo orden que la masa del Higgs, $\sqrt{s}
\gsim {\cal O} (M_H)$. El tama\~no de la secci\'on eficaz se ilustra
en la Fig.~\ref{esp-fg:eehx} para energ\'{\i}as $\sqrt{s}=500$ GeV del
colisionador lineal $e^+e^-$ como funci\'on de la masa del Higgs.
Como la masa $Z$ de retroceso en la reacci\'on de dos cuerpos
$e^+e^- \to ZH$ es mono-energ\'etica, la masa del bos\'on de Higgs
puede ser reconstruida a partir de la energ\'{\i}a del bos\'on $Z$,
$M_H^2 = s -2\sqrt{s}E_Z + M_Z^2$, sin la necesidad de analizar el
producto de los decaimientos del bos\'on del Higgs. Para los
decaimientos lept\'onicos del $Z$, t\'ecnicas de p\'erdida de masa
proporcionan una se\~nal muy clara, como se demuestra en la
Fig.~\ref{esp-fg:zrecoil}.

\vspace*{3mm}
\noindent {\it (b) \underline{Fusi\'on $WW$}}

\vspace*{1mm}
\noindent
Tambi\'en la secci\'on eficaz del proceso de
fusi\'on (\ref{esp-eq:wwfusion}) puede ser puesta impl\'{\i}citamente en
una forma compacta:
\begin{eqnarray}
\sigma (e^+e^-\to\bar \nu_e \nu_e H) & = & \frac{G_F^3
M_W^4}{4\sqrt{2}\pi^3}
\int_{\kappa_H}^1\int_x^1\frac{dx~dy}{[1+(y-x)/\kappa_W ]^2}f(x,y)
\\ \nonumber \\
f(x,y) & = & \left( \frac{2x}{y^3} - \frac{1+3x}{y^2} +
\frac{2+x}{y} -1 \right) \left[ \frac{z}{1+z} - \log (1+z) \right] +
\frac{x}{y^3} \frac{z^2(1-y)}{1+z} ~, \nonumber
\end{eqnarray}
con $\kappa_H=M_H^2/s$, $\kappa_W=M_W^2/s$ and
$z=y(x-\kappa_H)/(\kappa_Wx)$.

Dado que el proceso de fusi\'on es un proceso de intercambio del
canal-$t$, el tama\~no est\'a establecido por la longitud de onda de
Compton de $W$, suprimida, sin embargo, con respecto al
Higgs-strahlung por la tercera potencia del acoplamiento
electrod\'ebil, $\sigma \sim \alpha_W^3/M_W^2$. Como resultado, la
fusi\'on de $W$ se convierte en el proceso de producci\'on principal
para las part\'{\i}culas de Higgs a energ\'{\i}as altas. A
energ\'{\i}as asint\'oticas la secci\'on eficaz se simplifica a
\begin{equation}
\sigma (e^+e^- \to \bar \nu_e \nu_e H) \to \frac{G_F^3
M_W^4}{4\sqrt{2}\pi^3} \left[ \log\frac{s}{M_H^2} - 2 \right] ~.
\end{equation}
En este l\'\i mite, la fusi\'on de $W$ a bosones de Higgs puede ser
interpretado como un proceso de dos pasos: los bosones $W$ son
radiados como part\'{\i}culas cuasi-reales  a partir de los
electrones y positrones, $e \to \nu W$, con un tiempo de vida del
estado dividido del orden de $E_W / M^2_W$;
con la formaci\'on subsecuente de los bosones de Higgs en los rayos
$W$ que colisionan. Las correcciones electrod\'ebiles de orden m\'as
alto est\'an bajo control \cite{38C}.

En la Fig.~\ref{esp-fg:eehx} se compara el tama\~no de la secci\'on
transversal de fusi\'on con el proceso de Higgs-strahlung. A
$\sqrt{s}=500$ GeV  las dos secciones eficaces son del mismo orden,
no obstante, el proceso de fusi\'on se vuelve cada vez m\'as
importante con el incremento en la energ\'{\i}a.

\vspace*{4mm}
\subsection{El Perfil del Higgs en el ME}
\vspace*{2mm}
\noindent
 Para establecer experimentalmente el
mecanismo de Higgs, debe explorarse la naturaleza de esta
part\'{\i}cula midiendo todas sus caracter\'{\i}sticas, su masa y
tiempo de vida, los n\'umeros cu\'anticos externos
esp\'{\i}n-paridad, los acoplamientos a los bosones de norma y
fermiones, y por \'ultimo pero no menos importante, los
auto-acoplamientos del Higgs.  Mientras que parte de este programa
puede ser realizado en LHC \cite{12,xx}, el perfil completo de la
part\'{\i}cula puede ser reconstruido a trav\'es de todo el rango de
masa en colisionadores $e^+ e^-$ \cite{13}.

\vspace*{3mm}
\noindent {\it (a) \underline{Masa}}

\vspace*{1mm}
\noindent
La masa de la part\'{\i}cula de Higgs puede ser medida juntando los
productos de decaimiento de la part\'\i cula en colisionadores de hadrones
y de $e^+e^-$.  Valores al nivel del dos por mil pueden ser alcanzados
con \'este m\'etodo en LHC \cite{12,12A}.  M\'as a\'un, en colisiones $e^+e^-$
el Higgs-strahlung puede ser explotado para reconstruir muy
precisamente la masa a partir de la energ\'{\i}a de retroceso del $Z$
en los procesos de dos cuerpos $e^+e^-\to ZH$. Se puede esperar una
precisi\'on total de $\delta M_H \sim 100$ MeV \cite{13}.

\vspace*{3mm}
\noindent {\it (b) \underline{Anchura / Tiempo de vida}}

\vspace*{1mm} \noindent La anchura del estado, i.e. el tiempo de
vida de la part\'{\i}cula, puede directamente ser medido por arriba
del umbral de decaimiento de $ZZ$ donde la anchura crece
r\'apidamente. En la parte inferior del rango intermedio de masa el
ancho puede ser medido indirectamente \citer{12,13} combinando la
raz\'on de desintegraci\'on $H\to WW$ con la medida del ancho
parcial del $WW$, accesible a trav\'es de la secci\'on eficaz para
la fusi\'on del bos\'on $W$: $\Gamma_{tot} = \Gamma_{WW} / BR_{WW}$.
Por lo tanto, la anchura total para la part\'{\i}cula de Higgs puede
ser determinado a trav\'es de todo el rango de masa cuando los
resultados del LHC y los colisionadores $e^+e^-$ puedan ser
combinados.

\vspace*{3mm}
\noindent {\it (c) \underline{Esp\'{\i}n-paridad}}

\vspace*{1mm}
\noindent

%
\vspace*{1mm}
\noindent {\bf 1.$\,$} El esp\'in cero de la
part\'icula de Higgs puede ser determinado por la distribuci\'on
isotr\'opica de los productos del decaimientos \cite{Muhll,vdB}.
M\'as a\'un, la paridad puede medirse al observar las correlaciones de
esp\'in de los productos del decaimiento. De acuerdo con el teorema
de equivalencia, los \'angulos azimutales de los planos de
decaimiento en $H\to ZZ\to (\mu^+\mu^-) (\mu^+\mu^-)$ est\'an
asint\'oticamente descorrelacionados, $d\Gamma^+/d\phi_* \to 0$,
para una part\'icula $0^{+}$; esto deber\'a ser contrastado con
$d\Gamma^-/d\phi_* \; \to 1-\frac{1}{4} \cos 2\phi_*$ para la
distribuci\'on del \'angulo azimutal entre los planos de decaimiento
de una part\'icula $0^{-}$. La diferencia entre las distribuciones
angulares es una consecuencia de los diferentes estados de
polarizaci\'on de los bosones vectoriales en los dos casos. Mientras
se aproximan a los estados de polarizaci\'on longitudinal para los
decaimientos del Higgs escalar, est\'an transversalmente polarizados
para los decaimientos de la
part\'icula pseudoescalar.

En el intervalo bajo de masas, en el cual los decaimientos del Higgs
a pares de bos\'on $Z$ est\'an suprimidos, la distribuci\'on angular
azimutal  entre los jets de quarks que acompa\~nan en la fusi\'on
$WW$ puede ser explotada para medir la paridad \cite{Zeppaz}.
Mientras los jets esten cercanamente descorrelacionados para la
producci\'on del bos\'on de Higgs en el Modelo Est\'andar, la
correlaci\'on  es de car\'acter oscilatorio marcadamente diferente
para la producci\'on de una part\'icula pseudoescalar [{\it i.e.}
$\mathcal{CP}-non$] los jets apuntando preferentemente en direcciones
perpendiculares una de la otra.

\begin{figure}[hbt]
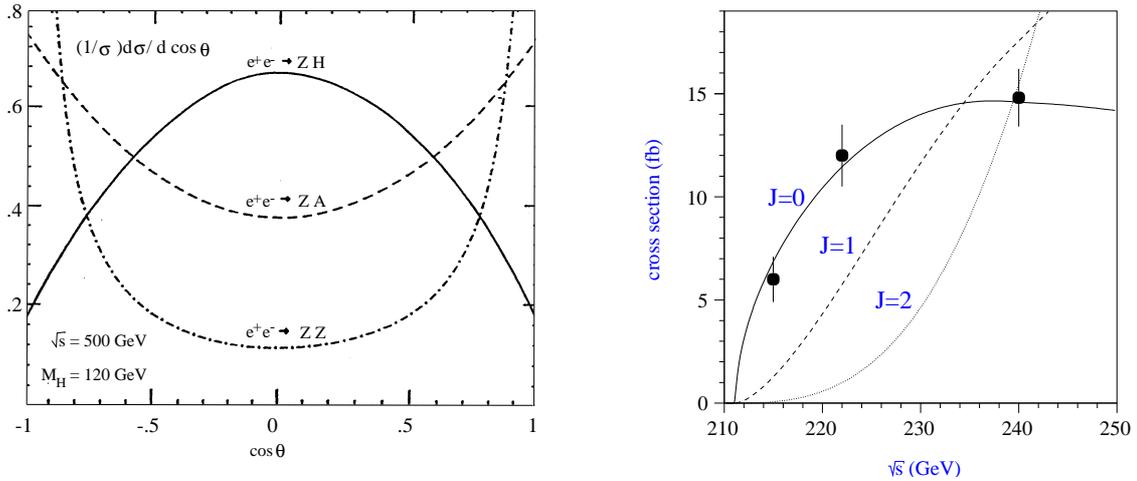


\vspace*{0.5cm} \hspace*{0.0cm} \epsfxsize=8cm \epsfbox{higgs_sm_spin_th.eps}
\vspace*{0.0cm}

\vspace*{-6.3cm} \hspace*{9.0cm}
\epsfig{figure=spinexp.eps,bbllx=0,bblly=0,bburx=560,bbury=539,width=6.5cm,clip=}
\vspace*{0.0cm}

\caption[]{\it \label{esp-fg:spinpar} Izquierda:  Distribuci\'on
angular de los bosones $Z/H$ en el Higgs-strahlung, comparada con
la producci\'on de part\'{\i}culas pseudo-escalares y los estados
finales de fondo $ZZ$;
 Ref.~\cite{41}.
Derecha: Umbral de excitaci\'on del  Higgs-strahlung que discrimina
el  spin=0 de otras asignaciones, Ref. \cite{MMM,exp}.}
\end{figure}

\vspace*{1mm} \noindent {\bf 2.$\,$} La distribuci\'on angular de
los bosones  $Z/H$ en el proceso de Higgs-strahlung es sensible al
esp\'{\i}n y paridad de la part\'{\i}cula de Higgs \cite{41}. Como
la amplitud de producci\'on est\'a dada por ${\cal A}(0^+) \sim
\vec{\epsilon}_{Z^*} \cdot \vec{\epsilon}_Z$, el bos\'on $Z$ es
producido en un estado de polarizaci\'on longitudinal a
energ\'{\i}as altas
 -- en concordancia con el teorema de equivalencia.
 Como resultado, la distribuci\'on angular
\begin{equation}
\frac{d\sigma}{d\cos\theta} \sim \sin^2 \theta + \frac{8M_Z^2}{\lambda s}
\end{equation}
se aproxima a la ley para esp\'{\i}n cero $\sin^2\theta$
asint\'oticamente. Esto puede ser contrastado con la distribuci\'on
$\sim 1 + \cos^2\theta$ para estados de paridad negativa, la cual
viene de amplitudes de polarizaci\'on transversal ${\cal A}(0^-)
\sim \vec{\epsilon}_{Z^*} \times \vec{\epsilon}_Z \cdot \vec{k}_Z$.
Es tambi\'en caracter\'\i sticamente diferente de la distribuci\'on
del proceso de fondo $e^+e^- \to ZZ$, el cual, como resultado del
intercambio de $e$ en los canales $t/u$, tiene un pico estrecho en
la direcci\'on adelante/atr\'as,
Fig.~\ref{esp-fg:spinpar}~izquierda.

Un m\'etodo diferente para determinar el esp\'{i}n del bos\'on de
Higgs es porporcionado por la exploraci\'on del inicio de la curva
de excitaci\'on en el Higgs-strahlung \cite{MMM} $e^+e^- \to ZH$.
Para el esp\'{\i}n del Higgs $S_H = 0$ la curva de excitaci\'on se
levanta abruptamente en el umbral $\sim \sqrt{s - (M_H + M_Z)^2}$.
Esta conducta es claramente diferente de las excitaciones de
esp\'{\i}n m\'as altas, las cuales se elevan con una potencia $> 1$
en el factor de umbral.  Una ambig\"{u}edad para estados con
esp\'{\i}n/paridad $1^+$ y $2^+$ puede ser resuelta evaluando
tambi\'en la distrubuci\'on angular del Higgs y el bos\'on $Z$ en el
proceso Higgs-strahlung. La precisi\'on experimental ser\'a
suficiente para discriminar la asignaci\'on del esp\'{\i}n-0 al
bos\'on de Higgs de otras asignaciones, como se muestra en la
Fig.~\ref{esp-fg:spinpar}~derecha.

\vspace*{3mm} \noindent {\it (d) \underline{Acoplamientos del
Higgs}}

\vspace*{1mm}
\noindent
Dado que las part\'{\i}culas fundamentales
adquieren masa a trav\'es de la interacci\'on con el campo de Higgs,
la intensidad de los acoplamientos del Higgs a fermiones y bosones
de norma est\'a establecida por las masas de las part\'{\i}culas.
Ser\'a una tarea experimental crucial medir estos acoplamientos, los
cuales son
predichos un\'\i vocamente por la naturaleza misma del mecanismo de Higgs.\\

\begin{figure}[t]
\hspace*{4.3cm} \epsfig{figure=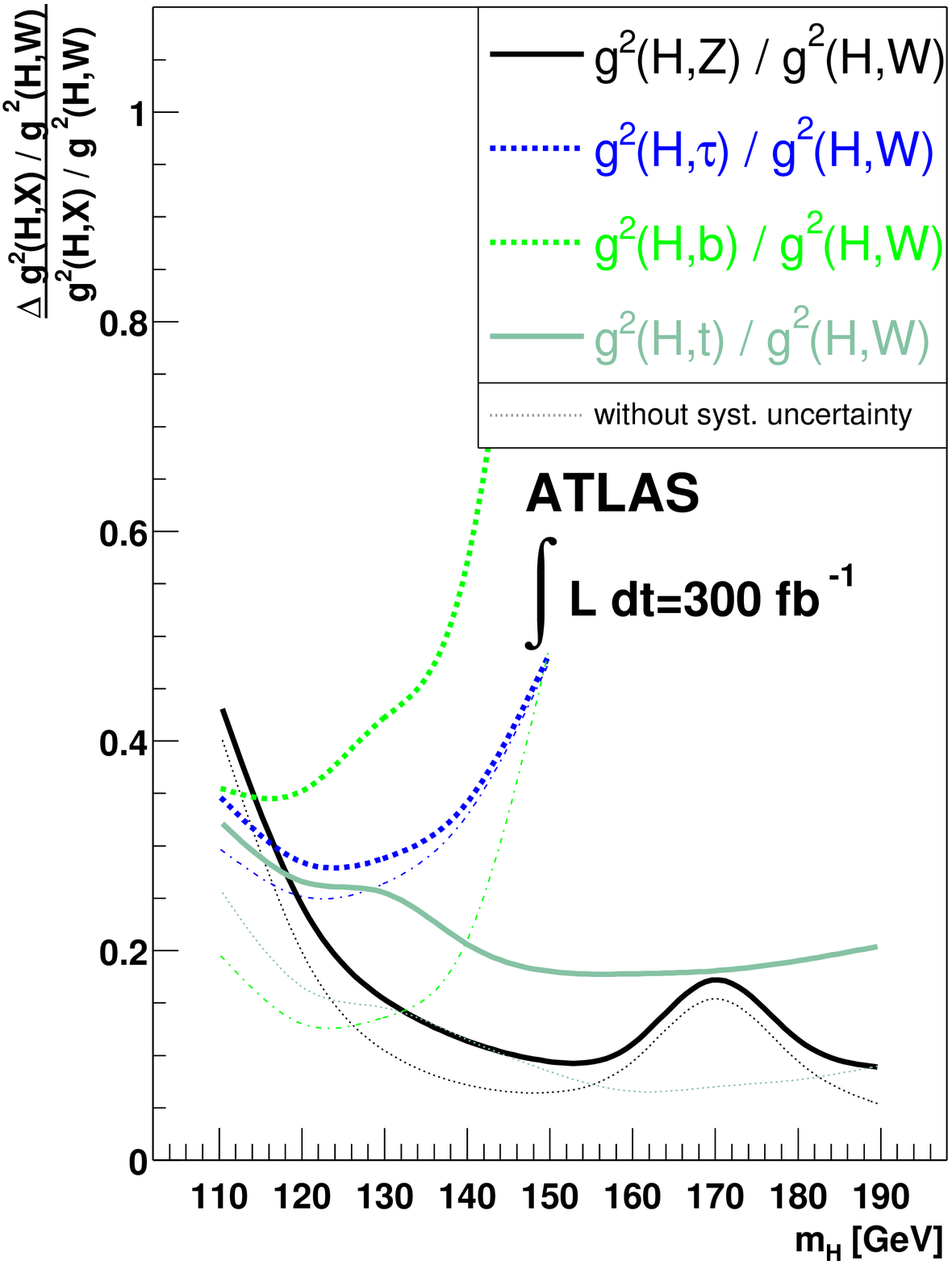,height=8.5cm,clip=}
\vspace*{-0.0cm}

\caption[]{\label{esp-fg:Hrat} \it Las exactitudes esperadas en las
mediciones de las razones de los acoplamientos del Higgs en el LHC.
Ref. \cite{Duhr}.}
\end{figure}

\vspace*{3mm} \noindent {\bf 1.$\,$} En el LHC s\'olo las razones de
los acoplamientos de Higgs pueden ser determinadas de forma
independiente del modelo en el intervalo intermedio de masa. Debido
a que \'unicamente puede ser medido el producto $\sigma_i \cdot BR_f
\sim \Gamma_i \Gamma_f / \Gamma_{tot}$, los anchos parciales
$\Gamma_{i,f}$ pueden ser reescalados y las variaciones
ba\-lan\-cea\-das en $\Gamma_{tot}$ por los canales de decaimiento
no identificados. La precisi\'on esperada para las razones de varios
canales se muestra en Fig.~\ref{esp-fg:Hrat} \cite{Duhr}.
Aparentemente se puede obtener en el LHC una primera visi\'on de la
regla fundamental
\begin{equation}
  g_i/g_j = m_i/m_j
\end{equation}
para varias part\'\i culas $i,j = W,Z,\tau, etc$.
\begin{figure}[hbt]
\begin{center}
\hspace*{-0.5cm}
\epsfig{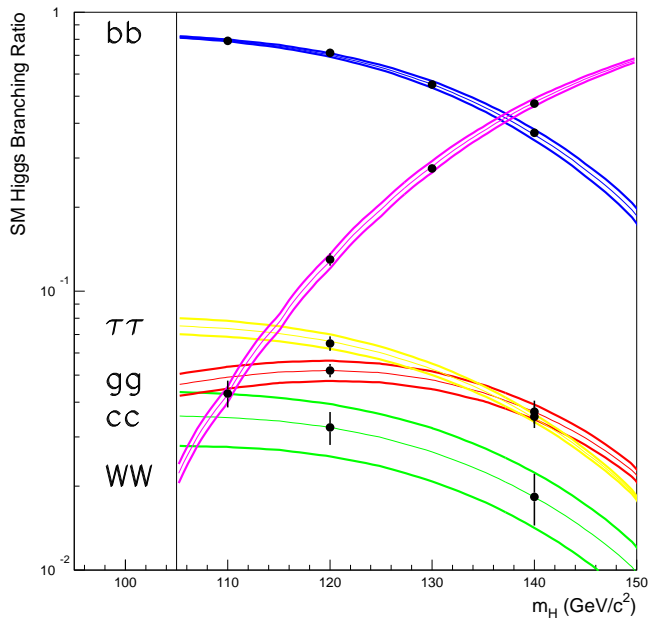}
\end{center}
\vspace*{-0.5cm}

\caption[]{\it \label{esp-fg:brmeas} Las razones de decaimiento del
bos\'on de Higgs predichas por el SM. Los puntos con barras de error
muestran  la precisi\'on experimental esperada, mientras que las
l\'{\i}neas muestran la incertidumbre estimada en las predicciones
del SM. Ref.~\cite{13}.}
\end{figure}

\vspace*{3mm} \noindent {\bf 2.$\,$} En los colisionadores  $e^+e^-$
puede ser medido el valor absoluto de los acoplamientos del Higgs de
forma independiente del modelo y con una alta precisi\'on.

Los acoplamientos del Higgs a bosones de norma masivos pueden ser
determinados de la secciones eficaces de producci\'on en el
Higgs-strahlung y la fusi\'on $WW,ZZ$, con una precisi\'on esperada
al nivel porcentual. Para bosones de Higgs suficientemente pesados
los anchos de decaimiento pueden ser explotados para determinar los
acoplamientos a bosones de norma electrod\'ebiles. Para
acoplamientos del Higgs a fermiones, las razones de desintegraci\'on
$H\to b\bar b, c\bar c, \tau^+\tau^-$ pueden ser usadas en la parte
inferior del rango intermedio de masa, cf.~Fig.~\ref{esp-fg:brmeas};
estas observables permiten la medici\'on directa de los
acoplamientos de Yukawa del Higgs.

Un acoplamiento particularmente interesante es el del Higgs a quarks
top. Dado que el quark top es por mucho el fermi\'on m\'as pesado del
Modelo Est\'andar, las irregularidades en la representaci\'on est\'andar del
rompimiento de la simetr\'{\i}a electrod\'ebil a trav\'es de un campo
fundamental de Higgs pueden ser aparentes primero en este
acoplamiento. Por lo tanto, el acoplamiento de Yukawa $Ht t$ podr\'\i a
eventualmente proveer claves esenciales de la naturaleza del mecanismo
de rompimiento de las simetr\'{\i}as electrod\'ebiles.
\begin{figure}[hbt]
\vspace*{-0.01cm} \hspace*{3.4cm} \epsfxsize=8.2cm
\epsfbox{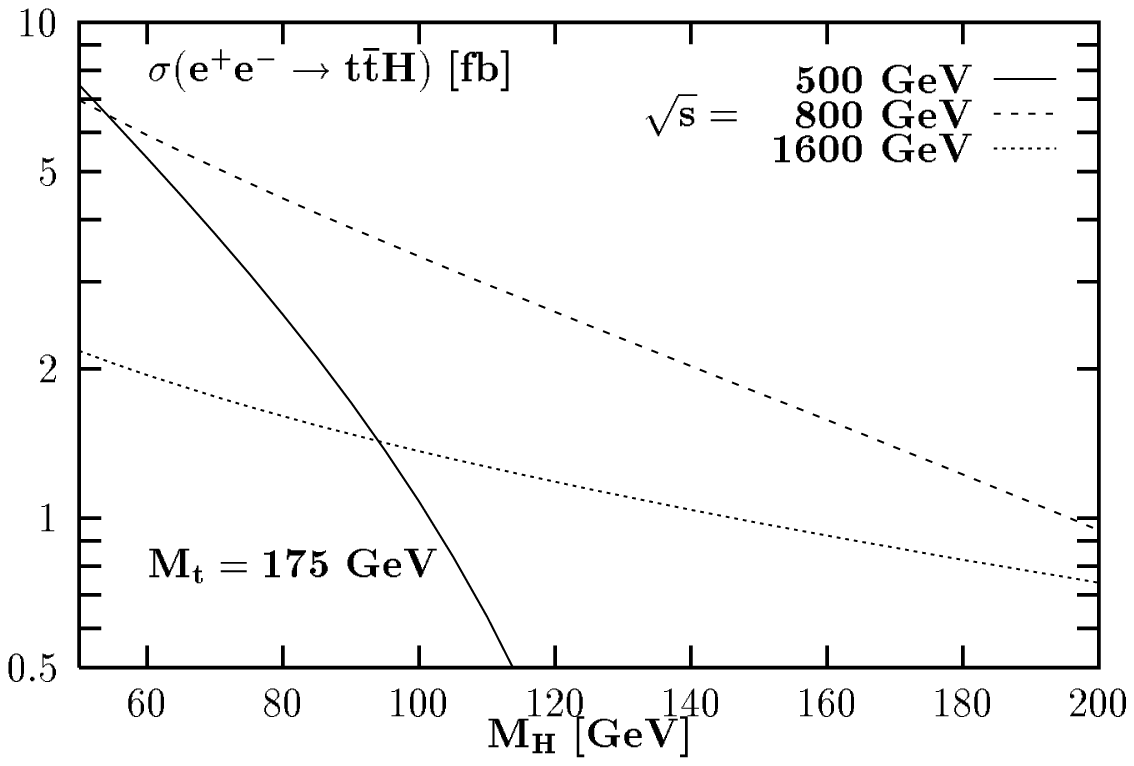} \vspace*{-0.3cm}

\caption[]{\it \label{esp-fg:eetth} La secci\'on eficaz para el
bremsstrahlung de bosones de Higgs del SM procedentes de quarks top
en los procesos de Yukawa $e^+e^-\to t\bar t H$. [La amplitud para
radiaci\'on procedente de la l\'{\i}nea intermedia del bos\'on $Z$
es peque\~na.] Ref. \cite{44}.}
\end{figure}
\begin{figure}[hbt]
\vspace*{-0.3cm} \hspace*{2.7cm} \epsfxsize=8.5cm
\epsfbox{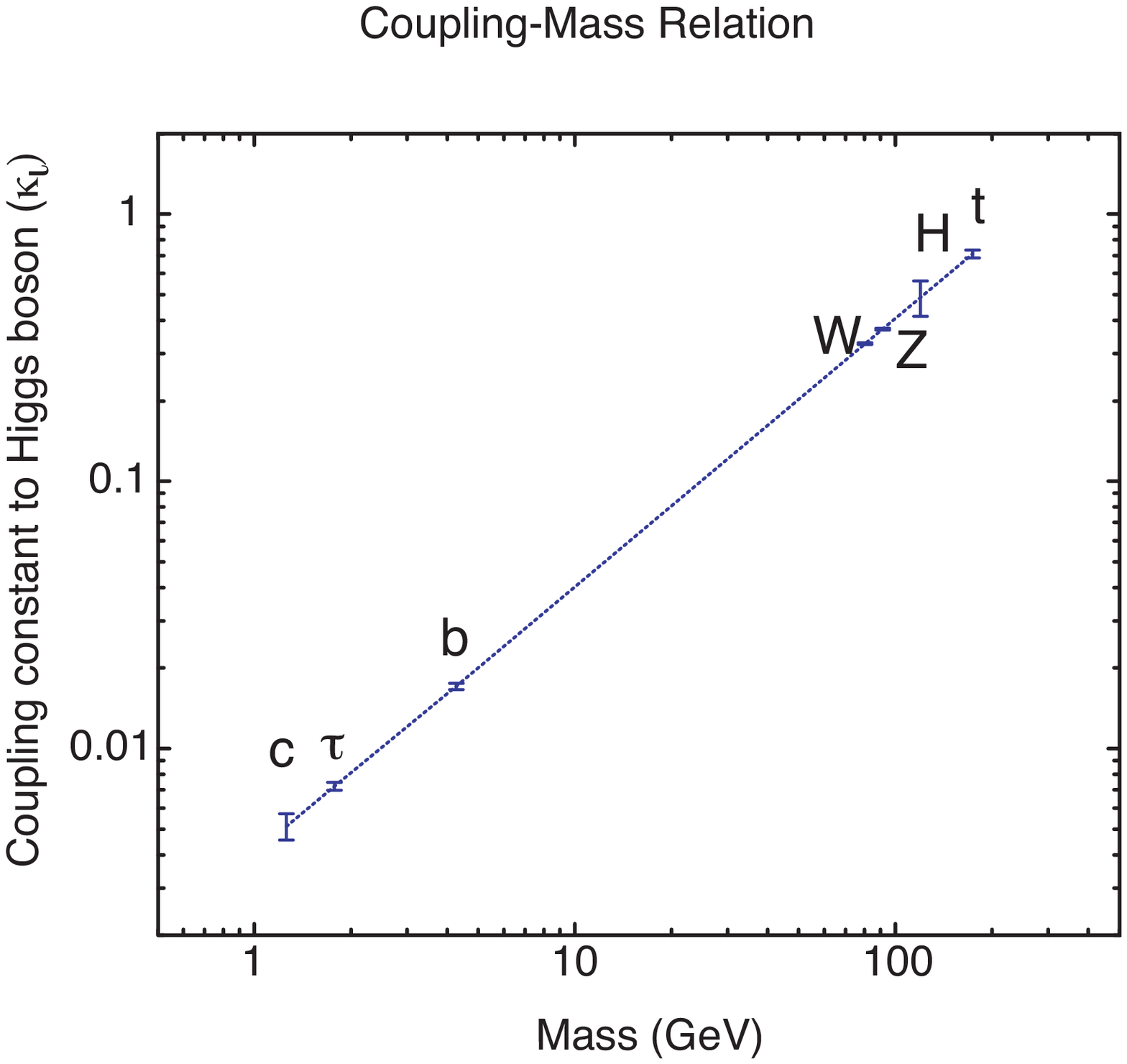} \vspace*{-0.3cm}

\caption[]{\it \label{esp-fg:Hcplg} Las constantes de acoplamiento
normalizadas predichas para el bos\'on de higgs del SM con los
fermiones pesados del SM y con los bosones vectoriales as\'\i  como
con el mismo bos\'on de Higgs como funci\'on de la correspondiente
masa de la part\'{\i}cula del SM. Los puntos con barras de error
muestran la precisi\'on experimental esperada, mientras que las
l\'{\i}neas muestran el crecimiento lineal de los acoplamientos con
las masas correspondientes. Ref.~\cite{Hcplg}.}
\end{figure}

Los lazos del top mediando los procesos de producci\'on
$gg\to H$ y $\gamma\gamma\to H$ (y los correspondientes canales de
decaimiento) dan lugar a secciones eficaces y anchos parciales,
los cuales son proporcionales a la ra\'{\i}z cuadrada de los
acoplamientos de Yukawa del Higgs--top. Este acoplamiento de
Yukawa puede ser medido directamente, para la parte inferior del
rango intermedio de masa, en los procesos de bremsstrahlung $pp\to
t\bar t H$ y $e^+e^- \to t\bar t H$ \cite{44}. El bos\'on de Higgs
es radiado, exclusivamente en el primer proceso, en el segundo
proceso predominantemente, de los quarks top pesados. A\'un cuando
estos experimentos son  dif\'{\i}ciles debido a que las secciones
eficaces son peque\~nas [cf. Fig.~\ref{esp-fg:eetth} para
colisionadores $e^+e^-$] y a la compleja topolog\'{\i}a del estado
final $b\bar bb\bar bW^+W^-$, este proceso es una herramienta
importante para explorar el mecanismo de rompimiento de la
simetr\'{\i}a electrod\'ebil. Para masas grandes del Higgs por arriba
del umbral $t\bar t$, el canal de decaimiento $H\to t\bar t$ puede
ser estudiado; en colisiones $e^+e^-$ la secci\'on eficaz de
$e^+e^- \to t\bar t Z$ aumenta a trav\'es de la reacci\'on $e^+e^-
\to ZH (\to t\bar t)$ \cite{45}. El intercambio de Higgs entre
quarks $t\bar t$ afecta tambi\'en la curva de excitaci\'on cerca del
umbral a un nivel de unos cuantos por ciento.


Las precisiones esperadas para algunos de los acoplamientos est\'an
reunidas en la Tabla~\ref{esp-tab:muehll}. El incremento lineal de los
acoplamientos del Higgs con las masas de las part\'{\i}culas es
claramente visible en la Fig.~\ref{esp-fg:Hcplg} en la cual la pendiente
esta un\'{\i}vocamente predicha dentro del Modelo Est\'andar. La
mezcla del bos\'on de Higgs con otras part\'\i culas escalares, como
radiones, puede cambiar estos acoplamientos de forma universal. Es
por esto necesario hacer un escrutinio no s\'olo de la dependencia
en la masa sino tambi\'en del valor absoluto de los acoplamientos.
\begin{table}[h]
\begin{center}
{\small
\begin{tabular}{|lll|}
\hline
Coupling & $M_H=120$~GeV & 140~GeV \\
\hline
$g_{HWW}$ & $\pm 0.012$ & $\pm 0.020$ \\
$g_{HZZ}$ & $\pm 0.012$ & $\pm 0.013$ \\
\hline
$g_{Htt}$ & $\pm 0.030$ & $\pm 0.061$ \\
$g_{Hbb}$ & $\pm 0.022$ & $\pm 0.022$ \\
$g_{Hcc}$ & $\pm 0.037$ & $\pm 0.102$ \\
\hline
$g_{H\tau\tau}$ & $\pm 0.033$ & $\pm 0.048$ \\
\hline
\end{tabular}
}
\end{center}
\caption{Las precisiones relativas en los acoplamientos de Higgs suponiendo
$\int\!{\cal L}=500$~fb$^{-1}$, $\sqrt{s}=500$~GeV ($\int\!{\cal L}=1$~ab$^{-1}$,
$\sqrt{s}=800$~GeV para $g_{Htt}$).} \label{esp-tab:muehll} \vspace*{-0.4cm}
\end{table}

\vspace*{3mm} \noindent {\it (e) \underline{Auto-acoplamientos del
Higgs}}

\vspace*{1mm}
\noindent
El mecanismo de Higgs, basado en un valor
diferente de cero del campo de Higgs en el vac\'{\i}o, debe
finalmente ponerse de manifiesto experimentalmente por la
reconstrucci\'on del potencial de interacci\'on que genera el campo
diferente de cero en el vac\'{\i}o. Este programa puede llevarse a
cabo midiendo la intensidad de los auto-acoplamientos trilineales y
cu\'articos de las part\'{\i}culas de Higgs:
\begin{eqnarray}
g_{H^3} & = & 3 \sqrt{\sqrt{2} G_F} M_H^2 \\ \non \\
g_{H^4} & = & 3 \sqrt{2} G_F M_H^2 ~.
\end{eqnarray}

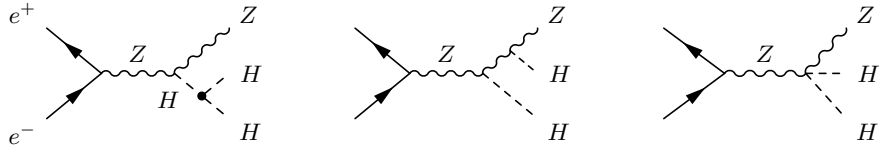
\begin{figure}[hbtp]
\begin{center}
\vspace*{0.25cm}
\begin{fmffile}{fd}
{\footnotesize \unitlength1mm
\begin{fmfshrink}{0.7}
\begin{fmfgraph*}(24,12)
  \fmfstraight
  \fmfleftn{i}{3} \fmfrightn{o}{3}
  \fmf{fermion}{i1,v1,i3}
  \fmflabel{$e^-$}{i1} \fmflabel{$e^+$}{i3}
  \fmf{boson,lab=$Z$,lab.s=left,tens=3/2}{v1,v2}
  \fmf{boson}{v2,o3} \fmflabel{$Z$}{o3}
  \fmf{phantom}{v2,o1}
  \fmffreeze
  \fmf{dashes,lab=$H$,lab.s=right}{v2,v3} \fmf{dashes}{v3,o1}
  \fmffreeze
  \fmf{dashes}{v3,o2}
  \fmflabel{$H$}{o2} \fmflabel{$H$}{o1}
  \fmfdot{v3}
\end{fmfgraph*}
\hspace{15mm}
\begin{fmfgraph*}(24,12)
  \fmfstraight
  \fmfleftn{i}{3} \fmfrightn{o}{3}
  \fmf{fermion}{i1,v1,i3}
  \fmf{boson,lab=$Z$,lab.s=left,tens=3/2}{v1,v2}
  \fmf{dashes}{v2,o1} \fmflabel{$H$}{o1}
  \fmf{phantom}{v2,o3}
  \fmffreeze
  \fmf{boson}{v2,v3,o3} \fmflabel{$Z$}{o3}
  \fmffreeze
  \fmf{dashes}{v3,o2}
  \fmflabel{$H$}{o2} \fmflabel{$H$}{o1}
\end{fmfgraph*}
\hspace{15mm}
\begin{fmfgraph*}(24,12)
  \fmfstraight
  \fmfleftn{i}{3} \fmfrightn{o}{3}
  \fmf{fermion}{i1,v1,i3}
  \fmf{boson,lab=$Z$,lab.s=left,tens=3/2}{v1,v2}
  \fmf{dashes}{v2,o1} \fmflabel{$H$}{o1}
  \fmf{dashes}{v2,o2} \fmflabel{$H$}{o2}
  \fmf{boson}{v2,o3} \fmflabel{$Z$}{o3}
\end{fmfgraph*}
\end{fmfshrink}
\\[0.5cm]
}
\end{fmffile}
\end{center}
\caption[]{\label{esp-fg:wwtohh} \it Generic diagrams contributing to
the double Higgs-strahlung process $e^+e^- \to ZHH$.}
\end{figure}

Esta es una tarea dif\'{\i}cil ya que los procesos a ser explotados
est\'an suprimidos por acoplamientos  y espacio fase peque\~nos. En
el LHC parece no ser posible determinar los auto-acoplamientos, no
obstante, se tiene alguna esperanza para la versi\'on de alta
luminosidad VLHC. Sin embargo este problema puede resolverse para el
acoplamiento trilineal $g_{H^3}$ en la fase de alta energ\'{\i}a en
colisionadores lineales $e^+e^-$ para luminosidades suficientemente
altas \cite{selfMMM}. La reacci\'on m\'as conveniente en los
colisionadores  $e^+e^-$ para la medida de los acoplamientos
trilineales de las masas del Higgs en el rango de masas preferido
te\'oricamente ${\cal O}(100~\mbox{GeV})$, es el proceso doble de
Higgs-strahlung
\begin{equation}
e^+e^- \to ZH \to ZHH
\end{equation}
en el cual, entre otros mecanismos, el estado final de dos Higgs es
generado por el intercambio de una part\'{\i}cula de Higgs virtual,
de modo que este proceso es sensible al acoplamiento trilineal $HHH$
en el potencial de Higgs, Fig.~\ref{esp-fg:wwtohh}. Como la secci\'on
eficaz es s\'olo una fracci\'on de 1 fb, es necesaria una
luminosidad integrada de $\sim$ 1 ab$^{-1}$ para aislar los eventos en
colisionadores lineales. Si se combina con mediciones de procesos de
fusi\'on doble del Higgs
\begin{equation}
 e^+e^- \to \bar{\nu} \nu H^\ast \to \bar{\nu} \nu H H
\end{equation}
se pueden esperar precisiones experimentales cercanas al 12\%
\cite{Rexp}. El acoplamiento cu\'artico $H^4$ parece ser accesible
s\'olo a trav\'es de efectos de lazos en un futuro
pr\'oximo.\\

\vspace*{2mm} \noindent \underline{Para resumir}: {\it Los elementos
esenciales del mecanismo de Higgs pueden establecerse
experimentalmente en el LHC y  colisionadores lineales $e^+e^-$ a
TeV.}

\vspace*{4mm}
\section{El Bos\'on de Higgs en Teor\'{\i}as Supersim\'etricas}
\vspace*{2mm}

\noindent {\bf 1.$\,$} Argumentos profundamente arraigados en el
sector de Higgs juegan un papel importante en la introducci\'on de
la supersimetr\'{\i}a como una simetr\'{\i}a fundamental de la
naturaleza \cite{14}. Esta es la \'unica simetr\'{\i}a que
correlaciona los grados de libertad bos\'onicos y fermi\'onicos, cf.
Ref.\cite{SUSY}.

\vspace*{2mm} {\bf (a)} La cancelaci\'on entre las contribuciones
bos\'onicas y fermi\'onicas a las correcciones radiativas de las
masas de los Higgs ligeros en teor\'{\i}as supersim\'etricas
prove\'e una soluci\'on al problema de la jerarqu\'{\i}a en el
Modelo Est\'andar. Si el Modelo Est\'andar se encastra en una
teor\'{\i}a de gran unificaci\'on, la gran brecha entre la escala
alta de gran unificaci\'on y la escala baja del rompimiento de la
simetr\'{\i}a electrod\'ebil se puede estabilizar de modo natural en
teor\'{\i}as con  simetr\'{\i}as bos\'on--fermi\'on \cite{15,601}.
Denotando la masa desnuda del Higgs por $M_{H,0}^2$, las
correcciones radiativas debidas a los lazos de bos\'on vectorial en
el Modelo Est\'andar por $\delta M_{H,V}^2$, y a las contribuciones
de los gauginos, los compa\~neros supersim\'etricos fermi\'onicos,
por $\delta M_{\tilde H,\tilde V}^2$, la masa f\'{\i}sica del Higgs
est\'a dada por la suma $M_H^2 = M_{H,0}^2 + \delta M_{H,V}^2 +
\delta M_{\tilde H,\tilde V}^2$. La correcci\'on del bos\'on
vectorial es cuadr\'aticamente divergente, $\delta M_{H,V}^2 \sim
\alpha [\Lambda^2 - M^2]$, de tal manera que para una escala de
corte $\Lambda \sim \Lambda_{GUT}$ ser\'a necesario un ajuste fino
extremo entre la masa desnuda intr\'{\i}nseca  y las fluctuaciones
cu\'anticas radiativas para generar la masa del Higgs del orden de
$M_W$. Sin embargo, debido al principio de Pauli, las contribuciones
del gaugino fermi\'onico adicional en teor\'{\i}as supersim\'etricas
son justo del signo opuesto, $\delta M_{\tilde H,\tilde V}^2\sim
-\alpha [\Lambda^2-\tilde M^2]$, de manera tal que los t\'erminos
divergentes se cancelan entre s\'{\i}\footnote{Las diferentes
estad\'{\i}sticas para bosones y fermiones son suficientes para
obtener la cancelaci\'on de las divergencias; sin embargo no
son necesarias. Las relaciones de simetr\'{\i}a entre los
acoplamientos, como se dan en Modelos de Higgs Peque\~no (Little
Higgs Models), tambi\'en pueden llevar a cancelaciones
individualmente entre amplitudes bos\'on-bos\'on o
fermi\'on-fermi\'on.}. Como $\delta M_H^2\sim\alpha [\tilde
M^2-M^2]$, se evita cualquier ajuste fino para masas de las
part\'{\i}culas supersim\'etricas $\tilde M \lessim {\cal O}(1$
TeV). As\'{\i}, dentro de este esquema de simetr\'{\i}a el sector de
Higgs es estable en el rango de bajas energ\'{\i}as $M_H\sim M_W$
incluso en el contexto de escalas de alta energ\'{\i}a de GUT. Este
mecanismo conduce de manera natural a la supersimetr\'{\i}a de bajas
energ\'{\i}as.

\vspace*{2mm} {\bf (b)} El concepto de supersimetr\'{\i}a est\'a
apoyado fuertemente por la exitosa predicci\'on del \'angulo de
mezcla electrod\'ebil obtenido en la versi\'on m\'\i nima de esta
teor\'{\i}a \cite{16}. El espectro extendido de part\'{\i}culas
lleva a la evoluci\'on del \'angulo de mezcla electrod\'ebil del
valor de GUT de 3/8  hasta $\sin^2\theta_W = 0.2336 \pm 0.0017$ a
bajas energ\'\i as, donde el  error  incluye contribuciones
desconocidas de umbral en las escalas de baja y alta masa
supersim\'etrica. La predicci\'on coincide con la medida
experimental $\sin^2\theta_W^{exp} = 0.23120 \pm 0.00016$ dentro de
una incertibumbre te\'orica menor al 2 por mil.

\vspace*{2mm} {\bf (c)} Es conceptualmente muy interesante la
interpretaci\'on del mecanismo de Higgs en las teor\'{\i}as
supersim\'etricas como un efecto cu\'antico \cite{50A}. El
rompimiento de la simetr\'{\i}a electrod\'ebil $SU(2)_L \times
U(1)_Y$ puede inducirse radiativamente mientras se conservan la
simetr\'\i a de norma electromagn\'etica $U(1)_{EM}$  y la
simetr\'\i a de norma de color $SU(3)_C$ para masas del quark--top
entre 150 y 200 GeV. Iniciando con un conjunto de masas escalares
universales a la escala de GUT, el par\'ametro de la masa al cuadrado
del sector de Higgs toma valores negativos a la escala
electrod\'ebil, mientras que las masas al cuadrado de los
sleptones y squarks se mantienen positivas.
\vspace*{2mm}

\noindent {\bf 2.$\,$} El sector de Higgs de las teor\'{\i}as
supersim\'etricas difiere del Modelo Est\'andard en muchos aspectos
\cite{17}. Para preservar la supersimetr\'{\i}a y la invariancia de
norma,  se deben introducir al menos dos campos iso-dobletes,
dej\'andonos con un espectro de cinco o m\'as part\'{\i}culas de
Higgs f\'{\i}sicas. En la extensi\'on supersim\'etrica m\'{\i}nima
del Modelo Est\'andar (MSSM) las auto-interacciones del Higgs son
generadas por la acci\'on del campo de norma--escalar, de manera tal
que los acoplamientos cu\'articos est\'an relacionados a los
acoplamientos de norma en este escenario. Despu\'es de incluir
correcciones radiativas esto lleva a l\'\i mites fuertes \cite{19}
para la masa del bos\'on de Higgs m\'as ligero, que debe ser menor
que aproximadamente $140$ GeV. Si se supone  que el sistema se
mantiene d\'ebilmente interactuante hasta escalas del orden de GUT o
escala de Planck, la masa se mantiene peque\~na, por razones muy
parecidas a las encontradas en el Modelo Est\'andar, incluso en
teor\'{\i}as supersim\'etricas m\'as complejas que involucran campos
de Higgs e interacciones de Yukawa adicionales. Las masas de los
bosones de Higgs pesados se espera que est\'en dentro del rango de
la escala de rompimiento de la simetr\'\i a electrod\'ebil hasta un
orden de 1 TeV.

\vspace*{4mm}
\subsection{El Sector de Higgs del MSSM}

\vspace*{2mm}
\noindent {\bf 1.$\,$}  El espectro de part\'{\i}culas
del MSSM \cite{14} consiste de leptones, quarks y sus compa\~neros
supersim\'etricos escalares, y part\'{\i}culas de norma,
part\'{\i}culas de Higgs y sus compa\~neros de esp\'{\i}n-1/2. Los
campos de mater\'{\i}a y de fuerzas est\'an acoplados en acciones
supersim\'etricas e invariantes de norma:
\begin{equation}
\begin{array}{lrcll}
S = S_V + S_\phi + S_W: \hspace*{1cm}
& S_V    & = & \frac{1}{4} \int d^6 z \hat W_\alpha \hat W_\alpha
\hspace*{1cm} & \mbox{acci\'on de norma (gauge)} ~, \\ \\
& S_\phi & = & \int d^8 z \hat \phi^* e^{gV} \hat \phi
& \mbox{acci\'on de materia} ~, \\ \\
& S_W    & = & \int d^6 z W[\hat \phi] & \mbox{superpotencial} ~.
\end{array}
\end{equation}
Descomponiendo los supercampos en sus componentes fermi\'onicas y
bos\'onicas, e integrando sobre las variables de Grassmann en
$z\to x$, se pueden derivar los siguientes Lagrangianos, los
cuales describen las interacciones de norma, materia y
campos de Higgs:
\begin{eqnarray*}
{\cal L}_V & = & -\frac{1}{4}F_{\mu\nu}F_{\mu\nu}+\ldots+\frac{1}{2}D^2 ~, \\ \\
{\cal L}_\phi & = & D_\mu \phi^* D_\mu \phi +\ldots+\frac{g}{2} D|\phi|^2  ~, \\ \\
{\cal L}_W & = & - \left| \frac{\partial W}{\partial \phi} \right|^2 ~.
\end{eqnarray*}
El campo $D$ es un campo auxiliar el cual no se propaga en el
espacio-tiempo y puede eliminarse aplicando las ecuaciones de
movimiento: $D=-\frac{g}{2} |\phi|^2$. Reinsertado en el
Lagrangiano, el auto-acoplamiento cu\'artico del campo escalar de
Higgs es generado,
\begin{equation}
{\cal L} [\phi^4] = -\frac{g^2}{8} |\phi^2|^2 ~,
\end{equation}
en teor\'{\i}as, como MSSM, en  las cuales el superpotencial no
genera t\'erminos cu\'articos. As\'\i, el acoplamiento cu\'artico de
los campos de Higgs est\'a dado, en la teor\'{\i}a supersim\'etrica
m\'\i nima, por el cuadrado del acoplamiento de norma.  A diferencia
del caso del Modelo Est\'andar, el acoplamiento cu\'artico no es un
par\'ametro libre.
Adem\'as, este acoplamiento es d\'ebil.\\

\vspace*{2mm}
\noindent {\bf 2.$\,$} Dos campos dobletes de Higgs
independientes $H_1$ y $H_2$ deben introducirse en el
superpotencial:
\begin{equation}
W = -\mu \epsilon_{ij} \hat H_1^i \hat H_2^j + \epsilon_{ij} [f_1
\hat H_1^i \hat L^j \hat R + f_2 \hat H_1^i \hat Q^j \hat D - f_2'
\hat H_2^i \hat Q^j \hat U]
\end{equation}
para proporcionar masa a las part\'{\i}culas de tipo-down ($H_1$) y
a las par\'{\i}culas de tipo-up ($H_2$). A diferencia del Modelo
Est\'andar, el segundo campo de Higgs no puede ser identificado con
el conjugado de carga del primero, ya que $W$ debe ser
anal\'{\i}tico para preservar la supersimetr\'{\i}a. Adem\'as, los
campos del Higgsino asociados con un s\'olo campo de Higgs
generar\'{\i}an anomal\'{\i}as triangulares; \'estas se cancelan si
los dos dobletes conjugados se suman, y la invariancia de norma
cl\'asica de las interacciones no se destruye a nivel cu\'antico.
Integrando el superpotencial sobre las coordenadas de Grassmann se
genera la auto-energ\'{\i}a de Higgs supersim\'etrica $V_0 = |\mu|^2
(|H_1|^2 + |H_2|^2)$. El rompimiento de la supersimetr\'{\i}a puede
ser incorporado en el sector de Higgs al introducir t\'erminos de
masa bilineales $\mu_{ij} H_i H_j$. Sumados a la parte de la
auto-energ\'{\i}a supersim\'etrica $H^2$ y la parte cu\'artica $H^4$
generada por la acci\'on de norma, llevan al siguiente potencial de
Higgs
\begin{eqnarray}
V & = & m_1^2 H_1^{*i} H_1^i + m_2^2 H_2^{*i} H_2^i - m_{12}^2 (\epsilon_{ij}
H_1^i H_2^j + hc) \non \\ \non \\
& & + \frac{1}{8} (g^2 + g'^2) [H_1^{*i} H_1^i -
H_2^{*i} H_2^i]^2 + \frac{1}{2} |H_1^{*i} H_2^{i}|^2 ~.
\end{eqnarray}
El potencial de Higgs incluye tres t\'erminos  bilineales de masa,
mientras que la intensidad del acoplamiento cu\'artico se
determina por los acoplamientos de norma $SU(2)_L$ y $U(1)_Y$ al
cuadrado. Los tres t\'erminos de masa son par\'ametros libres.

El potencial desarrolla un m\'{\i}nimo estable para $H_1 \to
[v_1,0]$ y $H_2\to [v_2,0]$, si se reunen las siguientes
condiciones: $ m_1^2 +  m_2^2 >  2 | m^2_{12} |  \;\;{\rm and}\;\;
m_1^2    m_2^2  <  | m^2_{12} |^2 ~.$

\vspace*{2mm}
\noindent
 Al expander los campos alrededor de los
valores del estado base $v_1$ y $v_2$,
\begin{equation}
\begin{array}{rclcl}
H_1^1 & = & v_1 &\!\! +\!\! & [H^0 \cos \alpha - h^0 \sin \alpha + i A^0 \sin \beta - i G^0
\cos \beta ]/\sqrt{2}                                   \\ \\
H_1^2 & = & &\!\!\!\! & H^- \sin \beta - G^- \cos \beta
\end{array}
\end{equation}
y
\begin{equation}
\begin{array}{rclcl}
H_2^1 & = & &\!\!\!\! & H^+ \cos \beta + G^+ \sin \beta   \\ \\
H_2^2 & = & v_2 &\!\!+\!\! & [H^0 \sin \alpha + h^0 \cos \alpha + i A^0 \cos \beta + i G^0
\sin \beta ]/\sqrt{2} \,,
\end{array}
\end{equation}
los eigenestados de masa est\'an dados por los estados neutros
$h^0, H^0$ y $A^0$, lo cuales son pares e impares bajo
transformaciones de ${\cal CP}$; y por los estados cargados $H^\pm$;
los estados $G$ corresponden a los modos de Goldstone, los cuales
son absorbidos por los campos de norma para construir las
componenetes longitudinales. Despu\'es de introducir los tres
par\'ametros
\begin{eqnarray}
M_Z^2 & = & \frac{1}{2} (g^2 + g'^2) (v_1^2 + v_2^2) \non \\ \non \\
M_A^2 & = & m_{12}^2 \frac{v_1^2 + v_2^2}{v_1v_2} \non \\ \non \\
\tgb  & = & \frac{v_2}{v_1} ~,
\end{eqnarray}
la matriz de masa puede descomponerse en tres bloques de $2\times
2$, los cuales son f\'aciles de diagonalizar:
\begin{displaymath}
\begin{array}{ll}
\mbox{\bf masa pseudoescalar:} & M_A^2 \\ \\
\mbox{\bf masa cargada:} & M_\pm^2 = M_A^2+M_W^2 \\ \\
\mbox{\bf masa escalar:} & M_{h,H}^2 = \frac{1}{2} \left[ M_A^2 +
M_Z^2 \mp \sqrt{(M_A^2+M_Z^2)^2
- 4M_A^2M_Z^2 \cos^2 2\beta} \right] \\ \\
& \displaystyle \tg 2\alpha = \tg 2\beta \frac{M_A^2 + M_Z^2}{M_A^2 - M_Z^2}
\hspace*{0.5cm} \mbox{con} \hspace*{0.5cm} -\frac{\pi}{2} < \alpha < 0
\nonumber
\end{array}
\end{displaymath}

De las f\'ormulas de la masa, pueden derivarse dos desigualdades
importantes,
\begin{eqnarray}
M_h & \leq & M_Z, M_A \leq M_H \\ \nonumber \\
M_W & \leq & M_{H^\pm} ~,
\end{eqnarray}
las que, por construcci\'on, son v\'alidas en una aproximaci\'on a nivel \'arbol.
Como resultado, se predice que la masa del escalar de Higgs m\'as ligero
est\'a acotada por la masa del $Z$, modulo correcciones
radiativas. Estos l\'{\i}mites provienen del hecho de que el
acoplamiento cu\'artico de los campos de Higgs est\'an determinados en el
MSSM por el tama\~no del acoplamiento de norma
al cuadrado. \\

\vspace*{2mm} \noindent {\bf 3.$\,$} \underline {\it Correcciones
Radiativas en SUSY:}
Las relaciones a nivel \'arbol entre las masas de los Higgs son
fuertemente modificadas por correcciones radiativas que involucran
el espectro de part\'{\i}culas supersim\'etricas  del sector del top
\cite{50B}; cf.~Ref.~\cite{DJ,mhplot} para recopilaciones recientes.
Estos efectos son proporcionales a la cuarta potencia de la masa del
top y al logaritmo de la masa del stop. El origen de estas
correcciones son las cancelaciones incompletas entre los lazos del
top virtual y del stop, reflej\'ando el rompimiento de la
supersimetr\'{\i}a. M\'as a\'un, las relaciones de masa son
afectadas por la mezcla potencialmente grande entre $\tilde t_L$ y
$\tilde t_R$
debido al acoplamiento de Yukawa del top.\\

Al orden dominante en $M_t^4$ las correcciones radiativas se pueden
resumir en el par\'ametro
\begin{equation}
\epsilon = \frac{3G_F}{\sqrt{2}\pi^2}\frac{M_t^4}{\sin^2\beta}\log
\frac{\langle M^2_{\tilde{t}} \rangle}{M_t^2}
\end{equation}
con $\langle M^2_{\tilde{t}} \rangle = M_{\tilde{t_1}}
M_{\tilde{t_2}} \, .$ En esta aproximaci\'on, la masa del Higgs
ligero $M_h$ puede expresarse por $M_A$ y $\tgb$ en la siguiente
forma compacta:
\begin{eqnarray*}
M^2_h & = & \frac{1}{2} \left[ M_A^2 + M_Z^2 + \epsilon \right.
\non \\
& & \left. - \sqrt{(M_A^2+M_Z^2+\epsilon)^2
-4 M_A^2M_Z^2 \cos^2 2\beta
-4\epsilon (M_A^2 \sin^2\beta + M_Z^2 \cos^2\beta)} \right]
\end{eqnarray*}
Las masas de los Higgs pesados  $M_H$ y $M_{H^\pm}$ se obtienen
de las reglas de suma
\begin{eqnarray*}
M_H^2 & = & M_A^2 + M_Z^2 - M_h^2 + \epsilon \non \\
M_{H^\pm}^2 & = & M_A^2 + M_W^2 ~.
\end{eqnarray*}
Finalmente, el par\'ametro de mezcla $\alpha$, el cual diagonaliza
la matriz de masa ${\cal CP}$-par, est\'a dada por la relaci\'on
mejorada radiativamente:
\begin{equation}
\tg 2 \alpha = \tg 2\beta \frac{M_A^2 + M_Z^2}{M_A^2 - M_Z^2 +
\epsilon/\cos 2\beta} ~.
\label{esp-eq:mssmalpha}
\end{equation}

\begin{figure}[hbt]
\begin{center}
\hspace*{-0.3cm}
\epsfig{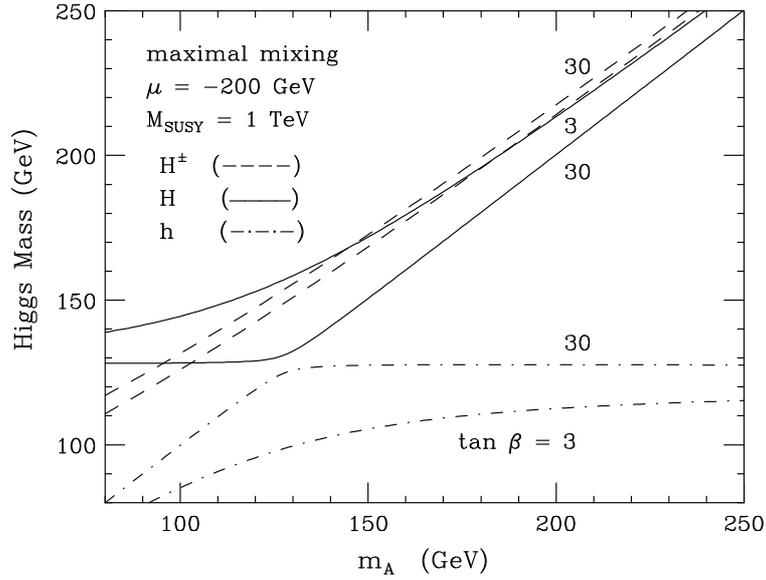}
\end{center}
\vspace*{-0.4cm} \caption[]{\label{esp-kdfig} \it Las masas de los
bosones de Higgs del MSSM CP-par y cargados como una funci\'on de
$m_A$ para $\tan\beta=3$ y $30$, incluyendo correcciones radiativas.
Ref.~\cite{66a}.}
\end{figure}

Para masas grandes de $A$, las masas de las part\'{\i}culas Higgs
pesadas coinciden aproximadamente, $M_A\simeq M_H \simeq
M_{H^\pm}$, mientras que la masa del Higgs ligero  se acerca a un
valor asint\'otico peque\~no. El espectro para valores grandes de
$\tgb$ es bastante regular: para  valores peque\~nos de $M_A$ uno
encuentra que $\{ M_h\simeq M_A; M_H \simeq \mbox{const} \}$
\cite{intense}; para valores grandes de $M_A$ se encuentra la
relaci\'on opuesta $\{ M_h\simeq \mbox{const}, M_H \simeq
M_{H^\pm}\simeq M_A \}$,
cf.~Fig.~\ref{esp-kdfig} la cual incluye correcciones radiativas.\\
\begin{figure}[hbt]
\begin{center}
\vspace*{7mm}
\vspace*{-2.0cm} \hspace*{-0.3cm}
\epsfig{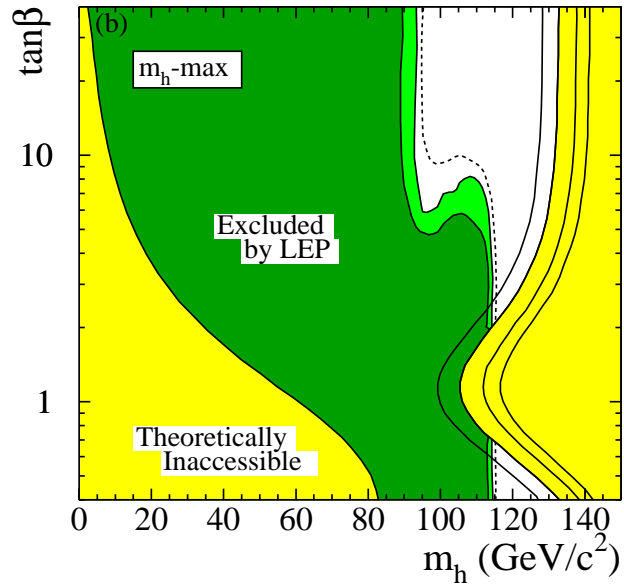}
\end{center}
\vspace*{-0.0cm} \caption[]{\label{esp-fg:mssmhiggs} \it Cotas
superiores a la masa del bos\'on de Higgs ligero como funci\'on de
$\tg\beta$ variando la masa del top y la regi\'on excluida por las
b\'usquedas negativas en los experimentos del LEP. Ref.
\cite{pdgplot}.}
\end{figure}

Mientras los efectos no dominantes de mezcla en las
relaciones de masa del Higgs son bastante complicados, el
impacto en la cota superior de la masa del Higgs ligero $M_h$ se
puede resumir de una forma sencilla:
\begin{equation}
M_h^2 \leq M_Z^2 +  \delta M_t^2 + \delta M_X^2 ~.
\end{equation}
La principal contribuci\'on dada por el top est\'a relacionada con
el par\'ametro $\epsilon$,
mientras la segunda contribuci\'on
depende del par\'ametro de mezcla en el sector del top escalar
\begin{equation}
M_tX_t = M_t \left[A_t - \mu~\ctgb \right] ~,
\end{equation}
el cual acopla estados de quiralidad izquierda y derecha en la
matriz de masa del stop:
\begin{equation}
M_h^2 \leq M_Z^2 + \frac{3G_F M_t^4}{2\sqrt{2}\pi^2} \, \left[
                   \log \frac{\langle M_{\tilde{t}}^2 \rangle}{M_t^2} +
                   \frac{X_t^2}{ \langle M_{\tilde{t}}^2 \rangle }
                   \left(1 - \frac{X_t^2}{\langle 12 M_{\tilde{t}}^2 \rangle}  \right) \right] \;.
\end{equation}

Contribuciones subdominantes pueden ser reducidas esencialmente a
efectos de QCD de orden m\'as alto. Estas contribuciones pueden ser
incorporadas efectivamente al interpretar el par\'ametro de masa del
top $M_t \to M_t(\mu_t)$ como la masa $\overline{\rm MS}$ del top
evaluada en la media geom\'etrica
entre las masas del top y el stop, $\mu_t^2 = M_t M_{\tilde t}$.\\

Las cotas superiores en la masa del Higgs ligero se muestran en
Fig.~\ref{esp-fg:mssmhiggs} como funci\'on de $\tg \beta$. Las curvas
son el resultado de los c\'alculos tomando en cuenta los efectos de
mezcla. Esto lleva a que la cota superior general para una mezcla
m\'axima est\'a dada por $M_h\lessim 140$ GeV, incluyendo valores
grandes de $\tgb$. De modo que el sector del Higgs ligero no puede
ser completamente cubierto por experimentos del LEP2 debido al
incremento que sufre el l\'{\i}mite de la masa con la masa del top.

\vspace*{4mm}
\subsection{Acoplamientos de las part\'\i culas de Higgs SUSY a las
  Part\'\i culas del SM}

\vspace*{2mm}
\noindent

El tama\~no de los acoplamientos de los Higgs del MSSM
con quarks, leptones y bosones de norma es similar al del Modelo
Est\'andar (SM), pero modificado por los \'angulos de mezcla
$\alpha$ y $\beta$. Normalizados a los valores del SM, estos
acoplamientos est\'an listados en la Tabla \ref{esp-tb:hcoup}. El bos\'on
de Higgs pseudoescalar $A$ no se acopla a bosones de norma a nivel
\'arbol, pero el acoplamiento, compatible con la simetr\'{\i}a de
${\cal CP}$, puede ser generado a nivel de lazos de m\'as alto
orden. Los bosones cargados de Higgs  se acoplan a los fermiones
up y down a trav\'es de las amplitudes quirales izquierda y
derecha $g_\pm = -
\left[ g_t (1 \mp \gamma_5) + g_b (1 \pm \gamma_5) \right]/\sqrt{2}$
donde $g_{t,b} = (\sqrt{2} G_F)^{1/2} m_{t,b}$.
\begin{table}[hbt]
\renewcommand{\arraystretch}{1.5}
\begin{center}
\begin{tabular}{|lc||ccc|} \hline
\multicolumn{2}{|c||}{$\Phi$} & $g^\Phi_u$ & $g^\Phi_d$ &  $g^\Phi_V$ \\
\hline \hline
SM~ & $H$ & 1 & 1 & 1 \\ \hline
MSSM~ & $h$ & $\cos\alpha/\sin\beta$ & $-\sin\alpha/\cos\beta$ &
$\sin(\beta-\alpha)$ \\
& $H$ & $\sin\alpha/\sin\beta$ & $\cos\alpha/\cos\beta$ &
$\cos(\beta-\alpha)$ \\
& $A$ & $ 1/\tg\beta$ & $\tg\beta$ & 0 \\ \hline
\end{tabular}
\renewcommand{\arraystretch}{1.2}
\caption[]{\label{esp-tb:hcoup}
\it Acoplamientos del Higgs a
fermiones y bos\'ons de norma [$V=W,Z$] en el MSSM relativos a los
acoplamientos del SM.}
\end{center}
\end{table}

Los acoplamientos modificados incorporan la renormalizaci\'on
debido a correcciones radiativas de SUSY, a primer orden en $M_t$,
si el \'angulo de mezcla $\alpha$ est\'a relacionado con $\beta$ y
con $M_A$ a trav\'es de la f\'ormula corregida
Eq.~(\ref{esp-eq:mssmalpha}). Para valores grandes de $M_A$, en la
pr\'actica $M_A\gsim 200$ GeV, los acoplamientos del bos\'on de
Higgs ligero $h$ a los fermiones y a los bosones de norma  se
aproximan asint\'oticamente a los valores del SM. Esta es la
esencia del \underline{teorema de desacoplamiento} en el sector
de Higgs \cite{66AA}: Part\'{\i}culas con masas grandes deben
desacoplarse del sistema de part\'{\i}culas ligeras como
consecuencia del principio de incertidumbre de la mec\'anica cu\'antica.
En el mismo l\'\i mite, el bos\'on de Higgs pesado $H$ se desacopla de los
bosones vectoriales y el acoplamiento a los fermiones tipo up es
suprimido por $1/{\tan\beta}$, mientras que el acoplamiento a los
fermiones tipo down es realzado por $\tan\beta$.  As\'\i, los acoplamientos
de los dos Higgs pesados degenerados $A,H$ son isomorfos en el l\'\i mite
de desacoplamiento.

\vspace*{4mm}
\subsection{Decaimientos de las Part\'{\i}culas de Higgs}

\begin{figure}[hbtp]

\vspace*{-2.5cm}
\hspace*{-4.5cm}
\begin{turn}{-90}%
\epsfxsize=16cm \epsfbox{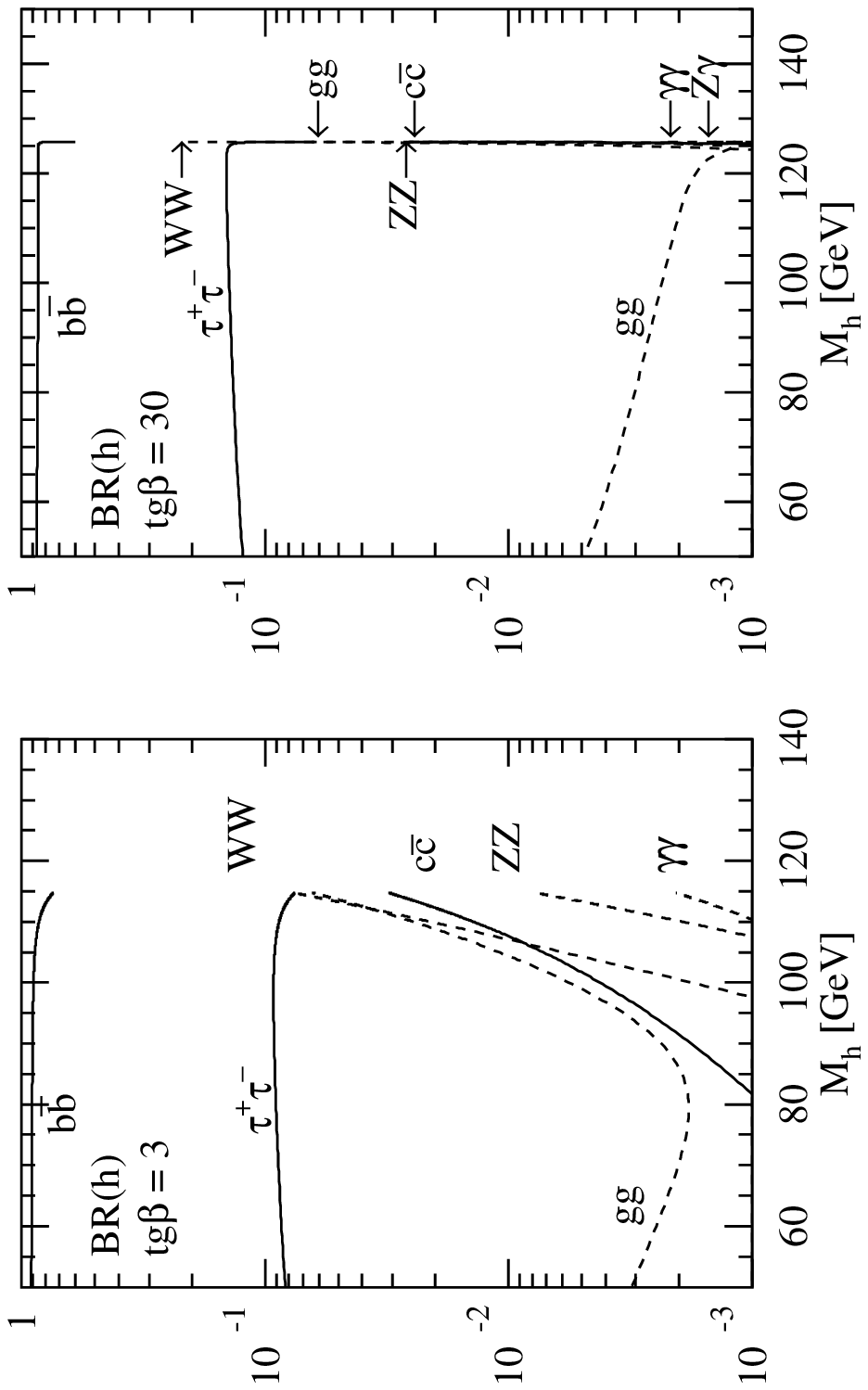}
\end{turn}
\vspace*{-4.2cm}

\centerline{\bf Fig.~\ref{esp-fg:mssmbr}a}

\vspace*{-2.5cm}
\hspace*{-4.5cm}
\begin{turn}{-90}%
\epsfxsize=16cm \epsfbox{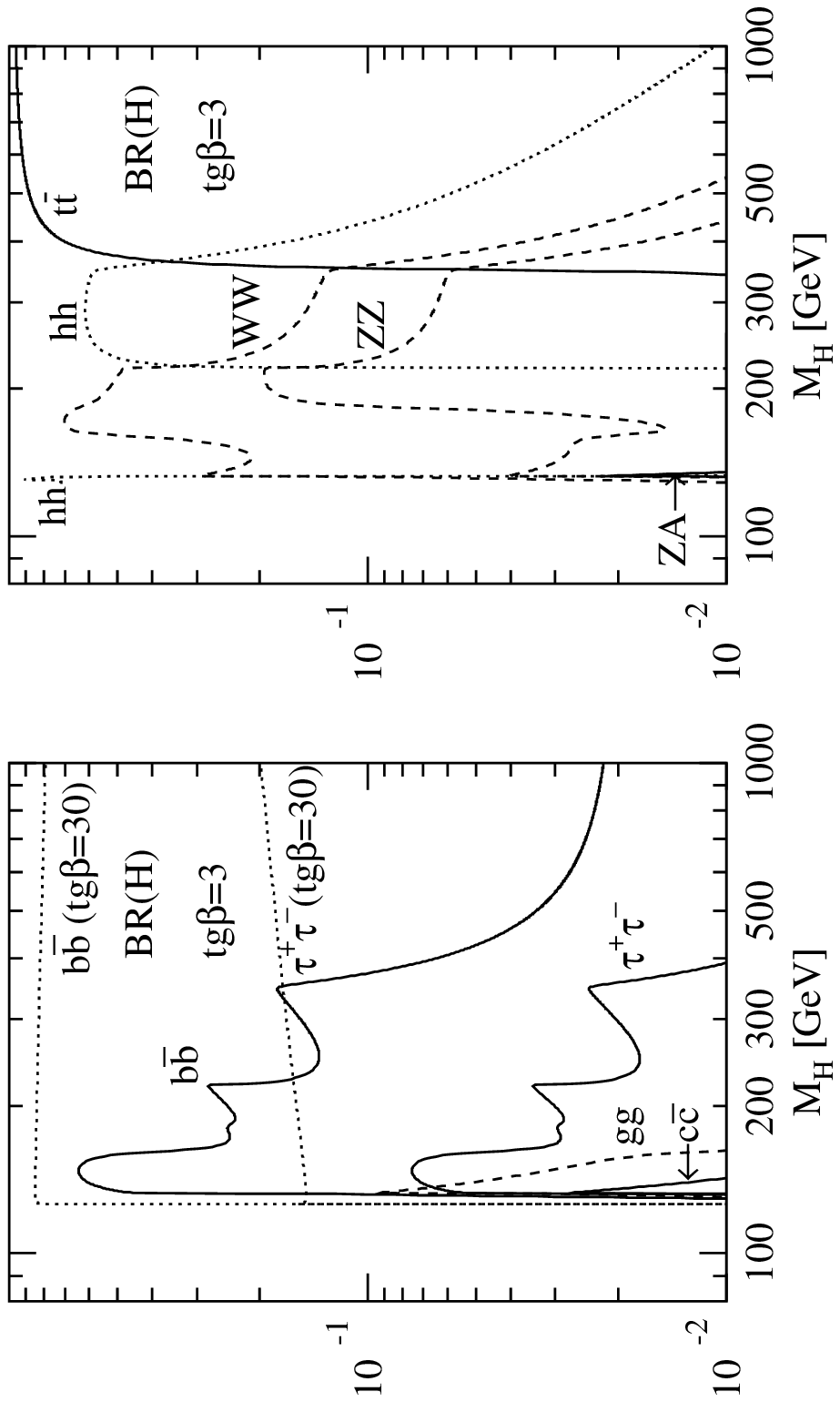}
\end{turn}
\vspace*{-4.2cm}

\centerline{\bf Fig.~\ref{esp-fg:mssmbr}b}

\caption[]{\label{esp-fg:mssmbr} \it Razones de desintegraci\'on de los
bos\'ones de Higgs del MSSM $h, H, A, H^\pm$, para decaimientos no
supersim\'etricos como funci\'on de las
masas, para dos valores de $\tgb=3, 30$ y sin considerar mezcla. Se ha
escogido la masa com\'un de los squark como $M_S=1$ TeV.}
\end{figure}

\vspace*{2mm}
\noindent El \underline{\it bos\'on de Higgs neutro} m\'as ligero $h$
decaer\'a principalmente a pares de fermiones ya que su masa es menor a
$\sim 140$ GeV, Fig.~\ref{esp-fg:mssmbr}a (cf.  \cite{613A} para un resumen
detallado). Este es, en general, tambi\'en el modo dominante de
decaimiento para el bos\'on pseudoescalar $A$. Para valores de $\tgb$
mayores a la unidad y para masas menores que $\sim 140$ GeV, los
principales modos de decaimiento de los bosones neutros de Higgs son
decaimientos a pares de $b\bar b$ y $\tau^+\tau^-$; las razones de
desintegraci\'on son del orden de $\sim 90\%$ y $8\%$, respectivamente.
Los decaimientos a pares de $c\bar c$ y gluones est\'an suprimidos,
especialmente para valores grandes de $\tgb$.  Para masas grandes de
los bosones neutros de Higgs, se abre el canal de decaimiento del top
$H,A \to t\bar t$; aunque para valores grandes de $\tgb$ este modo
permanece suprimido y los bosones neutros de Higgs decaen casi
exclusivamente en pares de $b\bar b$ y $\tau^+\tau^-$.  En contraste con
el bos\'on de Higgs pseudoescalar $A$, el bos\'on de Higgs pesado ${\cal
  CP}$-par $H$ puede en principio decaer en los bosones de norma
d\'ebiles, $H\to WW,ZZ$, si la masa es suficientemente grande. Sin
embargo, debido a que los anchos parciales de decaimiento son
proporcionales a $\cos^2(\beta - \alpha)$, en general est\'an
fuertemente suprimidos, y la se\~nal dorada $ZZ$ del
bos\'on pesado de Higgs en el Modelo Est\'andar se pierde en la
extensi\'on supersim\'etrica. Como resultado, los anchos totales
de los bosones de Higgs son mucho menores en teor\'{\i}as
supersim\'etricas que en el Modelo Est\'andar.

\addtocounter{figure}{-1}
\begin{figure}[hbtp]

\vspace*{-2.5cm}
\hspace*{-4.5cm}
\begin{turn}{-90}%
\epsfxsize=16cm \epsfbox{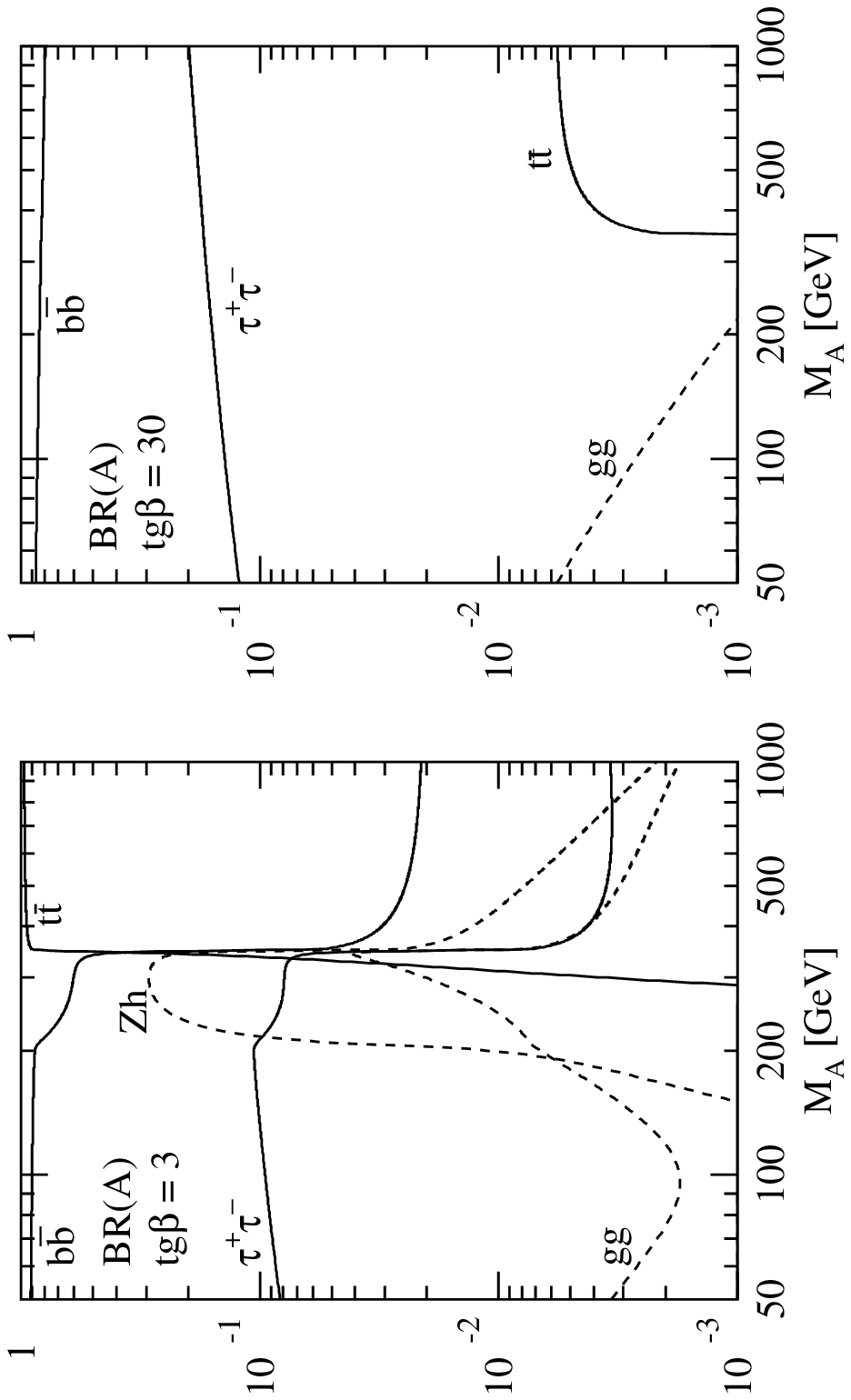}
\end{turn}
\vspace*{-4.2cm}

\centerline{\bf Fig.~\ref{esp-fg:mssmbr}c}

\vspace*{-2.5cm}
\hspace*{-4.5cm}
\begin{turn}{-90}%
\epsfxsize=16cm \epsfbox{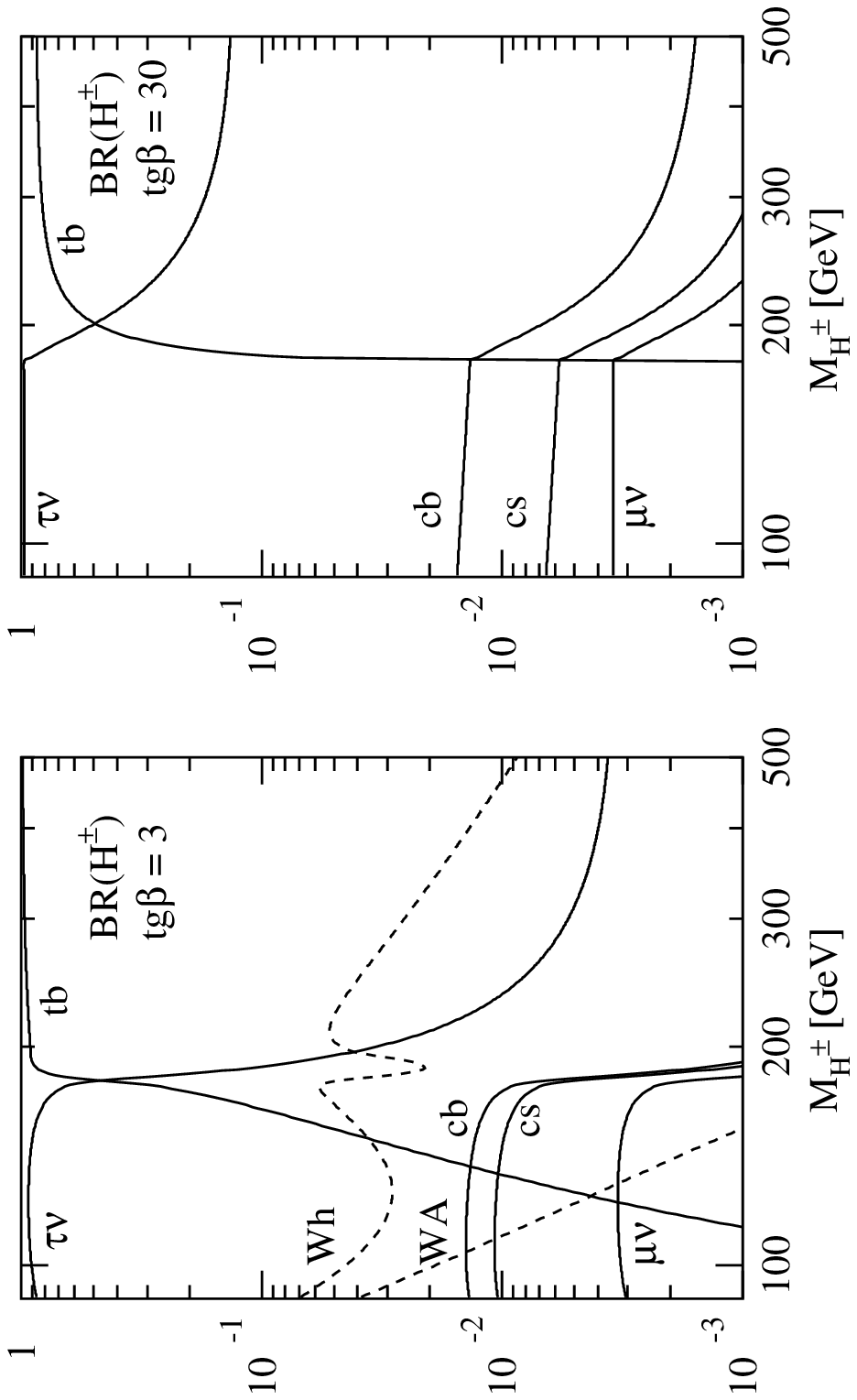}
\end{turn}
\vspace*{-4.2cm}

\centerline{\bf Fig.~\ref{esp-fg:mssmbr}d}

\caption[]{\it Continuaci\'on.}
\end{figure}

\begin{figure}[hbt]

\vspace*{-2.5cm}
\hspace*{-4.5cm}
\begin{turn}{-90}%
\epsfxsize=16cm \epsfbox{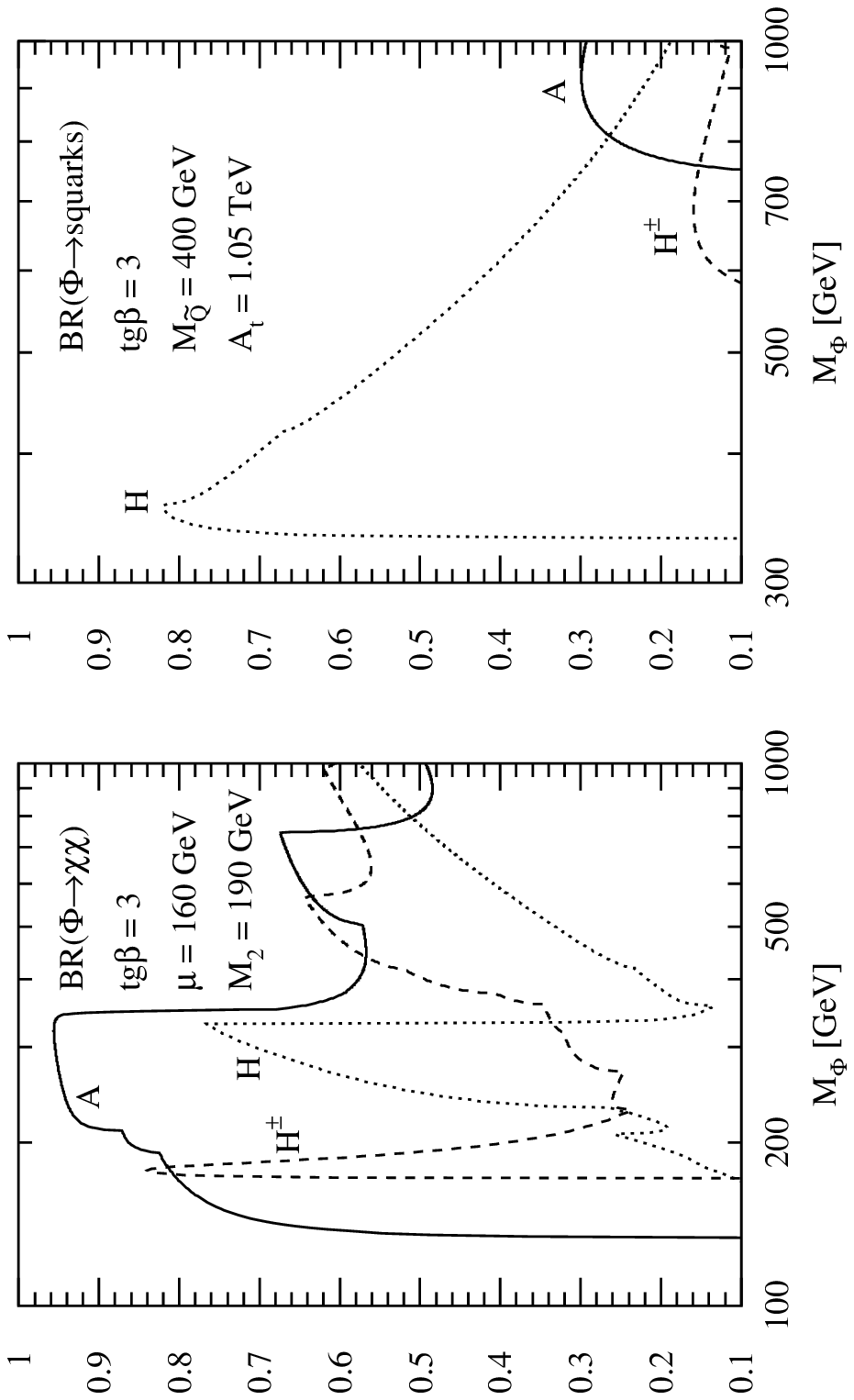}
\end{turn}
\vspace*{-4.2cm}

\caption[]{\label{esp-fg:hcharneutsq} \it Razones de desintegraci\'on de
los bosones de Higgs del MSSM $H,A,H^\pm$ en decaimientos a
charginos/neutralinos y squarks, en funci\'on de las masas del
Higgs para $\tgb=3$. Se han elegido los par\'ametros de mezcla
como $\mu=160$ GeV, $A_t=1.05$ TeV, $A_b=0$ y las masas de los
squarks de las primeras dos generaciones como
$M_{\widetilde{Q}}=400$ GeV. El par\'ametro de masa del gaugino se
toma como $M_2=190$ GeV.}
\end{figure}

El bos\'on de Higgs  pesado  neutro $H$ puede tambi\'en decaer en
dos bosones de Higgs ligeros. Otros posibles canales son:
decaimientos en cascada del Higgs y decaimientos a part\'{\i}culas
supersim\'etricas \citer{614,616}, Fig.~\ref{esp-fg:hcharneutsq}.
Adem\'as de los sfermiones ligeros, los decaimientos del bos\'on de
Higgs a charginos y neutralinos eventuralmente podr\'\i an ser
importantes. Estos nuevos canales son cinem\'aticamente
accesibles, al menos para los bosones de Higgs pesados $H,A$ and
$H^\pm$; de hecho, las razones de decaimiento pueden ser muy
grandes e incluso llegar a ser dominantes en algunas regiones del
espacio de par\'ametros del MSSM. Los decaimientos del $h$ a la
part\'{\i}cula supersim\'etrica m\'as ligera (LSP), neutralinos,
son tambi\'en importantes, excediendo el 50\% en algunas partes
del espacio  de par\'ametros. Estos decaimientos afectan
fuertemente las t\'ecnicas de investigaci\'on
experimentales.

\noindent
Las \underline{\it part\'{\i}culas de Higgs cargadas} decaen a
fermiones, pero tambi\'en, si es permitido cinem\'aticamente, al
Higgs neutro m\'as ligero y al bos\'on $W$. Abajo de los umbrales
de $tb$ y $Wh$, las part\'{\i}culas de Higgs cargadas decaer\'an
primordialmente a pares $\tau \nu_\tau$ y $cs$, siendo dominante
el primero para $\tgb>1$. Para valores grandes de $M_{H^\pm}$, el
modo de decaimiento  top--bottom $H^+\to t\bar b$ se vuelve
dominante. En algunas partes del espacio de par\'ametros de SUSY,
los decaimientos a part\'{\i}culas supersim\'etricas podr\'{\i}an
exceder el 50\%.
\vspace*{2mm}

Sumando los diversos modos de decaimiento, el ancho de todos los
cinco bosones de Higgs se mantiene muy estrecho, siendo del orden
de 10 GeV incluso para masas grandes.

\vspace*{4mm}
\subsection{La Producci\'on de Part\'\i culas de Higgs SUSY en Colisiones
  Hadr\'onicas}

\vspace*{2mm}
\noindent
Los procesos b\'asicos de producci\'on de part\'{\i}culas de
Higgs en colisionadores hadr\'onicos \cite{24A,32,620B} son
esencialmente los mismos que en el Modelo Est\'andar.  Diferencias
importantes son, no obstante, generadas por los acoplamientos
modificados, el espectro de part\'{\i}culas extendido y la paridad
negativa del bos\'on $A$. Para $\tgb$ grande el acoplamiento $hb\bar b$
es aumentado de tal manera que el lazo bottom-quark se vuelve
competitivo con respecto al lazo top-quark en el acoplamiento efectivo
$hgg$. M\'as a\'un, los lazos de squarks contribuir\'an a este
acoplamiento \cite{sqloop}.

\vspace*{2mm}
\noindent
{\bf 1.$\,$} La secci\'on eficaz part\'onica $\sigma(gg\to \Phi)$ para la
fusi\'on de gluones de las part\'{\i}culas de Higgs puede ser expresada
por  acoplamientos $g$, en unidades de los correspondientes
acoplamientos del SM, y por los factores de forma $A$; al orden m\'as
bajo \cite{32,sqloopqcd}:
\begin{eqnarray}
\hat\sigma^\Phi_{LO} (gg\to \Phi) & = & \sigma^\Phi_0 M_\Phi^2 \times
BW(\hat{s}) \\
\sigma^{h/H}_0 & = & \frac{G_{F}\alpha_{s}^{2}(\mu)}{128 \sqrt{2}\pi} \
\left| \sum_{Q} g_Q^{h/H} A_Q^{h/H} (\tau_{Q})
+ \sum_{\widetilde{Q}} g_{\widetilde{Q}}^{h/H} A_{\widetilde{Q}}^{h/H}
(\tau_{\widetilde{Q}}) \right|^{2} \nonumber \\
\sigma^A_0 & = & \frac{G_{F}\alpha_{s}^{2}(\mu)}{128 \sqrt{2}\pi} \
\left| \sum_{Q} g_Q^A A_Q^A (\tau_{Q}) \right|^{2} \nonumber \;.
\end{eqnarray}
Mientras los acoplamientos de quarks han sido definidos en la
Tabla \ref{esp-tb:hcoup}, los acoplamientos  de las part\'{\i}culas de
Higgs con squarks est\'an dados por
\begin{eqnarray}
g_{\tilde Q_{L,R}}^{h} & = & \frac{M_Q^2}{M_{\tilde Q}^2} g_Q^{h}
\mp \frac{M_Z^2}{M_{\tilde Q}^2} (I_3^Q - e_Q \sin^2 \theta_W)
\sin(\alpha + \beta) \nonumber \\ \nonumber \\
g_{\tilde Q_{L,R}}^{H} & = & \frac{M_Q^2}{M_{\tilde Q}^2} g_Q^{H}
\pm \frac{M_Z^2}{M_{\tilde Q}^2} (I_3^Q - e_Q \sin^2 \theta_W)
\cos(\alpha + \beta) \;.
\end{eqnarray}
S\'olo la no-invariancia de ${\cal CP}$ permite contribuciones
diferentes de cero de los squarks a la producci\'on del bos\'on
pseudoescalar $A$. Los factores de forma pueden ser expresados en
t\'erminos de las funciones de escala $f(\tau_i=4M_i^2/M_\Phi^2)$,
cf. Eq. (\ref{esp-eq:ftau}):
\begin{eqnarray}
A_Q^{h/H} (\tau) & = & \tau [1+(1-\tau) f(\tau)] \nonumber \\
A_Q^A (\tau) & = & \tau f(\tau) \nonumber \\
A_{\tilde Q}^{h/H} (\tau) & = & -\frac{1}{2}\tau [1-\tau f(\tau)] ~.
\end{eqnarray}
Para valores de $\tgb$ peque\~nos la contribuci\'on del lazo del top es
dominante, mientras que para $\tgb$ grandes el lazo del bottom aumenta
dr\'asticamente. Los lazos de squarks pueden ser muy
significativos para masas de squarks menores a $\sim 400$ GeV
 \cite{sqloopqcd,sqloopmass}.

\vspace*{2mm}
\noindent {\bf 2.$\,$} Otros mecanismos de producci\'on para bosones de
Higgs supersim\'etricos como fusi\'on de bosones vectoriales, Higgs-strahlung
por los bosones $W,Z$ y Higgs-bremsstrahlung por quarks top y bottom,
pueden tratarse en analog\'{\i}a con los correspondientes procesos del
SM.

De particular importancia es el proceso de fusi\'on de quarks $b$ para
valores grandes de  $\tan\beta$, cuando los acoplamientos del Higgs a $b$
son intensificados \cite{bH}.  Como los quarks  $b$ son moderadamente
ligeros, la escisi\'on del glu\'on $g \to \bar{b} b$ da lugar a haces de
alta energ\'\i a de  $b$-quark/anti-quark en protones/antiprotones
r\'apidos.  La fusi\'on de $\bar{b} + b \to h,H,A$ es, por lo tanto, una
fuente rica de bosones de Higgs en Tevatron y en LHC para $\tan\beta$ grande.

\begin{figure}[hbtp]

\vspace*{0.3cm}
\hspace*{1.0cm}
\begin{turn}{-90}%
\epsfxsize=8.5cm \epsfbox{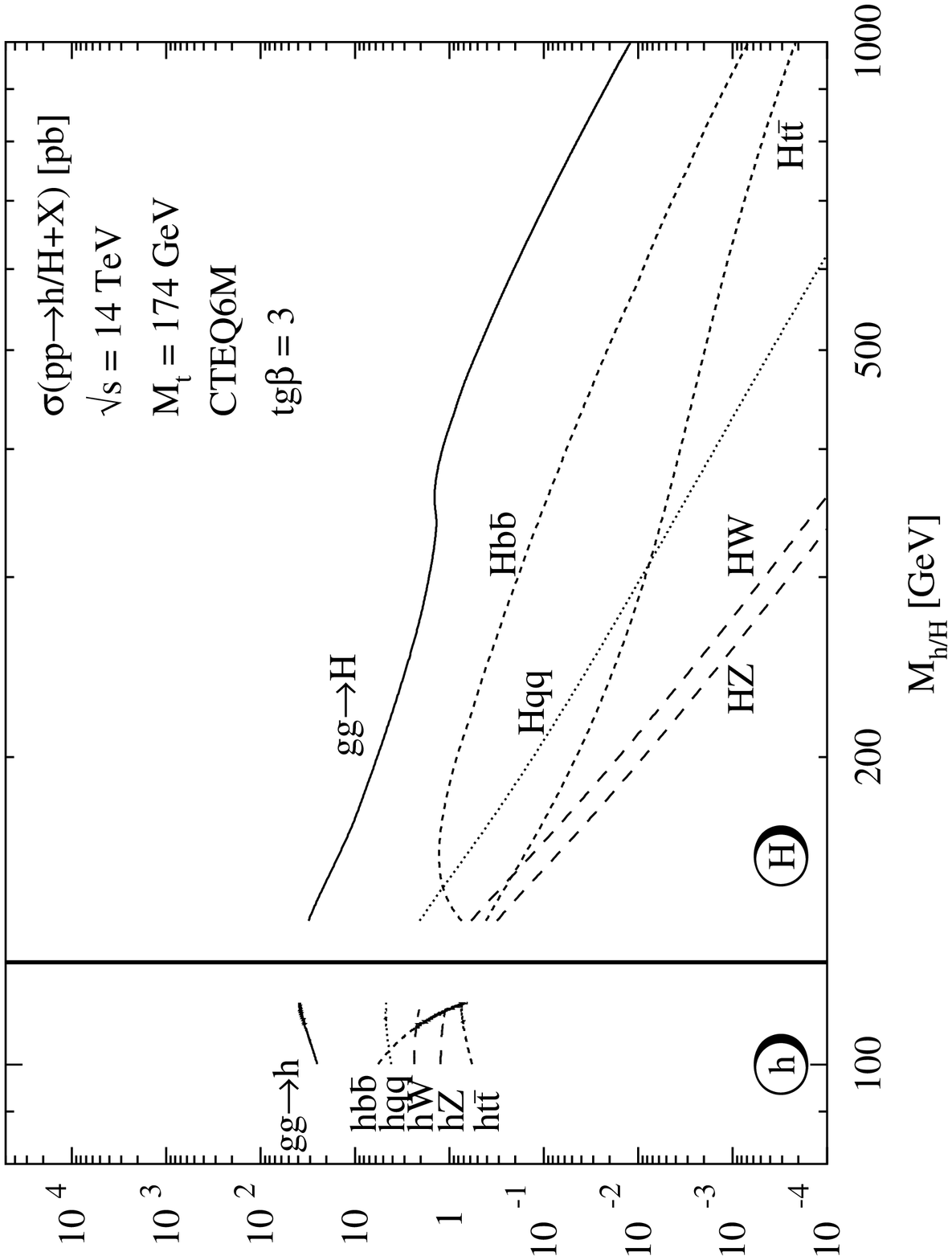}
\end{turn}
\vspace*{0.3cm}

\centerline{\bf Fig.~\ref{esp-fg:mssmprohiggs}a}

\vspace*{0.2cm}
\hspace*{1.0cm}
\begin{turn}{-90}%
\epsfxsize=8.5cm \epsfbox{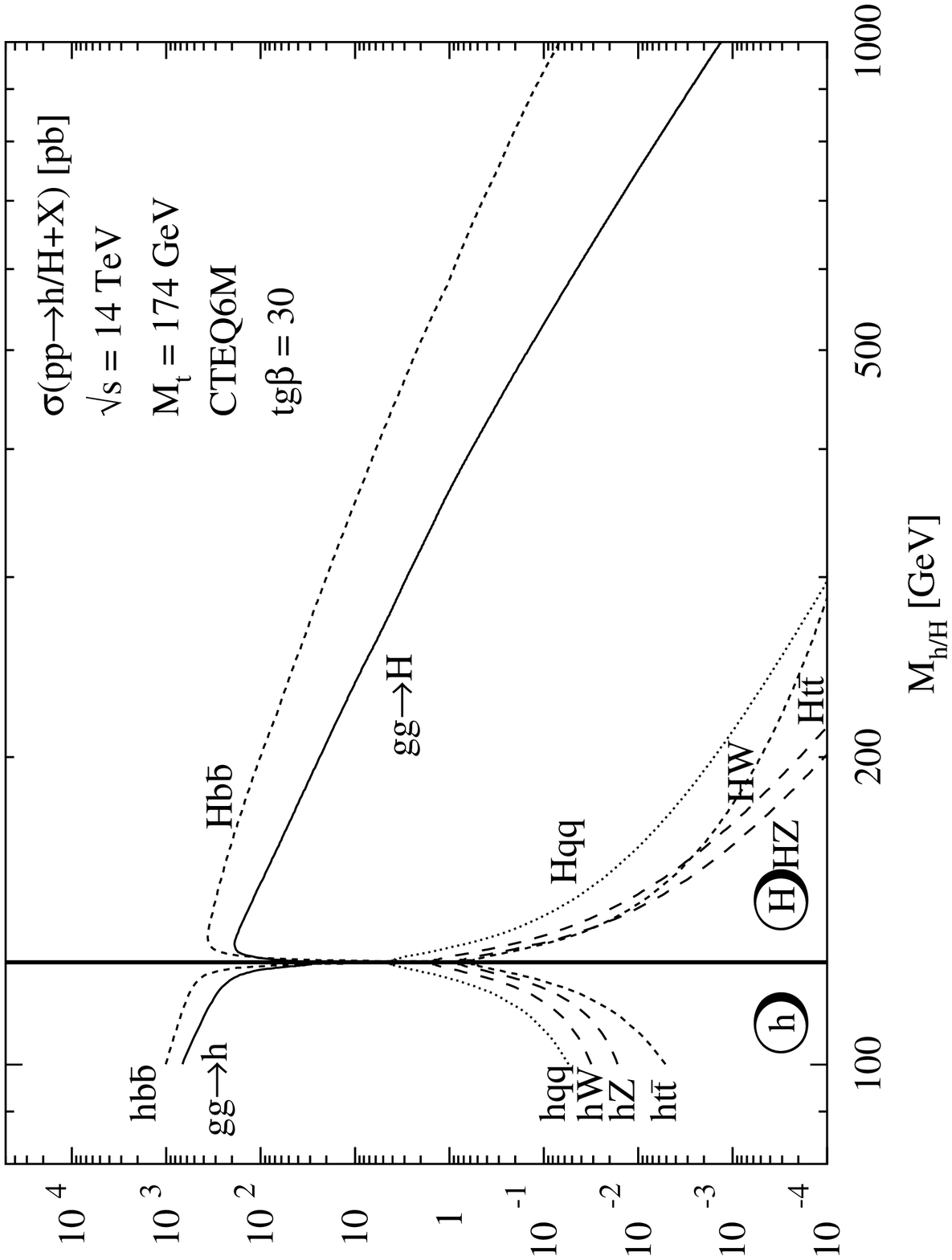}
\end{turn}
\vspace*{0.3cm}

\centerline{\bf Fig.~\ref{esp-fg:mssmprohiggs}b}

\caption[]{\label{esp-fg:mssmprohiggs} \it Secciones eficaces de la
producci\'on del Higgs del MSSM neutro en el LHC para fusi\'on de
glu\'on $gg\to \Phi$, fusi\'on del bos\'on vectorial $qq\to qqVV
\to qqh/ qqH$, Higgs-strahlung $q\bar q\to V^*
\to hV/HV$ y la producci\'on asociada $gg,q\bar q \to b\bar b \Phi/
t\bar t \Phi$, incluyendo todas las correcciones conocidas de QCD.
(a) producci\'on de $h,H$ para $\tgb=3$, (b) producci\'on de $h,H$
para $\tgb=30$, (c) producci\'on de $A$ para $\tgb=3$, (d)
producci\'on de $A$ para $\tgb=30$.}
\end{figure}
\addtocounter{figure}{-1}
\begin{figure}[hbtp]

\vspace*{0.3cm}
\hspace*{1.0cm}
\begin{turn}{-90}%
\epsfxsize=8.5cm \epsfbox{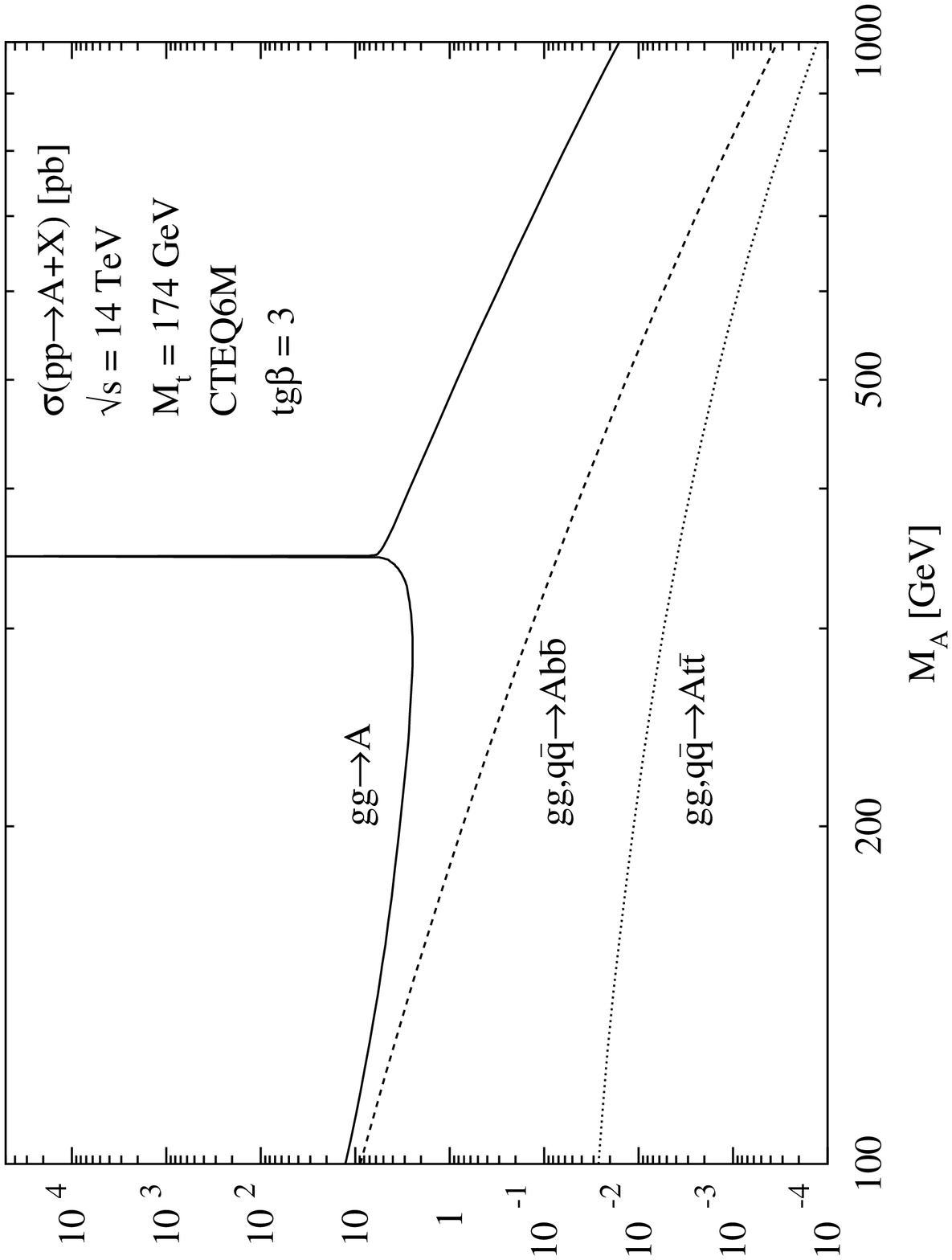}
\end{turn}
\vspace*{0.3cm}

\centerline{\bf Fig.~\ref{esp-fg:mssmprohiggs}c}

\vspace*{0.2cm}
\hspace*{1.0cm}
\begin{turn}{-90}%
\epsfxsize=8.5cm \epsfbox{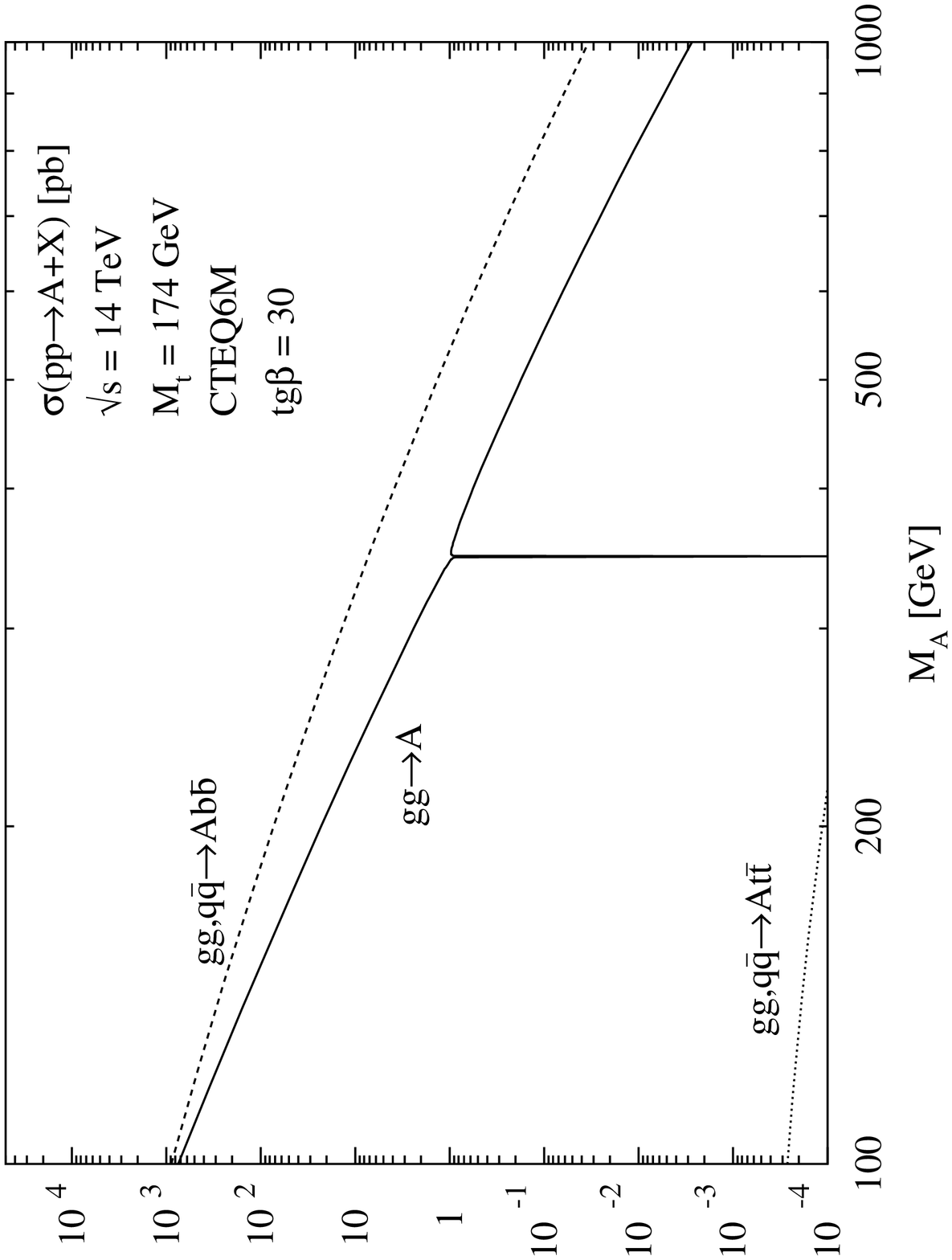}
\end{turn}
\vspace*{0.3cm}

\centerline{\bf Fig.~\ref{esp-fg:mssmprohiggs}d}

\caption[]{\it Continuaci\'on.}
\end{figure}

\vspace*{2mm}
\noindent
{\bf 3.$\,$} Los datos extra\'{\i}dos del Tevatron en al canal $p \bar p \to b
\bar b \tau^+ \tau^-$ han sido explotados para excluir
parte del espacio de par\'ametros supersim\'etrico del Higgs en el
plano $[ M_A, \tgb]$. En el rango interesante de $\tgb$ de entre
30 y 50, las masas para el pseudoescalar $M_A$ de hasta 150 a 190
GeV parecen estar excluidas.

\vspace*{2mm}
\noindent {\bf 4.$\,$} Las secciones eficaces de los diversos
mecanismos de producci\'on del Higgs del MSSM en el LHC se muestran en
las Figs.  \ref{esp-fg:mssmprohiggs}a--d para dos valores
representativos de $\tgb = 3$ y 30, como funci\'on de la
correspondiente masa del Higgs. Las densidades del part\'on CTEQ6M han
sido tomadas con $\alpha_s(M_Z)=0.118$; las masas del top y del bottom se
han fijado en los valores $M_t=174$ GeV y $M_b=4.62$ GeV. Para el
bremsstrahlung del Higgs pseudoescalar por los quarks $t,b$, $pp \to
Q\bar Q A +X$, se han usado las densidades dominantes del part\'on
CTEQ6L1. Para valores peque\~nos y moderados de $\tgb\lessim 10$ la
secci\'on eficaz de la fusi\'on-glu\'on provee la secci\'on eficaz del
producci\'on dominante para la regi\'on completa de masa del Higgs de
hasta $M_\Phi\sim 1$ TeV. Sin embargo, para $\tgb$ grandes, Higgs
bremsstrahlung por quarks bottom, $pp\to b\bar b \Phi+X$, domina sobre el
mecanismo de fusi\'on-glu\'on ya que los acoplamientos de Yukawa del
bottom aumentan dr\'asticamente en este caso.

\begin{figure}[hbtp]
\begin{center}
\epsfig{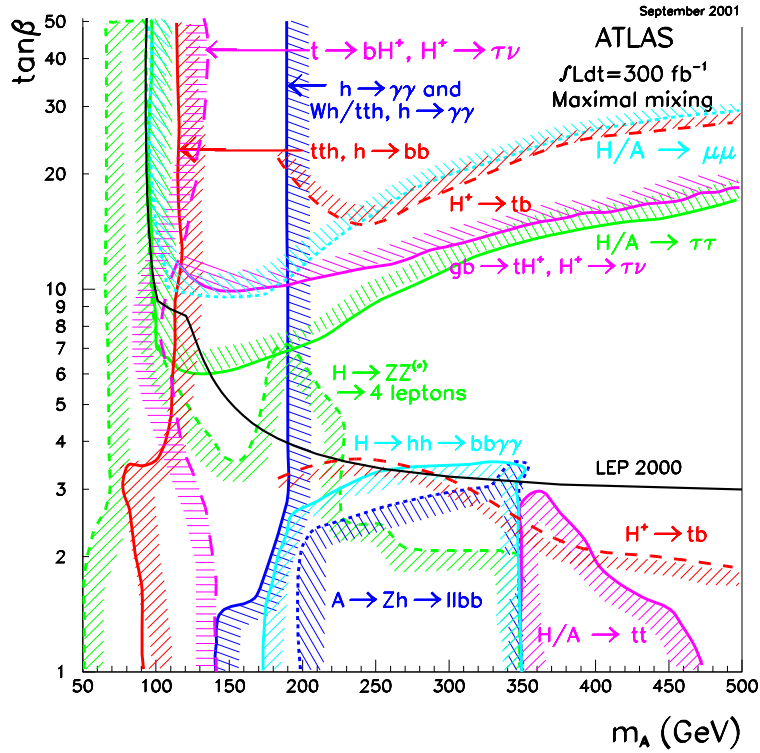}
\end{center}
\caption[]{\label{esp-fg:atlascms} \it La sensibilidad de ATLAS
para el descubrimiento del bos\'on de Higgs del MSSM en el caso de
 mezcla m\'axima. Las curvas de descubrimiento 5$\sigma$ se
muestran en el plano $(\tan\beta,m_A)$ para los canales
individuales y para una luminosidad integrada de 300 fb$^{-1}$.
Tambi\'en se muestra el correspondiente l\'{\i}mite del LEP.
Ref.~\cite{richter}.}
\end{figure}

La b\'usqueda del Higgs del MSSM en el LHC ser\'a m\'as complicada que la
b\'usqueda del Higgs del SM.  El resumen final se presenta en la
Fig.~\ref{esp-fg:atlascms}. Esta gr\'afica exhibe una regi\'on dif\'{\i}cil
para la b\'usqueda del Higgs del MSSM en el LHC. Para $\tgb \sim 5$ y $M_A
\sim 150$ GeV, es necesaria la luminosidad total y la muestra total de
datos de los experimentos ATLAS y CMS en el LHC para cubrir la regi\'on
problem\'atica de los par\'ametros \cite{richter}. Por otro lado, si no se
encontrara exceso de eventos de Higgs por encima de los procesos del
fondo del SM m\'as all\'a de 2 desviaciones est\'andar, los bosones de Higgs
del MSSM se pueden excluir en un 95\% C.L.

A\'un cuando  se espera que el espacio de par\'ametros
completo del Higgs supersim\'etrico sea finalmente cubierto por
experimentos del LHC, el conjunto total de los bosones de Higgs
individuales es accesible s\'olo en parte del espacio de
par\'ametros. En particular en el trozo de apertura ciega del espacio de
par\'ametros que est\'a aproximadamente en $M_A \sim$ 200 GeV y centrado alrededor de
$\tan\beta \sim 7$,  s\'olo el bos\'on de Higgs m\'as ligero  $h$ puede ser
descubierto, mientras que los bosones de Higgs pesados $A,H,H^\pm$ no
pueden ser encontrados en canales de decaimiento no-supersim\'etricos.
M\'as a\'un, la b\'usqueda de las part\'{\i}culas de
Higgs pesadas $H,A$ es muy
dif\'{\i}cil por el fondo continuo de $t\bar t$ para masas $\gsim
500$ GeV.

\vspace*{4mm}
\subsection{La Producci\'on de Part\'{\i}culas Higgs  Supersim\'etricas en
  Colisiones $e^+e^-$}

\vspace*{2mm}
\noindent
{\bf 1.$\,$}  La b\'usqueda de los bosones neutros de Higgs
supersim\'etricos en colisionadores lineales \ee  ser\'a una
extensi\'on directa de la b\'usqueda realizada en el LEP2, el cual
cubri\'o el rango de masas hasta los $\sim 100$~GeV para los
bos\'ones neutros de Higgs. Energ\'{\i}as mayores, con un exceso en
$\sqrt{s}$ de $250$~GeV, son requeridas para barrer todo el
espacio de par\'ametros del MSSM con valores desde moderados hasta
grandes de $\tgb$.\\

\noindent Los principales mecanismos de producci\'on de \underline{\it
los bosones neutros de Higgs} en colisionadores \ee \cite{19, 615,
617} son los procesos \Hs y producci\'on de pares asociados,
as\'{\i} como tambi\'en los procesos de fusi\'on:
\newpage
\begin{alignat}{3}
(a)& \ \ \mbox{Higgs--strahlung{\hspace*{0.7mm}}:} \hspace{1.5cm}& \epem &
\quad \stackrel{Z}{\longrightarrow} & Z+h/H
\nonumber  \\
(b)& \ \ {\rm Producci\acute{o}n \ de \ pares:}  & \epem &
\quad \stackrel{Z}{\longrightarrow} & A+h/H
\nonumber \\
(c)& \ \ {\rm Procesos \ de \ fusi\acute{o}n:}  & \epem &
\quad \stackrel{WW}{\longrightarrow} & \qquad \overline{\nu}_e \ \nu_e \ + h/H
 \nonumber  \\
& & \epem &
\quad \stackrel{ZZ}{\longrightarrow} & \qquad \epem + h/H \;. \nonumber
\end{alignat}

El bos\'on de Higgs ${\cal CP}$-impar $A$ no puede producirse en
procesos de fusi\'on a primer orden. Las secciones eficaces para
los cuatro \Hs y procesos de producci\'on de pares pueden ser
expresados como
\begin{eqnarray}
\sigma(\epem \ra Z + h/H) & =& \sin^2/\cos^2(\beta-\alpha) \ \sigma_{SM}
\nonumber \\
\sigma(\epem \ra A + h/H) & =& \cos^2/\sin^2(\beta-\alpha) \
\bar{\lambda} \  \sigma_{SM} ~,
\end{eqnarray}
donde $\sigma_{SM}$ es la secci\'on eficaz en el SM para el \Hs y
el coeficiente
$\bar{\lambda} \sim \lambda^{3/2}_{Aj} / \lambda^{\demi}_{Zj}$  que toma en cuenta la supresi\'on de onda-$P$
$Ah/H$ en las secciones eficaces cerca del umbral.

\STS Las secciones eficaces para el Higgs-strahlung y para la
producci\'on de pares, tanto como para la producci\'on de los
bosones de Higgs neutros, ligero y pesado, $h$ y $H$, son
complementarias, apareciendo con  coeficientes ya sea
$\sin^2(\beta-\alpha)$ \'o $\cos^2(\beta-\alpha)$.  Como resultado,
ya que $\sigma_{SM}$ es grande, al menos el bos\'on de Higgs
${\cal CP}$-par m\'as ligero debe ser detectado en los experimentos de
$e^+e^-$.

\begin{figure}[hbtp]
\begin{center}
\vspace*{5mm}
\hspace*{5mm}
\epsfig{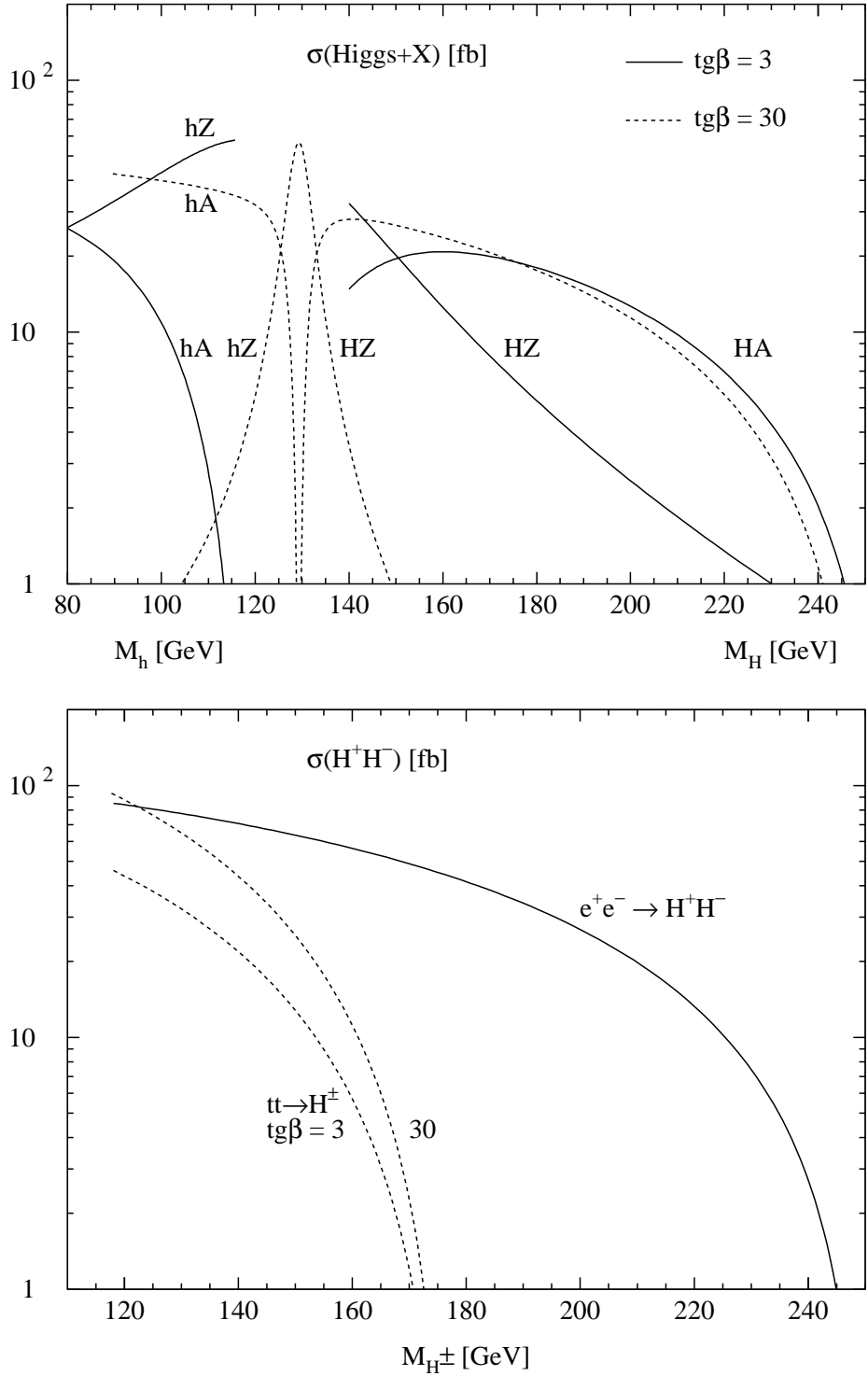}
\end{center}
\vspace{-2.5cm}
\caption[]{\it  Secciones eficaces de la producci\'on de bosones de Higgs
  en el MSSM a $\sqrt{s} = 500$~GeV: Higgs-strahlung y producci\'on de
  pares; parte superior: bosones de Higgs neutros; inferior: bosones de Higgs
  cargados.
  Ref. \protect\cite{613A}.  \protect\label{esp-f604}\label{esp-prodcs}}
\end{figure}

\STS Ejemplos representativos de secciones eficaces para
mecanismos de producci\'on de los bosones de Higgs neutros est\'an
dados en la Fig.~\ref{esp-f604}, como funci\'on de las masas de los
Higgs, para $\tgb= 3$ y 30.  La secci\'on eficaz para $hZ$ es
grande para $M_h$ cerca del valor m\'aximo permitido de $\tgb$; es
del orden de 50~fb, correspondiente a $\sim$ 2,500 eventos para
una luminosidad integrada de 50 fb$^{-1}$. En contraste, la
secci\'on eficaz para $HZ$ es grande si $M_h$ est\'a
suficientemente por abajo del valor m\'aximo [implicando que $M_H$ es
peque\~na]. Para $h$ y para una masa ligera de $H$, las se\~nales
consisten de un bos\'on $Z$ acompa\~nado por un par $b\bar{b}$ o
$\tau^+ \tau^-$. Estas se\~nales son f\'aciles de separar de las
se\~nales de fondo (background), el cual proviene principalmente
de la producci\'on de $ZZ$ si la masa del Higgs tiene un valor
cercano a $M_Z$. Para los canales asociados $\epem \to Ah$ y $AH$,
la situaci\'on es opuesta a la anterior: la secci\'on eficaz para
$Ah$ es grande para un $h$ ligero, mientras que la producci\'on
del par $AH$ es el mecanismo dominante en la regi\'on
complementaria para bosones $H$ y $A$ pesados. La suma de las dos
secciones eficaces decrece de $\sim 50$ a 10~fb si $M_A$ aumenta
de $\sim 50$ a 200~GeV en $\sqrt{s} = 500$~GeV.
En grandes partes del espacio de par\'ametros  las se\~nales consisten de cuatro
quarks $b$ en el estado final, requiriendo que se prevea un
eficiente caracterizaci\'on del quark-$b$. Las
restricciones en la masa ayudar\'an a eliminar las se\~nales de
fondo de estados finales de jets de QCD y $ZZ$. Para el mecanismo
de fusi\'on del $WW$, las secciones eficaces son mayores que para
el mecanismo de Higgs-strahlung, si la masa del Higgs es
moderadamente peque\~na -- menor que 160~GeV en $\sqrt{s} = 500$
GeV. Sin embargo, como el estado final no puede ser completamente
reconstruido, es m\'as dif\'{\i}cil extraer la se\~nal. Como en el
caso de los procesos de \Hs, la producci\'on de bosones de Higgs
ligero $h$ y pesado $H$ tambi\'en se complementan entre ellos en
la fusi\'on de $WW$.
 \\

\noindent  Los \underline{\it bosones de Higgs cargados}, si son m\'as
ligeros que el quark top, pueden producirse en decaimientos del
top, $t \ra b + H^+$, con una raz\'on de desintegraci\'on que
var\'{\i}a entre $2\%$ y $20\%$ en la regi\'on cinem\'aticamente
permitida. Debido a que la secci\'on eficaz para la producci\'on
de pares-top es del orden de 0.5 pb en $\sqrt{s} = 500$~GeV, esto
corresponde a una producci\'on de 1,000 a 10,000 bosones de Higgs
cargados a una luminosidad de 50~fb$^{-1}$. Dado que para $\tgb$
mayor a uno los bosones de Higgs cargados decaer\'an
principalmente a $\tau \nu_\tau$,  habr\'a un exceso de estados
finales $\tau$ sobre los estados finales $e, \mu$ en decaimientos
del $t$, un rompimiento aparente de la universalidad lept\'onica.
Para masas grandes del Higgs  el modo de decaimiento dominante es
el decaimiento al top  $H^+ \to t \overline{b}$. En este caso las
part\'{\i}culas de Higgs cargadas deben ser producidas en pares en
colisionadores \ee:
\[
              \epem \to H^+H^- ~.
\]
La secci\'on eficaz depende \'unicamente de la masa del Higgs
cargado. Es del orden de 100 fb para masas peque\~nas del Higgs en
$\sqrt{s} = 500$~GeV, pero decrece muy r\'apidamente debido a la
supresi\'on de onda-$P$ $\sim \beta^3$ cerca del umbral. Para
$M_{H^{\pm}} = 230$~GeV, la secci\'on eficaz cae a un nivel de
$\simeq 5\,$~fb. La secci\'on eficaz es considerablemente m\'as
grande para colisiones $\gamma \gamma$.

\vspace*{4mm} \noindent {\bf 2.$\,$} {\it \underline{ Estrategias de
B\'usqueda Experimental}:}
Las estrategias de b\'usqueda de los bosones de Higgs cargados y
neutros han sido descritas en la Ref. \cite{13}. La situaci\'on
experimental en su conjunto puede sintetizarse en los siguientes
dos puntos:

\vspace*{1mm}
\noindent
{\hspace*{3mm}}{\bf (i)}  La part\'{\i}cula de Higgs ${\cal
CP}$-par m\'as ligera, $h$ puede ser detectada en el rango
completo del espacio de par\'ametros del MSSM, ya sea v\'{\i}a
Higgs-strahlung $\epem \to hZ$ o v\'{\i}a producci\'on $\epem
\to hA$. Esta conclusi\'on se mantiene cierta incluso a
una energ\'{\i}a del c.m. de 250 GeV, independientemente del valor
de la masa de los squarks; es tambi\'en v\'alido si los
decaimientos a neutralinos invisibles y otras part\'{\i}culas SUSY
se dan en el sector de Higgs.

\vspace*{1mm}
\noindent
{\hspace*{3mm}}{\bf (ii)} El \'area del espacio de par\'ametros
donde {\it  todos los bosones de Higgs de SUSY} pueden ser
descubiertos en colisionadores \ee se caracteriza por $M_H, M_A
\lessim \frac{1}{2} \sqrt{s}$, independientemente de $\tgb$. Los
bosones de Higgs $h, H$ pueden producirse ya sea v\'{\i}a \Hs o en
producci\'on asociada de $Ah, AH$; los bosones de Higgs cargados
se producir\'an en pares
$H^+H^-$.  As\'\i{} la regi\'on ciega del LHC puede ser cubierto
para masas de los Higgses $A,H,H^\pm$ de hasta 500 GeV en el
colisionador de 1 TeV ILC, y hasta 1.5 TeV en el colisionador de 3 TeV
CLIC.  Si el colisionador $ee$ se convierte en un colisionador de
fotones de altas energ\'\i as por dispersi\'on hacia atr\'as de
Compton de luz de laser \citer{novo,MMMPMZ}, el rango de masas del
Higgs en experimentos de formaci\'on simple $\gamma \gamma \to A,H$ puede ser
extendido a 80\% de la energ\'\i a total $e^+e^-$, i.e. 800 GeV y 2.4TeV
en el ILC y CLIC, respectivamente, cf. \cite{Muhlgamgam,Krawgamgam}.

\vspace*{2mm}
La b\'usqueda del bos\'on de Higgs SUSY neutro m\'as ligero $h$ ha
sido uno de las tareas experimentales m\'as importantes del LEP2.
Valores de la masa del bos\'on pseudoescalar $A$ menor que
alrededor de 90 GeV ya han sido excluidos, independientemente de
$\tgb$, cf.~Fig.~\ref{esp-fg:mssmhiggs}. \\

\vspace*{4mm}
\subsection{Midiendo la Paridad de los Bosones de Higgs}

\vspace*{2mm}
\noindent  Una vez que los bosones de Higgs sean descubiertos, se
deben establecer las propiedades de las part\'{\i}culas. Adem\'as de
la reconstrucci\'on del potencial de Higgs supersim\'etrico
\cite{66A,selfMMM}, la cual ser\'a una tarea muy exigente, deben
establecerse los n\'umeros cu\'anticos externos, en particular la
paridad de las part\'{\i}culas
de Higgs escalar y pseudoescalar $H$ y $A$ \cite{618}.

\vspace*{2mm}
\noindent
{\bf 1.$\,$} Para masas grandes de $H,A$ los decaimientos $H,A\to t\bar t$ a
estados finales del top pueden ser usados para discriminar entre
las diferentes asignaciones de paridad \cite{618}. Por ejemplo,
los bosones $W^+$ y $W^-$ en los decaimientos $t$ y $\bar t$
tienden a emitirse antiparalelamente y paralelamente en el plano
perpendicular al eje  $t\bar t$:
\begin{equation}
\frac{d\Gamma^\pm}{d\phi_*} \propto 1 \mp \left( \frac{\pi}{4} \right)^2
\cos \phi_*
\end{equation}
para decaimientos $H$ y $A$, respectivamente.

De manera alternativa, la paridad $\mathcal{CP}$ de los bosones de
Higgs puede ser medida analizando la magnitud de la secci\'on eficaz
total y la polarizaci\'on del quark top en producci\'on asociada top-Higgs
en colisiones $e^+e^-$\cite{bup}. \\[-0.1cm]

\vspace*{2mm}
\noindent
{\bf 2.$\,$}  Para masas ligeras de $H,A$, las colisiones $\gamma\gamma$ parecen
proporcionar una soluci\'on viable \cite{618}. La fusi\'on de las
part\'{\i}culas de Higgs en rayos de fotones linealmente
polarizados depende del \'angulo entre los vectores de
polarizaci\'on. Para part\'{\i}culas escalares $0^+$ la amplitud
de producci\'on es diferente de cero para vectores de
polarizaci\'on paralela, mientras que las part\'{\i}culas
pseudoescalares $0^-$ requieren vectores de polarizaci\'on
perpendiculares:
\begin{equation}
{\cal M}(H)^+  \sim  \vec{\epsilon}_1 \cdot \vec{\epsilon}_2  \hspace*{0.5cm}
\mbox{y} \hspace*{0.5cm}
{\cal M}(A)^-  \sim  \vec{\epsilon}_1 \times \vec{\epsilon}_2 ~.
\end{equation}
El montaje experimental para dispersi\'on hacia atr\'as de Compton
de luz laser se puede ajustar de tal manera que la polarizaci\'on
lineal de los rayos de fotones duros se aproxime a valores cercanos
al 100\%. Dependiendo de la paridad $\pm$ de la resonancia
producida, la asimetr\'{\i}a medida  para fotones de
po\-la\-ri\-zaci\'on paralela y perpendicular,
\begin{equation}
{\cal A} = \frac{\sigma_\parallel - \sigma_\perp}{\sigma_\parallel +
\sigma_\perp} ~,
\end{equation}
es o positiva o negativa.

\vspace*{2mm} Observaciones excitantes en el sector de Higgs en
colisionadores de fotones son predichas en extensiones que violan
$\mathcal{CP}$.  Particularmente en el r\'egimen de desacoplamiento la
cuasi-degeneraci\'on de los estados $\mathcal{CP}$-par
$\mathcal{CP}$-impar da lugar a efectos potencialmente grandes,
rotando los eigenestados de corriente $ H,A $ a los eigenestados de
masa $ H_2,H_3 $.  En esta configuraci\'on grandes asimetr\'\i as,
\begin{equation}
{\cal A} = \frac{\sigma_{++} - \sigma_{--}}{\sigma_{++} +
\sigma_{--}} ~,
\end{equation}
pueden ser generadas en formaci\'on de un s\'olo Higgs  $\gamma \gamma$ entre
haces $\gamma$ con polarizaci\'on circular derecha e izquierda \cite{CKZ}.

\vspace*{4mm}
\subsection{Extensiones Supersim\'etricas No-M\'{\i}nimas}

\vspace*{2mm}
\noindent
{\bf 1.$\,$}  La extensi\'on supersim\'etrica m\'{\i}nima  del
Modelo Est\'andar puede parecer muy restrictiva para
teor\'{\i}as supersim\'etricas en general, en particular en el sector de
Higgs donde los acoplamientos cu\'articos se identifican con los
acoplamientos de norma. Sin embargo, resulta que el patr\'on de
masas del MSSM es bastante representativo si la teor\'{\i}a se
considera v\'alida hasta la escala de GUT -- la motivaci\'on para
supersimetr\'{\i}a {\it sui generis}. Este patr\'on general ha
sido estudiado concienzudamente dentro de la extensi\'on que sigue
a la m\'{\i}nima (next-to-minimal extensi\'on): el MSSM, que
incorpora dos iso-dobletes de Higgs, se extiende al introducir un
campo isosinglete adicional $N$. Esta extensi\'on lleva a un
modelo \citer{621,70A} que es generalmente referido como NMSSM.

El singlete de Higgs adicional puede resolver el llamado problema-$\mu$
($\mu$-problem) [i.e. $\mu \sim$ orden de $M_W$] al eliminar el par\'ametro
del higgsino $\mu$ del potencial y reemplazarlo por el valor esperado
(expectation value) del vac\'{\i}o del campo $N$, el cual
puede ser naturalmente relacionado con los valores esperados del
vac\'{\i}o usuales de los campos iso-dobletes de Higgs. En este
escenario el superpotencial involucra a los dos acoplamientos
trilineales $H_1 H_2 N$ y $N^3$.  Las consecuencias de este sector de
Higgs extendido se destacar\'an en el contexto de gran unificaci\'on,
incluyendo los t\'erminos de rompimiento suave universales de la
supersimetr\'{\i}a \cite{622,70A}.

\begin{figure}[hbt]
\begin{center}
\hspace*{-0.3cm}
\epsfig{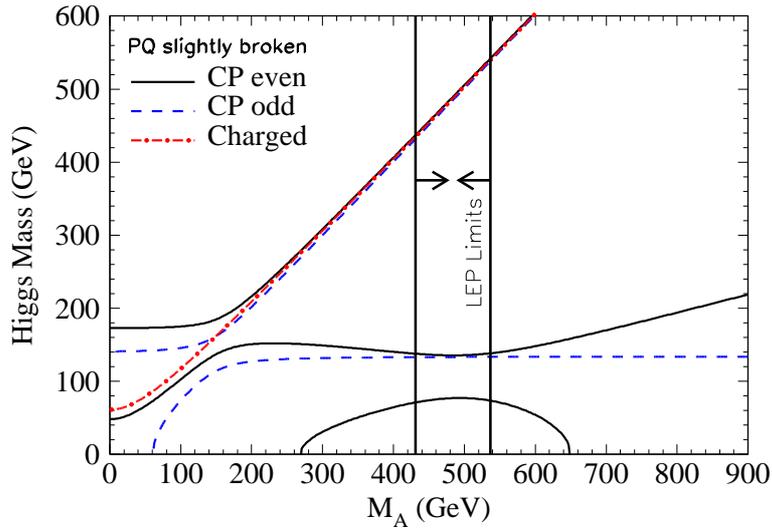}
\end{center}
\vspace*{-0.4cm}
\caption[]{\label{esp-fig:26} \it  Las masas del
bos\'on de Higgs a un lazo como funci\'on de $M_A$ para
$\lambda=0.3$, $\kappa=0.1$, $v_s=3v$, $\tan\beta=3$ y
$A_\kappa=-100$~GeV. Las flechas denotan la region permitida por
las b\'usquedas del LEP con un 95\% de confianza. Ref. \cite{70A}.}
\end{figure}

\vspace*{2mm}
 El espectro de Higgs del NMSSM incluye, adem\'as del conjunto
m\'{\i}nimo de part\'{\i}culas de Higgs, una part\'{\i}cula
adicional de Higgs escalar y una pseudoescalar:
\begin{eqnarray*}
  neutral\;CP=+              & &  H_1,\, H_2,\, H_3     \\ \non
  neutral\;CP=-              & &  A_1,\, A_2            \\ \non
  cargado{\hspace*{14.5mm}}  & &  H^\pm                 \, .
\end{eqnarray*}
Las
part\'{\i}culas de Higgs neutras son en general mezclas de
iso-dobletes que se acoplan a los bosones $W, Z$ y a los fermiones;
y del isosinglete, desacoplado del sector que no es de Higgs.
Las auto-interacciones trilineales contribuyen a las masas de las
part\'{\i}culas de Higgs; para el bos\'on de Higgs m\'as ligero de
cada especie:
\begin{eqnarray}
M^2 (H_1) & \leq & M^2_Z \cos^2 2\beta + \lambda^2 v^2 \sin^2 2 \beta \\
M^2 (A_1) & \leq & M^2 (A)   \nonumber \\
M^2 (H^{\pm}) & \leq & M^2 (W) + M^2 (A) - \lambda^2 v^2 \nonumber \;.
\end{eqnarray}
En contraste con el modelo m\'{\i}nimo, la masa de la
part\'{\i}cula de Higgs cargada podr\'{\i}a ser m\'as peque\~na
que la masa del \W. Un ejemplo del espectro de masas se muestra en
la Fig.~\ref{esp-fig:26}.  Dado que los acoplamientos
trilineales aumentan con la energ\'{\i}a, se pueden derivar cotas
superiores para la masa del bos\'on de Higgs m\'as ligero,
$h_1^0$, en analog\'{\i}a con el Modelo Est\'andar, y bajo la
suposici\'on de que la teor\'{\i}a es v\'alida hasta la escala de GUT:
$m(H_1) \lessim 140 $~GeV. As\'{\i}, a pesar de las
interacciones adicionales, el patr\'on distintivo de la extensi\'on
m\'{\i}nima permanece v\'alido incluso en escenarios
supersim\'etricos m\'as complicados. Si $H_1$ is (casi) puramente
isosinglete, se desacopla del sistema de bos\'on de norma y
fermiones y su papel lo toma la part\'{\i}cula de
Higgs siguiente con una componente de iso-doblete grande,
implicando, nuevamente, la
validez de la cota de la masa.
\\

Para un Higgs primordialmente isosinglete  $H_1$ el acoplamiento
$ZZH_1$ es peque\~no y la part\'{\i}cula
no puede producirse por Higgs-strahlung. Sin embargo, en este caso
$H_2$ es generalmente ligero y se
acopla con suficiente intensidad al bos\'on $Z$; de otra manera
$H_3$ juega este papel.

\vspace*{5mm} \noindent {\bf 2.$\,$} Una gran variedad de otras
extensiones m\'as all\'a del modelo supersim\'etrico m\'\i nimo han
sido a\-na\-li\-za\-das te\'oricamente.  Por ejemplo, si el sector
de bosones de norma se extiende con una simetr\'\i a Abeliana
adicional $U(1)'$ a altas energ\'\i as \cite{CHKZ,BL}, el campo
pseudoescalar de Higgs adicional es absorbido para generar la masa
del nuevo bos\'on $Z'$, mientras que la parte escalar del campo de
Higgs se puede observar como un nuevo bos\'on de Higgs adem\'as de
los del sector del MSSM. Si es generado por una simetr\'\i a
extendida como $E_6$, el sector de Higgs se expande en un conjunto
de estados nuevos \cite{E6} con propiedades bastante poco
convencionales.

\vspace*{2mm}
De manera bastante general, siempre y cuando los campos en las teor\'\i as
supersim\'etricas permanezcan d\'ebilmente interactuantes hasta la escala
de Planck can\'onica, la masa de los bosones de Higgs m\'as ligeros est\'a
acotada alrededor de 200 GeV, ya que los acoplamientos de Yukawa est\'an
restringidos a ser peque\~nos de la misma manera que el acoplamienteo
cu\'artico en el potencial de Higgs est\'andar. M\'as a\'un, el l\'\i mite de
140~GeV para la part\'\i cula de Higgs m\'as ligera se realiza en casi todas
las teor\'\i as supersim\'etricas \cite{623};
cf. Ref.~\cite{EspinQuiros} para expansiones de este l\'\i mite.
 Experimentos en colisionadores $e^+e^-$ no
est\'an en una situaci\'on de fracaso \cite{L141A} en la
detecci\'on de las part\'{\i}culas de Higgs en teor\'{\i}as
supersim\'etricas generales, incluso para energ\'{\i}as del c.m.
tan bajas como $\sqrt{s} \sim 500$ GeV.

\vspace*{4mm}
\section{Rompimiento Din\'amico de la Simetr\'{\i}a}

\vspace*{2mm}
\noindent El mecanismo de Higgs est\'a basado en el concepto te\'orico de
rompimiento espont\'aneo de la simetr\'\i a \cite{1}.  En la formulaci\'on
can\'onica, adoptada en el Modelo Est\'andar, un campo escalar {\it fundamental} de
cuatro-componentes es introducido, al cual se le dota con una
auto-interacci\'on tal que el campo adquiere un valor distinto de cero
en el estado base.  La direcci\'on espec\'\i fica en el iso-espacio, que
resulta elegida por la evoluci\'on del estado base, rompe la invariancia
del isosp\'\i n de la interacci\'on espont\'aneamente\footnote{Mantenemos
este lenguage usado comunmente en el contexto de teor\'\i as de norma, a
pesar de que la simetr\'\i a de norma no est\'a rota por el
mecanismo de Higgs en el sentido estricto.}  La interacci\'on de los campos de norma con el
campo escalar en el estado base genera las masas de estos campos.  Los
grados de libertad longitudinales de los campos de norma se construyen
por absorci\'on del los modos de Goldstone, los cuales est\'an asociados con el
rompimiento espont\'aneo de las simetr\'\i as electrod\'ebiles en el sector de
campo escalar.  Los fermiones adquieren masas a trav\'es de
interacciones de Yukawa con el campo en el estado base.  Mientras que
tres componentes escalares son absorbidas por los campos de norma, un
grado de libertad se manifiesta como una part\'\i cula f\'\i sica, el bos\'on de
Higgs.  El intercambio de esta part\'\i cula en amplitudes de dispersi\'on,
incluyendo  los campos de norma longitudinales y campos fermi\'onicos
masivos, garantiza la unitaridad de la teor\'\i a hasta energ\'\i as asint\'oticas.

\vspace*{2mm}
De forma alternativa, interpretar al bos\'on de Higgs como un bos\'on de
(pseudo-)Goldstone asociado con el rompimiento espont\'aneo de nuevas
interacciones fuertes ha sido una idea atractiva por mucho tiempo.

\vspace*{4mm}
\subsection{Modelos de Higgs Peque\~no}

\vspace*{2mm}
 El inter\'es en esta representaci\'on ha resurgido
dentro de escenarios de Higgs peque\~no \cite{2A} que
recientemente se han desarrollado para generar din\'amicamente el
rompimiento de la
simetr\'{\i}a electrod\'ebil por medio de nuevas interacciones fuertes.
Los modelos de Higgs peque\~no est\'an basados en un complejo
sistema de simetr\'{\i}as y mecanismos de rompimiento de
simetr\'{\i}a, para consultar una revisi\'on reciente ver
\cite{littlest}. Tres puntos son centrales para realizar
la idea:

\begin{itemize}
\item[(i)]  El campo de Higgs es un campo de Goldstone asociado con el
 rompimiento de una simetr\'{\i}a global $G$.  Las interacciones
 fuertes est\'an caracterizadas por una escala   $\Lambda \sim$ 10 a 30 TeV,
 mientras que la escala din\'amica de Goldstone se estima del orden de
$f \sim \Lambda / 4 \pi \sim$ 1 a 3 TeV;
\item[(ii)] En el mismo paso, la simetr\'{\i}a de norma $G_0 \subset G$ se
  rompe al  grupo de norma del Modelo Est\'andar $SU(2)
\times U(1)$, generando las masasa para los bosones vectoriales pesados y
los fermiones de tama\~no intermedio $M \sim g f \sim 1$ TeV;
\item[(iii)] Los bosones de Higgs adquieren finalmente masa por
  correcciones radiativas colectivas, i.e. a segundo orden, a la
  escala electrod\'ebil est\'andar  $v \sim g^2 f / 4 \pi \sim$ 100
  a 300 GeV.
\end{itemize}

As\'\i, en este modelo se encuentran tres escalas
caracter\'{\i}sticas: la escala de interacci\'on fuerte
$\Lambda_s$, la nueva escala de  masa $M$ y la escala de
rompimiento electrod\'ebil $v$, ordenadas en una cadena
jer\'arquica $\Lambda_s \gg M \gg v$. La masa del bos\'on de Higgs
ligero est\'a protegida a un valor peque\~no al requerir el
rompimiento colectivo de  dos simetr\'{\i}as. En contraste a la
simetr\'{\i}a bos\'on-fermi\'on que cancela las divergencias
cuadr\'aticas en supersimetr\'{\i}a, en los modelos de Higgs
peque\~no la cancelaci\'on opera individualmente en los sectores
bos\'onico y fermi\'onico, siendo asegurada por
las simetr\'{\i}as entre los acoplamientos de los campos
del SM y los nuevos campos, $\mathcal{O}$$(M)$, con los campos de Higgs.

\vspace*{4mm}
\noindent
{\it{\underline{Ejemplo: Modelo del Higgs M\'as Peque\~no}}}

\vspace*{2mm}
\noindent
 Un ejemplo interesante  en donde estos conceptos se llevan a
cabo es proporcionado por el ``Modelo del Higgs M\'as Peque\~no''
\cite{91A,91B}. El modelo est\'a formulado como un modelo sigma
no lineal con un grupo de simetr\'{\i}a global $SU(5)$. Este grupo
se rompe a $SO(5)$  por el valor esperado del
vac\'{\i}o distinto de cero
\beq
\Sigma_0 = crossdiag\,[{\scriptstyle '}\!\mathbb{I},1,
{\scriptstyle '}\!\mathbb{I}]
\eeq
del campo $\Sigma$.  Suponiendo que el subgrupo $[SU(2) \times U(1)]^2$  sea
normado (gauged),  el
rompimiento de la simetr\'{\i}a global lleva tambi\'en al
rompimiento de este grupo de norma al grupo
$[SU(2) \times U(1)]$. El rompimiento de la simetr\'{\i}a global
genera $24 - 10 = 14$ bosones de Goldstone, cuatro de los cuales
son absorbidos por los bosones de norma asociados con el grupo de
norma roto. Los 10 bosones de Goldstone restantes, incorporados
en el campo $\Sigma$
\beq
\Sigma = \exp[2i\Pi/f]: \quad \Pi = \left|\left|
\begin{array}{ccc} 0 & h^\dagger/\sqrt{2} &
\varphi^\dagger \\
h/\sqrt{2} & 0 & h^*/\sqrt{2} \\
\varphi & h^{\mathrm{T}}/\sqrt{2} & 0
\end{array} \right|\right|
\eeq
son identificados con un iso-doblete $h$ que se convierte
en el campo de Higgs ligero del Modelo Est\'andar y con un triplete de
Higgs $\varphi$ que adquiere una masa del orden de $M$.

\vspace*{2mm}
Los principios de construcci\'on fundamentales del modelo
deben ser ilustrados analizando cualitativamente los sectores
de norma y de Higgs. El sector del top, extendido por un nuevo
doblete pesado $[T_L, T_R]$, puede tratarse de un modo similar
despu\'es de introducir las interacciones apropiadas top-Higgs.\

\vspace*{4mm}
\noindent
{\it Sector del Bos\'on Vectorial:}

\vspace*{2mm}
\noindent
Al insertar los campos de norma $[SU(2) \times U(1)]^2$ en la
Lagrangiana de sigma,
\beq
{\cal L} = \frac{1}{2} \frac{f^2}{4} \mathrm{Tr} | {\cal D}_\mu \Sigma |^2
\eeq
con
\beq
{\cal D}_\mu \Sigma = \partial_\mu \Sigma - i \sum_{j=1}^2
[ g_j (W_j \Sigma + \Sigma W_j^\mathrm{T}) + \{U(1)\} ] \,,
\eeq
los cuatro bosones vectoriales de la simetr\'{\i}a de norma rota
$[SU(2) \times U(1)]$ adquieren masas
\beq
M[W_H,Z_H,A_H] \sim g f
\eeq
donde $W_H$ etc., denotan los campos de norma electrod\'ebil pesados.

\vspace*{2mm}
Notablemente, los bosones de norma  $W_H$ se acoplan con el signo
opuesto  al cuadrado del bos\'on de Higgs ligero comparado
con los bosones est\'andar $W$
\beq
{\cal L} &=& + \frac{g^2}{4} W^2 \,\mathrm{Tr} h^\dagger h \nonumber \\
&& - \frac{g^2}{4} W_H^2 \,\mathrm{Tr} h^\dagger h + ... \; .
\eeq
Es por esto que las divergencias cuadr\'aticas de los dos
diagramas de lazo cerrados W y W' que acompa\~nan al campo de Higgs
ligero, se cancelan entre ellas y, de manera similar a los grados de
libertad supersim\'etricos, los bosones vectoriales nuevos
deber\'an tener masas que no excedan de 1 a 3 TeV para evitar
ajustes finos excesivos.

\vspace*{2mm}
Los bosones de norma del Modelo Est\'andar se matienen
sin masa hasta este punto; adquieren masa despu\'es de que el mecanismo de
rompimiento electrod\'ebil est\'andar se pone en operaci\'on.

\vspace*{4mm}
\noindent
{\it Sector de Higgs:}

\vspace*{2mm}
\noindent
Hasta este nivel en la evoluci\'on de la teor\'{\i}a,
las simetr\'{\i}as globales impiden un potencial de Higgs
diferente de cero. S\'olo si las correcciones radiativas  se ponen
de manifiesto, entonces el mecanismo de Coleman-Weinberg genera el
potencial de Higgs que proporciona masas a los bosones de Higgs y
rompe la simetr\'{\i}a de norma del Modelo Est\'andar.

\vspace*{2mm}
D\'andole al potencial de Higgs la forma
\beq
V = m_\varphi^2 \,\mathrm{Tr} \varphi^\dagger \varphi - \mu^2 h h^\dagger +
\lambda_4 (h h^\dagger)^2
\eeq
el primer t\'ermino proporciona una
masa distinta de cero al bos\'on de Higgs $\varphi$, mientras que los
dos siguientes t\'erminos son responsables del rompimiento de la
simetr\'{\i}a en el sector de norma
del Modelo Est\'andar.

\vspace*{2mm}
\noindent
 -- Truncando las contribuciones al potencial de
Coleman-Weinberg cuadr\'aticamente divergentes en $\Lambda$, las
masas al cuadardo de los [ahora] bosones de pseudo-Goldstone
$\varphi$ son del orden de
\beq
m_\varphi^2 \sim g^2 (\Lambda / 4 \pi )^2 \sim g^2 f^2 \;.
\eeq
As\'\i{} que los bosones de Higgs
pesados adquieren masas del orden de la masa
de los bosones vectoriales pesados.

\vspace*{2mm}
\noindent
 -- El acoplamiento cu\'artico  del bos\'on de Higgs
ligero es del orden de $g^2$. Sin embargo, el coeficiente $\mu^2$
recibe \'unicamente contribuciones de las partes de un-lazo
logar\'{\i}tmicamente divergente y de dos-lazos cuadr\'aticamente
divergentes en el potencial de Coleman-Weinberg:
\beq
\mu^2 = \mu_1^2 + \mu_2^2 : && \mu_1^2 \sim (\Lambda/4 \pi)^2 \log
\left(\Lambda^2/f^2\right)/16\pi^2
\sim f^2 \log \left(\Lambda^2/f^2\right)/16\pi^2 \nonumber\\[0.3cm]
&& \mu_2^2 \sim \Lambda^2/(16\pi^2)^2 \sim f^2/16\pi^2 \;.
\eeq
Ambas contribuciones son naturalmente del orden de $f/4\pi$, i.e.
son  un orden de magnitud menor que la escala intermedia $M$ del
Higgs pesado y las masas vectoriales.
 \\

As\'\i,  un bos\'on de Higgs ligero con masa del orden de 100
GeV puede ser generado en modelos de Higgs Peque\~no como un
bos\'on pseudo-Goldstone asociado con el rompimiento espont\'aneo de
nuevas interacciones fuertes. La masa peque\~na est\'a protegida
contra correcciones radiativas grandes individualmente en los
sectores bos\'onico y fermi\'onico.

\vspace*{4mm}
\noindent
{\it{\underline{Fenomenolog\'{\i}a}}}

\vspace*{2mm}
\noindent
 De tales escenarios surgen muchas predicciones que
pueden verificarse experimentalmente. Lo m\'as importante, el espectro de los nuevos bosones vectoriales
pesados y fermiones deber\'an observarse con masas en el rango
intermedio de 1 a algunos TeV en el LHC o en colisionadores
lineales $e^+e^-$ TeV/multi-TeV.
Extensiones que van m\'as all\'a de la versi\'on m\'\i nima pueden generar
escalares adicionales con un fuerte impacto tambi\'en en el espectro del
sector de Higgs ligero.

\begin{figure}[hbt]
\begin{center}
\epsfig{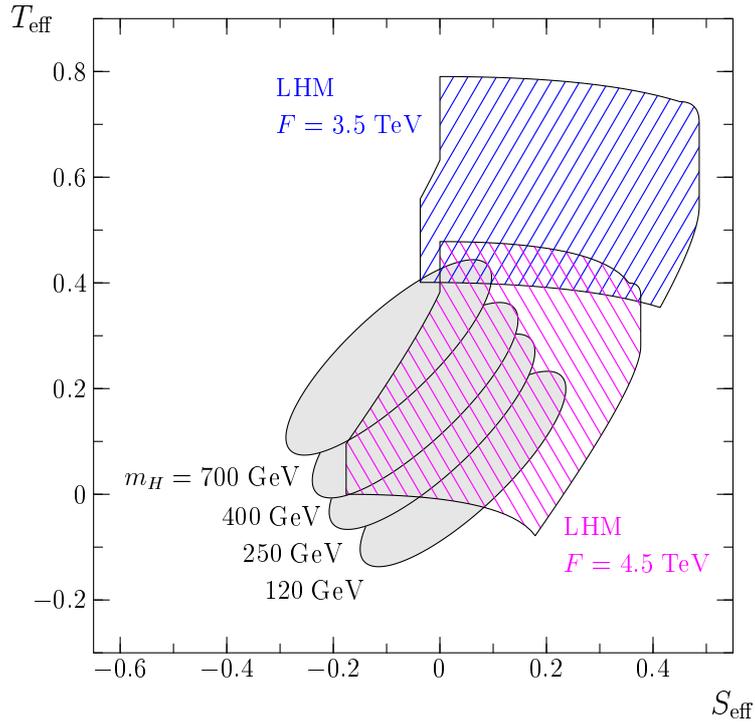}
\end{center}
\vspace*{-0.4cm}

\caption[]{\label{esp-fig:kilian} \it  Predicciones  de los
par\'ametros de precisi\'on $S,T$ para el modelo del Higgs M\'as
Peque\~no con asignaciones de carga est\'andares $U(1)$. Las
elipses sombreadas son el 68 \% de los contornos de exclusi\'on
los cuales provienen de los datos de precisi\'on electrod\'ebiles,
suponiendo cuatro masas diferentes de los Higgs. Las \'areas
rayadas son los rangos de los par\'ametros permitidos del modelo
del Higgs M\'as Peque\~no para dos valores diferentes de la escala
$F$. Los l\'{\i}mites de las interacciones de  contacto se han
tomado en cuenta.
Ref. \cite{kilfigure}.}
\end{figure}

\vspace*{2mm}
Sin embargo, el modelo puede ya ser verificado analizando los
datos de precisi\'on existentes extraidos del LEP y otros. El
impacto de los nuevos grados de libertad en los modelos de Higgs
Peque\~no debe mantenerse lo suficientemente reducido para no
arruinar el \'exito de las correcciones radiativas al incluir
s\'olo el bos\'on de Higgs ligero en la descripci\'on de los
datos.   Esto lleva a restringir el par\'ametro $f$ al orden de 3
a 5 TeV, Fig.~\ref{esp-fig:kilian}.  As\'{\i} la teor\'{\i}a es
compatible con los datos de precisi\'on actuales, pero s\'olo
marginalmente y con poca superposici\'on.

Extensiones del sistema en las cuales una nueva paridad \cite{Tpar} es
introducida, reducen las correcciones radiativas a lazos cerrados de
los nuevos grados de libertad, de tal manera que part\'\i culas ligeras nuevas
no est\'an excluidas.  Sin embargo, anomal\'\i as potenciales
\cite{HillHill} que rompen la invariancia bajo $T$ deben ser evitadas al
dise\~nar tales escenarios.

\vspace*{4mm}
\subsection{Bosones $W$ Fuertemente Interactuantes}

\vspace*{2mm}
\noindent
En escenarios alternativos \cite{2} un sistema de fermiones nuevos es
introducido,  el cual interact\'ua fuertemente a la
escala del orden de 1 TeV. En el estado base de dicho sistema un
condensado escalar de pares fermi\'on-antifermi\'on podr\'{\i}a
formarse. Generalmente se espera que tal proceso se lleve a cabo
en cualquier teor\'{\i}a de norma no Abeliana de las nuevas
interacciones fuertes [como en QCD].  Como el
condensado escalar rompe  la simetr\'{\i}a quiral del sistema
fermi\'onico, se formar\'an campos de Goldstone, y estos
pueden ser absorbidos por los campos de norma electrod\'ebiles
para construir las componentes longitudinales y las masas de los
campos de norma.
 Nuevas interacciones de norma deben ser
introducidas, las cuales acoplan los leptones y los quarks del
Modelo Est\'andar a los nuevos fermiones de tal manera que se
generen las masas de los leptones y quarks a trav\'es de las
interacciones con el estado base del condenesado
fermi\'on-antifermi\'on. En el sector de bajas energ\'{\i}as de la
teor\'{\i}a electrod\'ebil, la aproximaci\'on del campo de Higgs
fundamental y la alternativa din\'amica son equivalentes. Sin
embargo, las dos teor\'{\i}as son fundamentalmente diferentes a
energ\'{\i}as altas. Mientras que la unitaridad  de la
teor\'{\i}a de norma electrod\'ebil se garantiza por el
intercambio de un part\'{\i}cula de Higgs escalar en procesos de
dispersi\'on, la unitaridad se restablece a altas energ\'{\i}as en
la teor\'{\i}a din\'amica a trav\'es de interacciones fuertes no
perturbativas entre las part\'{\i}culas. Como las componentes de
los campos de norma longitudinales son equivalentes a los campos
de Goldstone asociados con la teor\'{\i}a microsc\'opica, sus
interacciones fuertes a altas energ\'{\i}as son transferidas a los
bosones de norma electrod\'ebiles. Puesto que por unitaridad, la
amplitud de dispersi\'on de la onda $S$ de los bosones $W, Z$
polarizados longitudinalmente en el canal iso-escalar $(2W^+W^- +
ZZ) / \sqrt{3}$, $a^0_0 = \sqrt{2} G_F s/ 16 \pi$, est\'a acotada
por 1/2, la escala caracter\'{\i}stica de las nuevas interacciones
fuertes debe estar cercana a 1.2 TeV. Entonces, cerca de la
energ\'{\i}a cr\'{\i}tica de 1 TeV, los bosones $W,Z$ interact\'uan
fuertemente entre ellos. Las teor\'{\i}as tecnicolor proporcionan
una forma elaborada de dichos escenarios.

\vspace*{4mm}
\subsubsection{Bases Te\'oricas}

\vspace*{2mm}
\noindent
{\bf 1.$\,$} Los escenarios f\'{\i}sicos del rompimiento din\'amico de la
simetr\'{\i}a pueden estar basados en teor\'{\i}as de nueva
interacci\'on fuerte, las cuales extienden el espectro de
part\'{\i}culas de materia y de las interacciones m\'as all\'a de
los grados de libertad tomados en cuenta en el Modelo Est\'andar.
Si las nuevas interacciones fuertes son invariantes bajo
transformaciones de un grupo de simetr\'{\i}a quiral $SU(2) \times
SU(2)$, la invariancia quiral generalmente es rota
espont\'aneamente al grupo de iso-esp\'{\i}n custodial diagonal
$SU(2)$. Este proceso est\'a asociado con la formaci\'on de un
condensado quiral en el estado base y con la existencia  de tres
bosones de Goldstone sin masa.
 \\[-1mm]

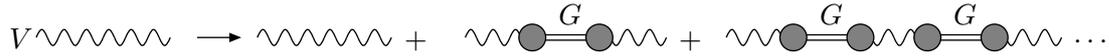
\begin{figure}[hbt]
\begin{center}
\begin{picture}(60,10)(90,30)
\Photon(0,25)(50,25){3}{6}
\LongArrow(60,25)(75,25)
\put(-10,21){$V$}
\end{picture}
\begin{picture}(60,10)(70,30)
\Photon(0,25)(50,25){3}{6}
\put(55,21){$+$}
\end{picture}
\begin{picture}(60,10)(55,30)
\Photon(0,25)(25,25){3}{3}
\Photon(50,25)(75,25){3}{3}
\Line(25,24)(50,24)
\Line(25,26)(50,26)
\GCirc(25,25){5}{0.5}
\GCirc(50,25){5}{0.5}
\put(80,21){$+$}
\put(35,30){$G$}
\end{picture}
\begin{picture}(60,10)(20,30)
\Photon(0,25)(25,25){3}{3}
\Photon(50,25)(75,25){3}{3}
\Photon(100,25)(125,25){3}{3}
\Line(25,24)(50,24)
\Line(25,26)(50,26)
\Line(75,24)(100,24)
\Line(75,26)(100,26)
\GCirc(25,25){5}{0.5}
\GCirc(50,25){5}{0.5}
\GCirc(75,25){5}{0.5}
\GCirc(100,25){5}{0.5}
\put(130,21){$\cdots$}
\put(35,30){$G$}
\put(85,30){$G$}
\end{picture}
\end{center}
\caption[]{\label{esp-fg:gaugemass} \it Generando masas de los bosones
de norma (V) a trav\'es de la interacci\'on con los bosones de
Goldstone (G).}
\end{figure}

\vspace*{1mm}
Los bosones de Goldstone pueden ser absorbidos por los campos de
norma, generando estados longitudinales y masas distintas de cero
para los bosones de norma, como se muestra en la
Fig.~\ref{esp-fg:gaugemass}. Sumando las series geom\'etricas de las
transiciones bos\'on de Goldstone-bos\'on vectorial en el
propagador, lleva a un cambio en el polo de la masa:
\begin{eqnarray}
\frac{1}{q^2} & \to & \frac{1}{q^2} + \frac{1}{q^2} q_\mu \frac{g^2 F^2/2}{q^2}
q_\mu \frac{1}{q^2} + \frac{1}{q^2} \left[ \frac{g^2 F^2}{2} \frac{1}{q^2}
\right]^2 + \cdots \nonumber \\
& \to & \frac{1}{q^2-M^2} \;.
\end{eqnarray}
El acoplamiento entre bosones de norma y bosones de Goldstone ha sido
definido como $ig F/\sqrt{2} q_\mu$. La masa generada por los
campos de norma est\'a relacionada a este acoplamiento por
\begin{equation}
M^2 = \frac{1}{2} g^2 F^2 ~.
\end{equation}
El valor num\'erico del acoplamiento $F$ debe coincidir con
$v/\sqrt{2}=174$ GeV.

\vspace*{2mm}
La simetr\'{\i}a custodial $SU(2)$ remanente garantiza que el
par\'ametro $\rho$, que es la intensidad relativa (al cuadrado) entre los
acoplamientos de las corrientes neutras y cargadas, sea uno.
Si se denota a  los elementos de la
matriz de masa $W/B$ por
\begin{equation}
\begin{array}{rclcrcl}
\langle W^i | {\cal M}^2 | W^j \rangle & = & \displaystyle \frac{1}{2} g^2
F^2 \delta_{ij}
& \hspace*{1cm} & \langle W^3 | {\cal M}^2 | B \rangle & = & \langle B |
{\cal M}^2 | W^3 \rangle \\ \\
\langle B | {\cal M}^2 | B \rangle & = & \displaystyle \frac{1}{2} g'^2 F^2 &
& & = & \displaystyle \frac{1}{2} gg' F^2
\end{array}
\end{equation}
la universalidad del acoplamiento $F$ conduce al cociente
  $M_W^2/M_Z^2 = g^2/(g^2+g'^2) = \cos^2\theta_W$
eigenvalores de masa, equivalente a $\rho=1$.

\vspace*{2mm}
\noindent
{\bf 2.$\,$} Puesto que las funciones de onda de los bosones vectoriales
polarizados longitudinalmente aumentan con la energ\'{\i}a, las
componentes del campo longitudinales son los grados de libertad
dominantes a altas energ\'{\i}as. Estos estados pueden, sin
embargo, para energ\'{\i}as asint\'oticas identificarse con los
bosones de Goldstone absorbidos. Esta equivalencia \cite{75} es
aparente en la norma de 't Hooft--Feynman donde, para
energ\'{\i}as asint\'oticas,
\begin{equation}
\epsilon_\mu^L W_\mu \to k_\mu W_\mu \sim M^2 \Phi ~.
\end{equation}
Por consiguiente, la din\'amica de los bosones de norma puede ser
identificada a altas energ\'\i as con la din\'amica de los
campos de Goldstone escalares. Una representaci\'on elegante de
los campos de Goldstone $\vec{G}$ en este contexto est\'a dada por
la forma exponencial
\begin{equation}
U = \exp [-i \vec{G} \vec{\tau}/v ] ~,
\end{equation}
la cual corresponde a un campo matricial $SU(2)$.

\vspace*{2mm}
El Lagrangiano de un sistema de bosones fuertemente interactuantes
consiste en tal escenario de la parte de Yang-Mills  ${\cal
L}_{YM}$ y de las interacciones de los campos de Goldstone ${\cal
L}_G$,
\begin{equation}
{\cal L}={\cal L}_{YM}+{\cal L}_G  \;.
\end{equation}
La parte de Yang--Mills es escrita en la forma usual
 ${\cal L}_{YM} = -\frac{1}{4} {\rm Tr} [W_{\mu\nu} W_{\mu\nu} +
B_{\mu\nu} B_{\mu\nu} ]$.
 La interacci\'on de los campos de
Goldstone pueden ser sistem\'aticamente expandida en teor\'{\i}as
quirales en las derivadas de los campos, correspondiendo a
expansiones en potencias de la energ\'{\i}a para amplitudes de
dispersi\'on \cite{76}:
\begin{equation}
{\cal L}_G = {\cal L}_0 + \sum_{dim=4} {\cal L}_i + \cdots \;.
\end{equation}
Denotando la derivada covariante del SM para los campos de
Goldstone por
\begin{equation}
D_\mu U = \partial_\mu U - i g W_\mu U + i g' B_\mu U
\end{equation}
el t\'ermino principal ${\cal L}_0$, el cual es de dimensi\'on = 2,
est\'a dado por
\begin{equation}
{\cal L}_0 = \frac{v^2}{4} {\rm Tr} [ D_\mu U^+ D_\mu U ] \;.
\end{equation}
Este t\'ermino genera las masas de los bosones de norma $W,Z$:
 $M_W^2 = \frac{1}{4} g^2 v^2$ y
$M_Z^2 = \frac{1}{4} (g^2+g'^2) v^2$.
 El \'unico par\'ametro en esta parte de la interacci\'on es
$v$, el cual sin embargo, se fija de manera \'unica por el valor
experimental de la masa del $W$; de modo que las amplitudes
predichas por el t\'ermino principal en la expansi\'on quiral
pueden efectivamente ser
consideradas como libres de par\'ametros.

\vspace*{2mm}

 La componente al siguiente orden
en la expansi\'on con dimensi\'on = 4 consiste de diez t\'erminos
individuales. Si la simetr\'{\i}a custodial $SU(2)$ se impone,
s\'olo quedan dos t\'erminos, los cuales no afectan a los
propagadores ni a los v\'ertices de 3 bosones, pero s\'{\i} a los
de 4 bosones. Introduciendo el campo vectorial $V_\mu$ como
\begin{equation}
V_\mu = U^+ D_\mu U
\end{equation}
estos dos t\'erminos est\'an dados por las densidades de
interacci\'on
\begin{equation}
{\cal L}_4  =  \alpha_4 \left[Tr V_\mu V_\nu \right]^2 \hspace*{0.5cm}
\mbox{y} \hspace*{0.5cm}
{\cal L}_5  =  \alpha_5 \left[Tr V_\mu V_\mu \right]^2 \;.
\end{equation}

Los dos coeficientes
$\alpha_4,\alpha_5$ son par\'ametros que caracterizan  a la teor\'\i a microsc\'opica
subyacente. En enfoques fenomenol\'ogicos deben ser ajustados
experimentalmente de los datos de la dispersi\'on  $WW$.

Ordenes mayores en la expansi\'on quiral dan lugar a una
expansi\'on de la energ\'{\i}a de las amplitudes de dispersi\'on
de la forma ${\cal A} = \sum c_n (s/v^2)^n$. Esta serie
divergir\'a a energ\'{\i}as para las cuales las resonancias de las
teor\'{\i}as de nuevas interacciones fuertes pueden formarse en
colisiones de $WW$: $0^+$ `tipo-Higgs', resonancias $1^-$
`tipo-$\rho$', etc. Las masas de estos estados de resonancia se
esperan en el rango $M_R \sim 4\pi v$ donde las expansiones de lazo
quiral divergen, i.e. aproximadamente entre 1 y 3 TeV's.

\vspace*{4mm}
\subsubsection{Un Ejemplo: Teor\'{\i}as Tecnicolor}

\vspace*{2mm}
Un ejemplo simple para tales escenarios es proporcionado por las
teor\'{\i}as de tipo tecnicolor, ver e.g. Ref.~\cite{94A}. \'Estas son
construidas con patrones similares a QCD pero caracterizadas por
una escala $\Lambda_{TC}$ en el rango de TeV, de manera que la
interacci\'on se vuelve fuerte ya a cortas distancias, del orden de $10^{-17}$~cm.

\vspace*{2mm} La simetr\'{\i}a quiral $SU(2)_L\times SU(2)_R$ de esta
teor\'{\i}a se rompe a la simetr\'{\i}a vectorial diagonal
$SU(2)_{L+R}$ por la formacion de condensados de vac\'{\i}o $\langle\bar{U}
U\rangle =\langle\bar{D} D\rangle = {\cal O}(\Lambda^3_{TC})$.  El rompimiento de la
simetr\'{\i}a quiral genera tres bosones de Goldstone sin masa $\sim
\bar{Q} i \gamma_5 \stackrel{\to}{\tau} Q$, que pueden ser absorbidos por
campos de norma del Modelo Est\'andar para construir los estados masivos
con $M_W \sim 100$~GeV. De la cadena
\beq
M_W = g F /\sqrt{2} \quad \mathrm{y} \quad F \sim \Lambda_{TC} / 4 \pi
\eeq
el par\'ametro $F$ es estimado abajo de 1 Tev, mientras que $\Lambda_{TC}$
debe estar en el rango de TeV.

\vspace*{2mm}

Mientras que el sector de norma electrod\'ebil puede ser formulado
consistentemente en esta marco, generar las masas de los fermiones
conduce a dificultades severas. Como las interacciones de norma
acoplan s\'olo las componentes del campo izquierdo-izquierdo y
derecho-derecho, un operador de masa de cambio de helicidad
izquierdo-derecho $\bar{f}_L f_R$  no es generado para los fermiones
del Modelo Est\'andar. Para resolver este problema, se deben introducir
nuevas interacciones entre fermiones del SM y fermiones del TC
[Tecnicolor Extendido ETC] de manera que la helicidad pueda cambiar a
trav\'es del condensado de ETC en el vac\'{\i}o.
 Las masas del SM
predichas de esta manera son del orden $m_f \sim g^2_E \Lambda^3_{ETC}/M_E^2$,
siendo $g_E$ el acoplamiento de la teor\'{\i}a de norma de Tecnicolor
extendida y $M_E$ la masa de los campos de norma del ETC. Sin
embargo, estimaciones de $M_E$ llevan a conflictos si uno intenta
reconciliar el tama\~no de la escala requerida para generar la masa del
top, del orden de TeV, con la supresi\'on de procesos de cambio de sabor, como
las oscilaciones $K\bar{K}$,
las cuales requieren un tama\~no del orden de PeV.

\vspace*{2mm}
De este modo, la realizaci\'on m\'as simple  de las teor\'{\i}as de
tecnicolor sufre de  conflictos internos en el sector
fermi\'onico. Modelos te\'oricos m\'as complicados son necesarios
para reconciliar estas estimaciones conflictivas \cite{94A}. No
obstante, la idea de generar el rompimiento de la simetr\'{\i}a
electrod\'ebil din\'amicamente es un escenario  te\'oricamente
atractivo e interesante.

\vspace*{4mm}
\subsection{Dispersi\'on $WW$ en colisionadores de Alta Energ\'{\i}a }

\vspace*{2mm}
\noindent
{\bf 1.$\,$}  Independientemente de la forma espec\'{\i}fica en que
se realiza el rompimiento din\'amico de la simetr\'{\i}a, se han
desarrollado herramientas te\'oricas que pueden ser
\'utiles para investigar  estos escenarios de una manera
bastante general. Las amplitudes de dispersi\'on (cuasi-)
el\'asticas de 2--2 $WW$ pueden expresarse a altas energ\'{\i}as
por una amplitud maestra $A(s,t,u)$, la cual depende de las tres
variables de Mandelstam de los procesos de dispersi\'on:
\begin{eqnarray}
A(W^+ W^- \to ZZ) & = & A(s,t,u) \\
A(W^+ W^- \to W^+ W^-) & = & A(s,t,u) + A(t,s,u) \nonumber \\
A(ZZ \to ZZ) & = & A(s,t,u) + A(t,s,u) + A(u,s,t) \nonumber \\
A(W^- W^- \to W^- W^-) & = & A(t,s,u) + A(u,s,t)  \nonumber \;.
\end{eqnarray}

\vspace*{2mm}
Al orden m\'as bajo en la expansi\'on quiral, ${\cal L} \to {\cal
L}_{YM} + {\cal L}_0$, la amplitud maestra est\'a dada, en la
forma libre de par\'ametros, por la energ\'{\i}a al cuadrado $s$:
\begin{equation}
A(s,t,u) \to \frac{s}{v^2} ~.
\end{equation}
Esta representaci\'on es v\'alida para energ\'{\i}as $s \gg M_W^2$
pero abajo de la nueva regi\'on de resonancia, i.e. en la
pr\'actica, a energ\'{\i}as de $\sqrt{s}={\cal O}(1~\mbox{TeV})$.
Denotando la longitud de dispersi\'on para el canal que lleva
iso-esp\'{\i}n $I$ y momento angular $J$ por $a_{IJ}$, los \'unicos
canales de dispersi\'on distintos de cero predichos por el t\'ermino
principal de la expansi\'on quiral corresponden a
\begin{eqnarray}
a_{00} & = & + \frac{s}{16\pi v^2} \\
a_{11}   & = & + \frac{s}{96\pi v^2} \nonumber \\
a_{20}   & = & - \frac{s}{32\pi v^2} ~.
\end{eqnarray}
Mientras que el canal ex\'otico $I=2$ es repulsivo, los canales
$I=J=0$ y $I=J=1$ son atractivos, indicando la formaci\'on de
resonancias no-fundamentales tipo Higgs y tipo $\rho$.

Tomando en cuenta los t\'erminos  siguientes al dominante en
la expansi\'on quiral (next-to-leading terms), la amplitud maestra
resulta ser \cite{24}
\begin{equation}
A(s,t,u) = \frac{s}{v^2} + \alpha_4 \frac{4(t^2+u^2)}{v^4}
+ \alpha_5 \frac{8s^2}{v^4} + \cdots ~,
\end{equation}
incluyendo los dos par\'ametros $\alpha_4$ y $\alpha_5$.

\vspace*{2mm}
Al incrementar la energ\'{\i}a, las amplitudes se aproximar\'an al
\'area de resonancia. En esa \'area, el car\'acter quiral de la
teor\'{\i}a no proporciona m\'as gu\'\i as en
la construcci\'on de amplitudes de dispersi\'on. En su lugar, se
deben introducir hip\'otesis {\it ad-hoc} para definir la
naturaleza de las resonancias; ver e.g. Ref. \cite{24a}.
 \\[2mm]

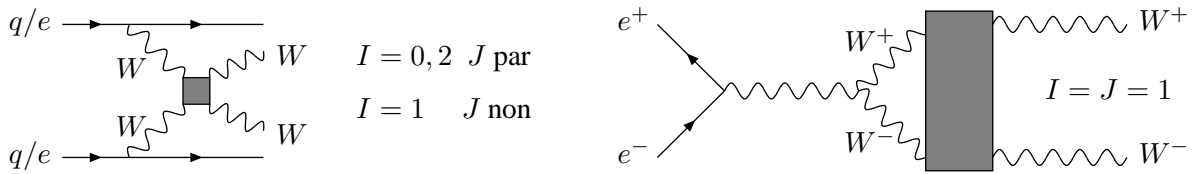
\begin{figure}[hbt]
\begin{center}
\begin{picture}(60,50)(140,0)
\ArrowLine(0,50)(25,50)
\ArrowLine(25,50)(75,50)
\ArrowLine(0,0)(25,0)
\ArrowLine(25,0)(75,0)
\Photon(25,50)(45,30){-3}{3}
\Photon(25,0)(45,20){3}{3}
\Photon(55,20)(75,10){-3}{3}
\Photon(55,30)(75,40){3}{3}
\GBox(45,20)(55,30){0.5}
\put(-20,48){$q/e$}
\put(-20,-2){$q/e$}
\put(20,8){$W$}
\put(20,30){$W$}
\put(80,35){$W$}
\put(80,5){$W$}
\put(110,35){$I=0,2~~J~\mbox{par}$}
\put(110,15){$I=1~~~~~J~\mbox{non}$}
\end{picture}
\begin{picture}(60,50)(-20,0)
\ArrowLine(25,25)(0,50)
\ArrowLine(0,0)(25,25)
\Photon(25,25)(75,25){3}{5}
\Photon(75,25)(100,50){3}{4}
\Photon(75,25)(100,0){3}{4}
\Photon(125,0)(175,0){3}{5}
\Photon(125,50)(175,50){3}{5}
\GBox(100,-5)(125,55){0.5}
\put(-15,48){$e^+$}
\put(-15,-2){$e^-$}
\put(70,2){$W^-$}
\put(70,40){$W^+$}
\put(180,48){$W^+$}
\put(180,-2){$W^-$}
\put(145,22){$I=J=1$}
\end{picture}
\end{center}
\caption[]{\label{esp-fg:qqtoqqww} \it  Dispersi\'on y redispersi\'on
de $WW$ a altas energ\'{\i}as en el LHC y en colisionadores TeV
lineales $e^+e^-$.}
\end{figure}

\begin{figure}[hbtp]
\begin{center}
\vspace*{-9.5cm}

\hspace*{-3.5cm}
\epsfig{file=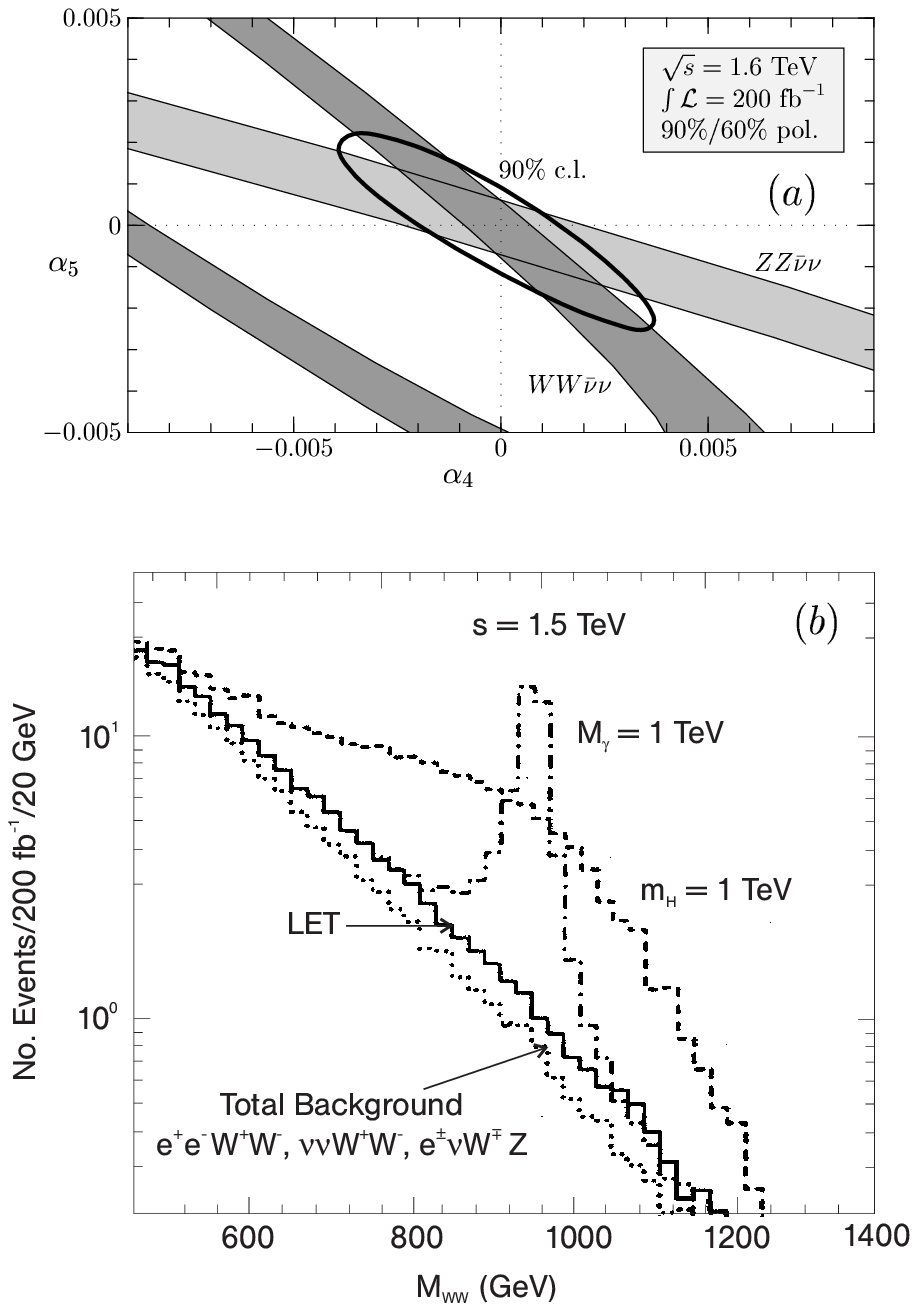,width=22.0cm,angle=0}
\vspace*{-8.5cm}

\end{center}
\caption[]{\it Parte superior: La sensibilidad a los par\'ametros de expansi\'on
  en modelos electrod\'ebiles quirales de la dispersi\'on $WW \to WW$ y $WW \to
  ZZ$ en el umbral de la interacci\'on fuerte;
  Ref. \protect\cite{24}.
Parte inferior: La distribuci\'on de la energ\'{\i}a invariante de
$WW$ en $e^+e^-
  \to \overline{\nu} \nu WW$ para modelos de resonancia escalares y
   vectoriales [$M_H, M_V$ = 1 TeV];
  Ref. \protect\cite{24a}.
\protect\label{esp-17tt}\label{esp-PKB}
}
\end{figure}

\vspace*{1mm}
\noindent {\bf 2.$\,$} La dispersi\'on de $WW$ puede estudiarse en el
LHC y en colisionadores TeV lineales $e^+e^-$.  A altas energ\'{\i}as,
haces de $W$ equivalentes acompa\~nan a los haces de quark y
electr\'on/positr\'on (Fig.~\ref{esp-fg:qqtoqqww}) en los procesos de
fragmentaci\'on $pp\to qq \to qqWW \, etc$ y $ee\to \nu\nu WW \, etc$; el
espectro de los bosones $W$ polarizados longitudinalmente est\'an dados
en la Eq.  (\ref{esp-eq:xyz}). En el ambiente hadr\'onico de LHC los bosones
de estado final $W \; etc$ s\'olo pueden ser observados en decaimientos
lept\'onicos y en decaimientos mixtos hadr\'onicos/lept\'onicos.
El ambiente limpio de los colisionadores de $e^+e^-$
permitir\'a la reconstrucci\'on de resonancias de decaimientos a pares de
jet. Los resultados de tres simulaciones experimentales se exhiben en
la Fig.~\ref{esp-PKB}. En la Fig.~\ref{esp-PKB}a se muestra la sensibilidad a
los par\'ametros $\alpha_4,\alpha_5$ de la expansi\'on quiral para la dispersi\'on
$WW$ en colisionadores $e^+e^-$ \cite{24}. Los resultados de estos
an\'alisis pueden reinterpretarse como sensibilidad a la predicci\'on de
par\'ametro libre de la expansi\'on quiral, correspondiendo a un
error de cerca del 10\% en el primer t\'ermino de la amplitud maestra
$s/v^2$.  Estos experimentos ponen a prueba el concepto b\'asico de rompimiento
de simetr\'{\i}a din\'amico a trav\'es del rompimiento espont\'aneo de la
simetr\'{\i}a. La producci\'on de una resonancia de bos\'on vectorial de masa
$M_V=1$ TeV se ejemplifica en la Fig.~\ref{esp-PKB}b \cite{24a}.\\

LHC permite la observaci\'on de resonacias $WW$ en estos canales hasta
un rango de masas de aproximadamente 1.5~TeV \cite{12}. \\

Un segundo poderoso m\'etodo  mide la dispersi\'on el\'astica $W^+W^-
\to W^+W^-$ en el canal $I=1, J=1$. La redispersi\'on de los
bosones $W^+W^-$ producidos en la aniquilaci\'on $e^+e^-$, cf.
Fig.~\ref{esp-fg:qqtoqqww}, depende, a energ\'{\i}as altas, de la fase
$\delta_{11}$ de dispersi\'on de $WW$ \cite{78}. La amplitud de
producci\'on $F = F_{LO} \times R$ es el producto del diagrama
perturbativo a m\'as bajo orden con la amplitud de redispersi\'on
de Mushkelishvili--Omn\`es ${\cal R}_{11}$,
\begin{equation}
{\cal R}_{11} = \exp \frac{s}{\pi} \int \frac{ds'}{s'}
\frac{\delta_{11}(s')}{s'-s-i\epsilon} ~,
\end{equation}
la cual est\'a determinada por el cambio en la fase $\delta_{11}$
de $WW$, obtenida de $I = J = 1$. La eficacia de este m\'etodo se
deriva del hecho de que toda la energ\'{\i}a del colisionador
$e^+e^-$ es transferida al sistema $WW$ [mientras que una
fracci\'on mayor de la energ\'{\i}a se pierde en la
fragmentaci\'on de $e \to \nu W$ si la dispersi\'on de $WW$ es
estudiada en el proceso $ee\to \nu\nu WW$]. Simulaciones
detalladas \cite{78} han demostrado que este proceso es sensible a
las masas del bos\'on vectorial de hasta
cerca de $M_V \lessim 6$ TeV en teor\'{\i}as tipo tecnicolor.

\vspace*{2mm}
El an\'alisis experimental de los par\'ametros $\alpha$ en el
colisionador lineal $e^+e^-$ en la primera fase con energ\'{\i}as
de hasta $\sim 1$~TeV puede ser reinterpretado de la siguiente
forma. Asociando los par\'ametros $\alpha$ con nuevas escalas de
interacci\'on fuerte, $\Lambda_\star \sim M_W/\sqrt{\alpha}$, cotas superiores para
  $\Lambda_\star$ de $\sim 3$~TeV pueden ser exploradas en la dispersi\'on $WW$.
 As\'{\i} pues, este instrumento
permite cubrir toda la regi\'on del umbral $\lessim 4\pi v \sim 3$~TeV
de las nuevas interacciones fuertes. En el canal de producci\'on
 $W^+W^-$ de colisiones $e^+e^-$ un rango de hasta  10~TeV puede ser
 probado de manera indirecta.  Si se descubriera una
nueva escala $\Lambda_\star$ de hasta $\sim 3$~TeV, se
podr\'{\i}an buscar nuevas resonancias $WW$ en el LHC mientras que
CLIC podr\'{\i}a investigar  nuevos estados de
resonancia con  masas de hasta 3 TeV y estados virtuales mucho m\'as all\'a
de \'estas.

\vspace*{4mm}
\section{Sinopsis}

\vspace*{2mm}
\noindent El mecanismo de rompimiento de la simetr\'{\i}a
elctrod\'ebil puede quedar establecido en los colisionadores
 actuales o en la nueva generaci\'on de colisionadores hadr\'onicos y
 lept\'onicos :
\begin{itemize}
\item[$\star$] Determinar la existencia de un bos\'on de Higgs fundamental ligero;
\item[$\star$]  El perfil de la part\'{\i}cula de Higgs puede ser reconstruido,
revelando as\'{\i} la naturaleza f\'{\i}sica
del subyacente mecanismo del rompimiento de la simetr\'{\i}a electrod\'ebil;
\item[$\star$] Pueden realizarse an\'alisis sobre la dispersi\'on WW fuerte
  si el rompimiento de la simetr\'{\i}a es generado din\'amicamente por
  nuevas interacciones fuertes.
\end{itemize}
M\'as a\'un, dependiendo de la respuesta experimental a estas
preguntas, el sector electrod\'ebil proporcionar\'a la plataforma
para extrapolaciones en \'areas f\'{\i}sicas m\'as all\'a del
Modelo Est\'andar: ya sea al sector supersim\'etrico de bajas
energ\'{\i}as; o  a una  teor\'{\i}a de nuevas
interacciones fuertes a una escala caracter\'{\i}stica del orden
de 1 TeV y m\'as all\'a; o, alternativamente, a dimensiones espaciales extra.

\vspace*{4mm}
\section*{Acknowledgments}

  P.M.Zerwas est\'a muy agradecido con los organizadores por la
  invitaci\'on a la 4th CERN Latin American School on High Energy Physics,
  Vi\~na del Mar, Chile, 2007.  La escritura de este trabajo fue apoyada
  parcialmente por el proyecto PAPIIT-IN115207.

\newpage

\appendix

\vspace*{4mm}
\section{The O(3) $\sigma$ Model}

\vspace*{2mm}
\noindent Un modelo transparente pero, a la vez, suficientemente
complejo para estudiar todos los aspectos del rompimiento de la
simetr\'{\i}a electrod\'ebil, es el modelo $\sigma$ O(3). Al
empezar por la versi\'on est\'andar, de un n\'umero de variantes
se puede desarrollar la idea del rompimiento espont\'aneo de la
simetr\'{\i}a y el teorema de Goldstone, mientras que normando
(gauging) la teor\'{\i}a se llegua al fen\'omeno del Higgs. Esta
evoluci\'on ser\'a descrita
paso a paso en las siguientes tres subsecciones.

\vspace*{2mm}

El modelo $\sigma$ O(3) incluye un triplete de componentes de
campo:
\beq
\sigma = (\sigma_1,\sigma_2,\sigma_3) \;.
\eeq
Si el potencial de auto-interacci\'on del campo depende
\'unicamente de la intensidad global del campo, la teor\'{\i}a,
descrita por el Lagrangiano
\beq
{\cal L} = \frac{1}{2} (\partial \sigma)^2 - V(\sigma^2)
\eeq
es invariante rotacional O(3). Estas
iso-rotaciones son generadas por la transformaci\'on
\beq
\sigma \to e^{i\alpha t}\sigma \quad \mathrm{con} \quad
(t^i)_{jk} = i \epsilon_{ijk}
\eeq
con par\'ametros de rotaci\'on  $\alpha =
(\alpha_1,\alpha_2,\alpha_3)$.
 Eligiendo una
interacci\'on cu\'artica para el potencial, la teor\'{\i}a es
renormalizable y por lo tanto bien definida.

\vspace*{4mm}
\subsection{Teor\'\i a Normativa:}

\begin{figure}[hbt]
\begin{center}
\epsfig{figure=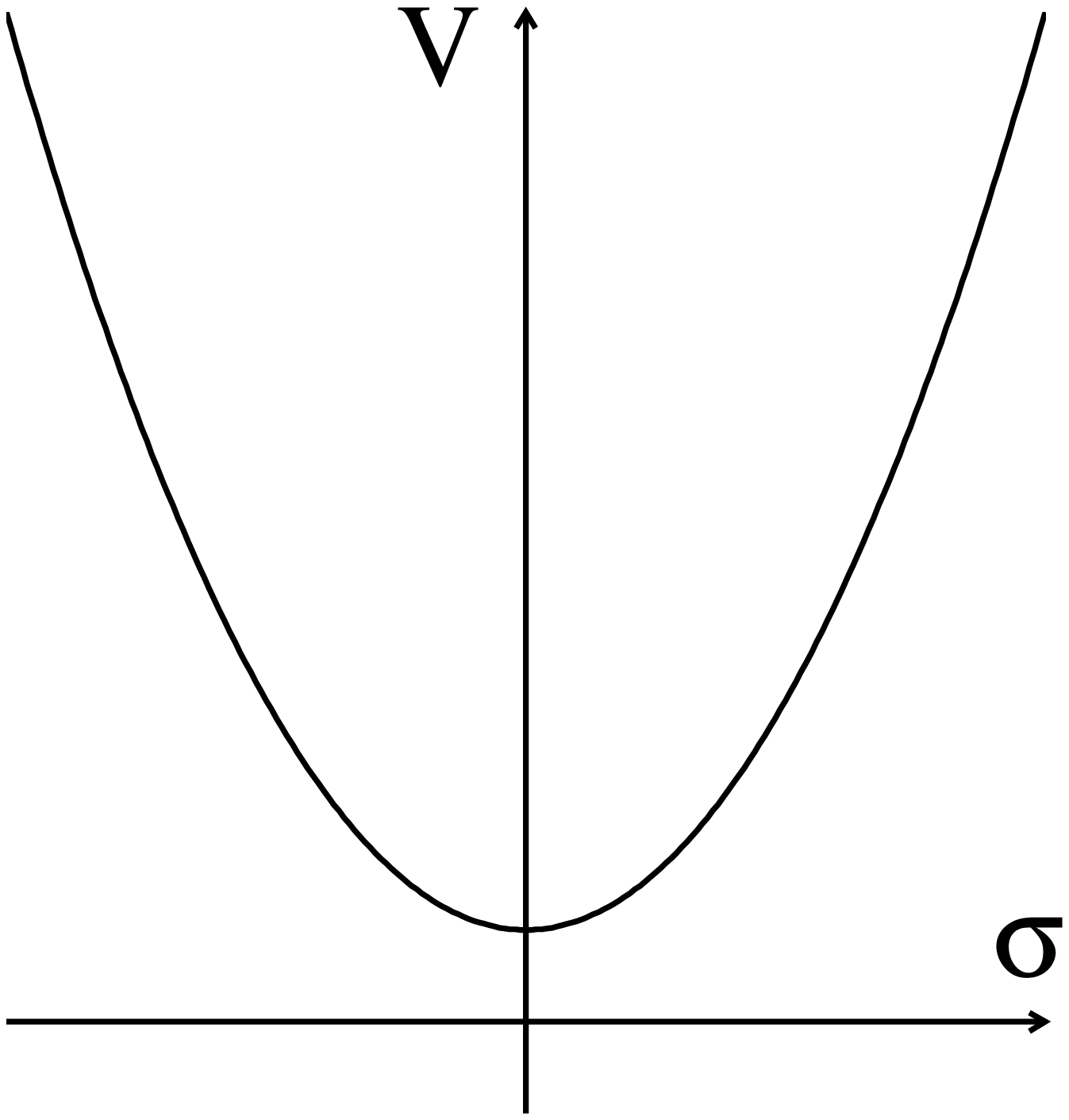,width=4cm}
\end{center}
\vspace*{-0.4cm}
\caption{\it \label{esp-fig:pot} }
\end{figure}

\vspace*{2mm}
\noindent
 Si el potencial cu\'artico $V$ se escoge a que sea
cf.~Fig.~\ref{esp-fig:pot},
\beq
V(\sigma^2) = \lambda^2 (\sigma^2 + \mu^2)^2
\eeq
el espectro de part\'{\i}culas y las interacciones
pueden f\'acilmente derivarse a partir de la siguiente forma
\beq
V(\sigma^2) = 2\lambda^2\mu^2\sigma^2 + \lambda^2 \sigma^4 + \mathrm{const.}
\;.
\eeq

\noindent El t\'ermino de campo bilineal describe tres masas
degeneradas
\beq
m(\sigma_1) = m(\sigma_2) = m(\sigma_3) = 2\lambda\mu
\eeq
que corresponden  a tres grados de libertad de
part\'{\i}culas f\'{\i}sicas. Los campos interact\'uan a trav\'es
del segundo t\'ermino cu\'artico. El estado base del sistema se
obtiene para la intesidad cero del campo:
\beq
\sigma^0 = (0,0,0) \;.
\eeq
Esta teor\'{\i}a describe un sistema de part\'{\i}culas est\'andar
en el cual el estado base preserva la invariancia rotacional del
Lagrangiano. Por lo que el Lagrangiano y la soluci\'on a la
ecuaci\'on del campo obedecen el mismo grado de simetr\'{\i}a.

\vspace*{4mm}
\subsection{Rompimiento Espont\'aneo de la Simetr\'{\i}a y el Teorema de
  Goldstone:}

\begin{figure}[hbt]
\begin{center}
\epsfig{figure=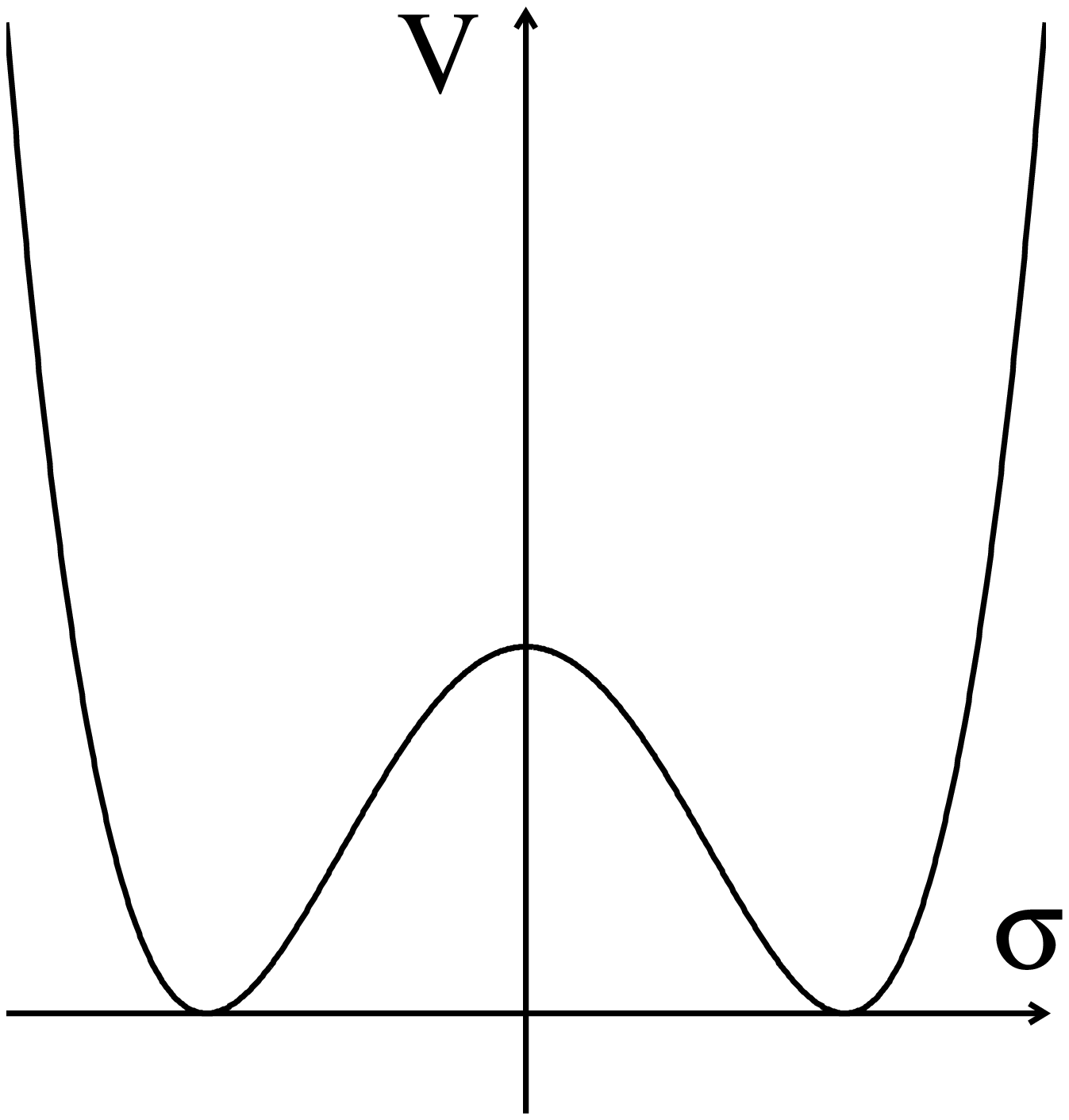,width=4cm}
\end{center}
\vspace*{-0.4cm}
\caption{\it \label{esp-fig:pothiggs} }
\end{figure}

\vspace*{2mm}
\noindent Sin embargo, si el signo del par\'ametro de masas el
potencial cambia a valores negativos,
\beq
V(\sigma^2) = \lambda^2(\sigma^2-\mu^2)^2
\eeq
el estado base es un estado de
intensidad de campo diferente de cero,
cf.~Fig.~\ref{esp-fig:pothiggs}. Fijando el eje del estado base de tal
manera que
\beq
\sigma^0 = (0,0,v) \quad \mathrm{con} \quad v=\mu
\eeq
la invariancia rotacional original O(3) del Lagrangiano ya no
es respetada por la soluci\'on del estado base, la cual distingue
una direcci\'on espec\'{\i}fica del iso-espacio. Sin embargo, no
hay un principio que determine la direcci\'on arbitraria del
estado base en el iso-espacio. Dicho fen\'omeno, en el cual las
soluciones de las ecuaciones de campo no obedecen la simetr\'{\i}a
del Lagrangiano, es generalmente llamado
``rompimiento espont\'aneo de la simetr\'{\i}a''.\\

Expandiendo el campo $\sigma$ alrededor del estado base,
\beq
\sigma = (\sigma_1', \sigma_2', v+\sigma_3')
\eeq
emerge una teor\'{\i}a efectiva para los nuevos grados de libertad
din\'amicos $\sigma_1', \sigma_2'$ y $\sigma_3'$. Evaluando el
potencial para los nuevos campos,
\beq
V = 4v^2\lambda^2\sigma_3^{'2} + 4 v \lambda^2 \sigma_3' (\sigma_1^{'2} +
\sigma_2^{'2} + \sigma_3^{'2}) + \lambda^2 (\sigma_1^{'2} + \sigma_2^{'2}
+ \sigma_3^{'2})^2 
\eeq
dos part\'{\i}culas sin masa m\'as una part\'{\i}cula masiva
corresponden a los t\'erminos de campo bilineales:
\beq
m(\sigma_1')=m(\sigma_2')=0 \nonumber\\
m(\sigma_3')=2\sqrt{2}\lambda v \neq 0 \;.
\eeq
A las dos part\'{\i}culas sin masa se les llama bosones de Goldstone,
Ref.~\cite{appcite}.\\

\vspace*{2mm} Adem\'as de los t\'erminos cu\'articos est\'andar, los bosones
de Goldstone y las part\'{\i}culas masivas interact\'uan entre ellas por
medio de t\'erminos trilineales del potencial efectivo.

\vspace*{2mm}
La simetr\'{\i}a de la teor\'{\i}a efectiva se reduce, de la
invariancia rotacional O(3) original a la invariancia O(2)
restringida a
rotaciones alrededor del eje del estado base.

\vspace*{4mm}
\noindent Este modelo $\sigma$ es s\'olo un simple ejemplo del m\'as general

\vspace*{2mm}
{\it \noindent\underline{Teorema de Goldstone:}\\[1mm]
\noindent
Si N es la dimensi\'on del grupo de simetr\'{\i}a del Lagrangiano
b\'asico, pero la simetr\'{\i}a de la soluci\'on del estado base
se reduce a una dimensi\'on M, entonces la teor\'{\i}a incluye
(N-M) bosones escalares de Goldstone sin masa.}

\vspace*{2mm}
Por cada grado de libertad de la simetr\'{\i}a que es destruido,
aparece en el espectro una part\'{\i}cula sin masa. Un ejemplo muy
famoso de este teorema son los tres piones casi sin masa que
emergen del rompimiento espont\'aneo de la simetr\'{\i}a de
iso-esp\'{\i}n quiral en QCD.

\vspace*{4mm}
\subsection{El mecanismo de Higgs}

\vspace*{2mm}
\noindent
El mecanismo de Higgs Ref.~\cite{1} proporciona masa a los bosones
vectoriales en teor\'{\i}as de norma sin destruir la
renormalizabilidad de la teor\'{\i}a. Si las masas fueran introducidas
a mano, la invariancia de norma que garantiza la renormalizabilidad,
se destruir\'{\i}a por los t\'erminos de masa {\it ad-hoc} en el
Lagrangiano.

\vspace*{2mm}
La simetr\'{\i}a global de iso-esp\'{\i}n del modelo $\sigma$ O(3)
puede ser extendida a una simetr\'{\i}a local al introducir un
iso-triplete $W$ de campos de norma acoplados de forma m\'{\i}nima
al campo $\sigma$. Introduciendo la derivada covariante
\beq
\partial_\mu \sigma \to \partial_\mu \sigma + igtW_\mu \sigma
\eeq
en el Lagrangiano,
\beq
{\cal L} = \frac{1}{2} [ (\partial + igtW)\sigma]^2 - V(\sigma^2) +
{\cal L}_{kin}(W)
\eeq
la teor\'{\i}a es invariante bajo transformaciones de norma local
\beq
\sigma \to e^{i\alpha t} \sigma \quad \mathrm{con} \quad
\alpha = \alpha(x)
\eeq
con la transformaci\'on de materia complementada por la
transformaci\'on usual del campo de norma no abeliano. El
Lagrangiano normado (gauged) incluye la parte cin\'etica de norma,
la parte cin\'etica de $\sigma$ y la interacci\'on $\sigma$-norma, as\'{\i} como
el potencial.

\begin{itemize}
\item[--] Si el potencial $\sigma$ es justo el potencial est\'andar,$V =
  \lambda^2 (\sigma^2 + \mu^2)^2$, la teor\'{\i}a es una teor\'{\i}a de norma de
  Yang-Mills no abeliana con un triplete de part\'{\i}culas $\sigma$ de
  masa degenerada, interactuando de la manera est\'andar con los campos
   de norma tripletes $W$ sin masa.
\\

\item[--] Sin embargo, si el potencial se escoge del tipo Mexicano,
  $V=\lambda^2 (\sigma^2-\mu^2)^2$, que en el modelo $\sigma$ conduce al rompimiento
  espont\'aneo de la simetr\'{\i}a, el contenido f\'{\i}sico
  campo/part\'{\i}cula de la teor\'{\i}a cambia dram\'aticamente [un
  fen\'omeno similar a la teor\'{\i}a no normada (non-gauged theory)].
\end{itemize}

Parametrizando el campo-triplete $\sigma$ a trav\'es de una
rotaci\'on del campo alrededor del eje del estado base,
\beq
\sigma = e^{i\Theta t/v} (\sigma^0 + \eta)
\eeq
con
\beq
\sigma^0 = (0,0,v)\,;\quad \eta = (0,0,\eta)\,; \quad
\Theta = (\Theta_1,\Theta_2,0)
\eeq
las componentes $\Theta$ de $\sigma$ perpendiculares al eje del
estado base pueden ser removidas por la transformaci\'on de norma
$\sigma \to exp[-i\Theta t/v] \sigma$ complementada por la
transformaci\'on correspondiente del campo de norma. Manteniendo
la notaci\'on original para los campos transformados de norma, el
nuevo Lagrangiano para los grados de libertad f\'{\i}sicos est\'a
dado por
\beq
{\cal L} = \frac{1}{2} [(\partial + igWt)(\sigma^0 +\eta)]^2 - V([\sigma^0 +
\eta]^2) + {\cal L}_{kin}(W) \;.
\eeq
Despu\'es de escribir el Lagrangian resultante de la teor\'{\i}a
efectiva como
\beq
{\cal L} = {\cal L}_{kin}(W) + \frac{1}{4} g^2 v^2 ( W_1^2 + W_2^2) +
\frac{1}{2} (\partial\eta)^2 - V + {\cal L}_{int} (\eta, W)
\eeq
el contenido f\'{\i}sico
part\'{\i}cula/campo se pone de manifiesto:

\begin{itemize}
\item[--]  un campo vectorial sin masa $W_3$ correspondiente a la invariancia
rotacional residual alrededor del eje-3 del estado base;\\

\item[--]  dos campos $W$ masivos $W_1$ y $W_2$ perpendiculares al eje
  del estado base con masas que est\'an determinadas por la intensidad
  del campo $v$ del estado base $\sigma$ y la constante de acoplamiento de norma
  $g$.  Estos dos campos masivos corresponden a los grados de libertad
  de la simetr\'{\i}a que fueron rotos espont\'aneamente en el modelo
  $\sigma$   no-normado;\\

\item[--] los bosones de Goldstone han desaparecido del espectro,
  absorbidos para formar los grados longitudinales de los bosones
  de norma masivos;\\

\item[--] un bos\'on de Higgs escalar real $\eta$.
\end{itemize}

\vspace*{4mm}
\noindent
Este ejemplo puede extenderse f\'acilmente, paralelamente al
teorema de Goldstone, para formular de forma general

\vspace*{2mm}
{\it \noindent \underline{El mecanismo de Higgs:} \\[1mm]
\noindent
Si N es la dimensi\'on del grupo de simetr\'{\i}a del
Lagrangiano original, M la dimensi\'on del grupo de simetr\'{\i}a
que deja invariante el estado base de los n campos escalares, entonces
la teor\'{\i}a f\'{\i}sica consiste en M campos vectoriales sin
masa, (N-M) campos vectoriales masivos, y n-(N-M) campos de Higgs
escalares.}

\newpage

\selectlanguage{english}


\end{document}